\documentclass[braunschweig,accepted]{thesis}

\usepackage{float}
\usepackage{graphicx}
\usepackage{tabularx}
\usepackage{txfonts}
\usepackage{epstopdf}
\usepackage{aas_macros}
\usepackage{rotating}
\usepackage{tablefootnote}
\usepackage{manyfoot}
\usepackage{ctable}
\usepackage{amsmath}
\usepackage[final]{pdfpages}

\DeclareNewFootnote{B}[fnsymbol]
\DeclareNewFootnote{C}[fnsymbol]
\DeclareNewFootnote{D}[fnsymbol]
\DeclareNewFootnote{E}[fnsymbol]

\title{Analysis and modeling of solar irradiance variations}
\author{Kok Leng Yeo}
\town{Singapur}

\refereea{Professor Dr. Sami K. Solanki}
\refereeb{Professor Dr. Karl-Heinz Gla\ss{}meier}
\submitteddate{22 April 2014}
\submittedyear{2014}
\examinationdate{11 Juli 2014}
\publicationyear{2014}
\isbn{}

\begin{document}

\newcommand{\bl}{\left\langle{}B_l\right\rangle}
\newcommand{\sigb}{\sigma_{\bl}}
\newcommand{\bmu}{\bl/\mu}
\newcommand{\icn}{I_{\rm C}}
\newcommand{\ilc}{I_{\rm L}}
\newcommand{\icnqs}{I_{\rm C,QS}}
\newcommand{\ilcqs}{I_{\rm L,QS}}
\newcommand{\sicnqs}{\sigma_{\icnqs}}
\newcommand{\silcqs}{\sigma_{\ilcqs}}
\newcommand{\mqsicn}{\left\langle{}\icnqs\right\rangle}
\newcommand{\mqsilc}{\left\langle{}\ilcqs\right\rangle}
\newcommand{\cicn}{C_{\icn}}
\newcommand{\cilc}{C_{\ilc}}
\newcommand{\bmut}{\left(\bmu\right)_{\rm TH}}
\newcommand{\g}{\:{\rm G}}
\newcommand{\bmua}{\bmut<\bmu\leq50\g}
\newcommand{\bmub}{50\g<\bmu\leq100\g}
\newcommand{\bmuc}{100\g<\bmu\leq180\g}
\newcommand{\bmud}{180\g<\bmu\leq280\g}
\newcommand{\bmug}{500\g<\bmu\leq640\g}
\newcommand{\bmuh}{640\g<\bmu\leq800\g}
\newcommand{\mua}{0.94<\mu\leq1.00}
\newcommand{\mug}{0.36<\mu\leq0.50}
\newcommand{\bmuf}{\frac{\bl}{\mu}}
\newcommand{\bmuqs}{\bmu\leq10\g}
\newcommand{\bmuz}{\bmu=0\g}
\newcommand{\bz}{\bl=0\g}
\newcommand{\umax}{\mu_{\rm max}}
\newcommand{\cicnmax}{C_{\icn,{\rm max}}}
\newcommand{\cilcmax}{C_{\ilc,\mu=1}}
\newcommand{\scicnmax}{\cicnmax/\left(\bmu\right)}
\newcommand{\scilcmax}{\cilcmax/\left(\bmu\right)}
\newcommand{\mbmu}{\left\langle\bmu\right\rangle}
\newcommand{\ir}{\hat{I}_{r}}
\newcommand{\ia}{\hat{I}_{\phi}}
\newcommand{\irm}{\left\langle\ir\right\rangle}
\newcommand{\iam}{\left\langle\ia\right\rangle}
\newcommand{\sir}{\int\ir{}dr}
\newcommand{\sia}{{\rm RMS}_{\ia-\iam}}
\newcommand{\psfone}{K_1}
\newcommand{\psftwo}{K_2}
\newcommand{\inorm}{I/I_0}
\newcommand{\vl}{\left\langle{}\vec{v}_{\rm l}\right\rangle}
\newcommand{\dvl}{\Delta\vl_{\rm QS}}
\newcommand{\vbl}{\left\langle{}B_{\rm l}\right\rangle}
\newcommand{\mbl}{\epsilon_{\bl}}
\newcommand{\noise}{\sigma_{\vbl}}
\newcommand{\mblmag}{\epsilon_{\bl{\rm ,mag}}}
\newcommand{\dbl}{\Delta\mbl}
\newcommand{\nmag}{n_{{\rm mag}}}
\newcommand{\dflux}{\Delta\left(\nmag\mblmag\right)}
\newcommand{\iqs}{\left\langle \inorm\right\rangle_{\rm QS}}
\newcommand{\iqp}{\left(\inorm\right)_{\rm QS,P}}
\newcommand{\ipu}{\left(\inorm\right)_{\rm P,U}}
\newcommand{\vobs}{\vec{v}_{\rm SDO}}
\newcommand{\vrot}{\vec{v}_{\rm rot}}
\newcommand{\mps}{{\rm ms^{-1}}}
\newcommand{\as}{{\rm arcsec}}
\newcommand{\kpvt}{{\rm KP}}
\newcommand{\kpone}{\kpvt_{512}}
\newcommand{\kptwo}{\kpvt_{\rm SPM}}
\newcommand{\bsat}{\left(\bmu\right)_{\rm sat}}
\newcommand{\bcut}{\left(\bmu\right)_{\rm cut}}
\newcommand{\wms}{{\rm Wm^{-2}}}
\newcommand{\blhmi}{\bl_{\rm HMI}}
\newcommand{\blmdi}{\bl_{\rm MDI}}
\newcommand{\nfac}{n_{\rm fac}}
\newcommand{\esat}{{\rm TSI}_{\rm mod}}
\newcommand{\eobs}{{\rm TSI}_{\rm obs}}
\newcommand{\fsat}{{\rm SSI}_{\rm mod}}
\newcommand{\fobs}{{\rm SSI}_{\rm obs}}
\newcommand{\dpmod}{{\rm DIARAD}_{\rm PMOD}}
\newcommand{\dirmb}{{\rm DIARAD}_{\rm IRMB}}
\newcommand{\dtsifac}{\Delta{\rm TSI_{fac}}}
\newcommand{\dtsispt}{\Delta{\rm TSI_{spt}}}
\newcommand{\gobs}{{\rm SSI}_{\rm obs,ref}}
\newcommand{\gsat}{{\rm SSI}_{\rm mod,ref}}

\maketitle


\chapter*{Vorver\"offentlichung der Dissertation\markboth{Vorver\"offentlichung der Dissertation}{Vorver\"offentlichung der Dissertation}}

Teilergebnisse aus dieser Arbeit wurden mit Genehmigung der Fakult\"at f\"ur Elektrotechnik, Informationstechnik, Physik, vertreten durch den Mentor der Arbeit, in folgenden Beitr\"agen vorab ver\"offentlicht:

\paragraph{Publikationen}
\begin{itemize}
	\item Yeo, K. L., Krivova, N. A., Solanki, S. K., Glassmeier, K. H., 2014, Reconstruction of total and spectral solar irradiance since 1974 based on KPVT, SoHO/MDI and SDO/HMI observations, Astron. Astrophys., 570, A85
	\item Thuillier, G., Schmidtke, G., Erhardt, C., Nikutowski, B., Shapiro, A. I., Bolduc, C., Lean, J., Krivova, N. A., Charbonneau, P., Cessateur, G., Haberreiter, M., Melo, S., Delouille, V., Mampaey, B., Yeo, K. L., Schmutz, W., 2014, Solar spectral irradiance variability in November/December 2012: comparison of observations by instruments on the International Space Station and models, Sol. Phys., online
	\item Yeo, K. L., Krivova, N. A., Solanki, S. K., 2014, Solar cycle variation in solar irradiance, \ssr, online
	\item Yeo, K. L., Feller, A., Solanki, S. K., Couvidat, S., Danilovic, S., Krivova, N. A., 2014, Point spread function of SDO/HMI and the effects of stray light correction on the apparent properties of solar surface phenomena, Astron. Astrophys., 561, A22
	\item Yeo, K. L., Solanki, S. K., Krivova, N. A., 2013, Intensity contrast of network and faculae, Astron. Astrophys., 550, A95
\end{itemize}

\paragraph{Tagungsbeitr\"age}
\begin{itemize}
	\item Yeo, K. L., Solanki, S. K., Krivova, N. A., Solar irradiance variability and the Earth's climate, SCOSTEP's 13th Quadrennial Solar-Terrestrial Physics Symposium, Xi'an, China, 13 to 18 October 2014 (invited talk)
	\item Yeo, K. L., Solanki, S. K., Krivova, N. A., Model reconstruction of total and spectral solar irradiance since 1974, AOGS 11th Annual Meeting, Sapporo, Japan, 28 July to 1 August 2014 (poster)
	\item Yeo, K. L., Krivova, N. A., Solanki, S. K., Reconstruction of TSI and SSI in the satellite era. EGU General Assembly 2014, Vienna, Austria, 27 April to 2 May 2014 (invited talk)
	\item Krivova, N. A., Yeo, K. L., Solanki, S. K., Dasi-Espuig, M., Ball, W., Modelling solar irradiance with SATIRE, 2014 SORCE Science Meeting, Cocoa Beach, Florida, USA, 28 to 31 January 2014 (invited talk)
	\item Yeo, K. L., Solanki, S. K., Krivova, N. A., SATIRE-S reconstruction of total and spectral solar irradiance from 1974 to 2013, International CAWSES-II Symposium, 18 to 22 November 2013, Nagoya, Japan (poster)
	\item Yeo, K. L., Krivova, N. A., Solanki, S. K., Cyclic variation in solar irradiance, `The Solar Activity Cycle: Physical Causes And Consequences' Workshop, Bern, Switzerland, 11 to 15 November 2013 (invited talk)
	\item Yeo, K. L., Solanki, S. K., Krivova, N. A., Network \& facular contribution to solar irradiance variation, Space Climate 5, Oulu, Finland, 15 to 19 June 2013 (poster)
	\item Yeo, K. L., Solanki, S. K., Krivova, N. A., Comparing irradiance reconstructions from HMI magnetograms with SORCE observations, 2011 SORCE Science Meeting, Sedona, Arizona, USA, 13 to 16 September 2011 (poster)
	\item Yeo, K. L., Solanki, S. K., Krivova, N. A., Reconstructing total solar irradiance from HMI/SDO observations, LWS/SDO-1 Workshop, Squaw Creek, California, USA, 1 to 5 May 2011 (poster)
\end{itemize}

\tableofcontents

\chapter*{Summary\markboth{Summary}{Summary}}
\addcontentsline{toc}{chapter}{Summary}

A prominent manifestation of the solar dynamo is the 11-year activity cycle, visible in indicators of solar activity, including the topic of this thesis, solar irradiance. Two quantities are of interest, total and spectral solar irradiance, TSI and SSI. They are defined as the total and wavelength-resolved solar radiative flux above the Earth's atmosphere, normalized to one AU. Excluding the interaction between solar radiation and the Earth's atmosphere, and changes in the Earth to Sun distance, TSI and SSI isolate the radiant property of the Earth-facing hemisphere of the Sun.

A relationship between solar activity and the brightness of the Sun had long been suspected. It was however, only directly observed when satellite measurements, free from the effects of atmospheric intensity fluctuations and stray light, became available. TSI and SSI (at least in the ultraviolet) have been measured on a regular basis by a succession of space missions, almost without interruption, since 1978. The measurement of solar irradiance from space is accompanied by the development of models aimed at describing the apparent variability in these observations by the intensity excess/deficit in the solar surface and atmosphere brought on by magnetic structures in the photosphere. While the body of satellite measurements is largely consistent at solar rotation timescales and show obvious solar cycle modulation, there is considerable scatter in the absolute radiometry, secular variation and the spectral dependence of the variation over the solar cycle, due to the challenge in accounting for instrumental influences. Consequently, models of solar irradiance serve as an important complement to direct observations, helping us understand the apparent variability and the physical processes driving them.

The more sophisticated models, termed semi-empirical, rely on the calculated intensity spectra of magnetic structures on the solar surface and in the solar atmosphere, generated with spectral synthesis codes from semi-empirical solar model atmospheres. An established example of such models is SATIRE-S (Spectral And Total Irradiance REconstruction for the Satellite era). Obviously, the robust reconstruction of solar irradiance depends on how realistic these intensity spectra are. There are two key sources of uncertainty. One, the account of departures from local thermodynamic equilibrium (LTE) in the spectral synthesis. Two, the fact that the radiant properties of network and faculae are neither fully understood nor adequately represented in current models by the use of plane-parallel model atmospheres (as opposed to three-dimensional model atmospheres). In SATIRE-S, this is responsible for the sole free parameter in the model. Semi-empirical models have achieved considerable success replicating the apparent variability in solar irradiance observations. The unambiguous account of the outstanding discrepancy between model and measurement will require however, an improvement in how non-LTE effects, and the influence of network and faculae on solar irradiance is included in semi-empirical models.

This thesis is the compilation of four publications, detailing the results of investigations aimed at setting the groundwork necessary for the eventual introduction of three-dimensional atmospheres into SATIRE-S. Also presented is an update of the SATIRE-S model, and a review of the current state of the measurement and modelling of solar irradiance.

We examined the intensity contrast of network and faculae in observations from the Helioseismic and Magnetic Imager onboard the Solar Dynamics Observatory (SDO/HMI), and estimated the point spread function (PSF) of the instrument. The derived intensity contrasts and PSF can be used, in future efforts, to constrain three-dimensional model atmospheres, key to improving the reliability of semi-empirical models. The results of these studies also offered new insights into the radiant behaviour of network and faculae (and their contribution to variation in solar irradiance), and the effects of stray light on the apparent properties of solar surface phenomena.

The SATIRE-S model had previously been applied to full-disc intensity images and magnetograms from the Kitt Peak Vacuum Telescope and the Michelson Doppler Imager (onboard the Solar and Heliospheric Observatory) to reconstruct TSI and SSI over the period of 1974 to 2009. On top of extending these preceding efforts to the present time with similar data from HMI, we made various refinements to the reconstruction method. The result is a daily reconstruction of TSI and SSI, covering 1974 to the present, that is more reliable and, in most cases, extended than similar reconstructions from contemporary models. The reconstruction is also highly consistent with observations from multiple sources, demonstrating its utility for solar irradiance and climate studies.

\chapter{Introduction}
\label{thesisintroduction}

\section{Measurements and models of solar irradiance}
\label{introductionmmsi}

\begin{cfig}
\includegraphics[width=\textwidth]{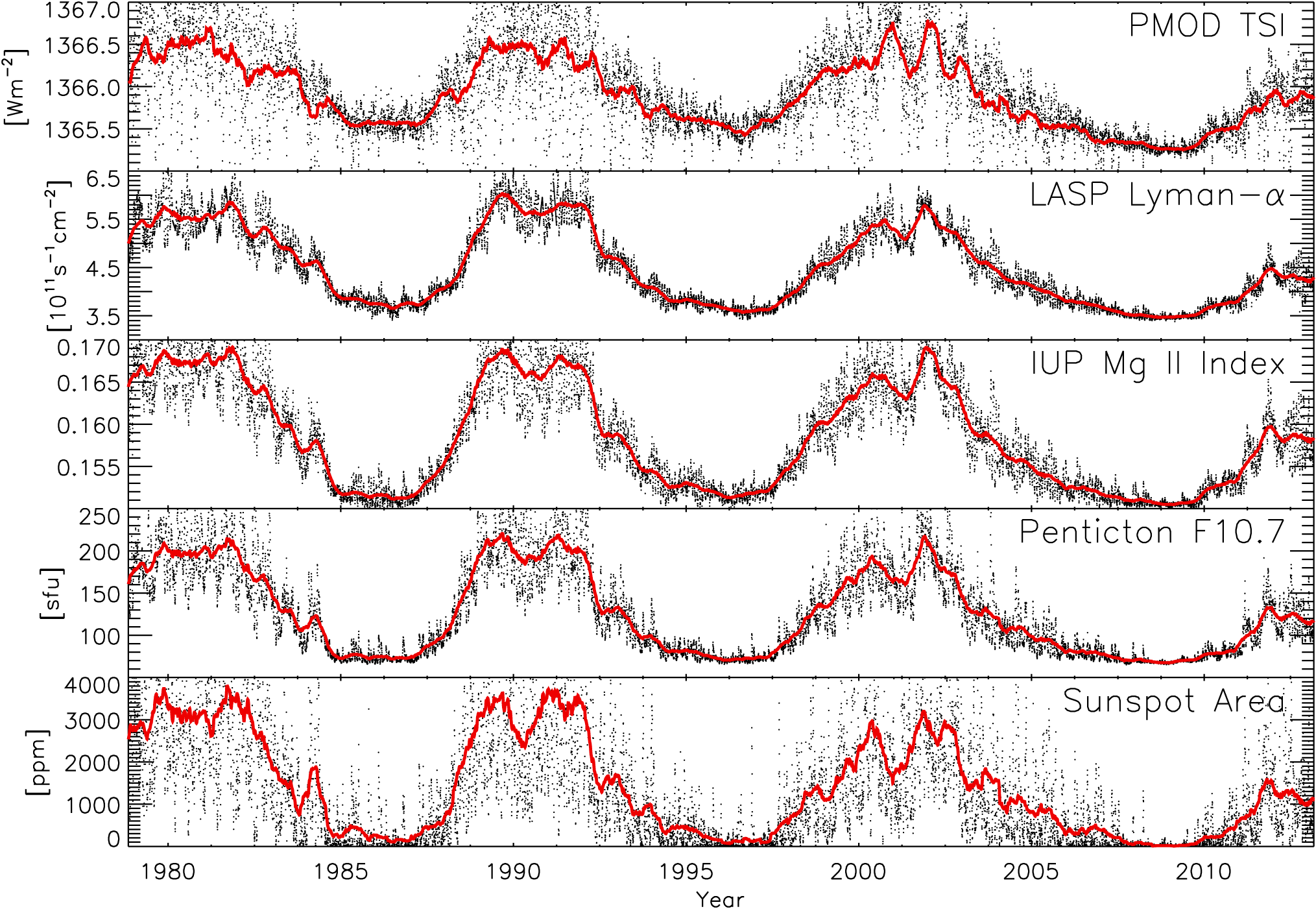}
\caption{Indices of solar activity. From top to bottom; the PMOD total solar irradiance composite \citep[version ${\rm d}41\_62\_1302$,][]{frohlich00}, the LASP Lyman-$\alpha$ irradiance composite \citep{woods00}, the IUP Mg II index composite \citep[version 4,][]{viereck99,skupin05b,skupin05c}, the Ottawa and Penticton adjusted 10.7 cm radio flux record \citep{tapping87,tapping13} and the projected sunspot area composite by \citealt{balmaceda09} (version 0613). The red curves follow the 181-day moving mean.}
\label{introproxy}
\end{cfig}

The 11-year activity cycle of the Sun, a manifestation of the solar dynamo, can be seen in indicators of solar activity \citep{hathaway10}. This includes, the sunspot area and number, chromospheric and coronal indices such as the 10.7 cm radio flux, Mg II index and X-ray flux, and in the topic of this thesis, solar irradiance (Fig. \ref{introproxy}). Solar irradiance, the radiative output of the Sun, is described in terms of what is termed total and spectral solar irradiance, TSI and SSI. They are defined the total and wavelength-resolved solar radiative flux above the Earth's atmosphere, normalized to one AU (units of power per unit area, and power per unit area and wavelength, respectively). Excluding the interaction between solar radiation and the Earth's atmosphere, and changes in the Earth-Sun separation, TSI and SSI follows the radiant property of the hemisphere of the Sun facing the Earth.

A relationship between solar activity and the radiative output of the Sun had long been speculated \citep{abbot23,smith75,eddy76}. This was however, not confirmed until satellite measurements, free from the effects of fluctuations in atmospheric transmittance, became available. TSI and SSI (at least in the ultraviolet) have been monitored regularly by a succession of space missions, almost without interruption, since 1978 \citep{hickey80,willson88,frohlich06,deland08,kopp12}. The early TSI observations quickly revealed a correlation between the apparent variability and the passage of active regions across the solar disc \citep{willson81,hudson82,oster82,foukal86}. Consequently, the measurement of solar irradiance from space is accompanied by the development of models aimed at describing the apparent variability in these observations by the intensity excess/deficit in the solar surface and atmosphere brought about by photospheric magnetism.

The action of magnetic concentrations in the photosphere on the thermal structure and therefore the radiant property of the solar surface/atmosphere is not the only mechanism mooted to explain the observed variability in solar irradiance, but it is by far the most established. Models relating variations in solar irradiance to the emergence and evolution of photospheric magnetism have achieved considerable success in replicating observations \citep{domingo09}. Other mechanisms, related to physical processes in the solar interior, have been proposed \citep{wolff87,kuhn88,cossette13} but there is as yet little direct evidence.

\begin{cfig}
\includegraphics[width=\textwidth]{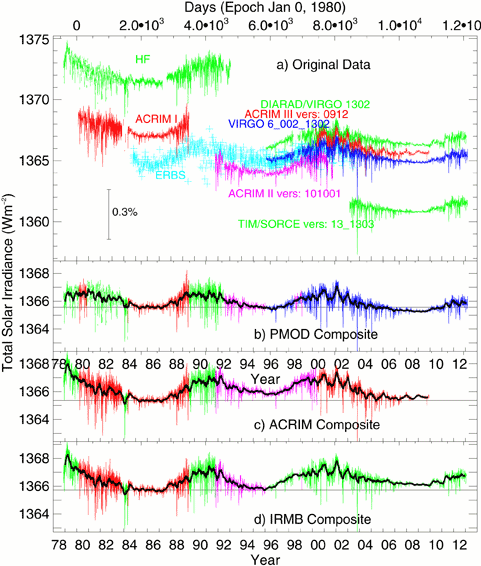}
\caption{a) The published TSI measurements from the various radiometers sent into orbit since 1978 and the competing composite records of TSI by b) PMOD, c) ACRIM and d) IRMB. These observations and composites are introduced in detail in Chap. \ref{tsimeasurements}. Courtesy of C. Fr\"ohlich (http://www.pmodwrc.ch/pmod.php?topic=tsi/composite/SolarConstant).}
\label{introfrohlich}
\end{cfig}

\begin{cfig}
\includegraphics[width=\textwidth]{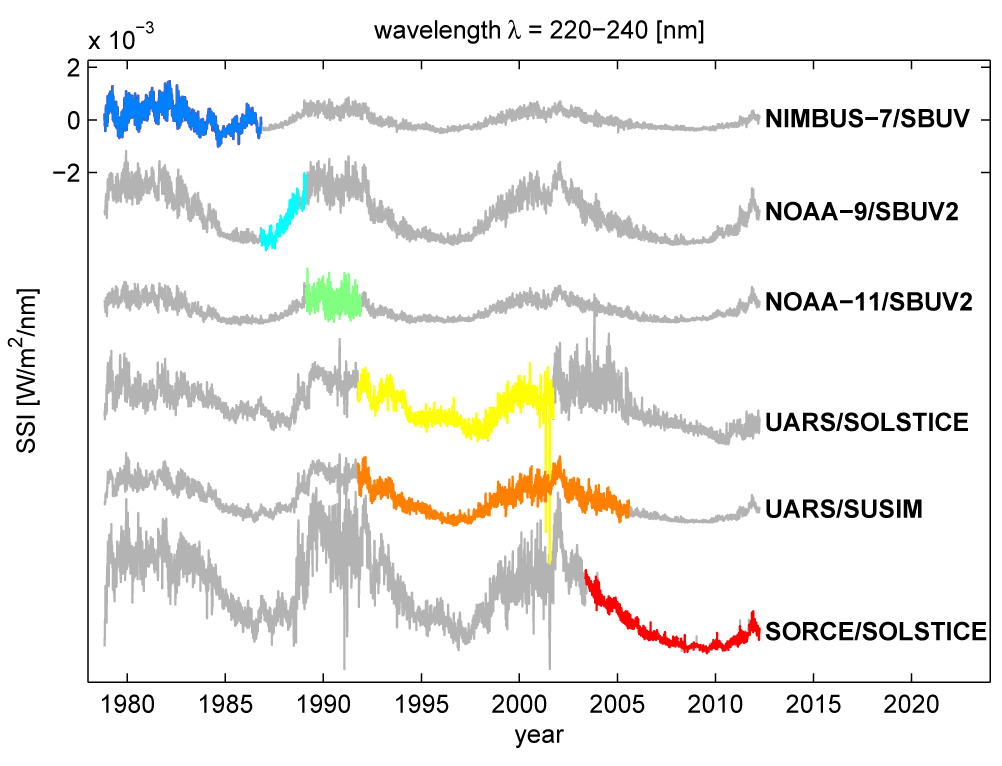}
\caption{Integrated solar irradiance between 220 and 240 nm from six instruments, offset from one another for the purpose of illustration. Each record is extrapolated backwards and forwards in time by the method set out in \cite{dudokdewit11}, drawn in grey. Taken from \cite{ermolli13}.}
\label{introermolli13}
\end{cfig}

While the measurements from the succession of solar irradiance monitors sent into orbit are largely consistent at solar rotation timescales and show obvious solar cycle modulation (though not without exception), there is considerable scatter in the absolute radiometry, secular variation and the spectral dependence of the variation over the solar cycle \citep[see Figs. \ref{introfrohlich} and \ref{introermolli13}, and][]{ermolli13,solanki13}. This is primarily due to the significant challenge in accounting for changes in instrument response from ageing and exposure \citep{hoyt92,lee95,dewitte04a,frohlich06,deland08}. Due to the uncertainties afflicting the direct observation of solar irradiance, models of solar irradiance based on photospheric magnetism have emerged as an important tool for understanding the apparent variability in these measurements and the associated physical processes.

The most straightforward way to model solar irradiance (adopted since the earliest models) is to reconstruct it by the regression of indices of solar activity, acting as proxies of the radiant effects of photospheric magnetism, to measurements.

The influence of sunspots and pores is typically represented by sunspot area or what is termed the photometric sunspot index \citep[PSI,][]{hudson82,frohlich94}, and network and faculae by chromospheric indices. (The PSI is the proportional deficit in solar irradiance, from the magnetically quiet Sun level, due to sunspot darkening. This can be calculated from the sunspot number, sunspot area or from full-disc intensity images.) While straightforward, such models depend on and are therefore limited by the availability of reliable measurements. They are also limited by uncertainties in the solar activity index data employed and offer little physical insight into the underlying relationship between solar irradiance and photospheric magnetism. Critically, these models usually assume a linear relationship between the proxies and solar irradiance, which is not true in the case of chromospheric indices \citep[][]{solanki04,foukal11}. Variation in solar irradiance is the sum effect of the influence of photospheric magnetism on the solar surface and enclosed atmosphere \citep{mitchell91,unruh99,preminger02}. This would obviously not be entirely captured in chromospheric indices, highlighting another fundamental limit of the approach.

A more physics-based approach has been in development by various groups over the past two decades \citep{fontenla99,fligge00,krivova03,penza03,haberreiter05}. The overall architecture of these models, referred to as semi-empirical, is similar. The solar disc is segmented by surface magnetic feature type. The intensity spectrum of each feature type is calculated from spectral synthesis codes with the respective semi-empirical model atmosphere\footnote{The model atmospheres describe the temperature and density within each feature type as a function of height. They are described as semi-empirical from the fact that they are constrained by observations.} as input. The solar spectrum is then recreated from the sum of these intensity spectra, weighted by the apparent surface coverage of each feature type. An established example of semi-empirical models is SATIRE-S \citep[Spectral And Total Irradiance REconstruction for the Satellite era,][]{fligge00,krivova03,krivova11a}. Prior to this thesis, it had been applied to full-disc observations from the Kitt Peak Vacuum Telescope \citep[KPVT,][]{livingston76,jones92} and the Michelson Doppler Imager onboard the Solar and Heliospheric Observatory \citep[SoHO/MDI,][]{scherrer95} to reconstruct TSI and SSI between 1974 and 2009 \citep{krivova03,krivova06,krivova09a,krivova11b,wenzler05b,wenzler06,wenzler09,unruh08,ball11,ball12,ball14}.

Evidently, the robust reconstruction of solar irradiance by the semi-empirical approach depends on how reliable/realistic are the intensity spectra of solar surface features utilized. A major source of uncertainty is the fact that the radiant behaviour of quiet Sun network and active region faculae is neither fully understood nor adequately represented in current models by the use of plane-parallel model atmospheres (as opposed to three-dimensional model atmospheres). The small-scale magnetic concentrations that make up network and faculae are, at present, still largely unresolved in available observations. The effect of atmospheric and instrumental scattered light on the apparent properties of these surface features is also not completely known. In spite of the current insufficiencies, semi-empirical models such as the SATIRE-S have been very successful in reproducing most of the apparent variability in solar irradiance observations. The unambiguous account of the outstanding discrepancy between model and measurement will require, amongst other things, an improvement in how the effects of network and faculae on solar irradiance is included in semi-empirical models.

This thesis is the compilation of four publications, detailing the results of investigations aimed at addressing the present limits of semi-empirical models (discussed above) and updating the SATIRE-S model, and also includes a review of the current state of the measurement and modelling of solar irradiance. These studies made use of full-disc observations from the Helioseismic and Magnetic Imager onboard the Solar Dynamics Observatory \citep[launched in 2010,][]{schou12}. In the following, we give a brief introduction to the HMI instrument (Sect. \ref{introductionsdohmi}) and the SATIRE-S model (Sect. \ref{introductionsatires}), before providing an outline of this thesis (Sect. \ref{introductionthesisoutline}).

\section{SDO/HMI}
\label{introductionsdohmi}

SDO/HMI \cite{schou12} is the follow-up to the highly successful SoHO/MDI \citep{scherrer95}, the first ever spaceborne magnetograph. HMI is designed to return continuous full-disc measurements of intensity, magnetic field vector and line-of-sight velocity from spectropolarimetry of the Fe I 6173 \AA{} line.

\begin{sidewaysfigure}
\centering
\includegraphics[width=\textwidth]{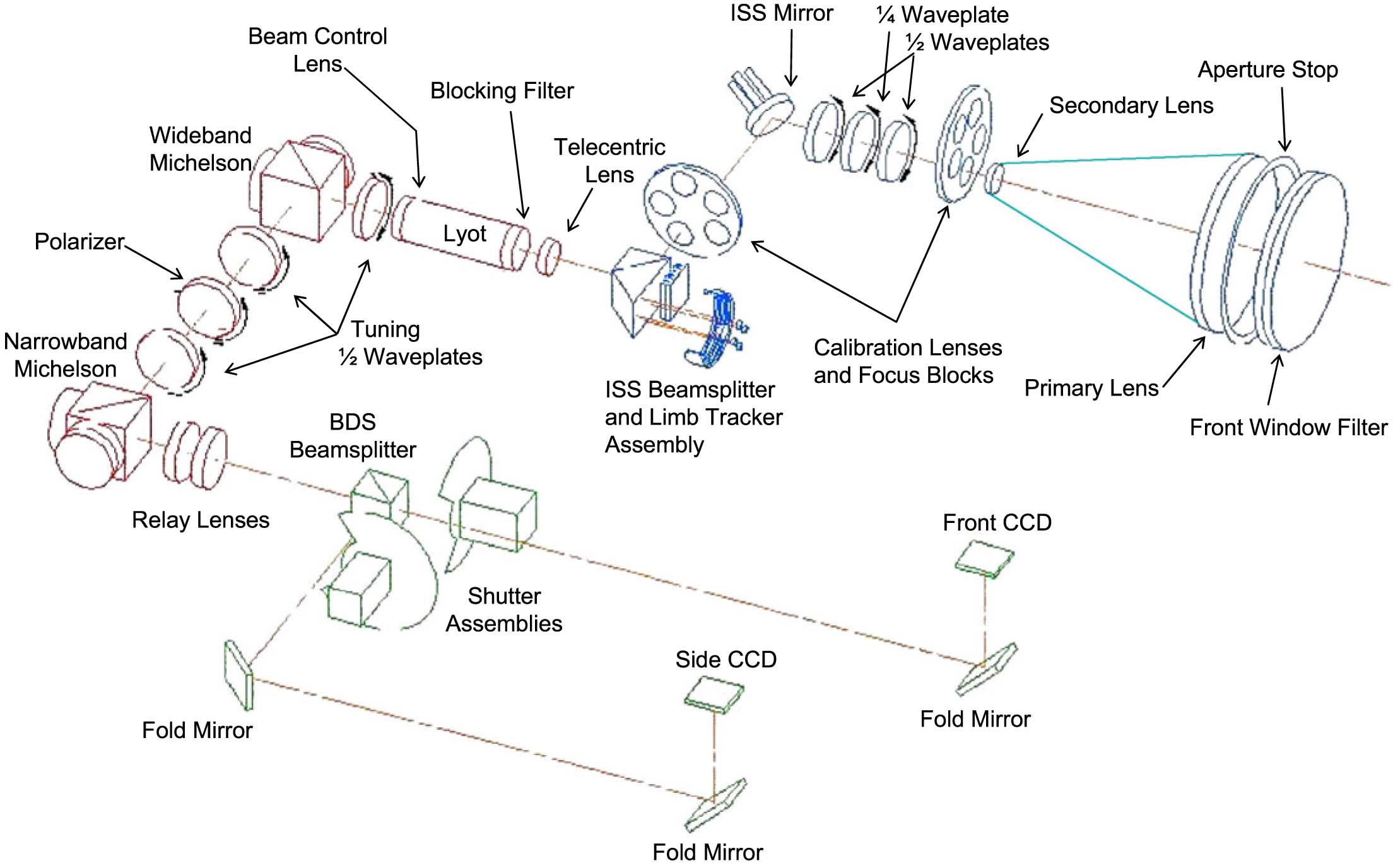}
\caption{HMI optical assembly (not to scale). Taken from \cite{schou12}.}
\label{introschou12}
\end{sidewaysfigure}

The instrument comprises of two $4096\times4096$ pixel CCD cameras, which share a common optical path, referred to as the side and front CCDs (Fig. \ref{introschou12}). The optical assembly includes a Lyot filter and two Michelson interferometers (both tunable), and a series of waveplates, which set the bandpass and polarization, respectively. The pixel scale is 0.505 arcsec and the diffraction-limited spatial resolution is 0.91 arcsec. Narrow (FWHM of 76 m\AA{}) bandpass images or filtergrams of the full solar disc are collected continuously, at 1.875 second intervals, on the two CCDs in turn. The observation sequence cycles through six positions within the Fe I 6173 \AA{} line (spaced 69 m\AA{} apart) and six polarizations (Stokes $I\pm{}Q$, $I\pm{}U$ and $I\pm{}V$).

A set of 12 filtergrams, covering the $I+V$ and $I-V$ polarizations at each line position is collected on the front CCD every 45 seconds. A set of 36 filtergrams, of each combination of polarization and line position, is registered on the side CCD every 135 seconds. The sequence of filtergrams from the front and from the side CCD are combined to yield longitudinal magnetograms, Dopplergrams and intensity images (continuum intensity, line depth and line width) at 45 and 720-second cadence, respectively. Full-Stokes parameters (i.e., the Stokes I, Q, U and V images) are also generated from the side CCD filtergram sequence, again at intervals of 720 seconds. These are inverted using the VFISV (Very Fast Inversion of Stokes Vector) Milne-Eddington inversion scheme by \cite{borrero11}, the primary output of which is the vector magnetogram (giving both magnetic flux density and pointing). For the work detailed in this thesis we preferred the longitudinal magnetogram data product over the vector magnetogram, even though it gives just the line-of-sight component of the magnetic flux density. Due to the significant noise level in the Stokes Q and U parameters, the VFISV algorithm produces artificial magnetic fields (order of 100 G in strength, largely horizontal and random in pointing) in the quiet Sun.

The temporal and spatial resolution of HMI is the highest of any full-disc spectromagnetograph. Being spaceborne, it is also free from the detrimental effects of atmospheric seeing. The noise level of HMI longitudinal magnetograms has been demonstrated to be much lower than in similar observations from MDI \citep{liu12}. This is, in all likelihood, also true of the other data products. The unprecedented quality of HMI magnetograms permits the resolution of small and/or weak magnetic features that would otherwise be hidden in similar data from other instruments. This grants us the ability to characterize the prevailing photospheric magnetism at never before accuracy, a boon for solar irradiance studies.

Also, while MDI intensity images, and magnetograms and Dopplergrams are generated from different filtergrams\footnote{The MDI data processing pipeline is different from that implemented in HMI, requiring unpolarized filtergrams for the intensity observables, and circularly polarized filtergrams for the line-of-sight magnetograms and Dopplergrams.}, these observables are generated from the exact same filtergrams in HMI, allowing perfectly co-spatial and co-temporal observations of intensity, magnetic field and line-of-sight velocity. This, together with the superior image quality, made HMI observations particularly suitable for the investigations detailed in this thesis.

\section{SATIRE-S}
\label{introductionsatires}

The SATIRE-S semi-empirical model of solar irradiance is one version of the SATIRE model \citep{fligge00,krivova03,krivova11a}. The key assumption of the model is that variations in solar irradiance, on timescales greater than a day, arise from photospheric magnetism alone. At timescales shorter than a day, fluctuations from flares, granulation and $p$-modes (i.e., acoustic oscillations) become significant \citep{hudson88,woods06,seleznyov11}. Variations in solar luminosity from thermal relaxation of the convection zone of the Sun and changes in the chemistry of the core occur at timescales exceeding $10^5$ years \citep{solanki13} and can therefore be safely ignored when considering variations over the 11-year activity cycle.

The solar disc is modelled as comprising of four components, quiet Sun, faculae, sunspot umbra and sunspot penumbra. The difference between SATIRE-S and the other variants of the model is the data used to determine the surface coverage by faculae and sunspots. SATIRE-T uses the sunspot number to reconstruct solar irradiance back to the 17th century \citep{krivova07}, and SATIRE-M\footnote{The suffixes denote their application to reconstructing solar irradiance over the period Telescopes (and therefore sunspot number records) are available, and over Millennial timescales.} cosmogenic isotope data covering the Holocene \citep{vieira11}. The SATIRE-S is the most accurate, employing spatially resolved full-disc observations of intensity and magnetic flux. Such observations allow the prevailing magnetism, including the disc position, to be determined with much greater precision than from sunspot number or cosmogenic isotope records. Apart from offering no information on the position of magnetic structures on the solar disc, the sunspot number is modulated by active region activity and cosmogenic isotope concentrations by the open magnetic flux alone. Neither index of solar magnetism constitutes a complete measure of prevailing magnetism. The only drawback with full-disc magnetograms is that they are only available for the last four decades. However, for the purpose of aiding the interpretation of satellite measurements of solar irradiance, which span a similar period, this is sufficient.

\begin{cfig}
\includegraphics[width=\textwidth]{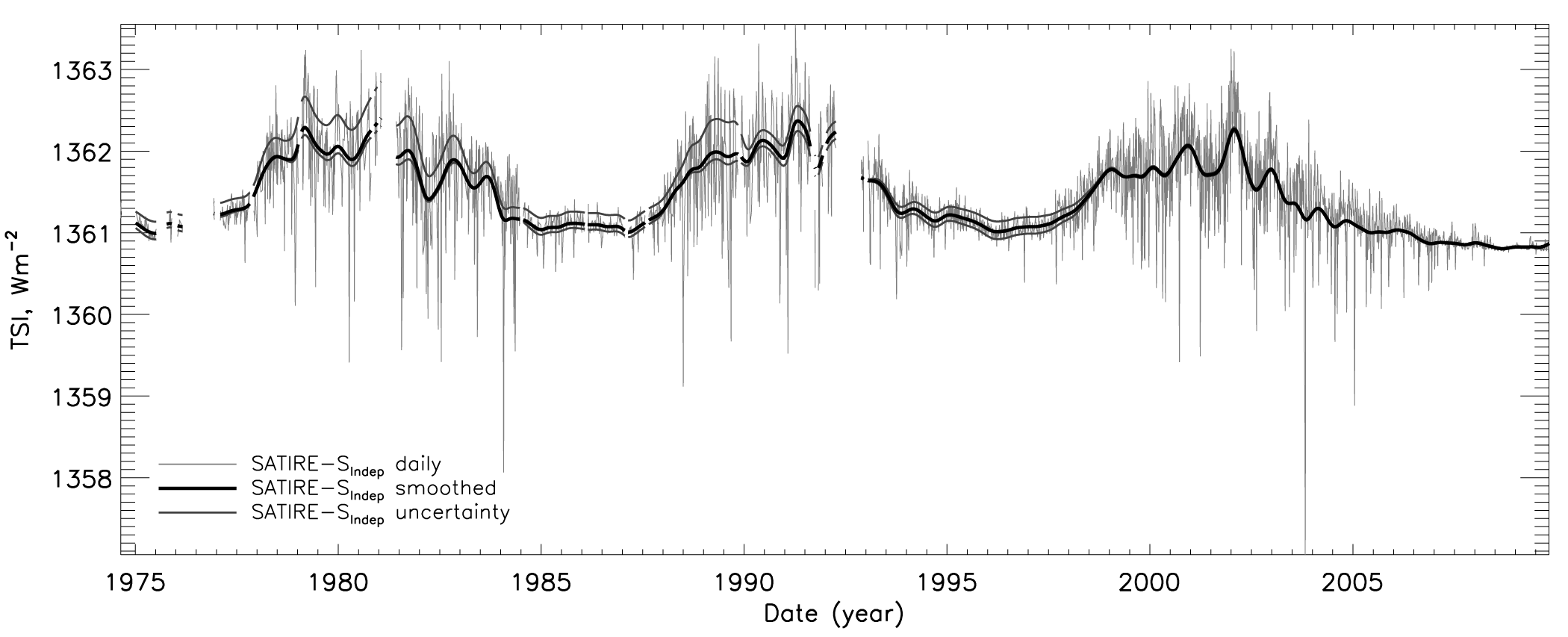}
\caption{SATIRE-S reconstruction of TSI based on KPVT and MDI full-disc longitudinal magnetograms and continuum intensity images. Taken from \cite{ball12}.}
\label{introball12}
\end{cfig}

As stated in Sect. \ref{introductionmmsi}, the model has previously been applied to longitudinal magnetograms and continuum intensity images from the KPVT and MDI. The TSI and SSI reconstructions from these studies extend over various periods between 1974 and 2009 (see example in Fig. \ref{introball12}). The KPVT ceased observations in 2003, and MDI in 2011. One of the objectives of this thesis was to extend the model to the present with similar observations from HMI.

\begin{cfig}
\includegraphics[width=\textwidth]{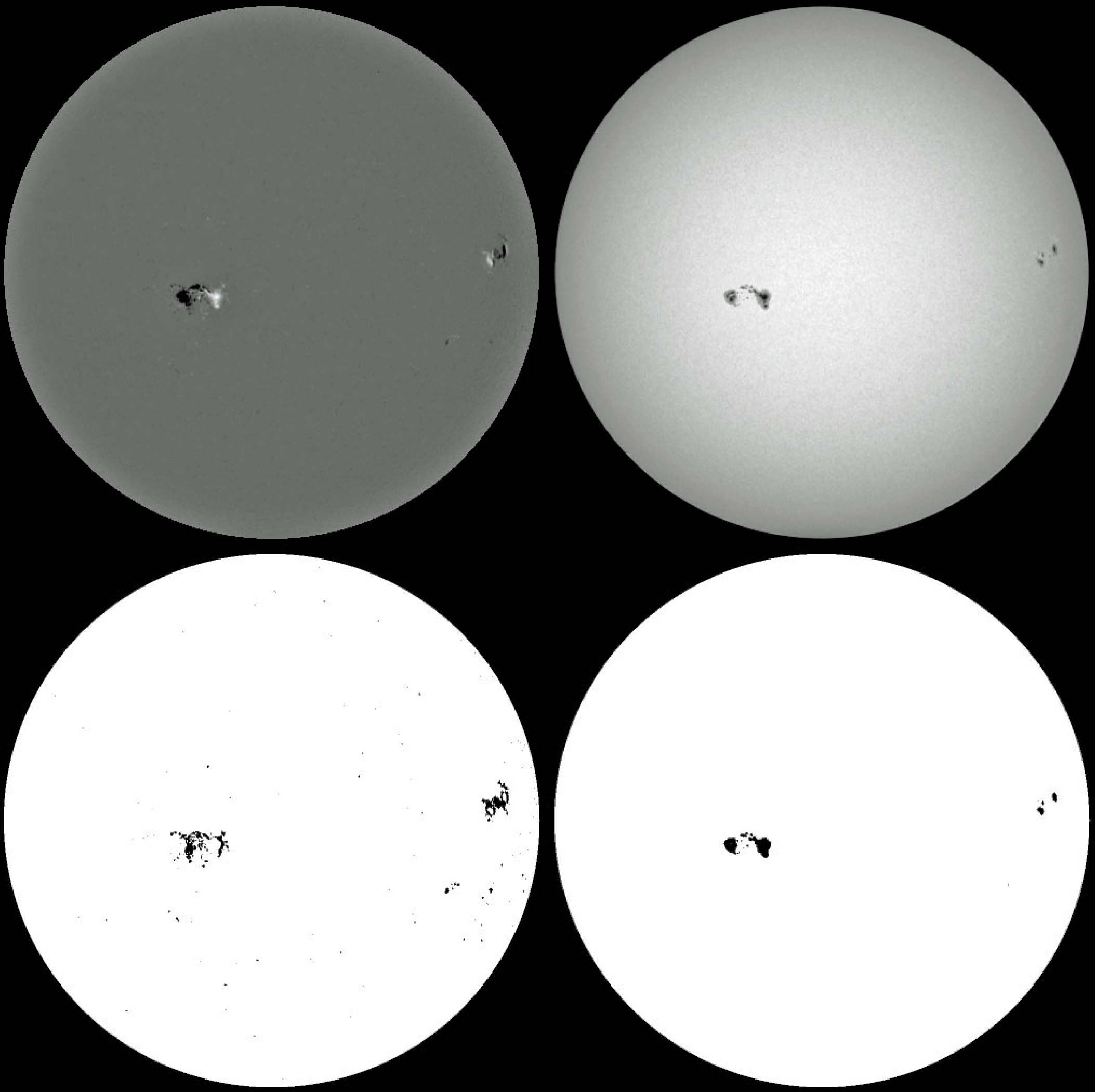}
\caption{Top: MDI longitudinal magnetogram (left) and continuum intensity image (right) from November 25, 1996. Bottom: Corresponding map indicating (in black) the pixels identified as corresponding to faculae (left) and to sunspots (right) by the image segmentation method employed in the SATIRE-S model. Taken from \cite{fligge00}.}
\label{introfligge00}
\end{cfig}

Sunspots are identified from the continuum intensity, and faculae by the longitudinal magnetogram signal. Points on the solar disc below threshold intensity levels representing the umbra-to-penumbra and penumbra-to-granulation boundaries are taken to correspond to umbra and penumbra, respectively. Image pixels above a threshold magnetogram signal (determined by the noise level) and not already classed as sunspots are identified as faculae. (This encompasses all bright magnetic features detectable by such an analysis, which includes both active region faculae and quiet Sun network.) An example is given in Fig. \ref{introfligge00}.

The small-scale magnetic concentrations that make up network and faculae are not fully resolved in available full-disc observations. This is roughly accounted for by scaling the faculae filling factor, the effective proportion of a given resolution element covered by faculae, with the magnetogram signal. The filling factor of each facular pixel is scaled linearly with $\bmu$, the ratio of the longitudinal magnetogram signal and the cosine of the heliocentric angle\footnote{The longitudinal magnetogram signal, $\bl$, represents the pixel-averaged line-of-sight magnetic flux density. Small-scale magnetic concentrations are largely orientated normal to the solar surface due to magnetic buoyancy. The ratio with $\mu$, $\bmu$ is therefore an approximation of the pixel-averaged magnetic flux density.}, saturating at unity at what is termed $\bsat$. The $\bmu$ level at which the faculae filling factor saturates, $\bsat$, is the only free parameter in the model, determined by comparing the reconstruction to measured TSI.

The apparent surface coverage by faculae and sunspots is converted to solar irradiance by means of the intensity spectra of quiet Sun, faculae, umbra and penumbra by \cite{unruh99}. The quiet Sun model atmosphere is given by the ATLAS9 standard solar model \citep{kurucz93}, and the sunspot penumbra and umbra model atmospheres by the standard stellar models corresponding to effective temperatures of 5450 K and 4500 K from the ATLAS9 grid of stellar models. The faculae model atmosphere, introduced by \cite{unruh99}, is a modification of the FAL P model by \cite{fontenla93}.

The intensity spectrum of each surface component at varying heliocentric angles was generated with the ATLAS9 spectral synthesis code. The code assumes local thermodynamic equilibrium, LTE (see Chap. \ref{locallte}). This assumption breaks down in the ultraviolet, formed in the upper photosphere and lower chromosphere, due to the increasingly collisionless condition. As a result, the output from the code is too weak below approximately 300 nm. Prior to the work of this thesis, this was accounted for by rescaling the 115 to 270 nm segment of reconstructed solar irradiance to the measurements from UARS/SUSIM\footnote{The Solar Ultraviolet Spectral Irradiance Monitor onboard the Upper Atmosphere Research Satellite \citep{brueckner93,floyd03}.} \citep{krivova06}. We introduced an updated correction, described in Chap. \ref{uvfix}.

SSI is reconstructed by assigning to each image pixel on the solar disc the appropriate surface component intensity spectrum, and summing the result over the entire solar disc. The wavelength range of the reconstructed solar spectra is 115 to 160000 nm, basically given by the ATLAS9 spectral synthesis code and the wavelength range of the spectroscopic data used to correct the ultraviolet segment. TSI is derived taking the integral under the reconstructed solar spectra.

\section{Thesis outline}
\label{introductionthesisoutline}

The main body of this thesis comprises of four publications, presented as individual chapters in chronological order (Chaps. \ref{paper1} to \ref{paper4}). Before that, we will first discuss background knowledge relevant to the investigations detailed in these publications in Chap. \ref{background}.

In Chap. \ref{paper1}, we examined the intensity contrast of quiet Sun network and active region faculae in HMI data. The aim was to gain insights into the complex radiant behaviour of these magnetic features and their contribution to variation in solar irradiance. In Chap. \ref{paper2}, we derived an estimate of the point spread function, PSF of the HMI instrument. We also investigated the effect of correcting HMI observations for stray light (using this PSF), including the apparent surface coverage and magnetic field strength of network and faculae (relevant to the use of such data in semi-empirical models of solar irradiance). An overarching objective with these two studies is the derivation of information that can be used, in future efforts, to constrain three-dimensional model atmospheres, key to improving the reliability of semi-empirical models of solar irradiance. Specifically, the relationship between intensity contrast, and disc position and magnetic field strength in HMI data from the first study, and the PSF of the instrument from the second will allow a precise quantitative comparison of the intensity contrast in HMI observations with that in artificial solar images synthesized from three-dimensional model atmospheres (based on magnetohydrodynamics or MHD simulations, see Chap. \ref{muram}).

In Chap. \ref{paper3}, we present a daily reconstruction of TSI and SSI, with the SATIRE-S model, based on full-disc observations from the KPVT, MDI and HMI. The reconstruction spans 1974 to 2013. On top of extending earlier efforts with the model based on KPVT and MDI data to the present time with HMI observations, we made various refinements to the reconstruction method. The most important improvement being how the model output based on observations from the various instruments are combined into a single, consistent time series. The aim was to provide a reliable, extended daily reconstruction of TSI/SSI for solar irradiance and climate studies (climate models require solar irradiance input). Then, in Chap. \ref{paper4}, we have a review of the current state of solar irradiance measurements and models, including a discussion of the key challenges in reconciling the outstanding discrepancies between the two. This review also sets the results presented in Chaps. \ref{paper1} and \ref{paper3} in the wider context of the field of study.

Finally, a summary of Chaps. \ref{paper1} to \ref{paper4}, including an outlook of future work based on the results presented, is provided in Chap. \ref{summaryoutlook}.

\chapter[Background]{Background}
\label{background}

\section{Physical origin of solar radiation}

\begin{cfig}
\includegraphics[width=\textwidth]{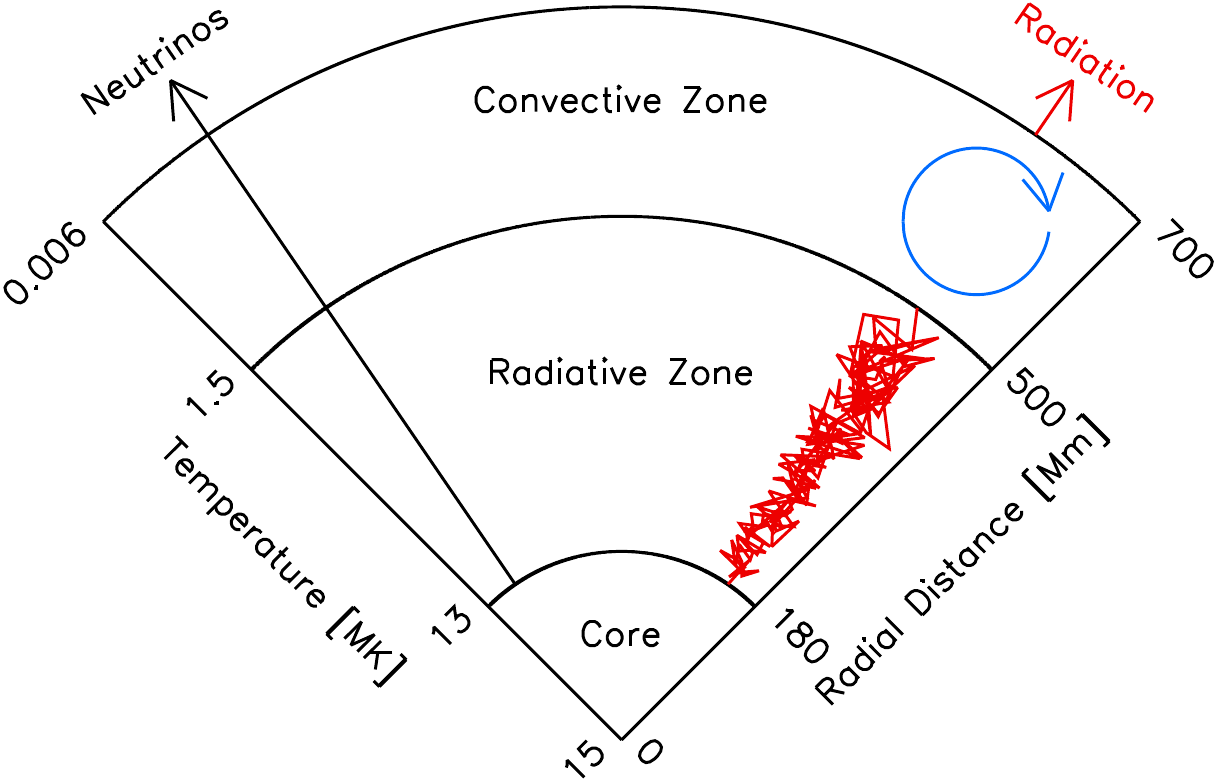}
\caption{Schematic of the solar interior. Neutrinos produced in the core pass largely unhindered out of the Sun. The bulk of the energy generated is transported by radiation (red) and then by convection (blue) to the solar surface.}
\label{solarinterior}
\end{cfig}

\begin{cfig}
\includegraphics[width=\textwidth]{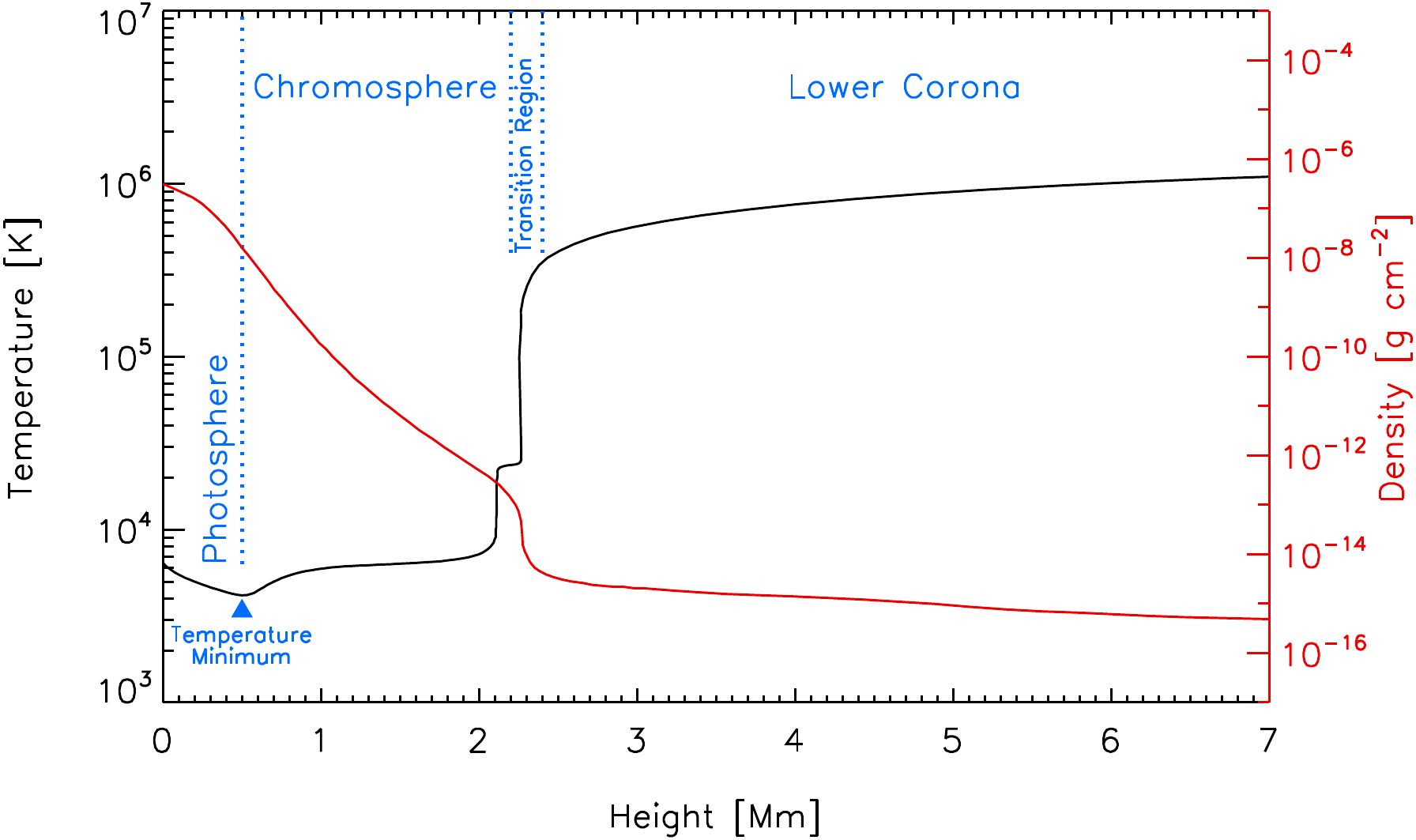}
\caption{Temperature (black, left axis) and density (red, right axis) profile of the solar atmosphere \citep[based on][]{reeves77,vernazza81,avrett92}. Here, as typically done, geometric height of zero is defined as at where the optical depth at 5000 \AA{} is unity. Adapted from \cite{jafarzadeh13}.}
\label{solaratmosphere}
\end{cfig}

In classical theory, the Sun is treated as a radially stratified body. In this convention, it is described as a succession of spherical layers, which track the graduation in physical property. Going out from the centre, we have the core, radiative zone, convective zone, photosphere, chromosphere, transition region and corona, the general properties of which are summarized in Figs. \ref{solarinterior} and \ref{solaratmosphere}.

\subsection{Energy transport in the solar interior}
\label{energytransportsolarinterior}

The energy flux radiated by the Sun, comprising of both corpuscular and electromagnetic radiation (the focus of this thesis), originates in the thermonuclear core. Nuclear fusion in the core is dominated by the proton-proton chain reaction, which releases neutrinos and gamma photons. Neutrinos, being weakly interacting, pass out of the Sun largely unhindered. Photons, on the other hand, are repeatedly scattered in the dense plasma medium. Estimates vary, but according to \cite{mitalas92}, the overall mean free path in the core and radiative zone is $9\times10^{-4}\:{\rm m}$ and the diffusion timescale, the time it takes for a photon to reach the top of the radiative zone, is $1.7\times10^{5}\:{\rm years}$.

From the top of the convective zone to the bottom, hydrogen is increasingly ionized, with the effect of enhancing the opacity (and therefore the extent to which the plasma medium is absorbing radiation and heating up) with depth. This enhanced opacity gradient pushes the temperature gradient above what is termed the adiabatic temperature gradient (c.f., Schwarzschild's criterion for convective stability). Under such conditions, a vertically displaced parcel of fluid will keep ascending/descending until it reaches the top/bottom of the convective zone (or is dissipated by diffusive processes). In the convective zone, energy is no longer transported by radiation but by convection\footnote{Hydrogen opacity is the dominant but not sole effect driving the convection. The partial ionization also introduces latent heat. Let us assume that rising plasma behaves adiabatically (i.e., no heat exchange with its environment). To remain in pressure balance with its increasingly less dense and cooler surroundings, it must expand and cool. Since the latent heat makes it harder for the plasma to cool, it must instead expand more, enhancing the buoyancy. The partial ionization of helium in the convective zone also plays a similar albeit smaller role (being far less abundant).}. This is a far more efficient process; again estimates vary but the time it takes for a parcel of fluid to travel from the bottom of the convective zone to the top is in the order of about a month \citep{eggleton11}. The upwelling of heated plasma from deeper layers and the cool downflow produces the convection cell pattern visible on the solar surface (defined below). They show up at two distinct spatial scales, what is termed granulation (around 1 Mm in diameter) and supergranulation \citep[10 to 30 Mm,][]{hirzberger08}.

\subsection{Formation of the solar spectrum}
\label{formationsolarspectrum}

Above the convective zone, the density is low enough that photons can escape unhindered. The opacity of the plasma medium and therefore the height where this occurs varies with wavelength. The bulk of solar radiative flux is released in the photosphere, where the continuum is formed. The position of optical depth unity (in other words, where the plasma goes from being opaque to transparent) is deepest in the visible and near-infrared, at the lower photosphere. As this is the deepest into the Sun one can observe directly, the lower photosphere is a natural candidate for the solar surface (being a plasma body, the Sun has no `bona fide' surface in the conventional sense), marking the boundary between the solar interior and atmosphere.

The solar atmosphere is by no means spherically symmetrical. The symmetry is broken by the combined influence of the Sun's rotation and magnetism (see Sect. \ref{solarmagnetism}). A plane-parallel description of the solar atmosphere, while not strictly correct, is nonetheless convenient and still informative. The temperature and density stratification of the solar atmosphere, from such a consideration \citep{reeves77,vernazza81,avrett92}, is depicted in Fig. \ref{solaratmosphere}. While density decreases monotonically with height, temperature declines from around 6000 to 4000 K across the photosphere, before increasing again, eventually to several $10^6$ K in the corona, most of the gain coming in the transition region. Due to efficient thermal conductivity, the decline in coronal temperature with radial distance is slow enough that the plasma eventually overcomes gravity and escape as solar wind (largely electrons and photons).

\begin{cfig}
\includegraphics[width=\textwidth]{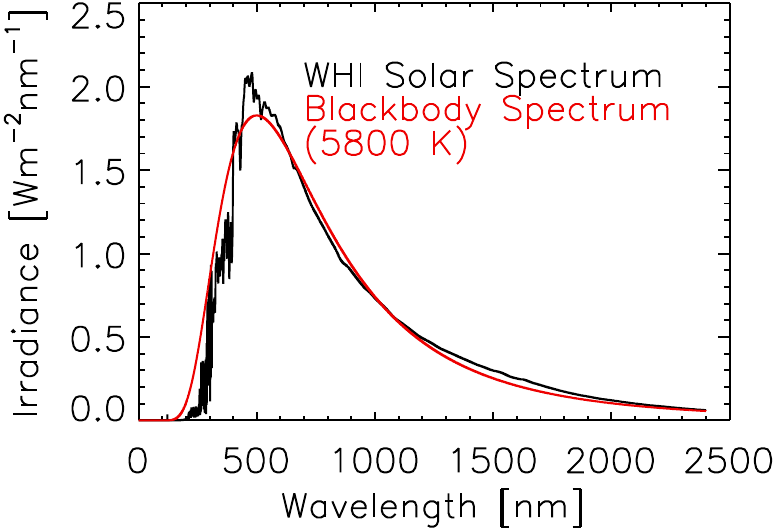}
\caption{The Whole Heliospheric Interval (WHI) reference solar spectrum (for low activity conditions) by \cite{woods09} and the solar spectrum if the Sun were a blackbody with a temperature of 5800 K.}
\label{solarspectrum}
\end{cfig}

Up to the photosphere, the plasma is collision-dominated, such that it behaves approximately like a blackbody emitter. For this reason, the solar spectrum broadly resembles that of a blackbody at its effective temperature, which is about 5800 K (Fig. \ref{solarspectrum}). The departures from the blackbody spectrum arises mainly from the combined action of the wavelength dependence of the opacity and the vertical temperature gradient (Fig. \ref{solaratmosphere}); the radiation at different wavelengths is formed at varying heights and therefore temperatures.

\begin{cfig}
\includegraphics[width=\textwidth]{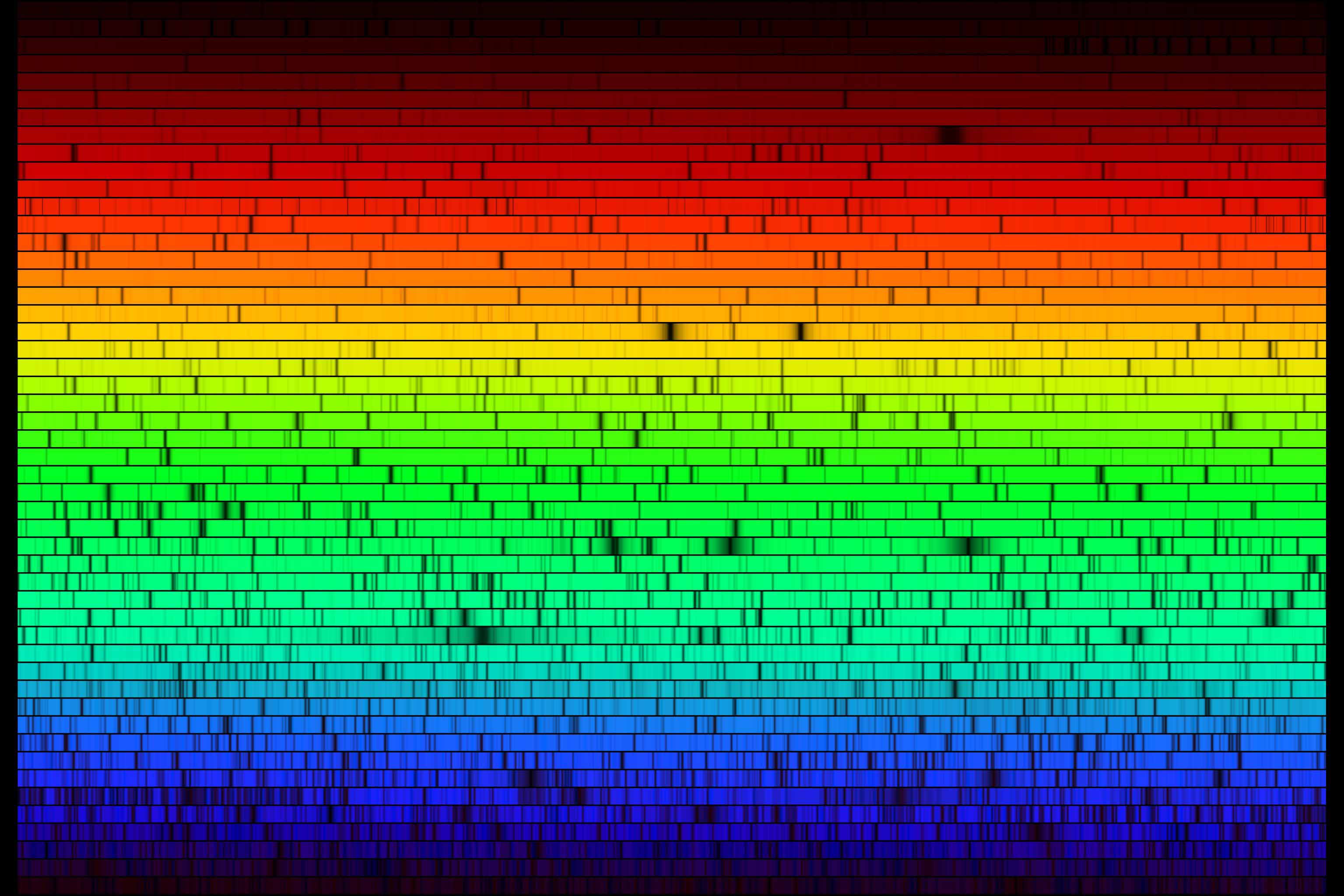}
\caption{Echelle spectrum constructed from the digital atlas of the solar spectrum by the National Optical Astronomy Observatory. From bottom to top, each row corresponds to 6 nm, covering 400 to 700 nm (i.e., the visible spectral range). The dark features correspond to absorption lines, formed dominantly in the photosphere. Courtesy of N. A. Sharp, NOAO/NSO/Kitt Peak FTS/AURA/NSF.}
\label{solarvisible}
\end{cfig}

\begin{cfig}
\includegraphics[width=\textwidth]{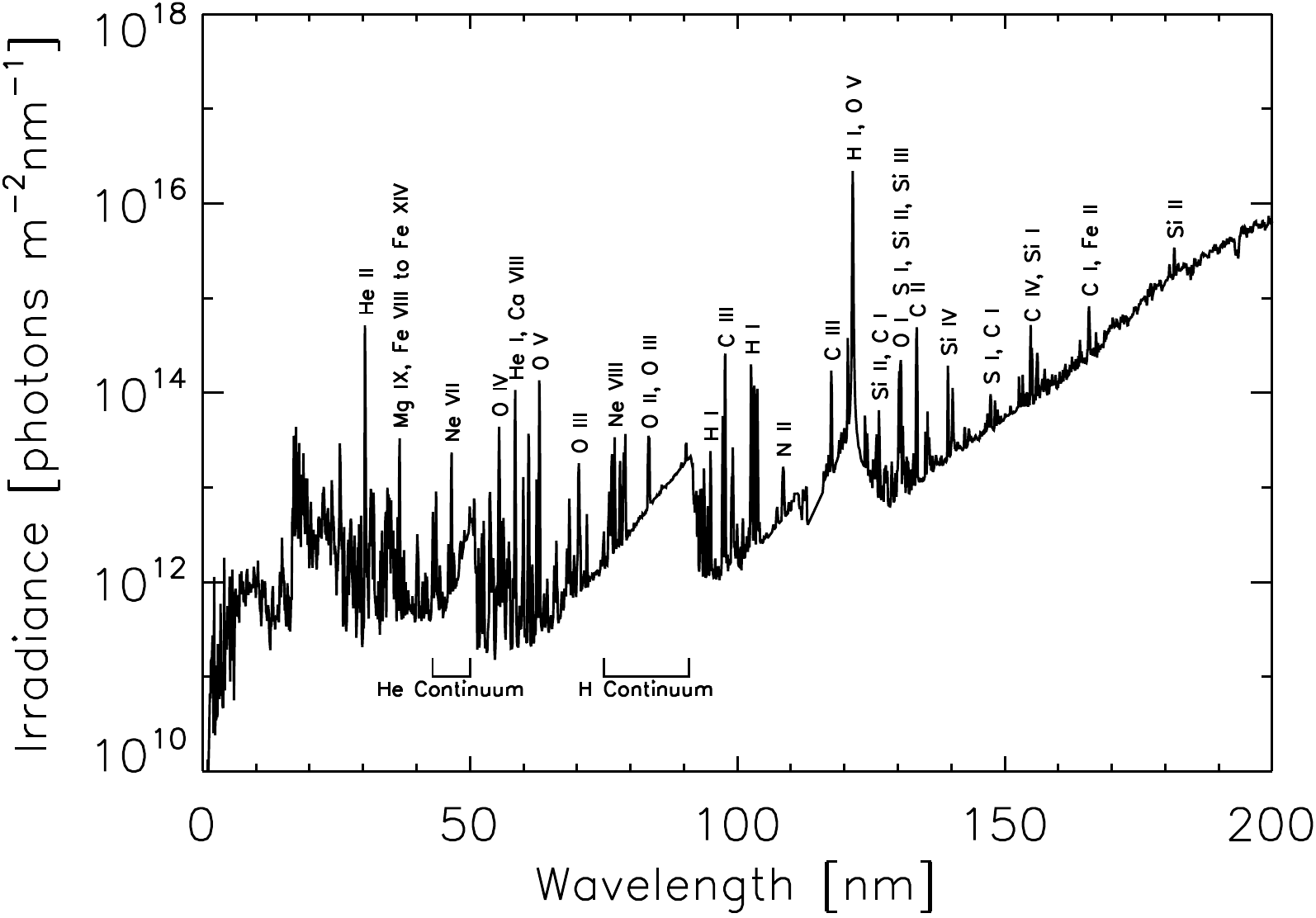}
\caption{Solar vacuum ultraviolet spectrum (from the WHI reference spectrum for low activity conditions, Fig. \ref{solarspectrum}). The emission lines are formed in the chromosphere, transition region and corona. The wedge shaped continua correspond to emission from hydrogen and helium bound-free recombination. The vertical edge correspond to the minimum energy from such an interaction, the ionization potential.}
\label{solaruv}
\end{cfig}

Continuum radiation is formed by free-free and bound-free interactions, and spectral lines by bound-bound interactions\footnote{Free-free and bound-free interactions refer to the absorption and emission processes that move an electron between two free states, and between a bound and a free state (i.e., ionization and recombination). Since the energy of a free electron is not discrete but continuous, interactions involving free states produce the smoothly varying (with wavelength) radiation we identify as continuum. Electron transitions between the discrete energy levels in an atom/molecule (bound-bound) give rise to spectral lines.}. At a given wavelength, the solar atmosphere is, above the continuum formation height, too sparse for free-free and bound-free interactions to take hold but bound-bound interactions (if there are present at that wavelength) are still relevant. At the wavelengths corresponding to bound-bound interactions in the solar atmosphere, the plasma medium remains optically thick up to where photons no longer interact with the responsible species and escape. Spectral lines are formed above the continuum radiation at similar wavelengths, and at different heights, depending on the abundance and location of the respective species. Whether a particular spectral line appears as an absorption or emission feature then depends on the property of the solar atmosphere at the formation height, explained in Sect. \ref{atlas9}. An echelle spectrum representation of the visible solar spectrum is depicted in Fig. \ref{solarvisible}, and a plot of the vacuum ultraviolet (<200 nm) spectrum in Fig. \ref{solaruv}.

\section{The 11-year activity cycle of the Sun and solar magnetism}
\label{solarmagnetism}

\begin{cfig}
\includegraphics[width=\textwidth]{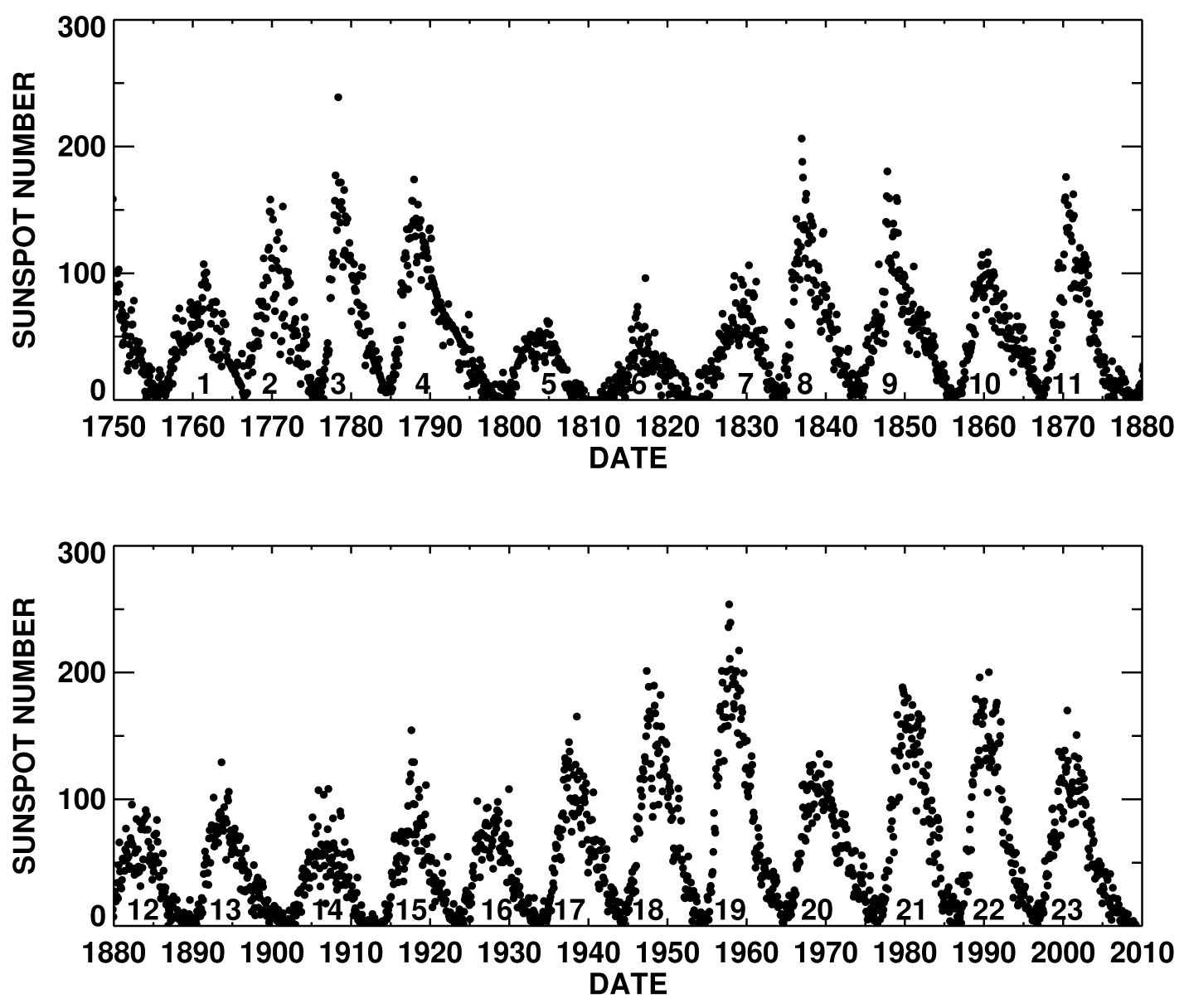}
\caption{Monthly mean of the international sunspot number. Solar cycle number is indicated. Taken from \cite{hathaway10}.}
\label{isn}
\end{cfig}

As pointed out in Chap. \ref{introductionsatires}, acoustic oscillations and convective motions on the solar surface are short-lived phenomena (lifetimes of minutes to around a day), whereas the thermal and nuclear timescale of the Sun exceeds $10^5$ years. Except at these very short and very long timescales, the dynamics of the Sun is dominated by solar magnetism, the most pronounced variability of which is the 11-year cycle. As a consequence, signatures of the 11-year magnetic cycle show up in just about every measurable indication of solar activity, including solar irradiance.

\subsection{Solar cycle variation in solar magnetism}

The most prominent configuration of emergent magnetic flux on the solar surface is the formation of bipolar active regions. A bipolar active region is a region of intense magnetic activity with two distinct zones of opposite magnetic polarity. There are active regions with more complex morphologies, the result of bipolar active regions forming near or within an existing active region \citep{bumba65}. At full development, the larger active regions feature sunspots, pores and faculae.

The 11-year activity cycle of the Sun was first discovered by Heinrich Schwabe in 1843, in the daily number of sunspot groups, and in the number of spotless days over an 18-year period. A few years after, Rudolf Wolf devised what is now termed the international sunspot number, $R$, given by
\begin{equation}
R=k(10g+n),
\end{equation}
where $g$ and $n$ denotes the number of sunspot groups and individual sunspots visible on the solar disc, respectively. Systematic differences between the measurements made by different observers are accounted for by the correction factor, $k$. He also incorporated earlier observations, extending the record back in time to 1749. Still tabulated today, the international sunspot number is one of the longest daily record of solar activity available\footnote{Only the sunspot group number record compiled by \cite{hoyt97}, which extends from 1610 to 1995, is longer.}. While the definition of the international sunspot number is somewhat arbitrary, it turns out to be highly correlated and so a good proxy of sunspot area, which has a more readily appreciable physical meaning \citep[see Figs. 6 and 7 in][]{hathaway10}. The minimum in 1755 (Fig. \ref{isn}) is designated as the start of solar cycle 1, and each successive minimum marks the start of the following cycle. Also apparent in the figure, solar cycle amplitude and length \citep[$10.9\pm1.2$ years,][]{hathaway02} fluctuates considerably.

\begin{cfig}
\includegraphics[width=\textwidth]{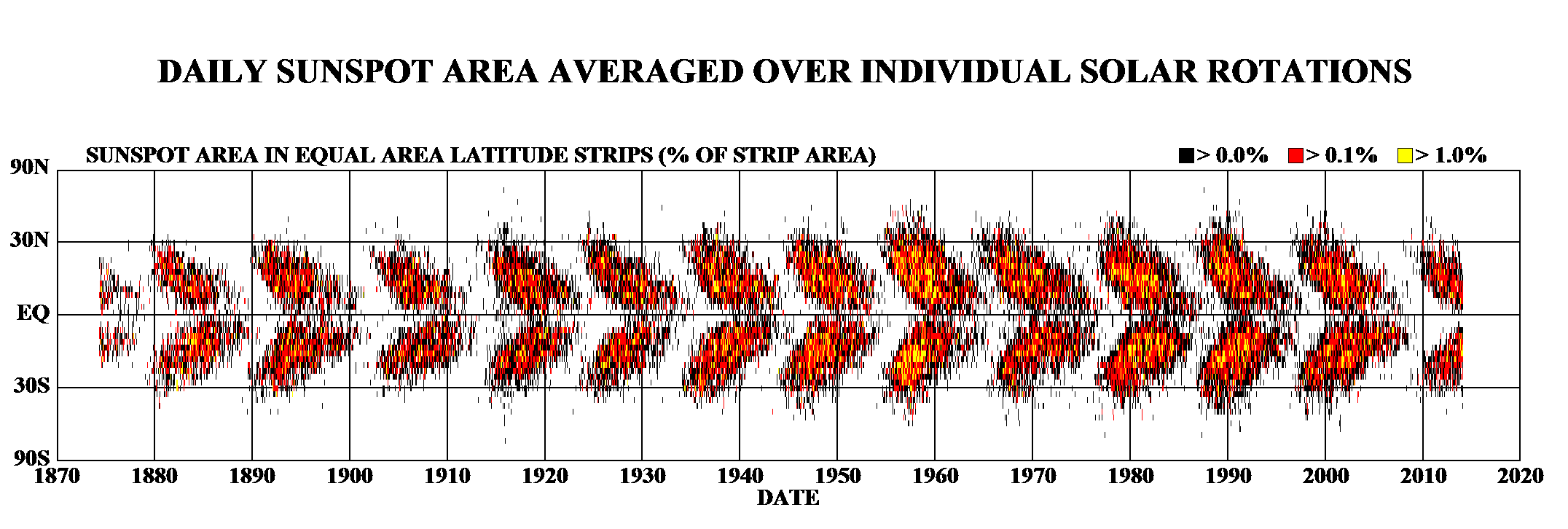}
\caption{The variation in sunspot area with time and latitude. Courtesy of D. H. Hathaway, NASA/MSFC.}
\label{sunspotarea}
\end{cfig}

\begin{cfig}
\includegraphics[width=\textwidth]{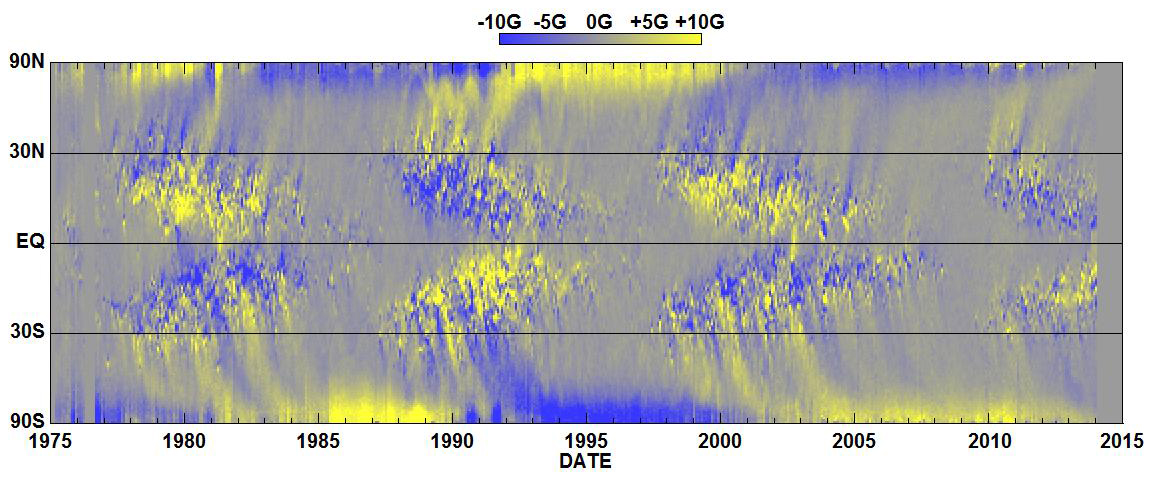}
\caption{Longitudinally-averaged signed magnetic flux density as a function of time and latitude. Courtesy of D. H. Hathaway, NASA/MSFC.}
\label{magbfly}
\end{cfig}

Apart from the 11-year periodicity in overall emergence apparent in the sunspot number (and similar measures), solar magnetism also exhibits the following large-scale behaviour.
\begin{itemize}
	\item Sp\"orer's law of zones: Near the onset of each solar cycle, active regions emerge between around $15^{\circ}$ and $30^{\circ}$ latitude, in both the northern and southern hemispheres. Over the course of the cycle, they appear at lower and lower latitudes, eventually close to the equator. This migration produces the `butterfly wings' pattern in time series representations of the latitudinal distribution of magnetic activity on the solar disc (not surprisingly, these are termed butterfly diagrams, two examples of which are given in Figs. \ref{sunspotarea} and \ref{magbfly}).
	\item Joy's law \citep{hale19}: Bipolar active regions are orientated such that the leading (in the direction of the Sun's rotation) polarity is closer to the equator. Overall, the higher the latitude, the more pronounced this tilt between the lagging and leading polarity. Taken together with Sp\"orer's law, it indicates that tilt angles decline over the solar cycle.
	\item Hale's polarity law \citep{hale25}: The broad orientation of bipolar active regions (which polarity is leading/lagging) within either the northern or southern hemisphere over a given cycle is constant but opposite between the two hemispheres and alternates from cycle-to-cycle. So, while there is an 11-year modulation in the emergence of solar magnetism, the magnetic cycle is really 22-years.
	\item Polar flux: The cyclic variation in active region activity at low and mid-latitudes is accompanied by an apparent anti-phase fluctuation (i.e., strongest/weakest around cycle minima/maxima) in the amount of magnetic flux near the poles. Polar flux in the northern and southern hemispheres have opposite dominant polarities which reverses mid-cycle.
\end{itemize}
Most of these large-scale features of solar magnetism are visible in the butterfly diagram of the longitudinally-averaged signed magnetic flux density (Fig. \ref{magbfly}). The magnetic polarity of the inner and outer edge of each `butterfly wing' arises from the combination of Joy's law and Hale's polarity law. As suggested by the `streams' leading from the wings towards higher latitudes, polar flux correspond to magnetic flux from since decayed active regions transported polewards by meridional flows\footnote{This is the more commonly accepted, but not only, interpretation of the apparent poleward drift. It has been suggested that this could indicate the presence of high latitude dynamo processes that somehow do not result in the formation of active regions \citep{gilman89,petrovay99}.}. Therefore, the mid-cycle reversal in the polarity of polar flux is ultimately related to the cycle-to-cycle alternation in the magnetic orientation of bipolar active regions (see next section). Meridional flows are slow (few $10\:\mps$), giving the half-cycle lag or apparent anti-phase relation between the solar cycle and the variation in the amount of polar flux.

\subsection{The solar dynamo}
\label{solardynamo}

The 11-year/22-year magnetic cycle, discussed above, is believed to be driven by dynamo processes\footnote{The processes by which the magnetic field of an astrophysical object is sustained by the inductive action of the motion of electrically conducting fluids.} in the convective zone and photosphere. Under the assumption that the plasma here is highly-conducting and flows are non-relativistic, Maxwell's equations can be combined into a single equation, which is termed the induction equation,
\begin{equation}
\frac{\partial{}\vec{B}}{\partial{}t}=\nabla\times(\vec{v}\times\vec{B})+\eta\Delta\vec{B},
\label{induction}
\end{equation}
which describes the time evolution of the magnetic field, $\vec{B}$. The first term on the right hand side gives the change from advection or bulk motion ($\vec{v}$ represents velocity), and the second term the change from diffusion ($\eta$ is the magnetic diffusivity, given by $\eta=1/\mu_{0}\sigma$, where $\mu_{0}$ and $\sigma$ denote the magnetic constant and electrical conductivity). Due to the high electrical conductivity, except at extremely small spatial scales (< 1 km, well below the present limits of observation), the advection term is dominant in the convective zone and photosphere (c.f., magnetic Reynolds number). In other words, diffusion is inefficient here. (For example, it will take sunspots on the order of a few thousand years to dissipate by diffusion alone, far greater than the observed lifetime of days to weeks.) This also means that magnetic flux is effectively frozen into the plasma, such that advecting plasma will carry the enclosed magnetic field along with it. The apparent variation in solar magnetism must, as stated earlier, involve the induction effect of bulk motion in the convective zone and photosphere (i.e., a dynamo).

The high electrical conductivity of the plasma medium is also the reason why the magnetic concentrations that make up network and faculae are mostly nested inside intergranular lanes, the cool dark downflows on the boundary of convection cells. Emergent magnetic field, frozen into the plasma, is expelled by convection flows towards the intergranular lanes \citep{parker63,weiss66,tao98}. 

\begin{cfig}
\includegraphics[width=\textwidth]{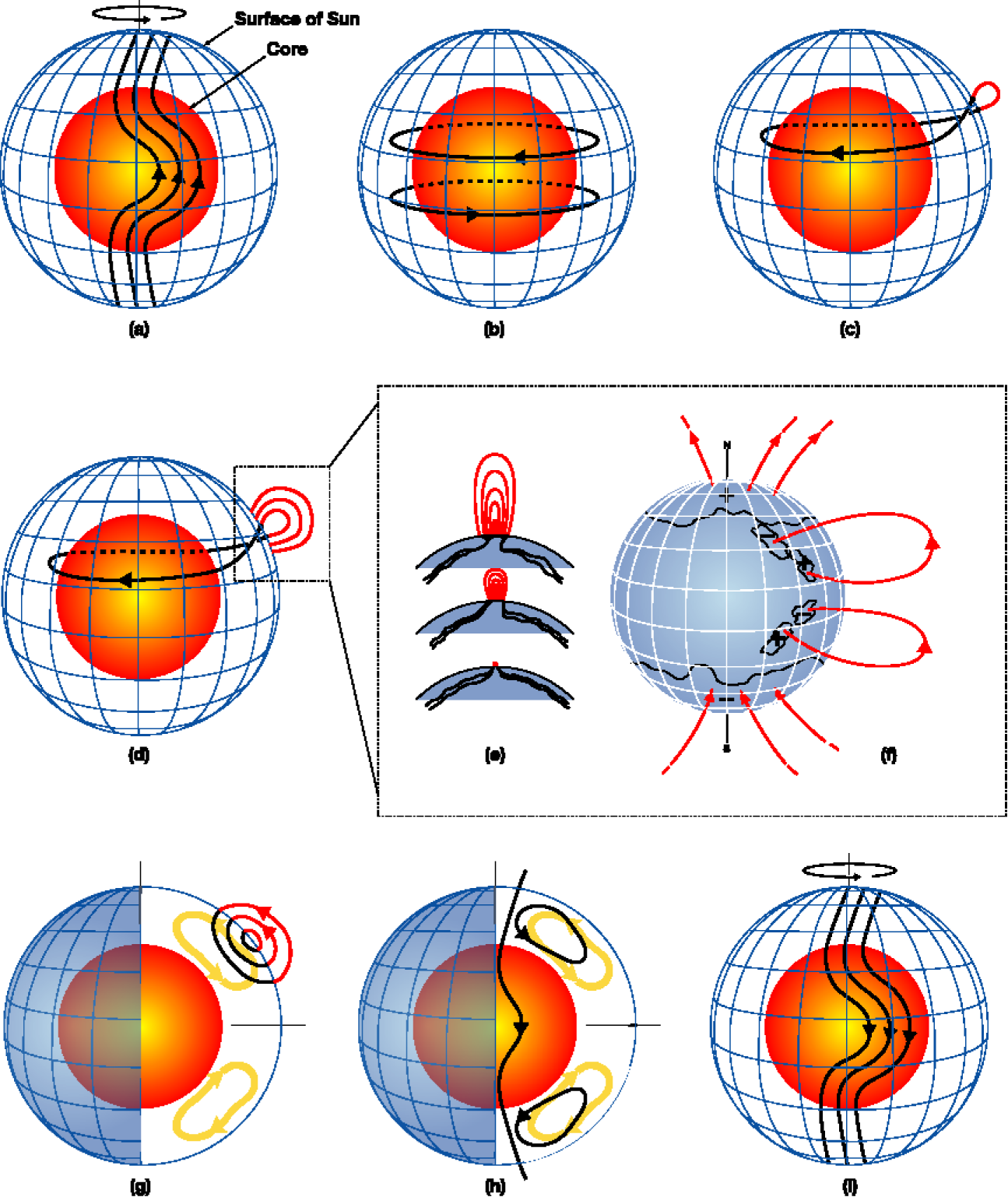}
\caption{Schematic of a `typical' surface flux transport model (see text for explanation). Taken from \cite{dikpati09}.}
\label{sftm}
\end{cfig}

While the exact workings of the solar dynamo is still debated \citep[see the review by][]{charbonneau10}, there is some degree of consensus over certain key features, illustrated in Fig. \ref{sftm}.
\begin{itemize}
	\item Omega-effect (Figs. \ref{sftm}a and \ref{sftm}b): In the radiative zone, the Sun is largely a rigid rotator. In the convective zone however, the rotational frequency varies significantly with latitude and radial distance (differential rotation). As a result, poloidal magnetic field in the convective zone is stretched and wound as the plasma it encloses rotates around the Sun at different speeds, producing toroidal magnetic field of opposite magnetic orientation in the northern and southern hemispheres (c.f., Hale's polarity law).
	\item Alpha-effect (Figs. \ref{sftm}c to \ref{sftm}f): For the magnetic ($B^{2}/2\mu_{0}$) and gas pressure within magnetic flux tubes to balance external gas pressure, the enclosed plasma must have a lower density. Magnetic flux tubes are therefore buoyant. Due to the action of the Coriolis force, magnetic loops (from the toroidal magnetic field) twist as they emerge, producing the tilt between the leading and lagging polarity of the resulting active regions (c.f., Joy's law).
	\item Meridional flow (Figs. \ref{sftm}g to \ref{sftm}i): As noted earlier in this section, this brings the decay products of active regions polewards, eventually reversing the polarity around mid-cycle. (Since the leading polarities from the two hemispheres have a greater tendency to meet and annihilate due to the equator-ward tilt, polar flux is dominated by the lagging polarity.) The return flow brings polar flux down into the convective zone, forming poloidal magnetic field with an opposite magnetic orientation to that at the beginning of the sequence (Fig. \ref{sftm}a).
	\item The solar dynamo is believed to be situated near the bottom of the convective zone. Due to the convective instability, the convective zone cannot store magnetic fields long enough for the omega-effect to take hold or sustain 11-year cycles. The radiative zone, lacking differential rotation, cannot support the omega-effect. Also, magnetic flux tubes here will not be able to rise to the surface for the same reasons that convection cannot take place (lack of a strong temperature gradient). It is proposed that magnetic flux is transported and confined to the layers just below the convective zone by convective downflows \citep{brandenburg96,tobias98,dorch01}. Here, in the interface between the radiative and convective zones, rotational shear is still present, permitting the omega-effect, and toroidal magnetic fields can, when sufficiently amplified, still rise up the convective zone.
\end{itemize}
These features are incorporated into the class of dynamo models termed surface flux transport models, which have found reasonable success reproducing the main features of the magnetic cycle summarized earlier in Sect. \ref{solarmagnetism} \citep{dikpati09}. 

The presence of magnetic concentrations on the solar surface has a profound effect on the temperature structure and consequently the radiant properties of the solar atmosphere, reviewed in Chap. \ref{models1}. This is believed to be the main driver of variations in solar irradiance, supported by the success of models of solar irradiance based on this mechanism \citep{domingo09}. Other mechanisms have been proposed to explain the observed solar cycle modulation in solar irradiance. For instance, slow decaying global oscillations driven by the Coriolis force \citep{wolff87}, thermal shadowing related to the toroidal magnetic field \citep{kuhn88} and magnetic modulation of convective flow patterns \citep{cossette13}. However, as stated in Chap. \ref{introductionmmsi}, direct evidence is still wanting.

\section{Solar model atmospheres and intensity spectra}

As we will review in Chap. \ref{paper4}, semi-empirical models of solar irradiance are the most sophisticated available. The robust reconstruction of solar irradiance through such models relies on two things. One, an accurate estimation of the prevailing solar magnetism at the sampled points in time. Two, realistic intensity spectra of the solar surface components. The latter is provided by numerical models aimed at returning the stratification of the solar atmosphere (what is usually termed `model atmospheres') and the emergent intensity spectrum through the solution to the radiative transfer equation, RTE (introduced next). The ATLAS9 code by \cite{kurucz93} is one such model. As noted in Chap. \ref{introductionsatires}, in SATIRE-S, the sunspot and quiet Sun model atmospheres are based on ATLAS9 model atmospheres, and the intensity spectra of solar surface components are generated from the respective model atmospheres using the ATLAS9 code.

\subsection{The radiative transfer equation}
\label{atlas9}

The specific intensity, $I_{\nu}(\vec{r},\vec{n},t)$ at frequency $\nu$, position $\vec{r}$, in direction $\vec{n}$ at time $t$ is defined the energy transported as radiation $dE_{\nu}$, per unit frequency, solid angle $d\omega$ and time passing through a unit area normal to $\vec{n}$. The energy passing through an area $dA$ is given by
\begin{equation}
dE_{\nu}=I_{\nu}(\vec{r},\vec{n},t)\:d\nu{}\:d\omega{}\:dt\:\mu{}dA,
\end{equation}
where $\mu$ is the cosine of the angle between the normal to $dA$ and $\vec{n}$. A useful property of the specific intensity is that, in the absence of sources (emission) and sinks (absorption), it is invariant along a pencil of light. This allows us to describe the change in specific intensity in a pencil of light from emission and absorption by
\begin{equation}
\mu{}dI_{\nu}(s)=j_{\nu}(s)ds-\alpha_{\nu}(s)I_{\nu}(s)ds,
\label{rte_before}
\end{equation}
where $s$ denotes path length, $j_{\nu}(s)$ the emission coefficient and $\alpha_{\nu}(s)$ the absorption coefficient. The quantity $1/\alpha_{\nu}(s)$ is also the local photon mean free path. A more natural way to describe distance than is by the optical distance, $\tau_{\nu}$ where $d\tau_{\nu}$=$\alpha_{\nu}(s)ds$. This allows us to rewrite equation \ref{rte_before} as
\begin{equation}
\mu\frac{dI_{\nu}}{d\tau_{\nu}}=S_{\nu}-I_{\nu},
\end{equation}
what is termed the radiative transfer equation (RTE), where $S_{\nu}$ is the source function, given by
\begin{equation}
S_{\nu}=\frac{j_{\nu}}{\alpha_{\nu}}.
\label{rte}
\end{equation}
The source function characterizes the combined action of emission and absorption on the radiation field. Whether a given spectral line manifests as an absorption or emission feature depends on the source function at that wavelength. As noted in Sect. \ref{formationsolarspectrum}, spectral lines are formed above the continuum at similar wavelengths. If the source function is weaker at the line formation height than at where the continuum at adjacent wavelengths is formed (i.e., absorption is dominant), the intensity here is weaker than in the nearby continuum, producing what we see as an absorption feature in the spectrum. Conversely, if the source function is stronger at the line formation height, an emission feature is formed.

It can be shown that, under conditions of local thermodynamic equilibrium, LTE (explained in Sect. \ref{locallte}) and in the absence of scattering, the source function at a given height is given by Planck's function at the local temperature. The LTE assumption is largely valid in the photosphere. As the temperature in the photosphere declines with height up to the temperature minimum (Fig. \ref{solaratmosphere}), so does the source function, explaining why photospheric lines are, with few exceptions, absorption lines (Fig. \ref{solarvisible}).

Apart from the frequency/wavelength dependence, the emission and absorption coefficients, and therefore the optical distance and source function, depend on temperature, pressure, atomic/molecular abundances and the corresponding occupation numbers\footnote{For a given species, the distribution between the various bound and ionized states.}. The goal then is to solve the RTE for the emergent intensity from a solar atmosphere, retrieving along the way the temperature and pressure stratification of said atmosphere. The problem is intractable without any simplifications. Classical radiative transfer models such as ATLAS9 are set up with the following assumptions.
\begin{itemize}
	\item Plane-parallel atmosphere: The thickness of the solar atmosphere is small enough compared to the radius of the Sun that we can treat it as a series of homogenous plane-parallel layers, reducing the problem down to one dimension. The limitations of this assumption are discussed in Chap. \ref{summaryoutlook}. In this geometry, $\mu$ is equivalent to the cosine of the heliocentric angle.
	\item Semi-infinite atmosphere: The solar atmosphere is sufficiently thick that the optical distance (zero at the observer's end, increasing with depth into the atmosphere) can be taken to extend to infinity.
	\item Radiative equilibrium: All energy transport is by radiation. This is valid since hydrogen ionization (and therefore convection) in the Sun occurs below the solar surface (Sect. \ref{energytransportsolarinterior}). It can be shown that under this condition, total flux is constant with depth.
	\item Local thermodynamic equilibrium: The infinitesimal slab of material at any given depth (i.e., between $\tau_{\nu}$ and $\tau_{\nu}+d\tau_{\nu}$), is taken to be in a state of thermodynamic equilibrium, therefore behaving like a blackbody at the local temperature. This allows one to calculate the occupation numbers from temperature and electron density (see next section).
	\item Steady state: The solar atmosphere does not vary with time. This requires that the occupation numbers are constant, a condition that is satisfied by the LTE assumption.
	\item Hydrostatic equilibrium: There are no velocity fields. The pressure stratification is given by the balance between gas pressure and gravity.
\end{itemize}
From an initial guess of the temperature stratification, the pressure and electron density profile can be calculated, and from that the optical distance and the source function. This is iterated, adjusting the temperature stratification until the solution to the RTE satisfies the radiative equilibrium condition (i.e., total flux is invariant with depth).

\subsection{Local thermodynamic equilibrium}
\label{locallte}

Occupation numbers are governed by radiative and collisional bound-bound and bound-free processes. In this context, by the term `radiative' we refer to transitions that are spontaneous or triggered by radiation, and by `collisional' transitions effected by collisions. Local thermodynamic equilibrium refers to the condition where the plasma is collision-dominated enough for the radiation field to relax to the blackbody radiation at the local temperature. The specific intensity of a blackbody at temperature $T$, $B_{\nu}(T)$, given by Planck's law, is
\begin{equation}
B_{\nu}(T)=\frac{2h\nu^2}{c^2}\frac{1}{\exp{\left(\frac{h\nu}{k_{B}T}\right)}-1},
\label{planckbbf}
\end{equation}
where $h$ is Planck's constant, $c$ the speed of light and $k_B$ Boltzmann's constant. Let $n$ represent number density and $E$ the energy at a given level, the distribution of atoms with electrons in either of two bound states (denoted $i$ and $j$) follow Boltzmann's distribution,
\begin{equation}
\frac{n_{i}}{n_{j}}=\frac{g_{i}}{g_{j}}\exp{\left(-\frac{E_{i}-E_{j}}{k_{B}T}\right)},
\label{boltzmann}
\end{equation}
where $g=2J+1$, a statistical weight to account for degenerate sub-levels ($J$ is the angular momentum quantum number). The analogue for atoms in either of two ionized state is given by the Saha equation,
\begin{equation}
\frac{n_{i}}{n_{j}}=\frac{2}{n_{e}}\left( \frac{2\pi{}m_{e}k_{B}T}{h^2} \right)^{\frac{3}{2}}\frac{g_{i}}{g_{j}}\exp{\left(-\frac{E_{i}-E_{j}}{k_{B}T}\right)},
\label{saha}
\end{equation}
where $n_{e}$ is the electron density and $m_{e}$ the electron mass. Notice, from equations \ref{boltzmann} and \ref{saha}, that the occupation numbers are functions of temperature and electron density alone.

As stated in the last section, the assumption of steady state requires that the occupation numbers must be constant. The mathematical expression of the requirement that the rate of population change from all relevant radiative and collisional processes sum to zero is what is termed the equations of statistical equilibrium. In the case collisional processes are not sufficiently dominant that we can assume LTE, occupation numbers must instead come from the solution to the equations of statistical equilibrium. This is difficult, as the rate of radiative processes is, by definition, influenced by the prevailing radiation field. Assuming LTE allows us to express occupation numbers as a function of temperature and electron density alone, which reduces the problem down to that of optimizing the temperature stratification to satisfy radiative equilibrium. If LTE cannot be assumed, one would have to solve the equations of statistical equilibrium for all depths and the RTE simultaneously in a self-consistent manner, a far more tedious problem. (Particularly as this would require complete knowledge of the rates of all relevant radiative and collisional processes.)

In the solar atmosphere, the plasma medium is increasingly sparse and therefore collisionless with height. LTE, while a convenient simplification, is not entirely realistic. The disparity between intensity spectra from ATLAS9 and measured solar spectra arising from the assumption of LTE is discussed in Chap. \ref{uvfix}.

\subsection{Line blanketing and opacity}
\label{opacitysampling}

The solar spectrum, especially in the visible and ultraviolet, is dense with spectral lines (Figs. \ref{solarvisible} and \ref{solaruv}). Formed above the continuum at similar wavelengths (Sect. \ref{formationsolarspectrum}), spectral lines block radiation from below, which in turn warms up the deeper layers, causing it to radiate at higher temperatures (`backwarming'). Consequently, the emergent intensity, not just at the spectral lines but also at continuum wavelengths, is spectrally redistributed. The formation of spectral lines in the higher layers of the solar atmosphere, by facilitating radiation, also causes it cool, an effect referred to as surface cooling. This, together with backwarming, enhances the vertical temperature gradient. These effects of spectral lines on stellar atmospheres and the emergent intensity spectra are collectively termed line blanketing. The inclusion of line blanketing is crucial to recovering accurate solutions of the solar atmosphere and intensity spectra. Ideally, this would be achieved by performing the computation described in Sect. \ref{atlas9} at every frequency with a spectral line. However, the spectral lines in the solar spectrum number in the millions, making this computationally unfeasible.

Spectral lines are represented in the RTE by their contribution to the absorption coefficient. The challenge is to include the opacity\footnote{There are varying definitions. For this discussion, by opacity we refer to the product of the absorption coefficient and the number density.} of the immense number of spectral lines present as completely as possible while solving the RTE at a manageable number of frequencies. There are various ways around this problem, each with its advantages and disadvantages.
\begin{itemize}
	\item Mean opacities: The idea here is to assume that the atmosphere is what is termed `grey', that is to say, the opacity does not vary with frequency. The problem is then reduced to calculating a suitably weighted mean opacity. One key advantage is that the RTE can be solved exactly for a grey atmosphere. However, the spectral distribution in the emergent intensity spectra turns out to be far from realistic. As line opacities are averaged out, the spectra will be missing spectral lines. The continuum will be affected as well as the decline in temperature with height will also be underestimated. In recent years, this approach has been used more to provide an initial estimate of the temperature stratification than as a solution in itself. Although, due to the computational simplicity, it is sometimes employed in MHD simulations to describe the action of radiative energy transfer.
	\item Opacity distribution function (ODF): This is the approach taken by the ATLAS9 code. The frequency range is divided into intervals. Within each interval, the absorption coefficient, $\alpha_{\nu}(s)$ is evaluated at narrow frequency steps to capture the rapid variation with frequency from the presence of multiple spectral lines. Following that, the opacity distribution function, describing the proportion of the interval where $\alpha_{\nu}(s)\geq\alpha$ as a function of $\alpha$, is constructed. When solving the RTE, the ODF within each interval is sampled at a few $\alpha$ points to estimate a representative opacity\footnote{As compared to sampling $\alpha_{\nu}(s)$ directly, very few sampling points are needed since ODFs are smooth, monotonic functions.}. ODFs are computationally expensive to build but it is a one-off task. They allow line blanketing to be included efficiently with reasonable accuracy by greatly limiting the number of frequency points that need to sampled. The use of ODFs however, implicitly assumes that spectral lines do not move in/out of a given interval and relative line strengths do not change with depth and temperature \citep{mihalas78}.
	\item Opacity sampling (OS): The principle of opacity sampling, first proposed by \cite{peytremann74}, is to restrict the RTE computation to the optimal set of frequency points necessary to capture most of the underlying physics. Various sampling schemes have been reported in the literature. For example, sampling at regularly spaced intervals \citep{sneden76,jorgensen92}, sampling randomly selected frequencies \citep{carbon84} and adopting a Planck function weighted sampling point distribution \citep{peytremann74,ekberg86}. Generally, the effect of introducing more sampling points on the output model atmosphere diminishes with the total number of sampling points, eventually making negligible different to the outcome (i.e., `saturation'). Taking this to imply that OS-based models approaches the `ideal' model atmosphere from complete sampling asymptotically with increased sampling, the output model atmosphere at saturation is the closest to the ideal solution practically achievable. Opacity sampling is computationally more expensive than using ODFs\footnote{For example, the ATLAS12 code, which is identical to ATLAS9 but for adopting the OS instead of the ODF approach, requires one to two orders of magnitude more computing time \citep{castelli05}.} but has the advantage that it makes no assumptions about the stability of the distribution of $\alpha_{\nu}(s)$ with depth and temperature.
\end{itemize}

\subsection{Three-dimensional model atmospheres}
\label{muram}

One of the limitations of present-day semi-empirical models of solar irradiance is the use of intensity spectra generated from plane-parallel model atmospheres. They do not capture all the complexities of the radiant behaviour of small-scale magnetic concentrations (reviewed in Chap. \ref{discussionsemiempirical}). Magnetic flux tubes are not isotropic structures; the apparent intensity and its centre-to-limb variation is not just a function of the vertical stratification of the confined atmosphere but also of the viewing geometry (c.f., flux tube model, Chap. \ref{models1}).

As stated in Chap. \ref{introductionthesisoutline}, one of the objectives of the investigations presented in Chaps. \ref{paper1} and \ref{paper2} is to set some of the groundwork necessary for the incorporation of intensity spectra of solar surface components from three-dimensional model atmospheres into SATIRE-S. One of our future plans (see Chap. \ref{summaryoutlook}) is to employ the results from these studies as observational constraints on three-dimensional model atmospheres based on MHD simulations.

Central to this aspiration are the three dimensional MHD simulations of the upper convective zone and solar atmosphere from the MURaM\footnote{In full, the \underline{M}ax-Planck Institute for Solar System Research/\underline{U}niversity of Chicago \underline{Ra}diative \underline{M}HD code.} code \citep{vogler03,vogler05,vogler05b}, briefly introduced here for information. The MURaM code describes the time variation of the density $\rho$, momentum density $\rho\vec{v}$, energy density $e=e_{\rm int}+\rho|\vec{v}|^2/2+|\vec{B}|^2/8\pi$ (sum of the internal, kinetic and magnetic energy) and magnetic field strength in a three-dimensional Cartesian grid. This is achieved by solving the following non-ideal MHD equations; the continuity equation (representing mass conservation),
\begin{equation}
\frac{\partial\rho}{\partial{}t}+\nabla{}\cdot(\rho\vec{v})=0,
\end{equation}
the equation of motion,
\begin{equation}
\frac{\partial\rho\vec{v}}{\partial{}t}+\nabla\cdot\left[\rho\vec{v}\vec{v}+\left( p+\frac{|\vec{B}|^2}{8\pi}\right)\underline{I}-\frac{\vec{B}\vec{B}}{4\pi}\right]=\rho\vec{g}+\nabla\cdot\underline{\tau}
\end{equation}
($p$ represents pressure, $\underline{I}$ the $3\times3$ identity matrix and $\underline{\tau}$ the viscous stress tensor), the energy equation,
\begin{gather}
\frac{\partial{}e}{\partial{}t}+\nabla\cdot\left[\vec{v}\left(e+p+\frac{|\vec{B}|^2}{8\pi}\right)-\frac{\vec{B}(\vec{u}\cdot\vec{B})}{4\pi}\right]=\nonumber\\
\frac{\nabla\cdot(\vec{B}\times\eta\nabla\times\vec{B})}{4\pi}+\nabla\cdot(\vec{u}\cdot\underline{\tau})+\nabla.(K\nabla{}T)+\rho(\vec{g}\cdot\vec{v})+Q_{\rm rad}
\label{energyequation}
\end{gather}
($K$ denotes thermal conductivity and $Q_{\rm rad}$ the radiative heating rate) and the induction equation (discussed in Sect. \ref{induction}). The solar simulation is set up by the following.
\begin{itemize}
	\item The system of MHD equations is closed by an equation of state. This equation of state, a set of grids relating temperature and pressure to density and internal energy, is constructed taking in the solution to Saha's equation (equation \ref{saha}) for the first ionization state of the 11 most abundant elements in the solar photosphere.
	\item Radiative energy transport, dominant above the convective zone, is included through the radiative heating rate term ($Q_{\rm rad}$) in the energy equation (equation \ref{energyequation}). This is calculated from the solution to the RTE (assuming LTE). Line opacities are included in an approximate manner by rearranging the entire frequency range into four subsets, grouping frequencies with similar opacities together, which are then each represented by a bin-averaged opacity \citep{vogler04}. This is a very crude implementation of the ODF approach, forced by the requirement to solve the RTE in multiple directions at each gridpoint and simulation timestep.
\end{itemize}
As demonstrated by \cite{unruh09} and \cite{afram11}, one can use the run of parameters along a line through a given MURaM simulation cube as a model atmosphere for generating intensity spectra with radiative transfer codes such as the ATLAS9. \cite{afram11} noted that the artificial solar images so generated from MURaM simulations match the observations from Hinode \citep{kosugi07} well at certain wavelengths (visible continuum) but less so others (CN and G-band). This prompted our interest to derive observational constraints on MURaM simulations from HMI data. As we will explain in Chap. \ref{summaryoutlook}, reconciling the intensity contrast of small-scale magnetic concentrations in HMI and MURaM will also allow us to relate the intensity spectra generated from MURaM-based model atmospheres to HMI magnetogram signal directly. This will obviate the empirical faculae filling factor and magnetogram signal relationship in SATIRE-S, described in Chap. \ref{introductionsatires}, and the free parameter it introduces to the model.

\chapter[Intensity contrast of solar network and faculae \\ \textit{\footnotesize{(The contents of this chapter are identical to the printed version of Yeo, K. L., Solanki, S. K., Krivova, N. A., 2013, Intensity contrast of network and faculae, Astron. Astrophys., 550, A95.)}}]{Intensity contrast of solar network and faculae}
\label{paper1}

\begin{flushright}
{\it Yeo, K. L., Solanki, S. K., Krivova, N. A.} \\
{\bf Astron. Astrophys., 550, A95 (2013)\footnoteB{The contents of this chapter are identical to the printed version of Yeo, K. L., Solanki, S. K., Krivova, N. A., 2013, Intensity contrast of network and faculae, Astron. Astrophys., 550, A95, reproduced with permission from Astronomy \& Astrophysics, \textcopyright{} ESO.}}
\end{flushright}

\section*{Abstract}

{\it Aims.} This study aims at setting observational constraints on the continuum and line core intensity contrast of network and faculae, specifically, their relationship with magnetic field and disc position.

\noindent
{\it Methods.} Full-disc magnetograms and intensity images by the Helioseismic and Magnetic Imager (HMI) on-board the Solar Dynamics Observatory (SDO) were employed. Bright magnetic features, representing network and faculae, were identified and the relationship between their intensity contrast at continuum and line core with magnetogram signal and heliocentric angle examined. Care was taken to minimize the inclusion of the magnetic canopy and straylight from sunspots and pores as network and faculae.

\noindent
{\it Results.} In line with earlier studies, network features, on a per unit magnetic flux basis, appeared brighter than facular features. Intensity contrasts in the continuum and line core differ considerably, most notably, they exhibit opposite centre-to-limb variations. We found this difference in behaviour to likely be due to the different mechanisms of the formation of the two spectral components. From a simple model based on bivariate polynomial fits to the measured contrasts we confirmed spectral line changes to be a significant driver of facular contribution to variation in solar irradiance. The discrepancy between the continuum contrast reported here and in the literature was shown to arise mainly from differences in spatial resolution and treatment of magnetic signals adjacent to sunspots and pores.

\noindent
{\it Conclusions.} HMI is a source of accurate contrasts and low-noise magnetograms covering the full solar disc. For irradiance studies it is important to consider not just the contribution from the continuum but also from the spectral lines. In order not to underestimate long-term variations in solar irradiance, irradiance models should take the greater contrast per unit magnetic flux associated with magnetic features with low magnetic flux into account.

\section{Introduction}
\label{introduction}

Photospheric magnetic activity is the dominant driver of variation in solar irradiance on rotational and cyclical timescales \citep{domingo09}. Magnetic flux in the photosphere is partly confined to discrete concentrations of kilogauss strengths, generally described in terms of flux tubes \citep{stenflo73,spruit83}. The brightness excess, or contrast relative to the Quiet Sun, of flux tubes is strongly modulated by their size and position on the solar disc \citep[see][for a review]{solanki93a}. Within these magnetic concentrations, pressure balance dictates an evacuation of the interior and consequent depression of the optical depth unity surface \citep{spruit76}. The horizontal extent influences the effect of radiative heating from the surroundings through the side walls on the temperature structure and contrast \citep{spruit81,grossmann94}. The position on the solar disc changes the viewing geometry, and therefore the degree to which the hot walls are visible and the apparent contrast \citep{steiner05}. Models describing the counteracting effects on solar irradiance of dark sunspots, and bright network and faculae, characterizing the latter by the magnetic filling factor (related to number density) and position have been successful in reproducing more than 90\% of observed variation over multiple solar cycles \citep{wenzler06,ball12}. Other factors, such as inclination, internal dynamics, phase of evolution and surrounding convective motions affect the brightness excess of a given flux tube, but become less important when considering the overall behaviour of an ensemble as is the case with such models \citep{fligge00}. The same is assumed of flux tube size, which enters these models only very indirectly, though it is known to have a significant effect on contrast.

Evidently, the robust reconstruction of solar irradiance variation from models based on photospheric magnetic activity is contingent, amongst other factors, on a firm understanding of the radiant behaviour of magnetic elements, in particular the variation with size and position on the solar disc. While the radiant behaviour of sunspots is relatively well known \citep{chapman94,mathew07} and sufficiently described by current models \citep{maltby86,collados94,unruh99}, the converse is true of network and faculae, and constitutes one of the main uncertainties in current solar irradiance reconstructions. This is primarily due to the difficulty in observing such small-scale features, the detailed structure of which are only starting to be resolved \citep{lites04,lagg10} with instruments such as the Swedish 1-m Solar Telescope, SST \citep{scharmer03} and the Imaging Magnetograph eXperiment, IMaX \citep{martinezpillet11} on-board SUNRISE \citep{solanki10,barthol11}. As such, the relationship between radiance and size cannot, as yet, be studied directly. It is however appropriate and more straightforward to consider instead the relationship between apparent intensity contrast and magnetogram signal. Apart from small-scale magnetic fields observed in the quiet Sun internetwork \citep{khomenko03,lites08,beck09}, magnetic concentrations carrying more than a minimum amount of flux exhibit similar field strengths regardless of size \citep{solanki93b,solanki99}. Flux tubes also tend towards vertical orientation due to magnetic buoyancy. As flux tubes exhibit a narrow range of magnetic field strengths and are largely vertical, on average the magnetogram signal at a given image pixel approximately scales with the proportion of the resolution element occupied by magnetic fields. Also, although the relationship between magnetogram signal and distribution of flux tube sizes is degenerate (a given magnetogram signal can, for example, correspond to either a single flux tube or a concentration of numerous smaller ones), flux tube size appears, on average, to be greater where the magnetogram signal is greater \citep{ortiz02}.

Relatively few studies examining network and faculae contrast variation with magnetogram signal and position on the solar disc have been reported in the literature. The majority of studies from the past two decades employed high-resolution ($<0.5\:{\rm arcsec}$) scans made with ground-based telescopes. For example, the efforts of \citet{topka92,topka97} and \citet{lawrence93} with the Swedish Vacuum Solar Telescope (SVST) and of \citet{berger07} with the SST. These studies suffer from variable seeing effects introduced by the Earth's atmosphere and poor representation of disc positions, a limitation imposed by the relatively narrow FOVs (field-of-view). \citet{ortiz02} and \citet{kobel11} repeated the work of \citet{topka92,topka97} and \citet{lawrence93} utilising observations from spaceborne instruments, and in so doing avoided seeing effects. \citet{ortiz02} employed full-disc continuum intensity images and longitudinal magnetograms from the Michelson Doppler Imager (MDI) on-board the Solar and Heliospheric Observatory (SoHO). While full-disc MDI observations presented a more complete coverage of disc positions, allowing the authors to derive an empirical relationship relating contrast to heliocentric angle and magnetogram signal, the spatial resolution is significantly poorer than in the SVST studies (4 arcsec versus $\gtrsim0.3$ arcsec). \citet{kobel11} examined the relationship between contrast and magnetogram signal near disc centre using spectropolarimetric scans from the Solar Optical Telescope (SOT) on-board Hinode \citep{kosugi07}. In this instance, the spatial resolution (0.3 arcsec) is comparable.

In this paper we discuss continuum and line core intensity contrast of network and facular elements from full-disc observations by the Helioseismic and Magnetic Imager (HMI) on-board the Solar Dynamics Observatory (SDO) spacecraft \citep{schou12}, and their relationship with heliocentric angle and magnetogram signal. The aim here is to derive stringent observational constraints on the relationship between intensity contrast, and position on the solar disc and magnetic field. This will be of utility to solar irradiance reconstructions, especially as HMI data will increasingly be used for this purpose.

This study partly echoes the similar studies discussed above, in particular that by \citet{ortiz02} utilising MDI observations. It presents a significant extension of the effort by \citet{ortiz02} in that we examined the entire solar disc not just in the continuum but also in the core of the HMI spectral line (Fe I 6173 \AA). This is of particular relevance to solar irradiance variation given the observation that spectral line changes appear to have a significant influence on such variations \citep{mitchell91,unruh99,preminger02}. Both here and in the study by \citet{ortiz02}, network and facular elements were distinguished from quiet Sun by the magnetogram signal, and sunspots and pores by the continuum intensity. HMI magnetograms are significantly less noisy than MDI magnetograms, allowing us to achieve a similar magnetogram signal threshold while averaging over a much shorter period (315 seconds versus 20 minutes). Network and facular features evolve at granular timescales \citep[$\sim10\:{\rm minutes,}$][]{berger96,wiegelmann13}. It is pertinent to keep the averaging period below this in order to avoid smearing and loss of signal. HMI also has a finer spatial resolution (1 arcsec compared to 4 arcsec), allowing weaker unresolved features to be detected at the same noise level than with MDI. The finer resolution however, also renders intensity fluctuations from small-scale phenomena such as granulation and filamentation more severe, which complicates the clear segmentation of sunspots and pores.

In Sect. \ref{data} we briefly present the HMI instrument, the observables considered and the data set. The data reduction process by which we identified and derived the intensity contrast of network and facular features is detailed in Sect. \ref{reduction}. Following that we describe the results of our analysis of these measured contrasts (Sect. \ref{results}). In Sect. \ref{discussion} we discuss our findings in the context of earlier studies and of their relevance to facular contribution to solar irradiance variation, before presenting our conclusions in Sect. \ref{conclusion}.

\section{Method}
\label{method}

\subsection{SDO/HMI data}
\label{data}

SDO/HMI \citep{schou12} is designed for the continuous, full-disc observation of velocity, magnetic field and intensity on the solar surface. The instrument comprises two $4096\times4096$ pixel CCD cameras observing the Sun at a spatial resolution of 1 arcsec (corresponding to two pixels).  By means of a tunable Lyot filter and two tunable Michelson interferometers, the instrument records 3.75-second cadence filtergrams at various polarizations and wavelengths across the Fe I absorption line at 6173 \AA{}. 45-second cadence Dopplergrams, longitudinal magnetograms and intensity (continuum, line depth and width) images are generated from the filtergram sequence. For this work we considered the longitudinal magnetic field, continuum intensity and line depth observables. HMI is full-Stokes capable, however, at time of study, only 720-second cadence Stokes IQUV parameters and Milne-Eddington inversions were available. As argued in Sect. \ref{introduction}, for this study it is important to keep the integration period of measurements below $\sim10\:{\rm minutes}$. We opted to utilise the 45-second longitudinal magnetograms also to keep in line with earlier studies, which examined intensity contrast variation with line-of-sight magnetic field \cite[e.g.][]{topka92,topka97,lawrence93,ortiz02,kobel11}. More details on the instrument can be found in \citet{schou12}.

The data set comprises simultaneous longitudinal magnetograms, continuum intensity and line depth images from 15 high activity days in the period May 2010 to July 2011. From each day, for each observable we took the average of the seven 45-second cadence images from a 315-second period, each rotated to the observation time of the middle image to co-register. Aside from signal-to-noise considerations, this averaging is to suppress variance from p-mode oscillations. The dates and times of the employed observations are listed in Table \ref{obs_date_time}.

\begin{table}
\caption{Observation date and time of the data set.}
\label{obs_date_time}
\centering
\begin{tabular}{cc}
\hline\hline
Date & Time \\
(year.month.date) & (hour.minute.second) \\
\hline
2010.05.04 & 00:03:00 \\
2010.06.11 & 00:00:00 \\
2010.07.24 & 00:05:15 \\
2010.08.11 & 00:00:00 \\
2010.09.02 & 00:00:00 \\
2010.10.25 & 00:00:00 \\
2010.11.13 & 00:00:00 \\
2010.12.04 & 00:00:00 \\
2011.01.01 & 00:00:00 \\
2011.02.14 & 00:00:00 \\
2011.03.08 & 00:00:00 \\
2011.04.15 & 00:00:00 \\
2011.05.30 & 00:00:00 \\
2011.06.02 & 00:00:00 \\
2011.07.18 & 00:12:00 \\
\hline
\multicolumn{2}{p{.5\textwidth}}{\textbf{Notes.} The time listed is the nominal time, in International Atomic Time (TAI), of the middle cadence in the sequence of seven considered from each day.}\\
\end{tabular}
\end{table}

Longitudinal magnetograms describe the line-of-sight component of the average magnetic flux density over each resolution element. To first order, for a given flux tube of intrinsic magnetic field strength $B$, the unsigned longitudinal magnetogram signal, $\bl$, is $\alpha{}B|\cos\gamma|$, scaled by $\alpha$, the magnetic filling factor and $\gamma$, the inclination of the magnetic field from the line-of-sight. As mentioned in the introduction (Sect. \ref{introduction}), flux tubes tend towards vertical orientation. In this study we examined the overall properties of an ensemble of magnetic elements. Under this condition it is reasonable to assume that on average $\gamma\approx\theta$, the heliocentric angle, allowing us to employ the magnetogram signal as a proxy for $\alpha{}B$ via the quantity $\bmu$, where $\mu=\cos\theta$. This quantity also represents a first-order approximation of the unsigned average magnetic flux density over each resolution element. At time of writing, HMI Dopplergrams and longitudinal magnetograms were generated from just the first Fourier component of the filtergram sequence resulting in a $\sqrt{2}$ factor increase in photon noise from the optimal level \citep{couvidat11,couvidat12}.

Each line depth image was subtracted from the corresponding continuum intensity image to yield the line core intensity image. Hereafter we will denote the continuum and line core intensity $\icn$ and $\ilc$ respectively. At time of writing, HMI data products are generated from the filtergram sequence assuming a Gaussian form to the Fe I 6173 \AA{} line and delta filter transmission profiles. The effect of these approximations on Doppler velocity and longitudinal magnetic field measurements are accounted for, but not completely for the intensity observables \citep{couvidat11,couvidat12}. The impact on this study is assumed minimal as we are only interested in contrast relative to the local mean quiet Sun level.

\subsection{Data reduction}
\label{reduction}
\subsubsection{Magnetogram noise level}
\label{magnetogramnoiselevel}

The noise level of 315-second HMI magnetograms as a function of position on the solar disc was determined. For this purpose we used 10 spot-free 315-second magnetograms recorded over a seven month period in 2010. First, we estimated the centre-to-limb variation, CLV of the noise level. The pixels within each magnetogram were ordered by distance from the disc centre and sampled in successive blocks of 5000 pixels. The blocks represent concentric rings (near the limb, arcs, as the circumference of the solar disc is greater than 5000 pixels) of pixels of similar distance from the disc centre. Within each block we computed the average distance from the disc centre and the standard deviation of the magnetogram signal (iteratively, with points outside three standard deviations from the mean excluded from the succeeding iteration till convergence). A fifth-order polynomial in $\mu$ was fitted to the noise CLV; the mean of the standard deviation versus distance profiles so derived from the magnetograms. The magnetograms were then normalized by the noise CLV fit. At each disc position, the standard deviation over a $401\times401$ pixel window centred on the pixel of interest was computed (iteratively as above) for each normalized magnetogram and the median taken \citep[following][]{ortiz02}. (Near the limb, the standard deviation was computed from just the image pixels that lie within the solar disc.) A sixth-order polynomial was fitted to the resultant surface (termed here the noise residue). This fit represents the noise level after the removal of the CLV as a function of position on the solar disc. The noise level, $\sigb$, shown in Fig. \ref{noise_surface}, is then the product of the noise CLV fit and the noise residue fit. The noise level is lowest near disc centre and increases radially up to the limb (mean of $4.9\g$ for $\mu>0.95$ and $8.6\g$ for $\mu<0.05$). The root-mean-square, RMS difference between the noise level and the noise CLV fit is $0.4\g$. The correlation with distance from disc centre and relatively small deviation from circular symmetry suggests photon noise is the dominant component. The noise level of 45-second magnetograms, derived by a like analysis, has a similar, albeit accentuated form. The ratio between the noise level, averaged over the solar disc, of 45-second and 315-second magnetograms is 2.7 (approximately $\sqrt{7}$). The noise level of 45-second magnetograms determined by \citet{liu12}, via a vastly different method, also exhibits a similar CLV.

\begin{cfig}
\includegraphics[width=\textwidth]{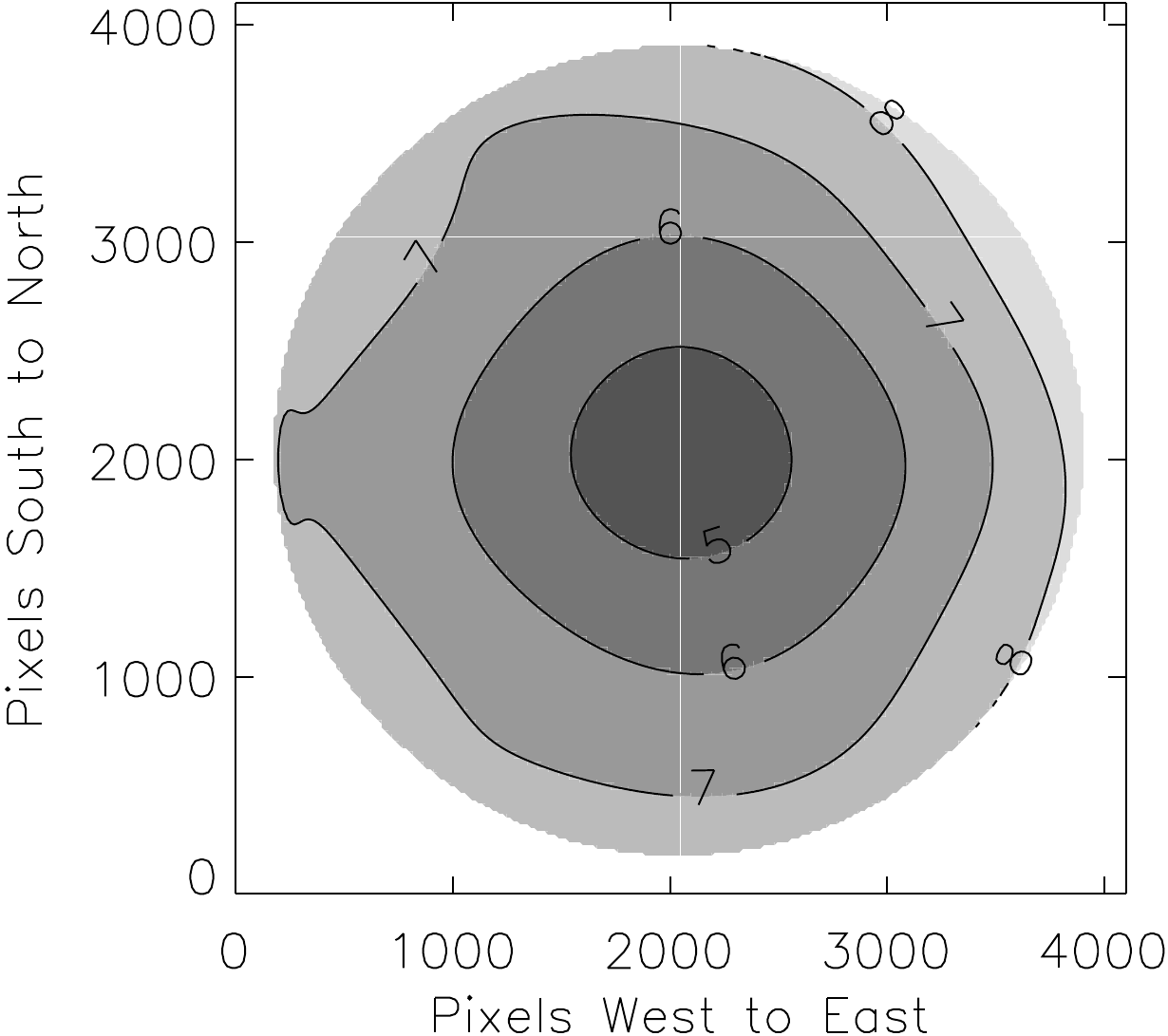}
\caption{Magnetogram noise level in units of Gauss, as a function of disc position, sampled at 16-pixel intervals in either direction on the CCD.}
\label{noise_surface}
\end{cfig}

\subsubsection{Quiet Sun intensity}
\label{qsintensity}

For this part of the data reduction process, where we examined how quiet Sun intensity and the noise level of the intensity images vary with position on the solar disc, we counted all pixels with $\bmuqs$ as corresponding to quiet Sun (QS).

The continuum and line core intensity images were normalized by the fifth-order polynomial in $\mu$ fit to the quiet Sun pixels to correct for limb-darkening \citep[following][]{neckel94}. In the case of the line core intensity images, this was also to correct for the centre-to-limb broadening of the Fe I 6173 \AA{} line \citep{norton06}. There are distortions in the intensity images such that after this normalization, the mean quiet Sun intensity is not constant at unity but varying with position on the solar disc. The mean quiet Sun intensity at the continuum and line core as a function of position on the solar disc, denoted $\mqsicn$ and $\mqsilc$ respectively, were determined for each of the selected days. A $401\times401$ pixel window was centred on each disc position and the mean continuum and line core intensity of all quiet Sun pixels inside the window computed. For each day represented in the data set, $\mqsicn$ and $\mqsilc$ were given by the fifth-order polynomial fits to the mean quiet Sun continuum and line core intensity surfaces so derived from the images from the day. This analysis had to be repeated for each selected day as we found the spatial distribution of $\mqsicn$ and $\mqsilc$ to vary significantly over the period of observation. The RMS value of $\mqsicn-1$ and $\mqsilc-1$, the scale of the image distortions, is on average 0.004 and 0.01 respectively.

The CLV of the standard deviation of quiet Sun intensity at continuum, $\sicnqs$ and line core, $\silcqs$ were determined, from $\icn/\mqsicn$ and $\ilc/\mqsilc$, by an analysis similar to the procedure used to derive the magnetogram noise CLV. In Fig. \ref{int_ld_noise_clv}, we express $\sicnqs$ and $\silcqs$, which carry information on the noise level of the intensity images and granulation contrast, as a function of $\mu$. Going from disc centre, $\sicnqs$ decreases gradually down to $\mu\sim0.15$ before increasing rapidly towards the limb. The monotonic decline from disc centre to $\mu\sim0.15$ resembles a similar trend in the CLV of granulation contrast reported by various authors \citep[][and references therein]{sanchezcuberes00,sanchezcuberes03}. $\silcqs$ exhibits a similar, though less accentuated, trend. For both $\sicnqs$ and $\silcqs$, the $\mu$-dependence from disc centre to $\mu\sim0.3$ is approximately linear, as highlighted by the linear fits (dotted lines). The elevation near limb is a direct consequence of limb-darkening; the diminishing signal-to-noise ratio translates into an escalating noise level in the normalized intensity. Given the gross scatter towards the limb, we excluded pixels outside $\mu=0.1$, representing about 1\% of the solar disc by area, from further consideration.

\begin{cfig}
\includegraphics[width=\textwidth]{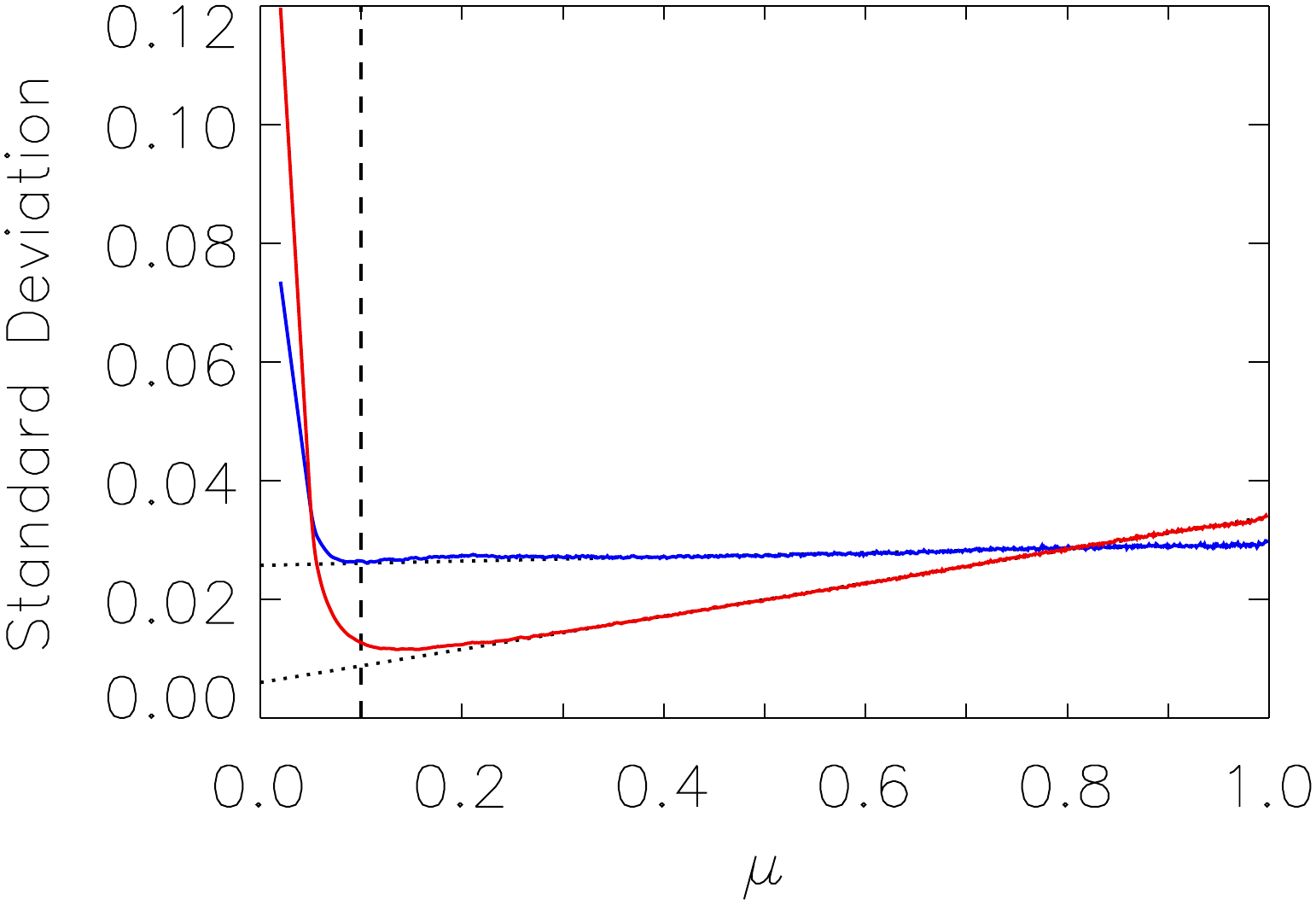}
\caption{Standard deviation of quiet Sun ($\bmuqs$) intensity at continuum, $\sicnqs$ (red) and line core, $\silcqs$ (blue) as a function of $\mu$. The dotted lines represents the linear fit to $\sicnqs$ and $\silcqs$ over the range $0.3\leq\mu\leq1.0$ (largely hidden due to the close agreement), extrapolated to $\mu=0$. The dashed line denotes the threshold ($\mu=0.1$) below which pixels were excluded from the rest of the study in view of the scatter in measured intensity.}
\label{int_ld_noise_clv}
\end{cfig}

\subsubsection{Identification of network and faculae}
\label{identificationofnetworkandfaculae}

Network and facular features, the subject of this work, were identified by first distinguishing them from quiet Sun by the magnetogram signal and from sunspots and pores by the continuum intensity. Pixels with $\bmu>3\sigb/\mu$ ($\sim14\g$ near disc centre, where $\sigb$ is lowest) were considered to harbour substantive magnetic fields. Isolated pixels above this threshold were assumed false positives and excluded. Hereafter we will denote $3\sigb/\mu$ as $\bmut$. Pixels with $\icn<0.89$ were counted as sunspots and pores. This value of the threshold is given by the mean of the minimum value of $\mqsicn-3\sicnqs$ from each selected day. It was so defined to distinguish sunspots and pores with minimal wrongful inclusion of intergranular lanes and magnetic features darker than the quiet Sun in the continuum. The continuum intensity versus $\mu$ scatter plot of network and facular pixels for one of the selected days (June 2, 2011) is shown in Fig. \ref{fac_cnp_spt_int} (top panel). The pixels identified as network and faculae lie clearly above the continuum intensity threshold.

\begin{cfig}
\includegraphics[width=\textwidth]{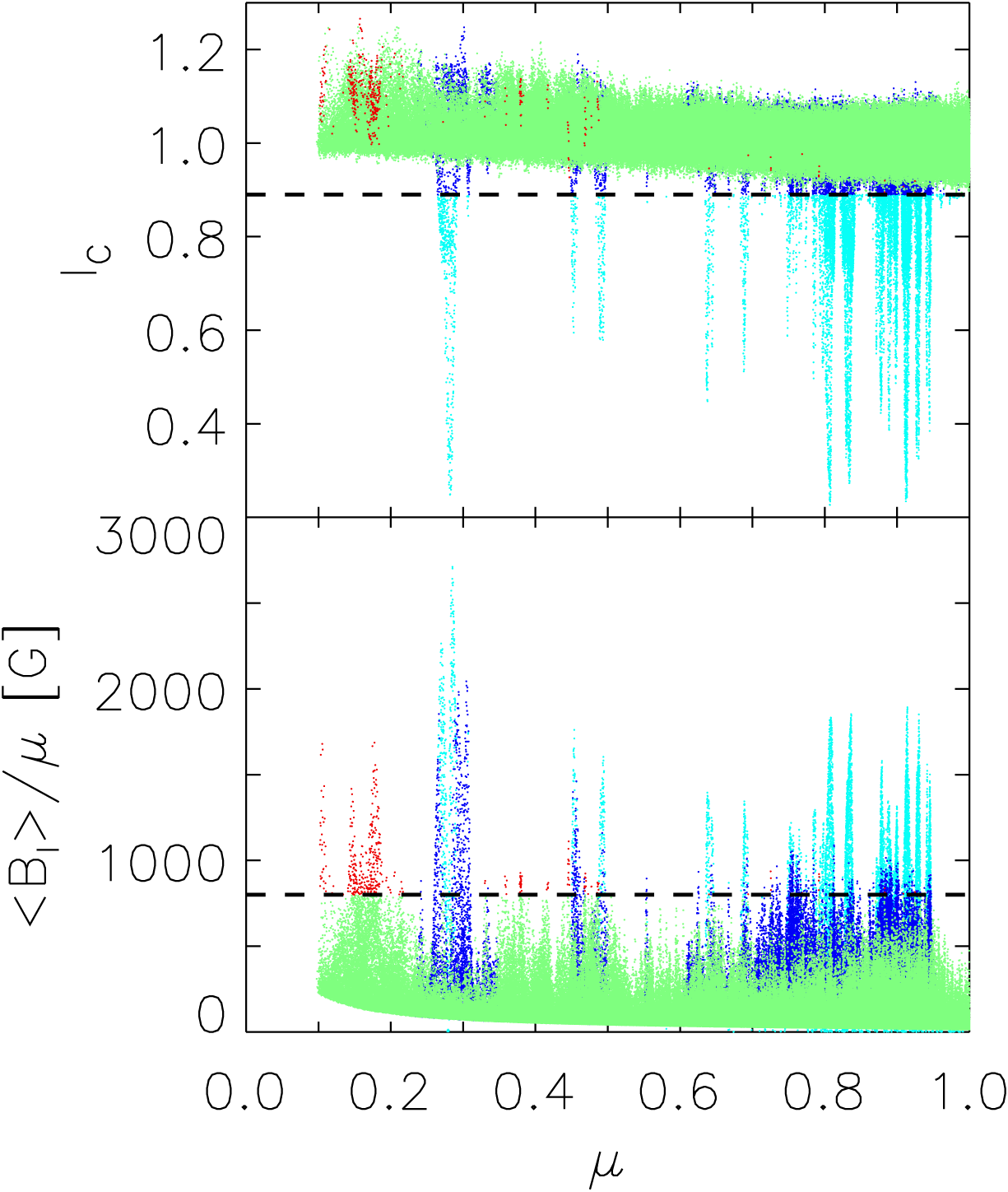}
\caption{Continuum intensity, $\icn$ versus $\mu$ (top) and $\bmu$ versus $\mu$ (bottom) scatter plots of pixels counted as network and faculae (green) from June 2, 2011. The pixels counted as sunspots and pores by the continuum intensity threshold and the magnetic extension removal procedure (see text) are represented by the cyan and blue dots respectively. The red dots represent network and faculae pixels that lie above the cutoff $\bmu$ level. The dashed lines denote the continuum intensity threshold (top) and the cutoff $\bmu$ level (bottom).}
\label{fac_cnp_spt_int}
\end{cfig}

In HMI magnetograms, magnetic signals produced by sunspots and pores extend beyond their boundary (in our analysis, the $\icn=0.89$ locus). This is illustrated for a sunspot near disc centre and another that is close to the limb on one of the selected days (July 18, 2011) in Fig. \ref{canopy}. The $\icn=0.89$ locus is plotted over the continuum intensity image and magnetogram of both sunspots (red contours) to highlight the extension of the magnetogram signal. This arises predominantly from the lateral expansion of their magnetic field with height (i.e., magnetic canopy) and partly also from the effect of straylight from sunspots and pores on nearby pixels. Towards the limb, these magnetic features become more extensive and bipolar due to the acute orientation of the largely horizontal magnetic canopy with the line-of-sight \citep{giovanelli80}, as illustrated by the near limb example (bottom panels). Figure \ref{canopy} also highlights the presence of bright sunspot structures (such as bright penumbral filaments) that lie above the continuum intensity threshold. All of these could easily be misidentified as network and faculae by the simple application of the magnetogram signal and continuum intensity thresholds described earlier.

\begin{cfig}
\includegraphics[width=.81\textwidth]{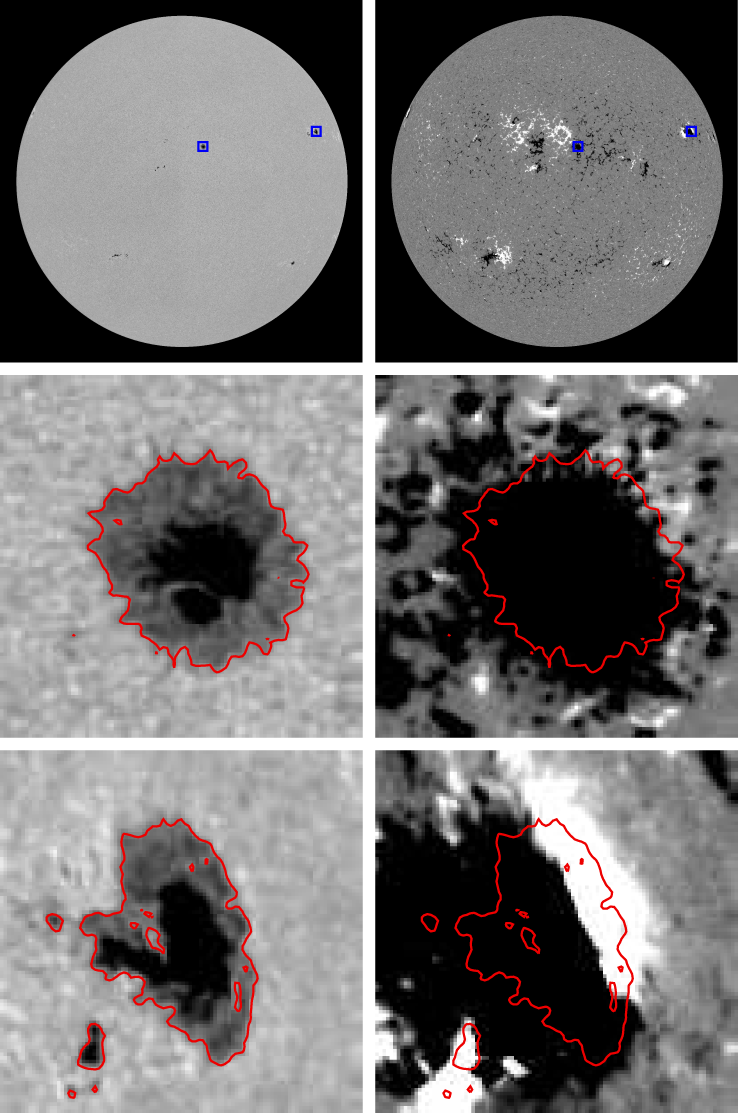}
\caption{Continuum intensity image (top left) and magnetogram (top right) from July 18, 2011 and the corresponding $50\times50\:{\rm arcsec}$ insets of the boxed sunspot features near disc centre (middle panels) and limb (bottom panels). The red contours represent the continuum intensity sunspot boundary (the $\icn=0.89$ locus). The continuum intensity image is saturated at 0.6 and 1.2, and the magnetogram at $\pm80\g$ (to highlight the extension of the magnetic signal from these sunspots beyond the intensity boundary).}
\label{canopy}
\end{cfig}

To account for the effects of straylight around sunspots and pores, magnetic canopies and bright sunspot structures, we expanded sunspots and pores to include adjoining magnetic signal. Pixels adjacent to sunspots and pores that lie above $\bmut$ are reassigned to these features. This was iterated till no more pixels could be added. Here we will refer to this process as magnetic extension removal. Simply adding only adjoining pixels within a threshold distance from sunspots and pores instead is not useful as the physical extent of magnetic canopies exhibits a broad dynamic range, dependent on the position and physical properties of the associated feature. \citet{kobel11}, in a similar study with Hinode/SOT scans, expanded pores to include adjoining pixels above a threshold magnetogram signal level of $200\g$ to account for their influence on surrounding pixels from telescope diffraction. While this technique appears to work for Hinode/SOT scans, here we observed that adding only adjoining pixels above a threshold magnetogram signal level results in the appearance of a knee in contrast versus $\bmu$ plots at this threshold level regardless of the value chosen. For these reasons we expanded sunspots and pores to include all adjoining magnetic signal, though this conservative approach inevitably assigns too many pixels, including legitimate faculae, to sunspots and pores.

All pixels identified as magnetic, and not as sunspots and pores or their extensions, were taken to correspond to network and faculae. Figure \ref{fac_cnp_spt_int} shows the continuum intensity versus $\mu$ (top panel) and $\bmu$ versus $\mu$ (bottom panel) scatter plots of pixels identified as network and faculae (green), and counted as sunspots and pores by the magnetic extension removal procedure (blue) from June 2, 2011. The pixels captured by the magnetic extension removal procedure are not well distinguished from network and faculae by the continuum intensity; largely hidden by network and faculae in the continuum intensity versus $\mu$ scatter plot. It is clear from the $\bmu$ versus $\mu$ scatter plots however that the two classes are significantly different magnetic populations. As noted earlier, this procedure is likely too severe and results in the exclusion of some true faculae. However, for the purpose of this study it is not necessary to identify all faculae present and far more important to avoid false positives.

Finally, network and facular pixels with $\bmu$ above a conservative cutoff level of $800\g$ were excluded from the subsequent analysis \citep[following][]{ball11,ball12}. They are mostly bright features concentrated near the limb (as illustrated for June 2, 2011 by the red dots in Fig. \ref{fac_cnp_spt_int}) associated with sunspots and pores. (The relatively high $\bmu$ values likely reflect nearly horizontal fields, for which $|\cos\gamma|\gg\mu$ towards the limb.) This is to account for non-facular magnetic signals that might have been missed by the continuum intensity threshold and the magnetic extension removal procedure.

The classification image, indicating the positions of the pixels classed as sunspot and pores, and network and faculae for another of the selected days (May 30, 2011) is shown in Fig. \ref{network_plage_mask}. In spite of the severe measures taken to minimise the influence of sunspots and pores, a fair fraction of active region faculae remains. In total, $7.6\times10^6$ pixels were identified as corresponding to network and facular features from the data set (i.e. $4.5\%$ of all solar disc pixels in the 15 continuum intensity image and magnetogram pairs examined).

\begin{cfig}
\includegraphics[width=\textwidth]{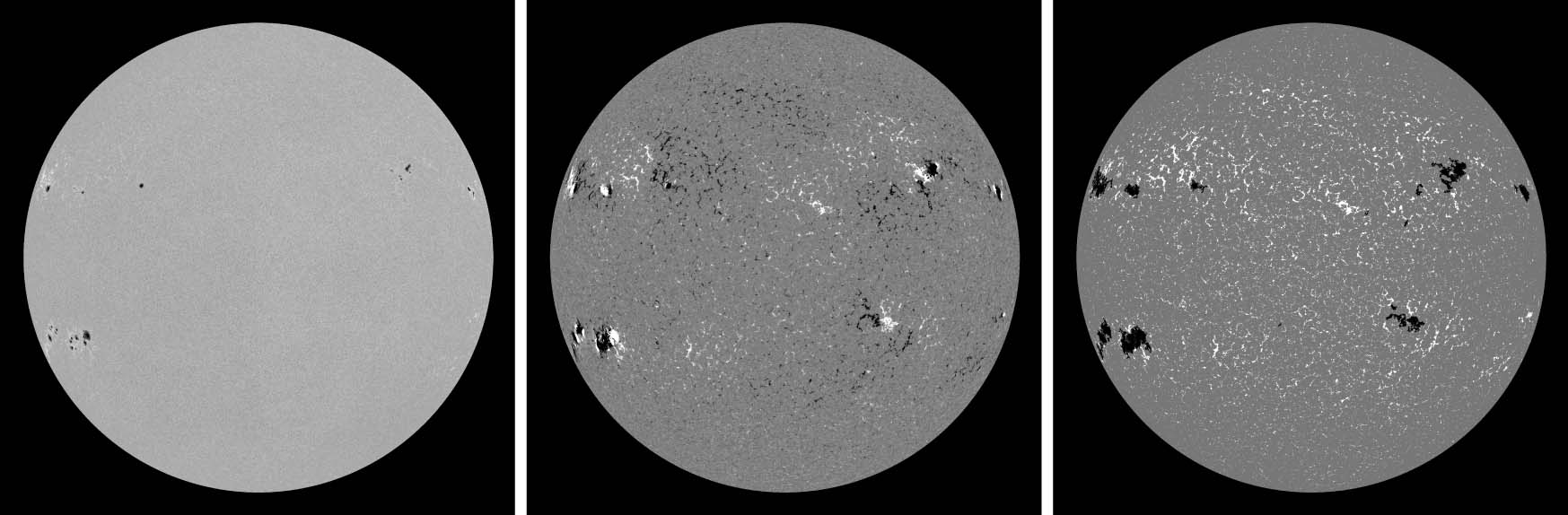}
\caption{HMI continuum intensity image (left), magnetogram (middle) and classification image (right) from May 30, 2011. The classification image indicates the positions of the pixels classed as network and faculae (white), and sunspots and pores (black). The latter includes magnetic signal adjoined to sunspots and pores, counted to them to avoid counting their magnetic canopy, possible bright structures within penumbrae and straylight as network and faculae erroneously. The continuum intensity image and magnetogram are scaled between 0.6 and 1.2, and $-80\g$ and $80\g$ respectively as in Fig. \ref{canopy}}
\label{network_plage_mask}
\end{cfig}

\subsubsection{Definition of intensity contrast}
\label{definitionofintensitycontrast}

The average continuum and line core intensity contrast over a given resolution element, $\cicn$ and $\cilc$ were defined as:
\begin{equation}
\cicn\left(x,y\right)=\frac{\icn\left(x,y\right)-\mqsicn(x,y)}{\mqsicn\left(x,y\right)}
\label{intensity_contrast}
\end{equation}
and
\begin{equation}
\cilc\left(x,y\right)=\frac{\ilc\left(x,y\right)-\mqsilc(x,y)}{\mqsilc\left(x,y\right)}
\label{linecore_contrast}
\end{equation}
respectively, where $\left(x,y\right)$ denote position on the CCD array. These two quantities, computed for each of the pixels identified as corresponding to network and facular features, represent the normalized difference between the continuum and line core intensities at a given pixel and the local mean quiet Sun levels as given by the mean quiet Sun continuum and line core intensity surfaces, $\mqsicn$ and $\mqsilc$, defined earlier in Sect. \ref{qsintensity}.

In summary, here we extracted an ensemble of $7.6\times10^6$ continuum and line core intensity contrast measurements corresponding to network and facular features covering as wide a range of heliocentric angles ($0.1\leq\mu\leq1.0$) and magnetogram signal ($\bmut<\bmu\leq800\g$) as reasonably possible from the data set for the succeeding analysis.

\section{Results}
\label{results}

\subsection{Variation with position and magnetogram signal}
\label{profile}

The positions of pixels identified as network and faculae, classified by $\bmu$, in a quiet region and an active region on one of the selected days (April 15, 2011) is shown in Fig \ref{qs_ar_fac_map}. As expected, at HMI's spatial resolution, magnetic signals with higher $\bmu$ are largely concentrated in active regions. Though magnetic signals with lower $\bmu$ are present in both quiet Sun and active regions, the fact that the solar disc is predominantly quiet Sun means these signals correspond largely to quiet Sun network and internetwork.

To elucidate the CLV of intensity contrast, we grouped the measured contrasts into eight intervals of $\bmu$ spanning the range $\bmut<\bmu\leq800\g$ and within each interval into $\mu$ bins 0.05 wide. As the distribution of magnetogram signal is skewed towards the lower bound \citep{wenzler04,parnell09}, the $\bmu$ intervals were defined such that the widths slide from about $36\g$ ($\bmua$) to $160\g$ ($\bmuh$) to ensure reasonable statistics in every interval. In grouping the measured contrasts into these broad $\bmu$ intervals we are effectively grouping the network and facular features by $\alpha{}B$ (Sect. \ref{data}), neglecting differences in quiet Sun network and active region faculae contrast \citep{lawrence93,kobel11}. This is reasonable since the lower intervals are mainly populated by quiet Sun network and the higher intervals by active region faculae. The bin-averaged continuum and line core intensity contrast as a function of $\mu$ and the cubic polynomial fit for each of the $\bmu$ intervals are expressed in Figs. \ref{intensity_contrast_scatter_bmu} and \ref{linecore_contrast_scatter_bmu} respectively. For brevity we will refer to these bin-averaged contrasts as the contrast CLV profiles.

\begin{cfig}
\includegraphics[width=\textwidth]{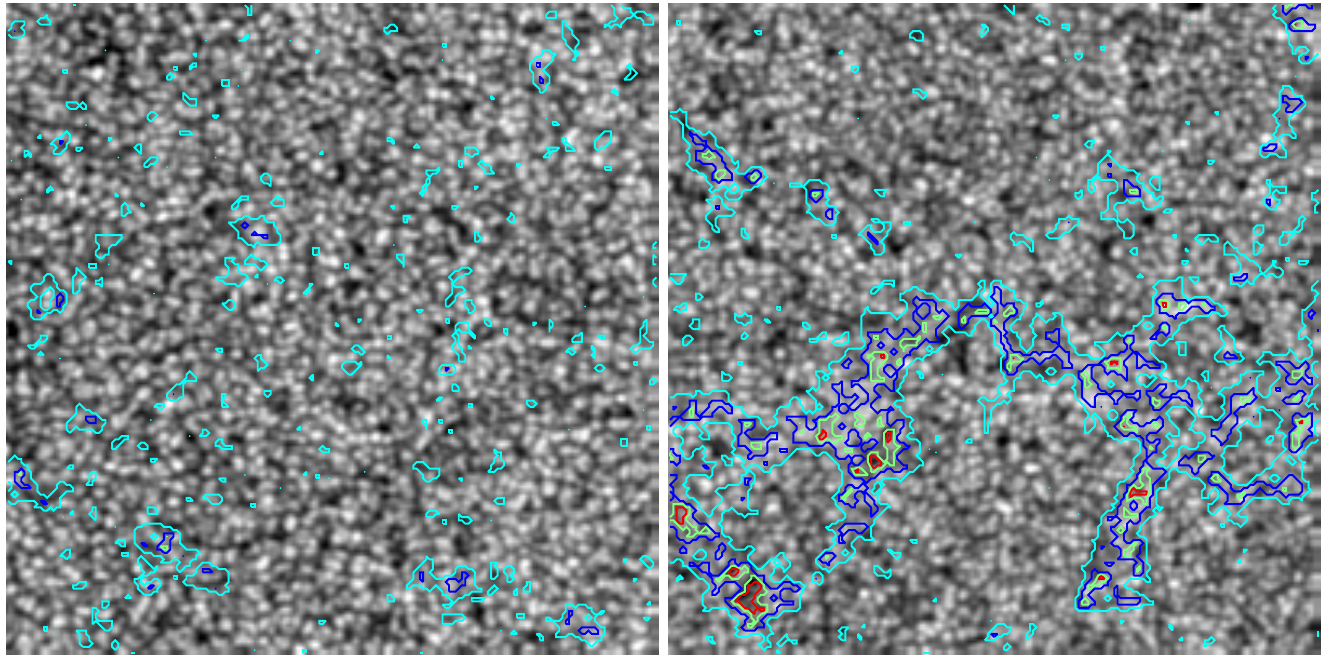}
\caption{$100\times100$ arcsec insets of a quiet region near disc centre ($\mu>0.99$, left) and active region NOAA 11187 ($0.82<\mu<0.91$, right) of the continuum intensity image from April 15, 2011. The cyan contours indicate the boundary of network and faculae. The blue, green and red contours correspond to $\bmu=100\g$, $280\g$ and $500\g$, respectively.}
\label{qs_ar_fac_map}
\end{cfig}

\begin{cfig}
\includegraphics[width=.98\textwidth]{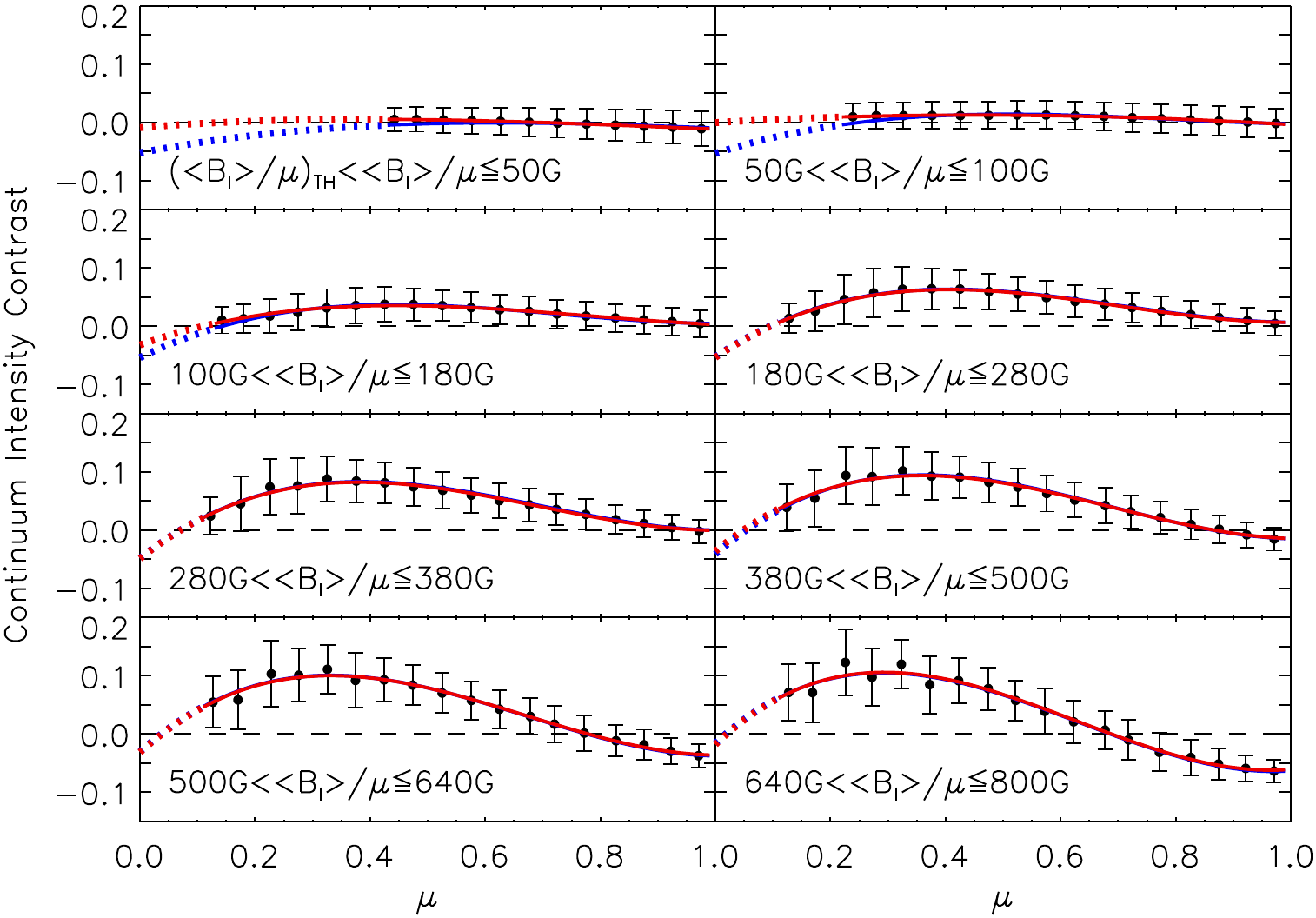}
\caption{CLV of continuum intensity contrast, $\cicn$ over eight $\bmu$ intervals. The filled circles and error bars represent the mean and standard deviation of $\cicn$ grouped by $\mu$ in bins 0.05 wide. The red curves are third-order polynomial fits to the filled circles and the blue curves are the cross-sections of the surface fit to $\cicn$ at the mean $\bmu$ within each interval (largely hidden due to the close agreement), extrapolated to $\mu=0$ (dotted segments). The horizontal dashed lines denote the mean quiet Sun level.}
\label{intensity_contrast_scatter_bmu}
\end{cfig}

\begin{cfig}
\includegraphics[width=.98\textwidth]{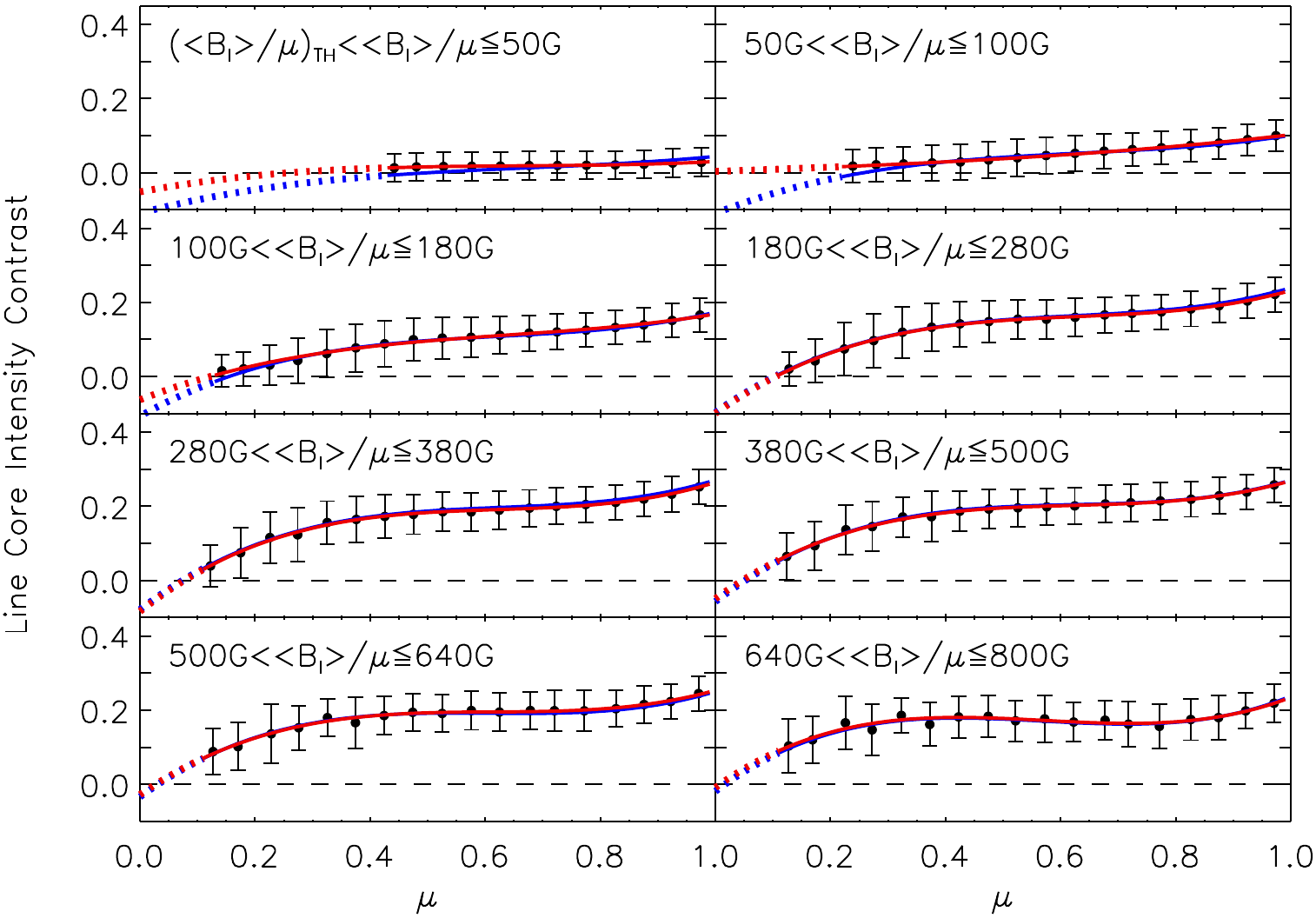}
\caption{Similar to Fig. \ref{intensity_contrast_scatter_bmu}, but for line core intensity contrast.}
\label{linecore_contrast_scatter_bmu}
\end{cfig}

The truncated $\mu$ coverage in the lower $\bmu$ intervals is due to foreshortening. As flux tubes are mainly vertical, going from disc centre to the limb the corresponding longitudinal magnetogram signal diminishes, eventually dropping below the threshold level which itself rises towards the limb (Fig. \ref{noise_surface}). The fluctuations near the limb, more pronounced in the higher $\bmu$ intervals, is due to the inhomogeneous distribution of active regions on the solar disc, where stronger magnetic signals are concentrated on the selected days. Diminishing statistics also play a role; there are comparatively fewer pixels in the higher $\bmu$ intervals.

To investigate the $\bmu$-dependence of intensity contrast, the measured contrasts were grouped into eight $\mu$ intervals spanning the range $0.1\leq\mu\leq1.0$ and within each interval into $\bmu$ bins $40\g$ wide. The $\mu$ intervals were defined such that they represent an approximately equal proportion of the solar disc by area. The bin-averaged continuum and line core intensity contrasts as a function of $\bmu$ and the cubic polynomial fit for each of the $\mu$ intervals are shown in Figs. \ref{intensity_contrast_scatter_mu} and \ref{linecore_contrast_scatter_mu} respectively. For brevity we will refer to these bin-averaged contrasts as the contrast versus $\bmu$ profiles. These profiles represent the variation of continuum and line core intensity contrast ranging from internetwork and weak network to active region faculae at different distances from disc centre.

\begin{cfig}
\includegraphics[width=.98\textwidth]{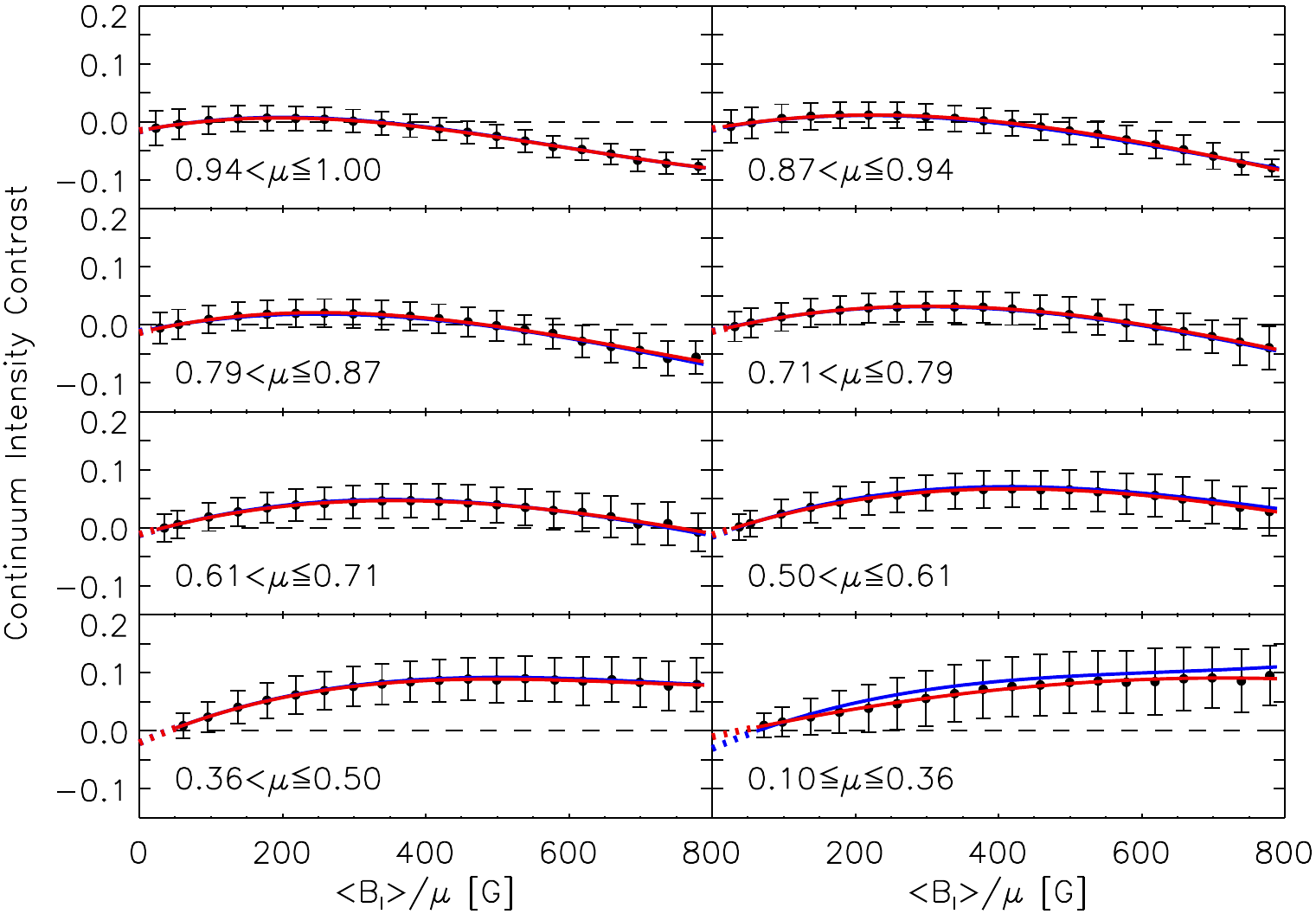}
\caption{Continuum intensity contrast, $\cicn$ as a function of $\bmu$ over eight $\mu$ intervals. The filled circles and error bars represent the mean and standard deviation of $\cicn$ grouped by $\bmu$ in $40\g$ bins. The red and blue curves and the horizontal dashed lines have the same meanings as in Fig. \ref{intensity_contrast_scatter_bmu}.}
\label{intensity_contrast_scatter_mu}
\end{cfig}

\begin{cfig}
\includegraphics[width=.98\textwidth]{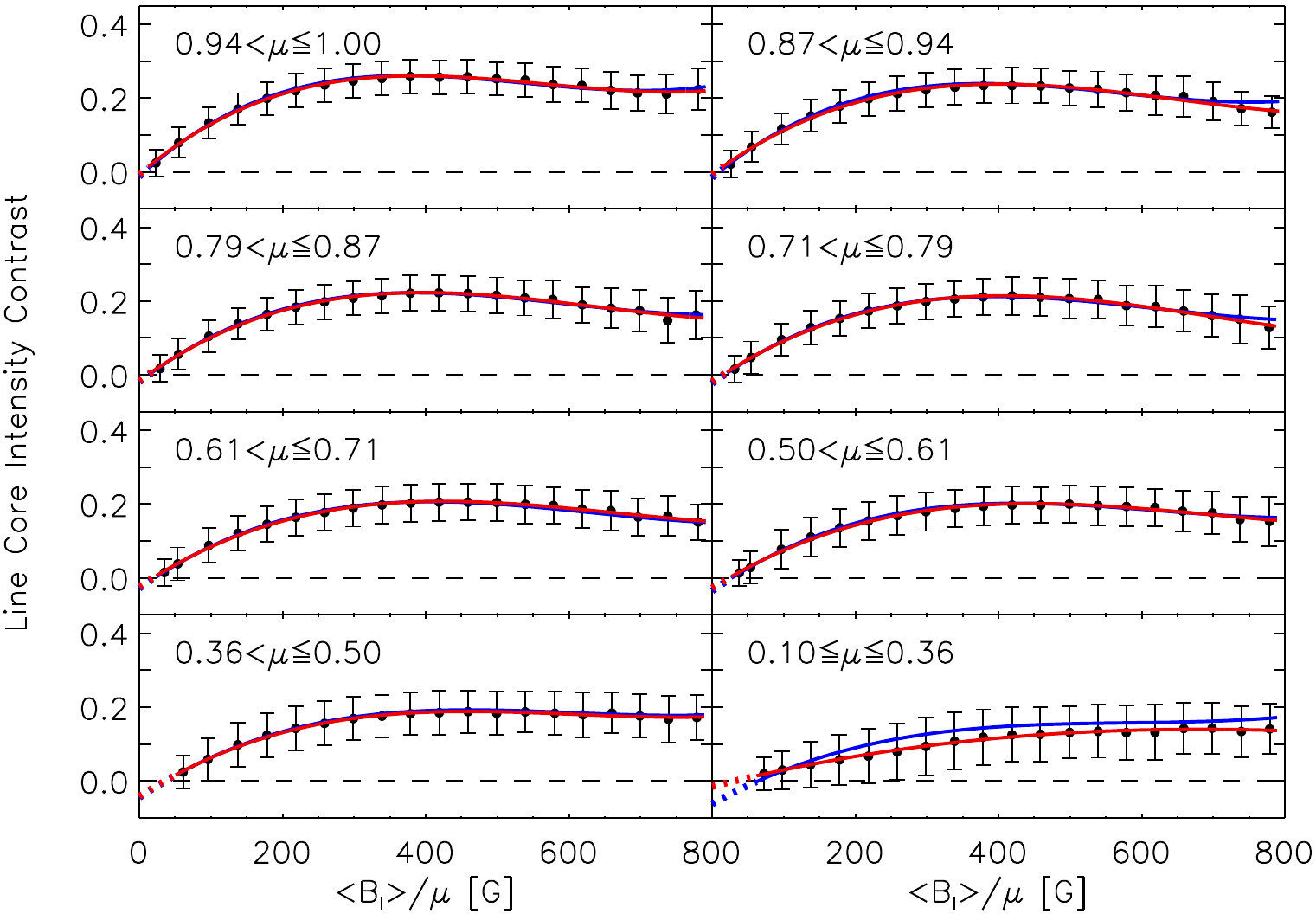}
\caption{Similar to Fig. \ref{intensity_contrast_scatter_mu}, but for line core intensity contrast.}
\label{linecore_contrast_scatter_mu}
\end{cfig}

The cubic polynomial fit to each contrast versus $\bmu$ profile included the zeroth-degree term. This produced better fits to data at low $\bmu$ than constraining the fits to pass through the origin by excluding it. Approaching $\bmuz$, the contrast versus $\bmu$ profiles decline to weak, broadly negative levels. Although the cubic polynomial fits express good agreement with measurements, they are too simple to accommodate this decline well without including the unphysical zeroth-degree term. For similar considerations we included the zeroth-degree term in the cubic polynomial fits to the contrast CLV profiles, so rendering them non-zero at $\mu=0$. These apparent offsets in the contrast CLV and contrast versus $\bmu$ profiles probably reflects the fact that magnetic elements are generally located in dark intergranular lanes.

In Fig. \ref{fishhook} we show a recomputation of the continuum intensity contrast versus $\bmu$ profile over $\mua$ from Fig. \ref{intensity_contrast_scatter_mu} where we included pixels below the magnetogram signal threshold and not identified as sunspots and pores (i.e., quiet Sun), and grouped the measurements by $\bmu$ in bins $10\g$ (instead of $40\g$) wide. Approaching $\bmuz$, contrast declines gradually to below the reference level before turning back up sharply towards the origin. Similar trends were reported by \citet{narayan10}, \citet{kobel11} and \citet{schnerr11}, who termed it the fishhook feature, based on SST and Hinode/SOT disc centre scans. \citet{schnerr11} demonstrated the resolution of granules and dark intergranular lanes, where magnetic flux concentrates, at the relatively fine spatial resolution of both instruments (0.15 and 0.3 arcsec respectively) to be the cause of the fishhook feature at low magnetogram signal levels. Though HMI has a coarser resolution (1 arcsec) than either Hinode/SOT or SST, the fishhook feature near $\bmuz$ in Fig. \ref{fishhook} indicates granulation is still sufficiently resolved to have a measurable impact on apparent contrast.

\begin{cfig}
\includegraphics[width=\textwidth]{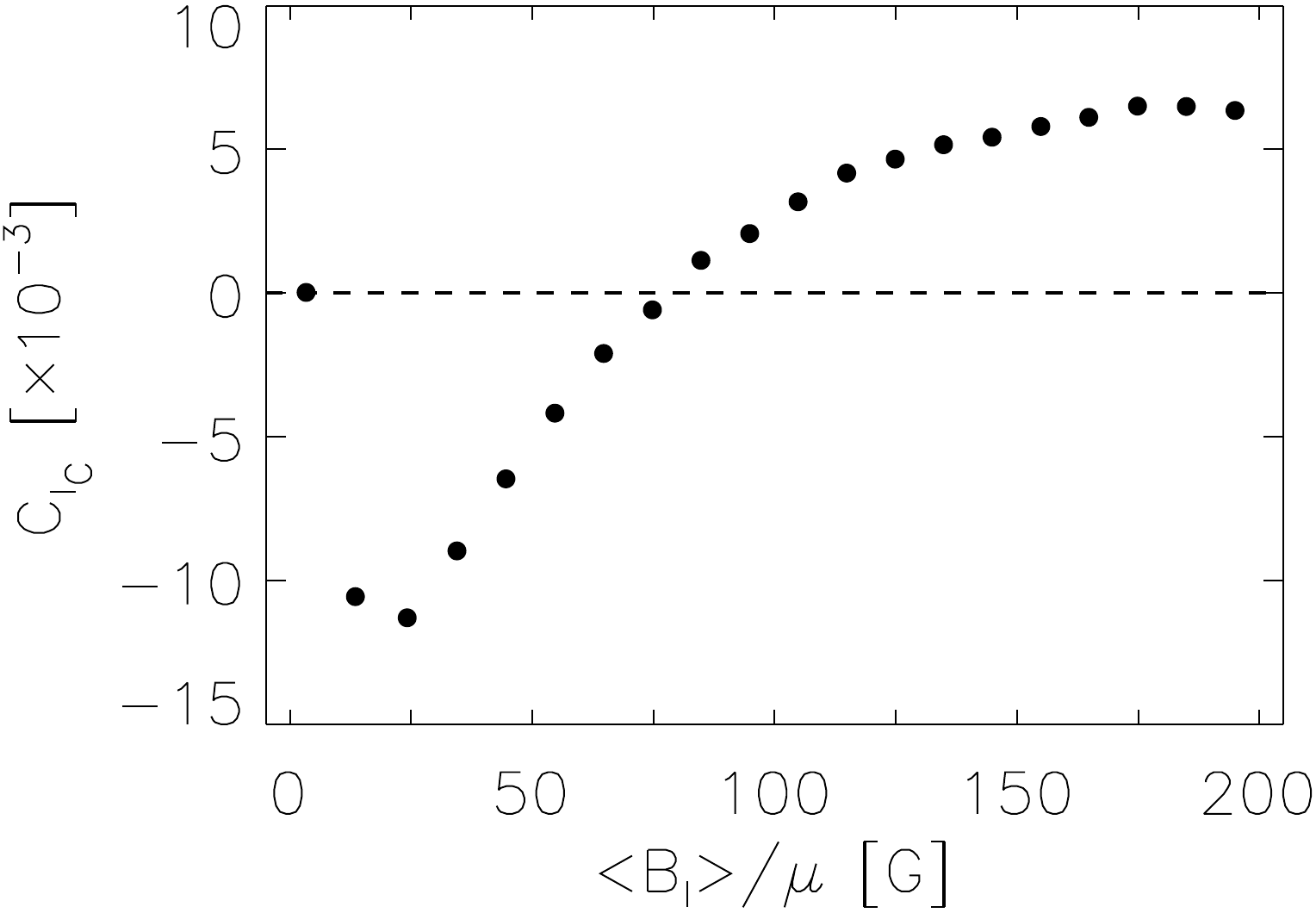}
\caption{Continuum intensity contrast, $\cicn$ of quiet Sun, network and faculae over $\mua$ as a function of $\bmu$. The filled circles represent the mean of $\cicn$ grouped by $\bmu$ in bins $10\g$ wide. The dashed line denotes the mean quiet Sun level.}
\label{fishhook}
\end{cfig}

Comparing the measured intensity contrast at continuum and line core derived here, the most distinct difference is the opposite CLV. Confining this discussion to broad trends in the measurements, continuum intensity contrast is weakest near disc centre and increases up to a maximum before declining quite significantly towards the limb (Fig. \ref{intensity_contrast_scatter_bmu}). Conversely, line core intensity contrast is strongest near disc centre and declines monotonically from disc centre to limb (Fig. \ref{linecore_contrast_scatter_bmu}). Line core intensity is modulated by line strength and shape, and continuum intensity. The centre-to-limb decline exhibited by the measured line core intensity contrast arises from variation in line strength and shape (if this were absent the continuum and line core intensity contrast would vary proportionally) and would be even more acute than reflected in the CLV profiles if not partially offset by the accompanying increase in continuum intensity. This will be demonstrated, along with a closer discussion of the diverging trends exhibited by both sets of measurements in Sect \ref{compare_lc}.

\subsection{Surface fits}
\label{surface}

The cubic polynomial fit to the contrast CLV and contrast versus $\bmu$ profiles reproduced the observations well (Figs. \ref{intensity_contrast_scatter_bmu} to \ref{linecore_contrast_scatter_mu}). Given this, we fit the entire set of measured network and faculae continuum and line core intensity contrast as functions of $\mu$ and $\bmu$ \citep[following][]{ortiz02}. The measured contrasts were grouped by $\mu$ and the natural logarithm of $\bmu$ into a grid of $36\times41$ bins. The grid spans $0.1\leq\mu\leq1.0$ and $2.6\leq\ln\left(\bmu\right)\leq6.7$ or about $14\g$ to $810\g$. Each bin represents an interval of 0.025 in $\mu$ and 0.1 in $\ln\left(\bmu\right)$. The logarithmic binning in $\bmu$ is to compensate for the bottom-heavy distribution of magnetogram signal, to ensure the distribution of points is not too concentrated in the lower magnetogram signal bins. At each grid element we considered the mean $\mu$, $\bmu$, continuum and line core intensity contrast of the points within the bin. In accord with the cubic polynomial fits to the individual contrast CLV and contrast versus $\bmu$ profiles, bivariate polynomials cubic in $\mu$ and $\bmu$ were fitted to the bin-averaged continuum and line core intensity contrast. The zeroth $\mu$ and $\bmu$ orders were included based on similar considerations as with the polynomial fits to the contrast CLV and contrast versus $\bmu$ profiles. In linear algebra notation, the surface fit to the bin-averaged continuum and line core intensity contrast are:
\begin{eqnarray}
\cicn\left(\mu,\bmuf\right)=\left[\begin{array}{c}
10^{-2}\left(\bmuf\right)^0 \\ 10^{-3}\left(\bmuf\right)^1 \\ 10^{-6}\left(\bmuf\right)^2 \\ 10^{-9}\left(\bmuf\right)^3
\end{array}\right]^T
\left[\begin{array}{rrrr}
-5.11 & 7.74 & 0.34 & -4.72 \\ -0.04 & 3.84 & -7.42 & 3.90 \\ 0.19 & -6.27 & 12.03 & -6.78 \\ -0.08 & 3.58 & -8.04 & 5.04
\end{array}\right]
\left[\begin{array}{c}
\mu^0 \\ \mu^1 \\ \mu^2 \\ \mu^3
\end{array}\right]
\label{intensity_sfit_kx}
\end{eqnarray}
and
\begin{eqnarray}
\cilc\left(\mu,\bmuf\right)=\left[\begin{array}{c}
10^{-1}\left(\bmuf\right)^0 \\ 10^{-3}\left(\bmuf\right)^1 \\ 10^{-5}\left(\bmuf\right)^2 \\ 10^{-8}\left(\bmuf\right)^3
\end{array}\right]^T
\left[\begin{array}{rrrr}
-1.08 & 2.09 & -1.78 & 0.66 \\ -0.08 & 6.84 & -11.22 & 6.34 \\ 0.07 & -1.50 & 2.52 & -1.48 \\ -0.06 & 1.03 & -1.90 & 1.18
\end{array}\right]
\left[\begin{array}{c}
\mu^0 \\ \mu^1 \\ \mu^2 \\ \mu^3
\end{array}\right]
\label{linecore_sfit_kx}
\end{eqnarray}
respectively. Since contrast is wavelength dependent and bright magnetic features are largely unresolved at HMI's spatial resolution, these relationships are valid only at the instrument's operating wavelength (6173 \AA) and spatial resolution (1 arcsec).

The surface fits are illustrated as surface and grey scale plots in Figs. \ref{intensity_contrast_sfit} and \ref{linecore_contrast_sfit}. Cross-sections to the surface fits are plotted in Figs. \ref{intensity_contrast_scatter_bmu} to \ref{linecore_contrast_scatter_mu} (blue curves) along the contrast CLV and contrast versus $\bmu$ profiles, and the corresponding cubic polynomial fits (red curves). The surface fits are in excellent agreement with the cubic polynomial fits almost everywhere. The agreement is so close that the surface fit cross-sections are completely hidden by the cubic polynomial fits in most places. The advantage with these surface fits is that they allow us to describe how the measured contrasts vary with $\mu$ and $\bmu$ almost equally well and with far fewer free parameters than by the cubic polynomial fit to each individual contrast CLV and contrast versus $\bmu$ profile (32 versus 128 free parameters; 4 from each of 32 profiles).

\begin{cfig}
\includegraphics[width=\textwidth]{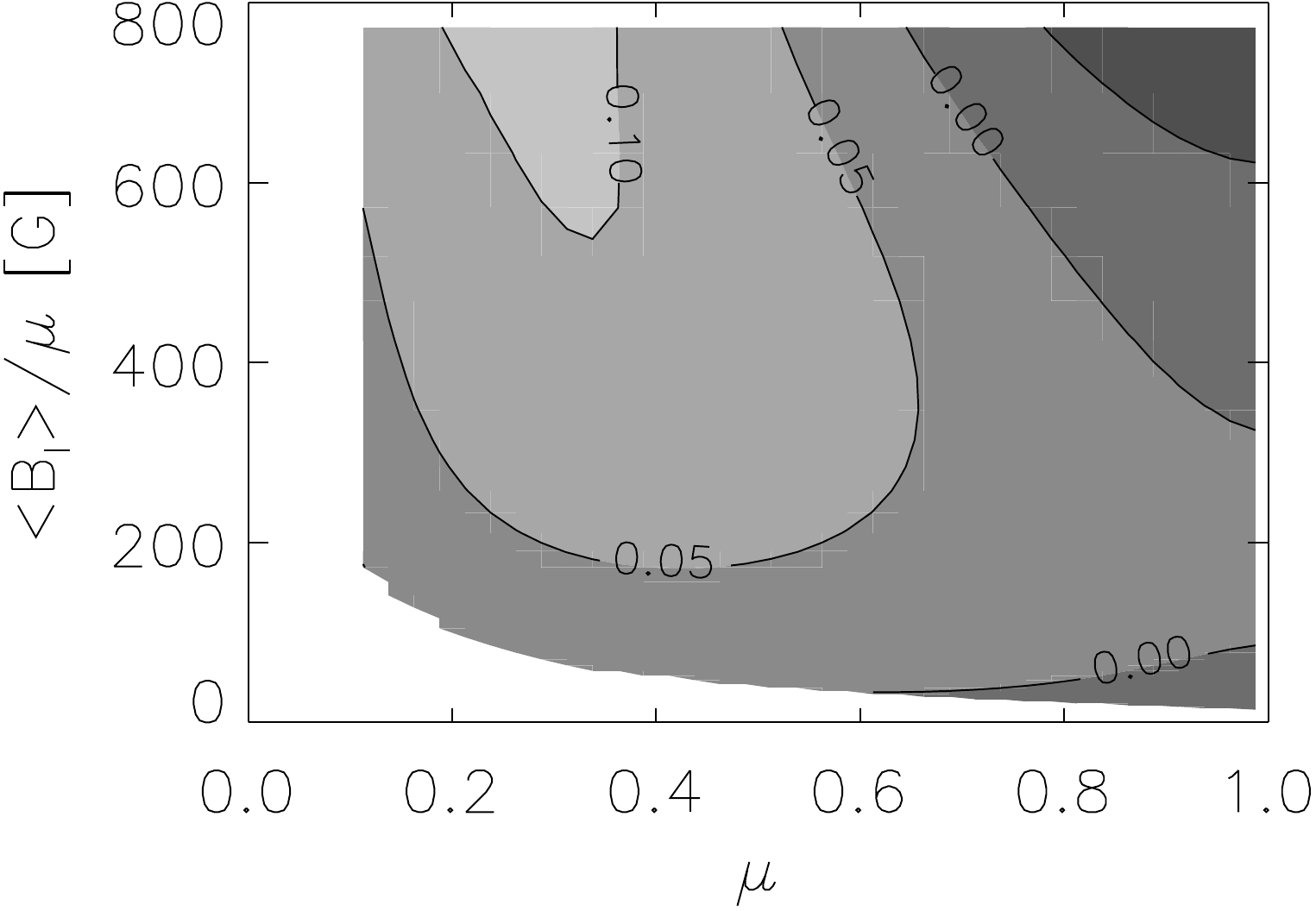}
\caption{Grey scale plot of the polynomial fit to bin-averaged continuum intensity contrast over the region where data exist, sampled at the mid-point of each bin in the $36\times41$ bins grid employed to compute the averages.}
\label{intensity_contrast_sfit}
\end{cfig}

\begin{cfig}
\includegraphics[width=\textwidth]{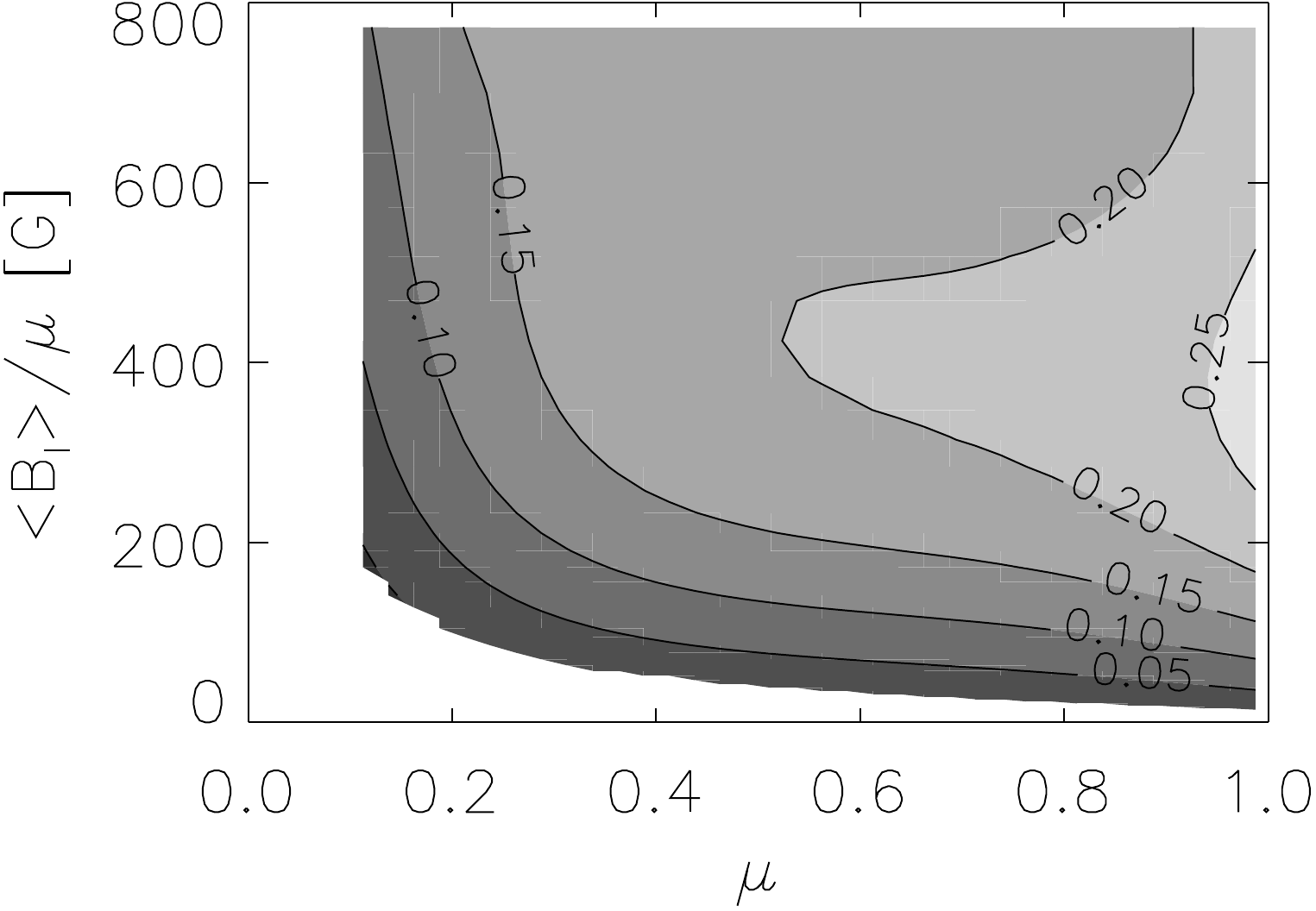}
\caption{Same as Fig. \ref{intensity_contrast_sfit}, but for line core intensity contrast.}
\label{linecore_contrast_sfit}
\end{cfig}

\subsection{Intrinsic contrast}
\label{intrinsic_contrast}

At each $\bmu$ interval, we estimated the maximum continuum intensity contrast, $\cicnmax$ and the heliocentric angle at which it is reached, $\umax$ from the cubic polynomial fit to the contrast CLV profile (Fig. \ref{intensity_contrast_scatter_bmu}). Following \citet{ortiz02} we term $\scicnmax$ the specific contrast; estimated here from the quotient of $\cicnmax$ and $\mbmu$, the mean $\bmu$ of all the points within a given $\bmu$ interval. The values of $\umax$, $\cicnmax$ and $\scicnmax$ are shown in Fig. \ref{intensity_contrast_analysis} as a function of $\bmu$ where the abscissa is given by $\mbmu$. Also plotted are the values obtained from the surface fit (Equation \ref{intensity_sfit_kx}). The uncertainty in $\cicnmax$ is given by the RMS difference between the contrast CLV profiles and their cubic polynomial fits. The uncertainty in $\scicnmax$ was estimated from the uncertainty in $\cicnmax$ and the standard error of $\mbmu$ employing standard propagation of errors.

While the position of the continuum intensity contrast CLV maximum varies with $\bmu$, line core intensity contrast at a given $\bmu$ is invariably strongest at disc centre (Fig. \ref{linecore_contrast_scatter_bmu}) as pointed out in Sect. \ref{profile}. Maximum line core intensity contrast (i.e., the value at $\mu=1$), $\cilcmax$ and specific contrast, $\cilcmax/\left(\bmu\right)$ derived similarly as above from the cubic polynomial fit to the contrast CLV profiles are expressed in Fig. \ref{linecore_contrast_analysis} as a function of $\bmu$, along with the values obtained from the surface fit (Eq. \ref{linecore_sfit_kx}). In both instances, there is some disparity between the values obtained from the cubic polynomial fits to the contrast CLV profiles and the surface fit below $\bmu\sim100\g$. This is likely due to the truncated coverage of disc positions at low $\bmu$ discussed in Sect. \ref{profile}.

\begin{cfig}
\includegraphics[width=\textwidth]{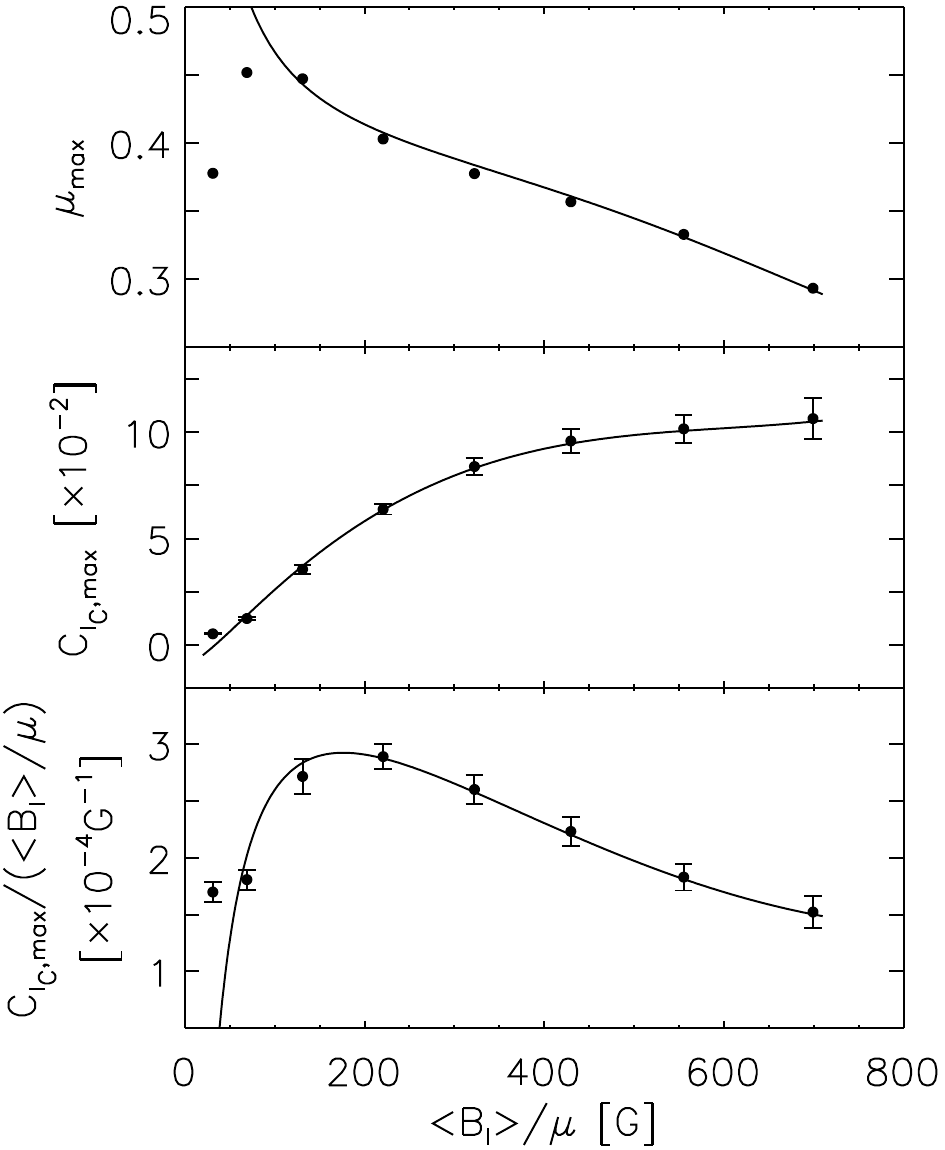}
\caption{Heliocentric angle at which continuum intensity contrast reaches its maximum, $\umax$ (top), as well as the contrast, $\cicnmax$ (middle) and specific contrast, $\scicnmax$ (bottom) there, as a function of $\bmu$. The filled circles represent the values derived from the cubic polynomial fit to the contrast CLV profiles (Fig. \ref{intensity_contrast_scatter_bmu}) and the error bars the uncertainty in $\cicnmax$ and $\scicnmax$. The curves follow the solution from the surface fit to measured continuum intensity contrast (Eq. \ref{intensity_sfit_kx}).}
\label{intensity_contrast_analysis}
\end{cfig}

\begin{cfig}
\includegraphics[width=\textwidth]{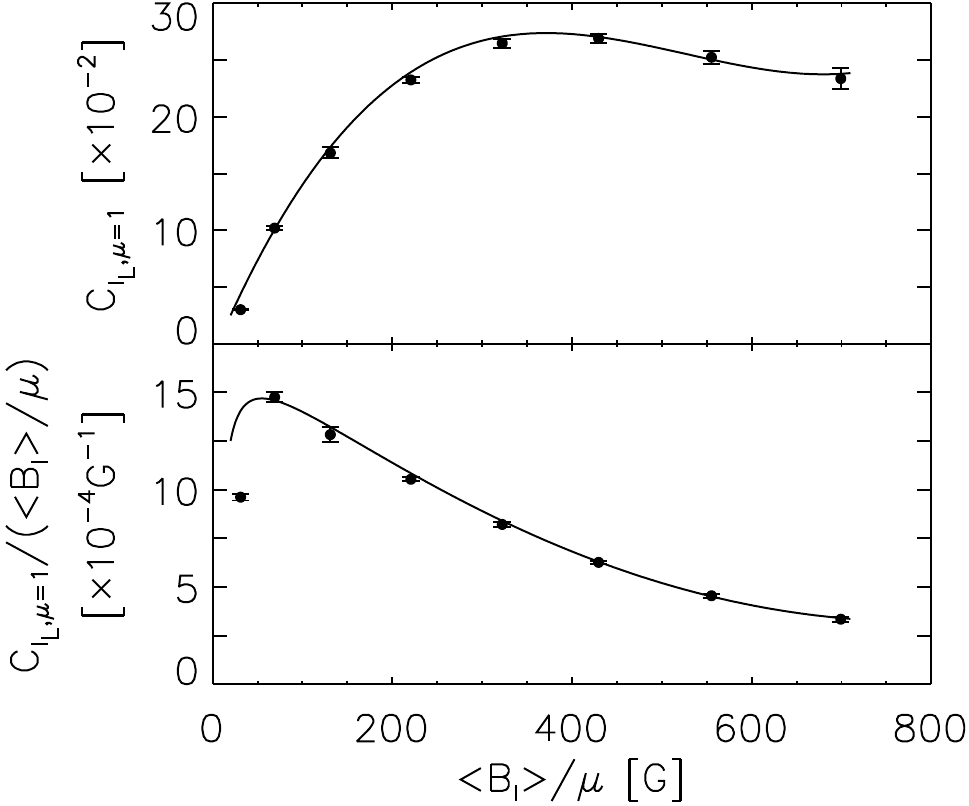}
\caption{Line core intensity contrast, $\cilcmax$ (top) and specific contrast, $\scilcmax$ (bottom) at disc centre as a function of $\bmu$. The filled circles represent the values derived from the cubic polynomial fit to the contrast CLV profiles (Fig. \ref{linecore_contrast_scatter_bmu}) and the error bars the uncertainty in $\cilcmax$ and $\scilcmax$. The curves follow the solution from the surface fit to measured line core intensity contrast (Eq. \ref{linecore_sfit_kx}).}
\label{linecore_contrast_analysis}
\end{cfig}

Network and facular features exhibit similar, kilogauss strength magnetic fields \citep{stenflo73,solanki84,rabin92,ruedi92}. This means, in general, the intrinsic magnetic field strength of magnetic elements, $B$ does not vary significantly with $\bmu$ and is also well above the range of $\bmu$ examined here, thus allowing us to assume that the magnetic filling factor, $\alpha$ never saturates. As mentioned in Sect. \ref{data}, the quantity $\bmu$ is representative of $\alpha{}B$. Taking these into account, within the limits of this study, $\bmu$ is representative of the magnetic filling factor. Therefore specific contrast, as it is defined and derived here, is a measure of the average intrinsic contrast or heating efficiency of magnetic features within a given $\bmu$ interval. Like the intrinsic field strength, the intrinsic contrast cannot be measured directly because most network and facular features are not resolved at HMI's spatial resolution and due to the effect of scattered light.

Above $\bmu\sim100\g$, the heliocentric angle at which continuum intensity contrast reaches its maximum decreases (i.e., the maximum is reached further away from disc centre) with increasing $\bmu$ (Fig. \ref{intensity_contrast_analysis}). A similar trend was reported by \citet{ortiz02} based on MDI continuum observations. Assuming the hot wall model \citep{spruit76} and a simple flux tube geometry, the authors demonstrated this to imply the average size of flux tubes increases with $\bmu$.

Both continuum and line core intensity specific contrast increases strongly with decreasing $\bmu$ up to a maximum (at around 200 and $50\g$ respectively) before gradually declining (Figs. \ref{intensity_contrast_analysis} and \ref{linecore_contrast_analysis}). The decline at low $\bmu$ aside, the observed trends suggests that brightness or temperature excess in both the lower and middle photosphere in flux tubes decreases with increasing magnetic filling factor \citep[the continuum and line core formation height of the Fe I 6173 \AA{} line is about 20 and 300 km respectively,][]{norton06}. This may partly be due to the increasing average size of flux tubes; larger magnetic elements appear darker as lateral heating is less efficient \citep{spruit76,spruit81,grossmann94} and magnetic suppression of surrounding convective energy transport is more severe \citep{kobel12}. The total radiation flux derived from MHD simulations by \citet{vogler05} exhibits a similar behaviour. This result also agrees with the observations that network elements appear hotter than facular elements in the lower and middle photosphere \citep{solanki84,solanki86,keller90} and G-band bright points are brighter in the quiet Sun than in active regions \citep{romano12}.

The decline in both continuum and line core specific contrast at low $\bmu$ suggests a diminishing heating efficiency of the smallest magnetic elements but is more likely due to the influence of intergranular lanes on apparent contrast. (As with the low contrast towards $\bmuz$ in the contrast versus $\bmu$ profiles discussed in Sect. \ref{profile}.) The probable impact of the truncated coverage of disc positions at low $\bmu$ on the quality of the fits to measured contrast here might have also played a role.

\section{Discussion}
\label{discussion}

\subsection{Comparison with \citet{ortiz02}}
\label{compare_ao}

A comparison of the results reported here for continuum intensity with those from the similar study based on full-disc MDI observations by \citet{ortiz02} reveal several notable differences.
\begin{itemize}
	\item The contrast reported here is generally higher, by as much as a factor of about two. The difference between the two studies becomes increasingly pronounced with $\bmu$ and distance from disc centre.
	\item Approaching disc centre, contrast appears to level off in our study (Fig. \ref{intensity_contrast_scatter_bmu}). In the earlier work, the contrast declines approximately linearly towards $\mu=1$ and there are also marked fluctuations about $\mu\sim0.95$ \citep[Fig. 3,][]{ortiz02}.
	\item Near $\bmuz$ contrast is negative here (Fig. \ref{intensity_contrast_scatter_mu}) but positive in the previous study \citep[Fig. 4,][]{ortiz02}.
	\item The specific contrast (a proxy of intrinsic contrast given by the quotient of contrast at the maximum point on the CLV profile and $\bmu$) presented here ascends with $\bmu$ up to $\bmu\sim200\g$ before descending monotonically thereafter (Fig. \ref{intensity_contrast_analysis}). \citet{ortiz02} found specific contrast to decline approximately linearly with $\bmu$ (Fig. 8 in their paper).
	\item For $\bmu\gtrsim200\g$, the specific contrast reported here is also nearly double that in the earlier work.
\end{itemize}
The lower contrast and specific contrast (for $\bmu\gtrsim200\g$), and difference in CLV towards disc centre reported by \citet{ortiz02} is likely, as we will show shortly, to be primarily due to the misidentification of the magnetic signal adjacent to sunspots and pores as network and faculae (discussed in Sect. \ref{reduction}) by those authors. Care was taken here to minimise such misidentification. The negative contrast towards $\bmuz$ found here, as argued in Sect. \ref{profile}, arises from the resolution of intergranular lanes at HMI's spatial resolution. We demonstrate below that this difference in spatial resolution also contributes to the opposite $\bmu$-dependence of specific contrast below $\bmu\sim200\g$ reported here and by \citet{ortiz02}.

To recreate the conditions of the study by \citet{ortiz02}, we recomputed continuum intensity contrast and specific contrast from the HMI data set employed here without applying the magnetic extension removal procedure, binning the data set spatially by $4\times4$ pixels to be consistent with MDI's spatial resolution and (very approximately) transforming measured contrast to the corresponding value at MDI's operating wavelength, 6768 \AA{}. This last transformation was carried out by describing quiet Sun, network and faculae as black bodies, and taking an effective temperature of 5800 K for the quiet Sun. This allowed us to crudely convert contrast measured at 6173 \AA{} by HMI into the corresponding contrast at 6768 \AA{}. It should be noted that this is a first-order approximation of the wavelength dependence of contrast which ignores the variation of the continuum formation height with wavelength \citep{solanki98a,sutterlin99,norton06}. In Fig. \ref{intensity_contrast_canopy} we depict the contrast CLV profile of network patterns ($\bmub$) and active region faculae ($\bmug$) after the application of the above procedure. Also plotted are the similarly treated contrast versus $\bmu$ profiles about disc centre ($\mua$) and near limb ($\mug$). The specific contrasts from this process are illustrated in Fig. \ref{intensity_contrast_analysis_bin}.

\begin{cfig}
\includegraphics[width=\textwidth]{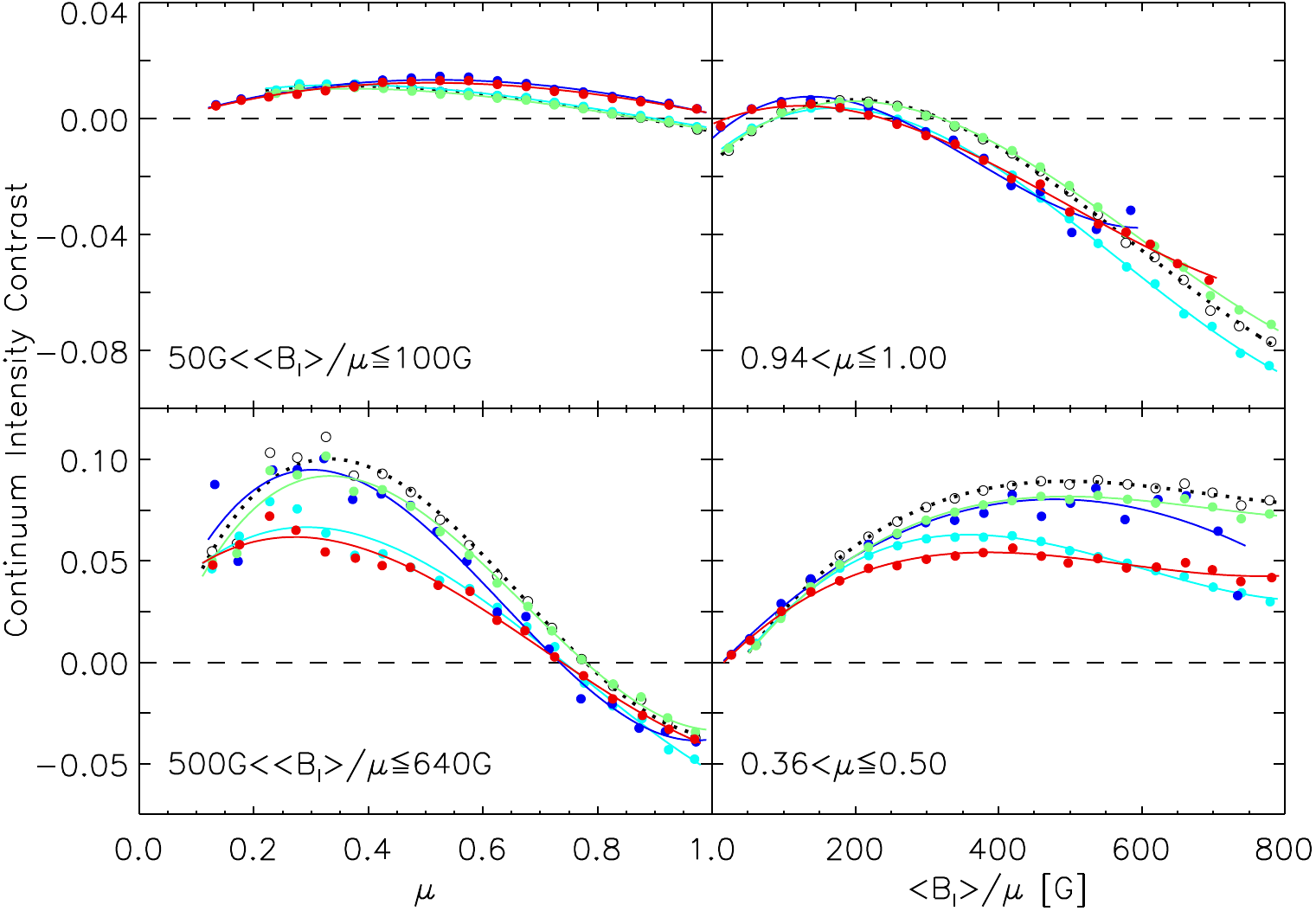}
\caption{Selected continuum intensity contrast CLV (left) and contrast versus $\bmu$ (right) profiles from Figs. \ref{intensity_contrast_scatter_bmu} and \ref{intensity_contrast_scatter_mu} (open circles and dotted curves). The selected profiles correspond to quiet Sun network (top left), active region faculae (bottom left), disc centre (top right) and near limb (bottom right). The cyan, blue and green series denote the profiles obtained by omitting the magnetic extension removal procedure, spatially binning the data set by $4\times4$ pixels and converting measured contrast to 6768 Å respectively. The red series indicates the results of taking into account all these three considerations. The circles represent the mean of measured contrast binned as in the referenced figures and the curves the corresponding third-order polynomial fits. The dashed lines mark the mean quiet Sun level.}
\label{intensity_contrast_canopy}
\end{cfig}

\begin{cfig}
\includegraphics[width=\textwidth]{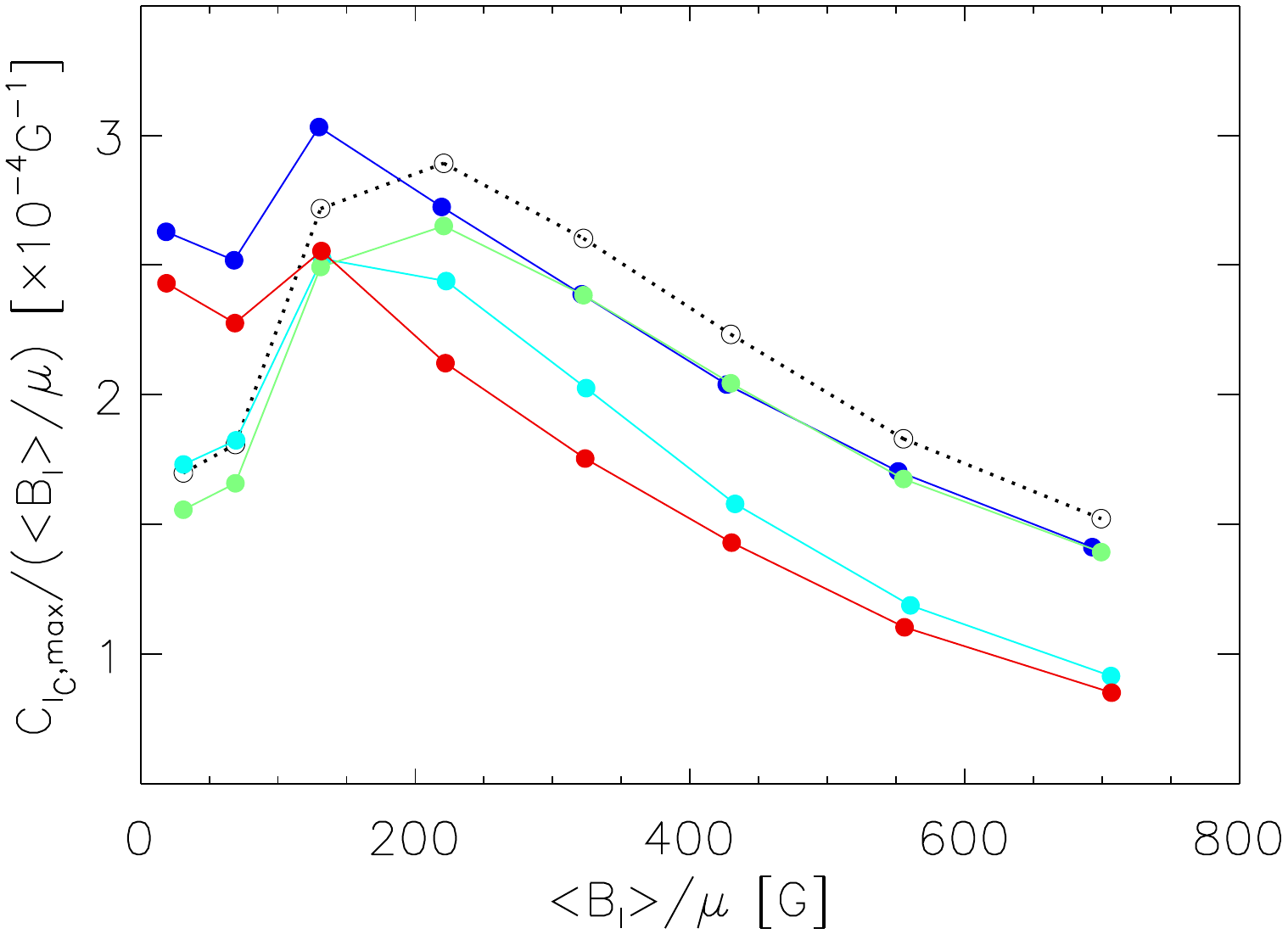}
\caption{Continuum intensity specific contrast as a function of $\bmu$ from Fig. \ref{intensity_contrast_analysis} (open circles). The cyan, blue and green filled circles represent the same quantity obtained by omitting the magnetic extension removal procedure, spatially binning the data set by $4\times4$ pixels and converting measured contrast to 6768 Å respectively. The red filled circles represent the result of taking into account all these three considerations. The points in each series are joined by straight lines to aid the eye.}
\label{intensity_contrast_analysis_bin}
\end{cfig}

Excluding the procedure to remove magnetic signal adjoined to sunspots and pores produced the largest effect. It produced an overall drop in contrast (cyan series, Fig. \ref{intensity_contrast_canopy}) and also reproduced the linear CLV near disc centre observed by \citet{ortiz02} ($\bmuh$ panel). The effects of the resizing of the data set (blue series) and the projection of measured contrasts to MDI's operating wavelength (green series) are relatively minor. The contrast profiles derived from the application of all three procedures (red series) are, in terms of both form and magnitude, in general agreement with the profiles covering similar disc positions and magnetogram signal levels reported by \citet{ortiz02} (Figs. 3 and 4 in their paper).

Similarly for specific contrast, the result of applying all three processes (red series, Fig. \ref{intensity_contrast_analysis_bin}) resembles the measurements presented by \citet{ortiz02} (Fig. 8 in their paper) and the greatest effect on magnitude came from the omission of the magnetic extension removal procedure. Binning the data set down to a MDI-like resolution also produced a significant increase in specific contrast at low $\bmu$. After combining with the other two processes, this rendered specific contrast approximately level below $200\g$. The influence of granulation on apparent contrast at HMI's spatial resolution plays a role in the observed decline in specific contrast from $\bmu\sim200\g$ towards $\bmuz$.

The largest effect is produced by the removal of magnetic signal adjoining to sunspots and pores. Hence it is worth considering this process more closely. The procedure by which we removed these signals inevitably discriminates against active region faculae, and various authors have noted lower contrast in active region faculae compared to quiet Sun network at similar magnetogram signal levels \citep{lawrence93,kobel11}, but we do not reckon this to be the reason for the greater contrast reported here. In spite of the severe steps taken to exclude magnetic signals adjoined to sunspots and pores, there remains a fair representation of active region faculae in the measured contrasts (Fig. \ref{network_plage_mask}). Also, at HMI's spatial resolution, network patterns are largely confined to the lower half of the $\bmu$ range considered (Fig. \ref{qs_ar_fac_map}) while the difference between the contrast reported here and by \citet{ortiz02} is more pronounced at higher $\bmu$.

Following the example of earlier studies \citep[e.g.][]{chapman80}, \citet{ortiz02} fit quadratic polynomials to continuum intensity contrast CLV profiles derived in their study. Here we fit cubic polynomials instead because of the different CLV near disc centre which we just demonstrated to arise from our treatment of magnetic signal adjoined to sunspots and pores.

We surmise that the fluctuations about disc centre in the continuum intensity CLV profiles reported by \citet{ortiz02} is due to the non-homogeneous distribution of active regions over the solar disc. Each unit $\mu$ represents a greater radial distance towards the disc centre rendering these fluctuations increasingly abrupt. Not accounting for magnetic signal adjoined to sunspots and pores probably accentuated these fluctuations.

\subsection{Continuum intensity contrast about disc centre}
\label{compare_dc}

In this section we compare the magnetogram signal dependence of continuum intensity contrast near disc centre reported here with that from other recent studies, summarized in Table \ref{disc_centre_compare}. Apart from the work of \citet{ortiz02}, the CLV of continuum intensity contrast has been examined in detail previously by \citet{topka92,topka97} and \citet{lawrence93}. While we and \citet{ortiz02} employed full-disc observations made at a single wavelength and spatial resolution, these authors collated telescope scans made at multiple wavelengths and resolutions. The magnetogram signal intervals represented by the contrast CLV profiles presented in these papers also differ considerably from those presented here and by \citet{ortiz02}. For these reasons it is not straightforward to make any quantitative comparisons with the contrast CLV reported by these studies, therefore the focus on results for near disc centre here.

\begin{sidewaystable}
\caption{Continuum intensity contrast dependence on magnetogram signal near disc centre obtained in the present and from earlier studies.}
\label{disc_centre_compare}
\begin{tabular}{lccccl}
\hline\hline
Reference & Instrument & Target$^b$ & Resolution [Arcsec] & Wavelength [\AA{}] & Results\\
\hline
Present & SDO/HMI & $\mu>0.94^c$ & $1$ & 6173 & Negative near $\bmuz$. Peak at $\bmu\sim200\g$.\\
1, 2, 3, 4 & SVST & AR & $\gtrsim0.3$ & Various$^d$ & Negative near $\bz$. Monotonic decline with $\bl$.\\
3 & SVST & QS & $\gtrsim0.3$ & 6302 & Negative near $\bz$. Peak at $\bl\sim400\g$.\\
 & SFO/SHG$^a$ & QS & $\gtrsim1$ & 6302 & Negative near $\bz$. Peak at $100\g\lesssim{}\bl\lesssim200\g$.\\
 & & AR & & & Negative near $\bz$. Monotonic decline with $\bl$.\\
5 & SoHO/MDI & $\mu>0.92^c$ & $4$ & 6768 & Positive near $\bmuz$. Peak at $\bmu\sim100\g$.\\
6 & SST & AR & $0.15$ & 6302 & Negative near $\bz$. Peak at $\bl\sim650\g$.\\
7 & Hinode/SOT & QS & $0.3$ & 6302 & Negative near $\bz$. Peak at $\bl\sim700\g$.\\
 & & AR & & & Negative near $\bz$. Peak at $\bl\sim700\g$.\\
8 & SST & QS & $0.15$ & 6302 & Negative near $\bz$. Saturate at $\bl\sim800\g$.\\
 & Hinode/SOT & QS & $0.3$ & 6302 & Negative near $\bz$. Peak at $\bl\sim500\g$.\\
\hline
\multicolumn{6}{p{\textwidth}}{\textbf{Notes.} $^{(a)}$ San Fernando Observatory (SFO) 28-cm vacuum telescope and vacuum spectroheliograph (SHG). $^{(b)}$ AR and QS denote active region and quiet Sun respectively. $^{(c)}$ Segment of full-disc observations considered. $^{(d)}$ 5250\AA{}, 5576\AA{}, 6302\AA{} and 6768\AA{}.\linebreak }\\
\multicolumn{6}{p{\textwidth}}{\textbf{References.} (1) \citet{title92}; (2) \citet{topka92}; (3) \citet{lawrence93}; (4) \citet{topka97}; (5) \citet{ortiz02}; (6) \citet{narayan10}; (7) \citet{kobel11}; (8) \citet{schnerr11}.}\\
\end{tabular}
\end{sidewaystable}

The negative contrast at low magnetogram signal levels found here and by the majority of the compared works counters expectations from thin flux tube models, which predict intrinsically bright magnetic features in this regime \citep{knoelker88}. As discussed in Sect. \ref{profile}, this is attributed here \citep[and in][]{title92,topka92,kobel11} to the influence of intergranular lanes, an assertion supported by various models \citep{title96,schnerr11}. Of the studies compared, only that by \citet{ortiz02} noted positive contrasts. Binning the data set here to a MDI-like resolution raised the disc centre end of the contrast CLV profile in the $\bmub$ interval into positive territory (Fig. \ref{intensity_contrast_canopy}), suggesting that the positive contrasts reported by \citet{ortiz02} arose from utilising data at a resolution (the lowest of the studies compared) where granulation is largely unresolved.

From SVST scans of active regions, \citet{title92}, \citet{topka92,topka97} and \citet{lawrence93} found contrast to decline monotonically with magnetogram signal. \citet{lawrence93} also noted the same with active region data acquired at the San Fernando Observatory (SFO). \citet{kobel11} demonstrated the comparatively poorer resolution and straylight from pores to be likely culpable for the monotonic decline observed in the SVST studies. In all the other compared works, contrast exhibits a peak, the position of which varied from $\bmu\sim100\g$ \citep{ortiz02} to $\bl\sim700\g$ \citep{kobel11}, except for the quiet Sun SST scan examined by \citet{schnerr11}, where contrast saturated at $\bl\sim800\g$. Generally, the finer the spatial resolution, the higher the position of the peak. Most of these studies were based on observations made at the Fe I 6302 \AA{} line. Even if we discount the studies made at other wavelengths, this broad pattern is still apparent, ruling out differences in wavelength as the major driver. \citet{kobel11} reported similar contrast peak positions for quiet Sun and active region scans. Therefore the difference between active region and quiet Sun contrast had likely little role in the spread in reported peak positions amongst the compared works. As shown in the $\mua$ panel of Fig. \ref{intensity_contrast_canopy}, both resizing the HMI data set to a MDI-like spatial resolution and omitting the magnetic extension removal procedure shifted the position of the contrast versus $\bmu$ profile maximum towards the origin.

The above comparison points to differences in spatial resolution and treatment of magnetic signal near sunspots and pores as the dominant factors behind the spread in reported dependence of contrast on magnetogram signal. Indeed, the recent MHD simulation work of \citet{rohrbein11} suggests that the contrast peak at intermediate magnetogram signal levels seen in direct observations, but not observed in MHD simulations, is a product of the limited spatial resolution. The bulk of the studies listed in Table \ref{disc_centre_compare} were based on higher spatial resolution data than utilised here. For this reason, a more quantitative comparison like we did in Sect. \ref{compare_ao} with the findings of \citet{ortiz02} is not workable here.

The recent works of \citet{berger07} and \citet{viticchie10}, examining the contrast of bright points in the G-band, bear tenuous relevance to our study. These studies utilised observations made in a molecular band (i.e., neither continuum nor line core) where the contrast of magnetic features is enhanced \citep[due to CH depletion,][]{steiner01,schussler03}. More importantly, while this study and the works cited in Table \ref{disc_centre_compare} considered each pixel a separate entity, \citet{berger07} and \citet{viticchie10} examined the overall contrast of each individual bright structure. As the body of magnetic features isolated by both approaches differ, the results are not directly comparable. \citet{berger07} did however also report a pixel-by-pixel consideration of G-band contrast. Scans at four disc positions were surveyed. Barring one that appeared anomalous, the contrast versus magnetogram signal profiles from each scan bear general resemblance to ours in terms of form. Notably, the profile from the disc centre scan, with a spatial resolution of $0.15\:{\rm arcsec}$, exhibits a peak at $\bl\sim700\g$, consistent with the broad pattern between spatial resolution and peak position described here.

\subsection{Line core intensity contrast}
\label{compare_lc}

As pointed out in Sect. \ref{profile}, the most notable difference between the continuum and line core intensity contrasts present here is the converse CLV. While continuum intensity contrast is weakest near disc centre and strengthens towards the limb, line core intensity contrast is strongest at disc centre and declines towards the limb. The divergent CLV exhibited by the two sets of measurements stem from their rather different physical sources. Continuum intensity is enhanced largely in the hot walls of magnetic elements, thus the centre-to-limb increase. The line core is formed in the middle photosphere, which is heated either by radiation from deeper layers \citep{knoelker91}, or by mechanical and Ohmic dissipations \citep{moll12}. Also stated in Sect. \ref{profile}, line core intensity is modulated by line strength and shape, and continuum intensity. Excluding variation related to continuum excess, line core intensity enhancement arises from the influence of the temperature excess in the middle photosphere and Zeeman splitting on line strength and shape. Given the relatively narrow, vertical geometry of flux tubes, as magnetic elements rotate from disc centre to limb, line-of-sight rays go from being largely confined to single flux tubes to increasingly passing into and out of multiple flux tubes, especially in densely packed facular regions \citep{bunte93}. Line core intensity contrast decreases towards the limb as the contribution to the spectral line from magnetic elements diminishes from absorption in the non-magnetic part of the atmosphere transversed by the rays \citep{solanki98b}. Another probable cause of the centre-to-limb decline is the spatial displacement of the line core with respect to the corresponding continuum towards the limb caused by the difference in formation height and oblique viewing geometry \citep{stellmacher91,stellmacher01}. The line core intensity enhancement arising from temperature excess in the middle photosphere and Zeeman splitting discussed here is not to be confused with that from the centre-to-limb broadening of the Fe I 6173 \AA{} line mentioned in Sect. \ref{qsintensity}, which arises from the viewing geometry independent of magnetic field.

We recomputed line core intensity contrast, this time normalizing the line core intensity images by the corresponding continuum intensity images prior to data reduction. The result, the line core residual intensity contrast, is essentially the component of line core intensity contrast arising from line weakening in magnetic features alone. Line core intensity and residual intensity contrast values can be compared directly. (Line core residual intensity contrast equates to the line core intensity contrast we would get from scaling the line core intensity of just the network and faculae pixels by $\mqsicn/\icn$.) The line core residual intensity contrast CLV profile of network patterns ($\bmub$) and active region faculae ($\bmug$), and contrast versus $\bmu$ profile about disc centre ($\mua$) and near limb ($\mug$) so derived are plotted along with the corresponding line core intensity contrast profiles from Figs. \ref{linecore_contrast_scatter_bmu} and \ref{linecore_contrast_scatter_mu} in Fig. \ref{line_weakening}. Line core residual intensity specific contrast is plotted along with the line core intensity specific contrast from Fig. \ref{linecore_contrast_analysis} in Fig. \ref{linecore_contrast_analysis_bin}.

\begin{cfig}
\includegraphics[width=\textwidth]{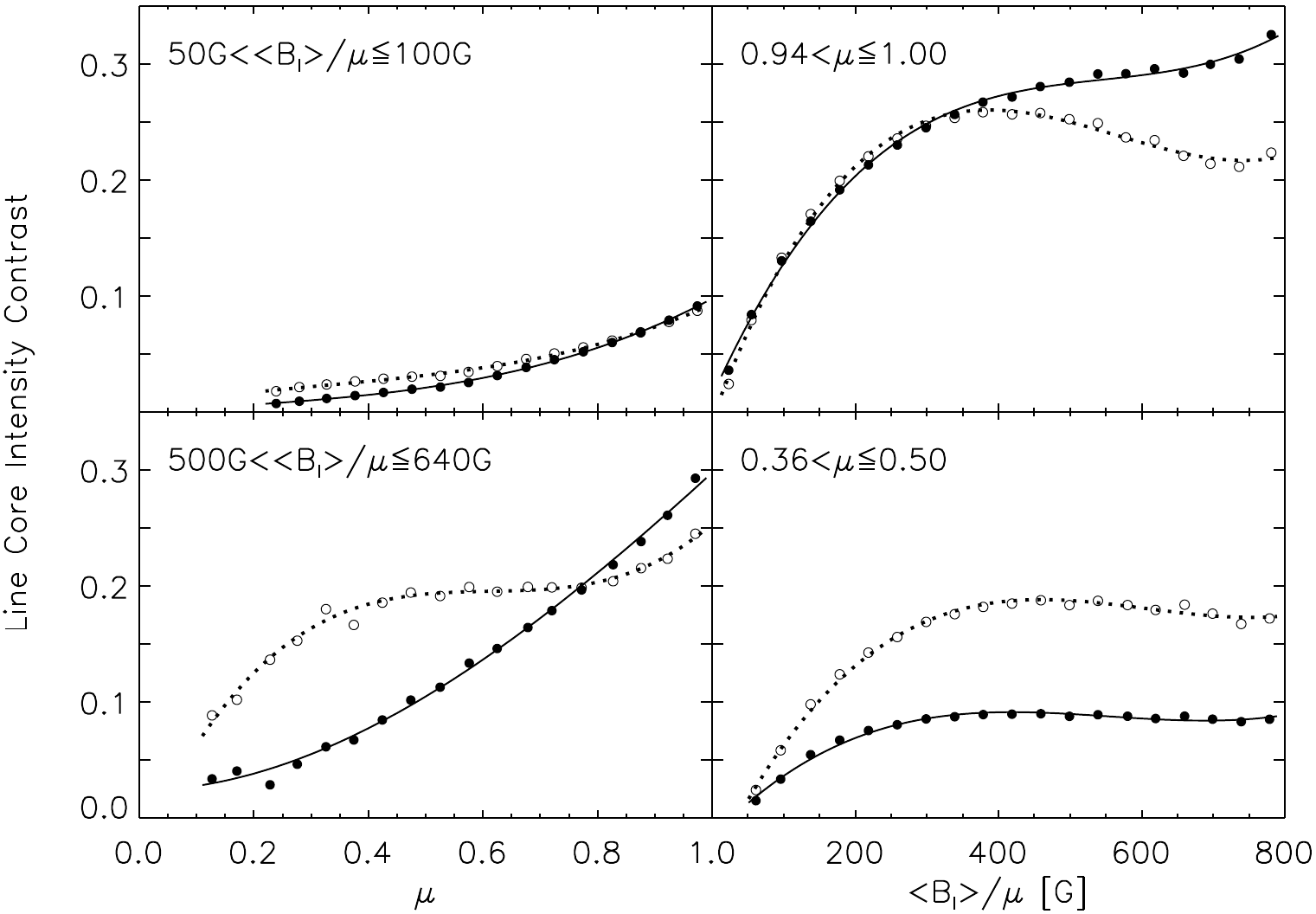}
\caption{Selected line core intensity (open circles and dotted curves) and corresponding residual intensity (filled circles and solid curves) contrast CLV (left) and contrast versus $\bmu$ (right) profiles. The circles represent the mean of measured contrast binned as in Figs. \ref{linecore_contrast_scatter_bmu} and \ref{linecore_contrast_scatter_mu} and the curves the corresponding cubic polynomial fits. The selected profiles correspond to quiet Sun network (top left), active region faculae (bottom left), disc centre (top right) and near limb (bottom right).}
\label{line_weakening}
\end{cfig}

\begin{cfig}
\includegraphics[width=\textwidth]{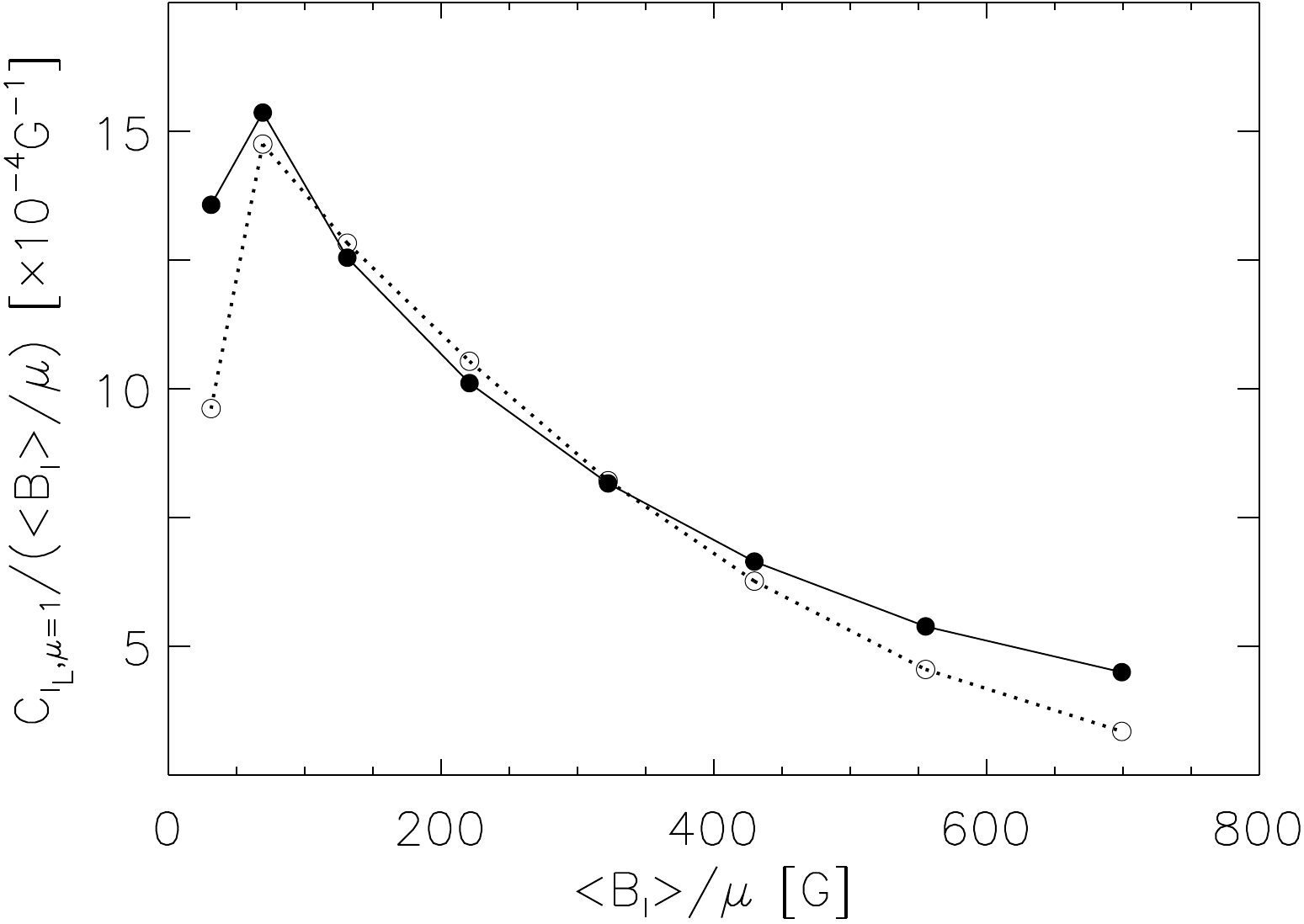}
\caption{Line core intensity (open circles) and corresponding residual intensity (filled circles) specific contrast as a function of $\bmu$. The circles are joined by straight lines to aid the eye.}
\label{linecore_contrast_analysis_bin}
\end{cfig}

As expected, excluding the contribution of continuum intensity enhancement to line core intensity contrast results in a more pronounced centre-to-limb decline ($\bmub$ and $\bmug$ panels, Fig. \ref{line_weakening}). This decline in the line core residual intensity contrast CLV profiles is also consistent with the spectroscopic observations of \citet{stellmacher79} and \citet{hirzberger05}. Comparing line core intensity and residual intensity contrast (Figs. \ref{line_weakening} and \ref{linecore_contrast_analysis_bin}), the broad similarity, especially near disc centre, imply that line core intensity contrast is dominated by the contribution from line weakening. Even towards the limb, where magnetic line weakening is at its weakest, it still comprises a significant proportion of observed line core intensity contrast. The potential implications of this result and the opposite CLV exhibited by continuum and line core intensity contrast on facular contribution to solar irradiance variations will be discussed in Sect. \ref{fac_tsi}.

Very few studies examining the disc position and magnetogram signal dependence of line core intensity contrast exist in the literature. \citet{frazier71}, \citet{lawrence91} and \citet{title92} measured intensity contrast at disc centre in the core of the Fe I 5250 \AA{}, Fe I 6302 \AA{} and Ni I 6768 \AA{} lines, respectively. Results from different lines are not directly comparable but it is still encouraging to see that in terms of magnitude and general trend with magnetogram signal, the results from these earlier studies express broad agreement with ours. The only notable exception is the steep monotonic decline at $\bl>600\g$ reported by \citet{title92}, which largely persisted even after the authors masked out pores. The decline coincides with a similar trend in intensity contrast measured at the nearby continuum, suggesting a greater relative influence of the continuum excess to magnetic line weakening at the Ni I 6768 \AA{} line compared to the Fe I 6173 \AA{} line. \citet{frazier71} also made measurements away from disc centre, but due to data scatter the author could do no better than to present a schematic representation of the CLV through visually fitted linear functions. \citet{walton87} reported measured line core residual intensities for eight magnetically insensitive lines at various disc positions. There were however only a small number of scattered measurements for each line, and a relatively narrow range of disc positions ($\mu>0.6$). To our knowledge, the present study is the first to examine line core intensity contrast employing full-disc observations. Together with the relatively fine resolution and low noise level, this allowed us to describe the magnetogram signal dependence and especially the CLV with greater accuracy and detail than in the previous efforts.

\subsection{Facular contribution to variation in solar irradiance}
\label{fac_tsi}

Our finding that intensity contrast in the line core is dominated by line weakening rather than continuum intensity enhancement and exhibits the opposite CLV as continuum intensity contrast has potential implications on facular contribution to total solar irradiance variation. Total solar irradiance (TSI) variation is the sum manifest of variation in the continuum and spectral lines. The converse CLV of the continuum and line core intensity contrasts reported here imply a different time variation of the contribution by the continuum and spectral lines to solar irradiance as magnetic features rotate across the solar disc.

In Fig. \ref{lone_ar}a we show the level 2.0 hourly TSI measurements from the DIARAD radiometer \citep{dewitte04a} on the SoHO/VIRGO instrument \citep{frohlich95} for the 22-day period of August 19 to September 9, 1996 (open circles). Active region NOAA 7986 rotated into view on August 23 and out of view on September 5, and was the only active region on the solar disc for the duration of its passage across the solar disc. The period is otherwise relatively quiet. This data therefore allows us to chart variations in TSI arising mainly from the passage of a single active region across the solar disc. The dip around August 29 corresponds to active region NOAA 7986 crossing disc centre. Even with darkening from sunspots present in the active region, the nadir of the dip is $\sim0.2\:{\rm Wm^{-2}}$ above the level before August 23 and after September 5. This suggests an overall positive contribution to TSI variation by the faculae in NOAA 7986 when it was near disc centre \citep[first pointed out by][]{fligge00}.

A schematic representation of facular contribution to variation in TSI during the passage of NOAA 7986 across the solar disc, depicted in Fig. \ref{lone_ar}e, was derived as follows.
\begin{itemize}
	\item The DIARAD TSI data was interpolated at 0.1 day intervals and the result smoothed \citep[via binomial smoothing,][]{marchand83}. Sunspot darkening was estimated from the Photometric Sunspot Index (PSI) by \citet{chapman94} based on full-disc photometric images acquired with the Cartesian Full Disk Telescope 1 (CFDT1) at SFO. A quadratic polynomial was fitted to the PSI values from the period of interest excluding the points where ${\rm PSI}=0$ (i.e., no sunspots in view). Taking a value of $1365.4\:{\rm Wm^{-2}}$ for the total irradiance of the quiet Sun, the fit was converted from units of parts per million to ${\rm Wm^{-2}}$ and subtracted from the DIARAD data. This value for the total irradiance of the quiet Sun is given by the average TSI at the last three solar minima stated in version d41\_62\_1204 (dated April 2, 2012) of the PMOD TSI composite \citep{frohlich00}. The DIARAD data, after this treatment, represents variation in TSI largely from faculae in NOAA 7986 alone. In Fig. \ref{lone_ar}a we plot the DIARAD data after interpolation and smoothing (red dotted curve), and after removing sunspot darkening (red solid curve) along the original measurements. In Fig. \ref{lone_ar}b we show the PSI (open circles) and the quadratic polynomial fit to the non-zero segment (curve).
	\item The trajectory of NOAA 7986 during its passage across the solar disc was estimated from 142 level 1.8 5-minute MDI magnetograms \citep{scherrer95} on which the active region was entirely in view (i.e. not only partially rotated into, or partially rotated off the solar disc). Taking the unsigned magnetogram signal, the magnetograms were binned spatially by $16\times16$ pixels. For each binned magnetogram, the position of NOAA 7986, in terms of $\mu$, was estimated from the mean position of the five pixels within the active region with the strongest signal. The trajectory of NOAA 7986 is then given by the quadratic polynomial fit to these estimates. In Fig. \ref{lone_ar}c we show the estimated position of NOAA 7986 in the magnetograms (open circles) and the quadratic polynomial fit (curve).
	\item Facular contribution to variation in TSI was very approximately modelled from the empirical relationships describing contrast as a function of $\mu$ and $\bmu$ derived in this study. Assuming a power law distribution of $\bmu$ with a scaling exponent of -1.85 \citep{parnell09}, we evaluated Eqs. \ref{intensity_sfit_kx} and \ref{linecore_sfit_kx}, scaled by $\left(\bmuf/15\right)^{-1.85}$ and integrated over $\bmu=15\g$ to $800\g$, at 0.1 day intervals taking $\mu$ from the trajectory of NOAA 7986 estimated earlier. The resulting time series were then scaled by $\mu$ (to correct for the CLV of projected area on the solar disc) and the limb darkening function from \citet{foukal04}. Given the approximate manner of this derivation, the results are non-quantitative. However for this analysis it is not the actual values but the temporal trends that is important. Here we had approximated the active region as a point object and so cannot include variation arising from the active region being only partially visible as it rotates into and off the solar disc. This derivation is therefore only valid, and confined to, the period where NOAA 7986 was entirely in view in MDI magnetograms.
\end{itemize}
The treated DIARAD data, giving TSI variation largely from faculae in NOAA 7986 alone, is plotted in Fig. \ref{lone_ar}d (red curve) along the conjectures based on the observed intensity contrast in the continuum (blue dashed curve) and line core (blue dotted curve). To compare how they varied with time, we subtracted the mean from and normalized each time series by the area bounded by the curve and the zero level. In Fig. \ref{lone_ar}e we show the multiple linear regression fit of the continuum and line core models to the DIARAD series (blue solid curve), which showed a much better agreement to it than either model. Facular contribution to variation in solar irradiance appears to be strongly driven by intensity contrast in both the continuum, and spectral lines, which derives largely from line weakening in magnetic elements. This complies with the observation that solar irradiance variations over the solar cycle seems to be significantly influenced by changes in spectral lines \citep{mitchell91,unruh99,preminger02}.

\begin{cfig}
\includegraphics[width=\textwidth]{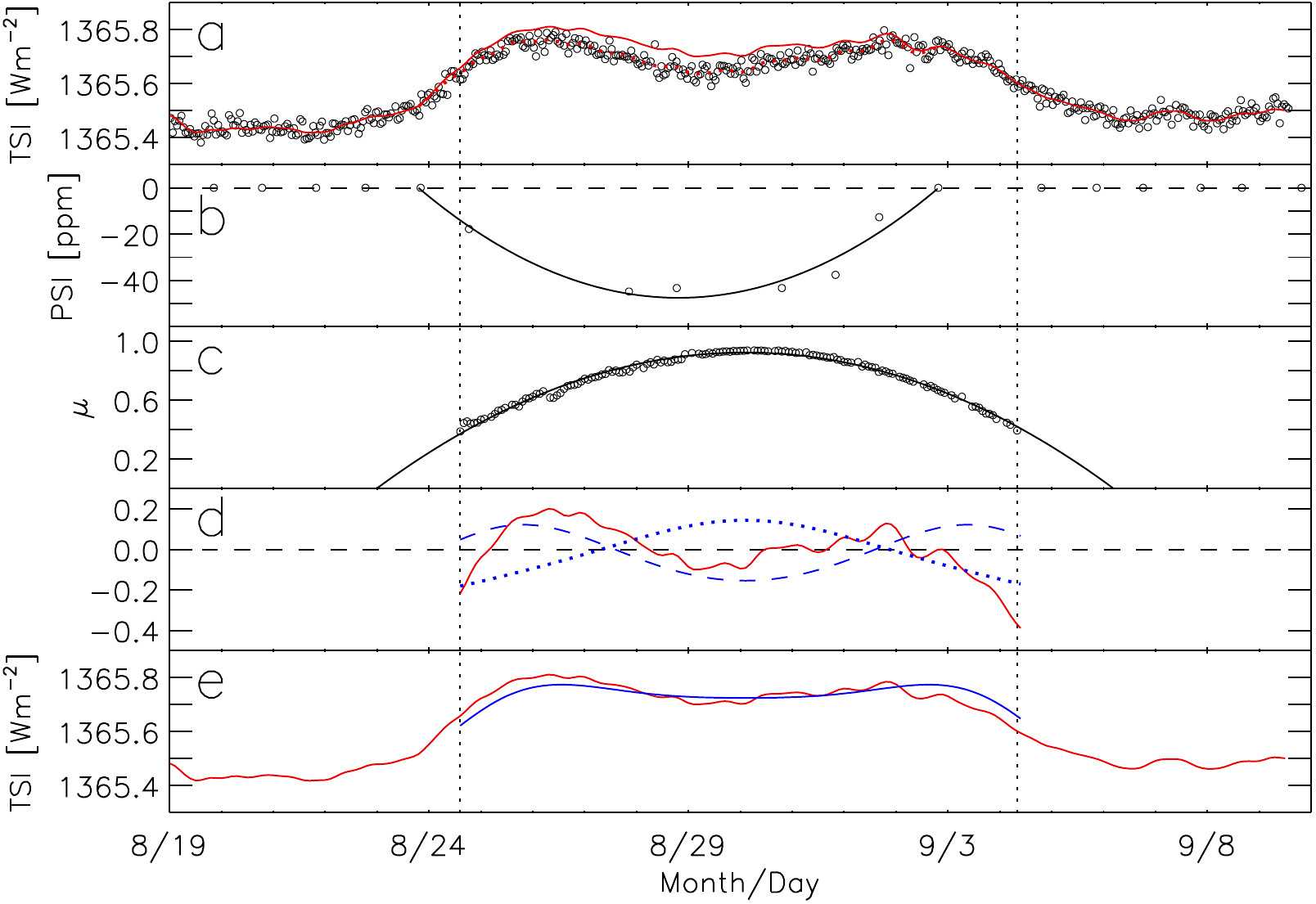}
\caption{a) Total solar irradiance, TSI from DIARAD on SoHO/VIRGO (open circles) for the period of August 19 to September 9, 1996. The dotted and solid red curves represent the interpolated and smoothed version before and after subtracting sunspot darkening. b) Photometric Sunspot Index, PSI (open circles) by \citet{chapman94} and the quadratic polynomial fit to the non-zero points (curve). The dashed line marks the zero level. c) Position of NOAA 7986 as estimated from MDI magnetograms (open circles) and the quadratic polynomial fit (curve). d) TSI minus sunspot darkening (red curve) and model of facular contribution to TSI based on observed intensity contrast in the continuum (blue dashed curve) and line core (blue dotted curve), each mean subtracted and normalized by the area bound between the time series and the zero level (black dashed line). e) TSI minus sunspot darkening (red curve) and the multiple linear regression fit of the continuum and line core models (blue solid curve). The dotted lines running down all the panels mark the period where NOAA 7986 was entirely on the solar disc in MDI magnetograms.}
\label{lone_ar}
\end{cfig}

Various studies examining the photometric contrast of faculae, as identified in calcium line images, reported positive values near disc centre \citep{lawrence88,walton03,foukal04}. The apparent divergence between the findings of these works, and the largely negative contrast near disc centre reported here (Fig. \ref{intensity_contrast_scatter_mu}) and in the similar studies listed in Table \ref{disc_centre_compare}, where network and faculae were characterized by the magnetogram signal, arises from the different selection methods \citep{ermolli07}. Significantly, the results of \citet{foukal04} were based on near total light broadband measurements. From the statistical analysis of an extensive catalogue of full-disc photometric observations obtained with the SFO/CFDT1, \citet{walton03} found contrast, in a continuum passband (i.e., negligible spectral line contribution) in the visible red, averaged over each facular region (as identified in the Ca II K) to be positive ($\sim0.005$) near disc centre. The authors took this result to indicate the distribution of flux tube sizes in facular regions is biased to the low end (i.e., more small bright flux tubes than large dark ones), which is reasonable considering the typical distribution of magnetogram signal \citep[i.e., much more weak signals than strong,][]{parnell09} and the observation that the average flux tube size is greater where magnetogram signal is greater \citep{ortiz02}. This suggests the continuum component of facular contribution to TSI variation is on average positive at disc centre in the visible red. From this and the analysis presented above, we conclude that the apparent overall positive contribution to TSI variation by the faculae in NOAA 7986 when it was near disc centre is, at least in part, due to the prevalence of smaller flux tubes and line weakening.

\section{Conclusion}
\label{conclusion}

Here we have presented measured network and facular intensity contrast in the continuum and in the core of the Fe I 6173 \AA{} line from SDO/HMI full-disc observations. We studied the dependence of the contrast on disc position and magnetogram signal, represented by $\mu$ and $\bmu$ respectively. Specifically, we derived empirical relationships describing contrast as a function of $\mu$ and $\bmu$, and specific contrast (contrast per unit $\bmu$, representative of intrinsic contrast) as a function of $\bmu$. This study exploits the unprecedented opportunity offered by the SDO mission to examine co-temporal full-disc observations of magnetic field and intensity at a constant intermediate spatial resolution (1 arcsec), relatively low noise, and without atmospheric interference. The quality of the data allowed us to examine intensity contrast for a larger sample at greater accuracy and detail than previous, similar studies, especially in the case of the line core. These results constitute stringent observational constraints on the variation of network and facular intensity with disc position and magnetogram signal in the low and middle photosphere. By constraining atmospheric models of network and faculae, these results should be of utility to solar irradiance reconstructions, especially as HMI data will increasingly be used for this purpose. Given this is the first study of its kind to examine the entire solar disc in both the continuum and line core, it should also be useful to reproduce these results in models of magnetic flux concentrations.

There are significant discrepancies in the continuum intensity contrast reported here and from earlier studies. In this study we had taken steps to account for magnetic signal in the periphery of sunspots and pores, arising from their magnetic canopies and the influence of straylight, which can easily be misidentified as network and faculae. From a comparison with the findings of past efforts, including a recomputation of the results obtained here recreating the conditions of the similar study by \citet{ortiz02}, we showed differences in resolution, and treatment of magnetic signal adjacent to sunspots and pores to be the likely main factors behind the spread in reported results. The apparent radiant behaviour of network and faculae elements is strongly modulated by spatial resolution \citep{title96,rohrbein11,schnerr11}. An understanding of its influence is necessary for the proper interpretation of direct measurements.

In terms of magnitude, trend with magnetogram signal and in particular the CLV, the results obtained here in the continuum and line core differ considerably. While continuum intensity contrast broadly ascends towards the limb, line core intensity contrast is greatest near disc centre and diminishes from disc centre to limb. The divergence between both sets of measurements arises dominantly from spectral line changes due to heating in the middle photosphere and Zeeman splitting in magnetic features, and the different mechanisms by which apparent contrast vary with viewing geometry going from disc centre to limb. From a simple model based on the empirical relationships between contrast, and $\mu$ and $\bmu$ derived here we confirmed that facular contribution to variation in solar irradiance is strongly driven by both continuum excess and spectral line changes.

The specific contrast in both the continuum and line core exhibit a marked decline with increasing magnetogram signal, confirming that network elements are, per unit magnetic flux, hotter and brighter than active region faculae. The different radiant behaviour of network and faculae, not accounted in present models of solar irradiance variations, would be an important factor to consider for more realistic modelling. This observation also implies that secular changes in solar irradiance may be considerably larger than what some models of solar irradiance variations have suggested, given the variation in the number of small-scale magnetic elements on the solar disc is a prime candidate driver of secular changes \citep{solanki02}. For example, the model employed by \citet{krivova07} to reconstruct variation in TSI from 1700 and by \citet{vieira11} for over the Holocene assumed the faculae contrast model by \citet{unruh99} for both network and faculae. This renders network and faculae with similar magnetic filling factors equally bright and therefore possibly underestimate the contribution by network to secular variation.

\chapter[Point spread function of SDO/HMI and the effects of stray light correction on the apparent properties of solar surface phenomena \\ \textit{\footnotesize{(The contents of this chapter are identical to the printed version of Yeo, K. L., Feller, A., Solanki, S. K., Couvidat, S., Danilovic, S., Krivova, N. A., 2014, Point spread function of SDO/HMI and the effects of stray light correction on the apparent properties of solar surface phenomena, Astron. Astrophys., 561, A22.)}}]{Point spread function of SDO/HMI and the effects of stray light correction on the apparent properties of solar surface phenomena}
\label{paper2}

\begin{flushright}
{\it Yeo, K. L., Feller, A., Solanki, S. K., Couvidat, S., Danilovic, S., Krivova, N. A.} \\
{\bf Astron. Astrophys., 561, A22 (2014)\footnoteC{The contents of this chapter are identical to the printed version of Yeo, K. L., Feller, A., Solanki, S. K., Couvidat, S., Danilovic, S., Krivova, N. A., 2014, Point spread function of SDO/HMI and the effects of stray light correction on the apparent properties of solar surface phenomena, Astron. Astrophys., 561, A22, reproduced with permission from Astronomy \& Astrophysics, \textcopyright{} ESO.}}
\end{flushright}

\section*{Abstract}

{\it Aims.} We present a point spread function (PSF) for the Helioseismic and Magnetic Imager (HMI) onboard the Solar Dynamics Observatory (SDO) and discuss the effects of its removal on the apparent properties of solar surface phenomena in HMI data.

{\it Methods.} The PSF was retrieved from observations of Venus in transit by matching it to the convolution of a model of the venusian disc and solar background with a guess PSF. We described the PSF as the sum of five Gaussian functions, the amplitudes of which vary sinusoidally with azimuth. This relatively complex functional form was required by the data. Observations recorded near in time to the transit of Venus were corrected for instrumental scattered light by the deconvolution with the PSF. We also examined the variation in the shape of the solar aureole in daily data, as an indication of PSF changes over time.

{\it Results.} Granulation contrast in restored HMI data is greatly enhanced relative to the original data and exhibit reasonable agreement with numerical simulations. Image restoration enhanced the apparent intensity and pixel averaged magnetic field strength of photospheric magnetic features significantly. For small-scale magnetic features, restoration enhanced intensity contrast in the continuum and core of the Fe I 6173 \AA{} line by a factor of 1.3, and the magnetogram signal by a factor of 1.7. For sunspots and pores, the enhancement varied strongly within and between features, being more acute for smaller features. Magnetic features are also rendered smaller, as signal smeared onto the surrounding quiet Sun is recovered. Image restoration increased the apparent amount of magnetic flux above the noise floor by a factor of about 1.2, most of the gain coming from the quiet Sun. Line-of-sight velocity due to granulation and supergranulation is enhanced by a factor of 1.4 to 2.1, depending on position on the solar disc. The shape of the solar aureole varied, with time and between the two CCDs. There are also indications that the PSF varies across the FOV. However, all these variations were found to be relatively small, such that a single PSF can be applied to HMI data from both CCDs, over the period examined without introducing significant error.

{\it Conclusions.} Restoring HMI observations with the PSF presented here returns a reasonable estimate of the stray light-free intensity contrast. Image restoration affects the measured radiant, magnetic and dynamic properties of solar surface phenomena sufficiently to significantly impact interpretation.

\section{Introduction}
\label{p2introduction}

Solar telescopes, like any real optical system, diverge from diffraction-limited behaviour due to optical aberrations and, in the case of ground-based instruments, the influence of the Earth's atmosphere. Optical aberrations arise from factors such as design and material constraints, imperfections in the fabrication, presence of impurities, thermal changes and jitter, and are practically unavoidable. Due to aperture diffraction and optical aberrations, radiation entering a given optical system is not entirely confined to the intended area on the focal plane but instead spread out, as described mathematically by the point spread function, PSF. This image blurring or loss of contrast is the so-termed stray light.

The apparent properties of solar phenomena is sensitive to stray light, accounting for its influence on solar imagery is necessary for proper interpretation and comparison with numerical models. This has been demonstrated repeatedly in the literature, for example, in studies looking at the limb darkening function \citep{pierce77,neckel94}, the intensity contrast of granulation \citep{sanchezcuberes00,danilovic08,wedemeyerbohm09,afram11}, and the brightness of small-scale magnetic concentrations \citep{title96,schnerr11} and sunspots \citep{albregtsen81,mathew07}.

Sophisticated techniques to correct solar observations for instrumental and atmospheric effects exist, the most common being speckle interferometry \citep{deboer92,vonderluhe93} and phase diversity \citep{gonsalves82,lofdahl94} methods. For spaceborne instruments, where variable atmospheric seeing is not a factor, a more conventional approach is often sufficient. Specifically, inferring the PSF from the distribution of intensity about the boundary between the bright and dark parts of partially illuminated scenes (such as, of the solar limb, and of transits of the Moon, Venus or Mercury) and restoring data by the deconvolution with it. Recent examples include the work of \cite{mathew07} with observations from SoHO/MDI\footnote{The Michelson Doppler Imager onboard the Solar and Heliospheric Observatory \citep{scherrer95}.}, \cite{wedemeyerbohm08}, \cite{wedemeyerbohm09} and \cite{mathew09} with Hinode/SOT/BFI\footnote{The Broadband Filter Imager of the Solar Optical Telescope onboard Hinode \citep{kosugi07}.}, and \cite{deforest09} with TRACE\footnote{The Transition Region And Coronal Explorer \citep{handy98}.} . 

In this paper we present an estimate of the PSF of the Helioseismic and Magnetic Imager onboard the Solar Dynamics Observatory, SDO/HMI \citep{schou12}. The PSF was derived from observations of Venus in transit. We also demonstrate the effects of correcting HMI data for stray light with this PSF on the apparent properties of various photospheric phenomena.

This study broadly follows the approach taken with the other spaceborne instruments listed above. It departs from these earlier efforts in that we constrain the PSF not only in the radial dimension but also in the azimuthal direction, recovering the anisotropy. This we will show to be crucial for accurate stray light removal (Sect. \ref{imagerestoration}).

The relationship between the radiance of magnetic features in the photosphere, and their size and position on the solar disc, is an important consideration in understanding and modelling the variation in solar irradiance \citep{domingo09}. HMI returns continuous, seeing-free, full-disc observations of intensity, Doppler shift and magnetic field at a constant, intermediate spatial resolution ($\sim1\:\rm{arcsec}$) and at relatively low noise. This renders it a suitable tool for constraining the radiant behaviour of photospheric magnetic features \citep{yeo13}. It is therefore of interest to enhance the quality of HMI observations by quantifying the stray light performance of the instrument. This would be of utility not only for the accurate examination of the radiant behaviour of magnetic features in HMI data but also any application that can benefit from stray light-free measurements of intensity, line-of-sight velocity and magnetic flux density.

The PSF presented here is, to our knowledge, the first on-orbit measurement of the stray light of HMI reported in the literature \citep[see][for the pre-launch measurement]{wachter12}. This is necessary given that the exact operating conditions of the sensor cannot be exactly simulated on the ground. Also, the stray light of the HMI might have changed from the time of the pre-launch calibration from changes in the condition of the instrument.

The HMI comprises of two identical $4096\times4096$ pixel CCD cameras, denoted `side' and `front'. The PSF was retrieved from images recorded on the side CCD during the transit of Venus on June 5 to 6, 2012. In addition to this transit of Venus, the HMI has also witnessed several partial lunar eclipses (seven, as of the end of 2012). In Sect. \ref{psl} we discuss the reasons for choosing the observations of Venus in transit over data from the partial lunar eclipses or of the solar limb for constraining the PSF, even though Venus has an atmosphere which had to be taken into account in deriving the PSF, introducing additional complexity to the task and uncertainty to the final estimate of the PSF.

In the following section, we detail the data selection (Sect. \ref{dataselection}), the PSF derivation (Sect. \ref{psfderivation}) and image restoration method (Sect. \ref{imagerestoration}), and how we accounted for the influence of Venus' atmosphere (Sect. \ref{venusatm}). Then, we verify the utility of the PSF presented here for image restoration, comparing the apparent granulation contrast in restored HMI observations and synthetic intensity maps generated from numerical simulation (Sect. \ref{gc}). We illustrate the result of image restoration on the intensity, Dopplergram and magnetogram data products of the instrument, looking at its effect on the following.
\begin{itemize}
	\item The intensity and magnetic field strength of small-scale magnetic concentrations (Sect. \ref{ssmcss1}).
	\item The intensity and magnetic field strength of sunspots and pores (Sect. \ref{ssmcss2}).
	\item The amount of magnetic flux on the solar surface (Sect. \ref{ssmcss3}).
	\item Line-of-sight velocity (Sect. \ref{ssmcss4}).
\end{itemize}
The retrieved PSF represents the stray light behaviour of the side CCD at the time of the transit of Venus, at the position in the field-of-view (FOV) occupied by the Venus disc. In Sect. \ref{psfdep} we examine the applicability of this PSF to other positions in the FOV, and to observations from the front CCD as well as from other times. Finally, a summary of the study is given in Sect. \ref{summary}.

\section{PSF derivation}
\label{p2method}

\subsection{Data selection}
\label{dataselection}

The HMI is a full-Stokes capable filtergram instrument. The instrument records full-disc polarimetric filtergrams continuously, at 3.75-s cadence, on the two identical CCDs. The filtergram sequence of the side CCD alternates between six polarizations ($I\pm{}Q$, $I\pm{}U$ and $I\pm{}V$) and six positions across the Fe I 6173 \AA{} line (at $\pm{}34$, $\pm{}103$ and $\pm{}172\:{\rm m\AA{}}$ from line centre). A set of 36 filtergrams, of each polarization at each line position, is collected every 135-s. The front CCD collects a set of 12 filtergrams, covering the Stokes $I+V$ and $I-V$ polarizations at the same line positions, every 45-s.

Dopplergrams, longitudinal magnetograms and intensity (continuum, and line depth and width) images, collectively termed the line-of-sight data products, are generated from the filtergram sequence of the side CCD at 720-s intervals, and from that of the front CCD at 45-s intervals. Stokes parameters (I, Q, U and V) and the corresponding Milne-Eddington inversion \citep{borrero11} are also produced, at 720-s cadence, from the filtergram sequence of the side CCD.

During the transit of Venus on June 5 to 6, 2012, the side CCD recorded filtergrams in the nearby continuum ($-344\:{\rm m\AA{}}$ from line centre), instead of the regular filtergram sequence. For the purpose of estimating the PSF of the instrument, we examined the $854\times854$ pixel crop, centred on the Venus disc, of 249 continuum filtergrams collected between second and third contact (i.e., the period the venusian disc was entirely within the solar disc). Care was taken to avoid filtergrams with pixels with spurious signal levels, a result of cosmic ray hits, on the venusian disc. The pixel scale was 0.504 arcsec/pixel. A summary description of this and the other HMI data employed in this study is given in Table \ref{hmidata}.

\begin{sidewaystable}
\caption{Summary description of the HMI data employed in this study and the sections in which their analysis is detailed.}
\label{hmidata}
\centering
\begin{tabularx}{\textwidth}{l>{\hsize=1.4\hsize}X>{\hsize=.6\hsize}Xl}
\hline\hline
Index & Description & UTC time of observation & Section(s)\\
\hline
1 & $854\times854$ pixel crop, centred on the venusian disc, of 249 continuum ($-344\:{\rm m\AA{}}$ from line centre) filtergrams recorded on the side CCD during the transit of Venus, between second and third contact. & Between 22:30, June 5 and 04:14, June 6, 2012. & \ref{dataselection}\\
2 & The mean of 42 of the 249 images in item 1, between which the spatial distribution of intensity on the venusian disc is relatively similar, termed the mean transit image. & Between 02:04 and 02:46, June 5, 2012. & \ref{dataselection}, \ref{psfderivation}, \ref{venusatm}\\
3 & One of the 42 images used to produce item 2, referred to as the test transit image. & 02:25:37, June 6, 2012. & \ref{imagerestoration}, \ref{venusatm}\\
4 & $854\times854$ pixel crop, centred on the venusian disc, of a continuum side CCD filtergram recorded just before second contact, denoted the ingress image. & 22:25:33, June 5, 2012. & \ref{venusatm}\\
5 & Continuum side CCD filtergram taken right after the venusian disc exited the solar disc completely, termed the test continuum filtergram. & 04:35:59, June 6, 2012. & \ref{gc}, \ref{psfdep1}\\
6 & 45-s longitudinal magnetogram from the front CCD closest in time ($<$ 1 minute) to item 5. & 04:35:22, June 6, 2012. & \ref{gc}, \ref{psfdep1}\\
7 & A set of simultaneous 720-s Dopplergram, longitudinal magnetogram, line depth and continuum intensity images from the side CCD, recorded about an hour after the end of the transit of Venus, when said CCD resumed collection of the regular filtergram sequence. & 05:35:32, June 6, 2012. & \ref{ssmcss}, \ref{psfdep1}\\
8 & A pair of filtergrams, one from each CCD, of similar bandpass ($-172\:{\rm m\AA{}}$ from line centre) and polarization (Stokes $I-V$), taken less than one minute apart of one another. & Around 05:36, June 6, 2012. & \ref{psfdep2}\\
9 & The continuum filtergram recorded on each CCD whenever the SDO spacecraft passes orbital noon and midnight. A total of 1866 filtergrams from each CCD, from when the HMI commenced regular operation to the time of the study. & Around 06:00 and 18:00 daily, between May 1, 2010 and June 30, 2013. & \ref{psfdep2}\\
\hline
\end{tabularx}
\end{sidewaystable}

When generating the various data products in the HMI data processing pipeline, filtergrams (from an interval of 1350-s for the 720-s data products and 270-s for the 45-s data products) are corrected for spatial distortion \citep{wachter12}, cosmic ray hits, polarization crosstalk \citep{schou12b} and solar rotation, and the filtergrams of similar polarizations averaged. These time-averaged filtergrams are then combined non-linearly to form the various data products \citep{couvidat12}. For the side CCD, these time-averaged filtergrams are outputted as the 720-s Stokes parameters product.

As a consequence of the correction for spatial distortion, the apparent PSF is different in the unprocessed and time-averaged filtergrams. The non-linearity of the algorithms used to derive the data products means they cannot be corrected for stray light by the deconvolution with the PSF. Instead, their restoration must go via restoring the unprocessed or time-averaged filtergrams.

The 249 continuum filtergrams considered (and all the other filtergram data utilised in the rest of the study) were corrected for spatial distortion. The retrieved PSF therefore represents stray light in undistorted HMI observations. This allows the generation of stray light-free data products through the deconvolution of the PSF from the time-averaged filtergrams. For the line-of-sight data products, this means correcting just the time-averaged Stokes $I+V$ and $I-V$ at each line position, a total of $2\times6=12$ images in each instance, instead of all the unprocessed filtergrams, which numbers 360 and 72 for the 720-s and 45-s data products respectively.

In this study we assumed Venus to be a perfect sphere \citep{archinal11}. Radial distance and azimuth are denoted $r$ and $\phi$, respectively. Azimuth is taken anti-clockwise from the CCD column axis such that zero is up.

The spatial distribution of intensity on the venusian disc, predominantly instrumental scattered light (aperture diffraction and stray light), varied significantly over the course of the transit (Fig. \ref{stable2}). The figure gives the intensity on the venusian disc in the 249 continuum filtergrams,
\begin{itemize}
	\item as a function of radial distance from the centre of the venusian disc, averaged over all azimuths and normalized to the level at the point of inflexion ($\ir$, top panel), and
	\item as a function of azimuth along the edge of the venusian disc as given by the point of inflexion on $\ir$, normalized to the mean level ($\ia$, bottom panel).
\end{itemize}
Also plotted are the mean $\ir$ and $\ia$ of all the filtergrams, $\irm$ and $\iam$ (red curves). The radius of the venusian disc as given by the point of inflexion on $\ir$ is, to 0.1 arcsec, constant at 29.5 arcsec. The fluctuation in the intensity on the venusian disc over the course of the transit arises from changes in the solar background from granulation, $p$-mode oscillations and limb darkening, as well as the variation of the PSF with position in the FOV.

\begin{cfig}
\includegraphics[width=.81\textwidth]{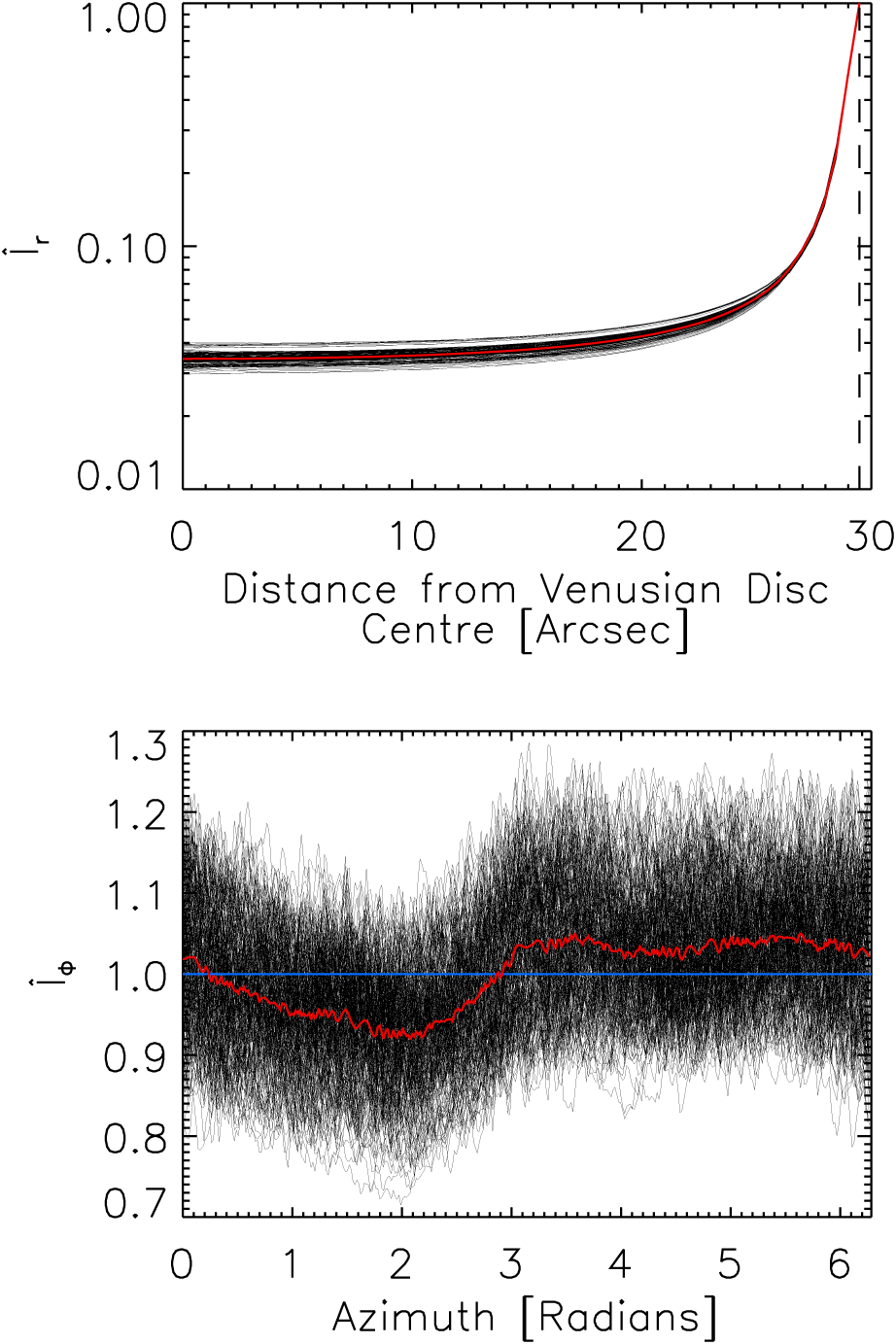}
\caption{Intensity on the venusian disc in the 249 continuum filtergrams from the transit of Venus. Top: Intensity as a function of distance from the centre of the venusian disc, averaged over all azimuths and normalized to the level at the point of inflexion (dashed line), $\ir$. Bottom: Intensity versus azimuth along the edge of the venusian disc as given by the point of inflexion on $\ir$, normalized to the mean level, $\ia$. The red curves follow the mean $\ir$ and $\ia$ of all the filtergrams, while the blue line represents $\ia=1$.}
\label{stable2}
\end{cfig}

To quantify the variation in the spatial distribution of intensity on the venusian disc over the course of the transit, and the influence of the changing solar background and the variation of the PSF with position in the FOV, we computed, for each of the 249 continuum filtergrams, the following two quantities.
\begin{itemize}
	\item The integral under $\ir$ from the centre of the venusian disc to the point of inflexion, $\sir$. The broader the PSF at a given position in the FOV, the brighter the venusian disc is relative to the level at its edge, and the greater this integral.
	\item The root-mean-square or RMS difference between $\ia$ and $\iam$, $\sia$. The closer the agreement in the trend with azimuth between $\ia$ and $\iam$, the smaller this quantity. $\sia$ reflects changes in the isotropy of the PSF (such as, from astigmatism and coma aberrations) and variation in the spatial distribution of intensity of the solar background.
\end{itemize}
$\sir$ and $\sia$ are plotted along the trajectory of the venusian disc, given in terms of the cosine of the heliocentric angle, $\mu=\cos\theta$ of the disc centre, as a function of time in Fig. \ref{stable1}.

\begin{cfig}
\includegraphics[width=.93\textwidth]{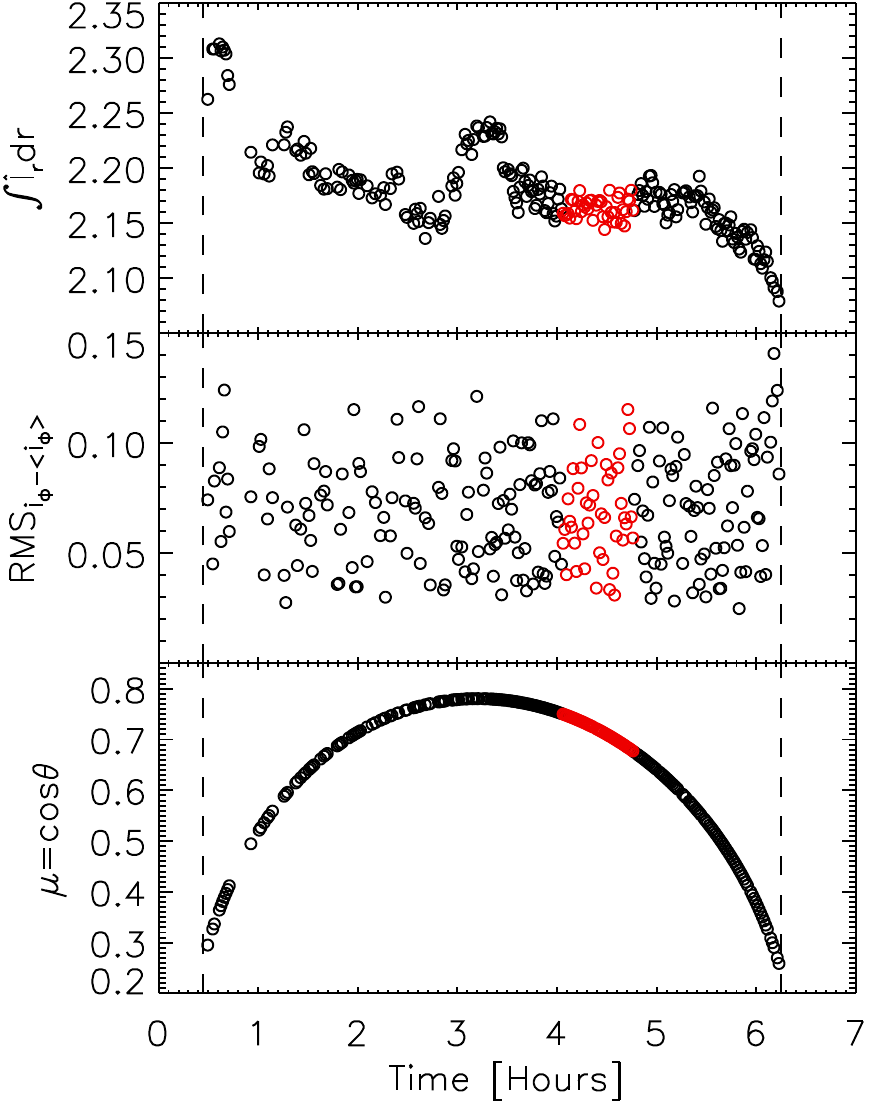}
\caption{The integral under $\ir$, $\sir$ (top), the RMS difference between $\ia$ and $\iam$, $\sia$ (middle) and the position of the centre of the venusian disc in terms of the cosine of the heliocentric angle, $\mu=\cos\theta$ (bottom) as a function of time from 22:00:00 UTC, June 5, 2012. The red circles highlight the values for the filtergrams used to derive the PSF. The dashed lines mark the approximate time of the second and third contacts; the period between which the venusian disc was entirely within the solar disc from SDO's position.}
\label{stable1}
\end{cfig}

$\sir$ changed slowly but notably over the course of the transit and the variation appears to be uncorrelated to distance from solar disc centre (and therefore, limb darkening). This suggests that the width of the PSF varies significantly along the path of Venus in the FOV. $\sia$ showed a marked point-to-point fluctuation but is otherwise relatively low ($\lesssim0.1$) and exhibits no obvious trend with time. This indicates that the azimuth dependence of $\ia$ and $\iam$ is mainly driven by a persistent PSF anisotropy along the path of Venus in the FOV rather than limb darkening (see also Sect. \ref{psfderivation}). The marked point-to-point fluctuation is likely from granulation and $p$-mode oscillations. This is supported by the comparatively smooth time variation of $\sir$, which is less sensitive to these small-scale intensity fluctuations in the solar background due to the averaging over all azimuths.

From the trend of $\sir$ and $\sia$ with time, we surmise that the width of the PSF varies significantly along the path of Venus in the FOV and while the anisotropy of the PSF is relatively stable, it is obscured by granulation and $p$-mode oscillations in individual filtergrams. Based on these considerations, we retained just the 42 filtergrams highlighted in Fig. \ref{stable1} (red circles), taken over a 40-minute period where $\sir$ is relatively stable, for the derivation of the PSF. The selected filtergrams were aligned by the venusian disc and the average, hereafter referred to as the mean transit image, taken. The objective is to derive an image of Venus in transit where the influence of granulation and $p$-mode oscillations is minimal by averaging filtergrams where the PSF at Venus' position in the FOV is fairly similar.

The 249 continuum filtergrams examined alternated between four polarizations (Stokes $I+Q$, $I-Q$, $I+U$ and $I-U$). We found no systematic differences in the spatial distribution of intensity on the venusian disc between filtergrams of different polarizations and so made no distinction between them here. We repeated the analysis described in this subsection on the filtergrams recorded at the six regular wavelength positions, on the front CCD during the transit of Venus. There are systematic differences (in $\sir$ and $\sia$) between different positions, even between positions at similar distance from but opposite sides of the line centre. This suggests spectral line changes from effects unrelated to the stray light behaviour of the instrument may exert an influence on intensity measured on the venusian disc. Specifically, Doppler shifts from the motion of the spacecraft (SDO is in a geosynchronous orbit) and the rotation of the Sun. For this reason we restricted ourselves to the continuum filtergrams from the side CCD.

\subsection{PSF derivation method}
\label{psfderivation}

A bivariate polynomial function was fitted to the mean transit image. We excluded the circular area of 50 arcsec radii centred on the venusian disc (about three times the area of the venusian disc) from the regression. The extrapolation of the surface fit over this excluded area represents an estimate of the intensity if Venus had been absent. We filled a circular area in the surface fit, corresponding to the venusian disc, with zeroes. We will refer to the result, essentially a model of the mean transit image in the absence of an atmosphere in Venus, aperture diffraction and stray light, as the artificial image.

The PSF was determined by minimizing the chi-square between the convolution of the artificial image with a guess PSF and the mean transit image, in the circular area of 50 arcsec radii centred on the venusian disc. For this we employed the implementation of the Levenberg-Marquardt algorithm (LMA) included in the IDL Astronomy User's Library, mpfit2dfun.pro.

Intensity in the circular area was sampled at equal intervals in the radial (0.504 arcsec, the pixel scale) and azimuthal (1/360 radians) dimensions. This is to give intensity measured at each radius from the centre of the venusian disc more equal weight in the LMA optimization. The result is a closer agreement between the convolution of the artificial image with the guess PSF and the mean transit image on convergence than achieved by comparing the circular area in the artificial and mean transit images directly.

We scaled the artificial image by a factor prior to convolution with the guess PSF. We allowed this factor and the radius of the disc of zeroes in the artificial image to be free parameters in the LMA optimization, taking an initial value of unity (i.e., no scaling) and 29.5 arcsec (the position of the point of inflexion on $\ir$). This is to minimize error from any misestimation of the surface fit to the mean transit image and the radius of the venusian disc\footnote{The point of inflexion on $\ir$ is not an accurate indication of the position of the edge of the venusian disc due to the influence of the venusian atmosphere.}.

Following the example of previous studies \citep{martinezpillet92,mathew07,mathew09}, we attempted to model the guess PSF as the linear sum of various combinations of Gaussian and Lorentzian functions. Except here we allowed the amplitude of each Gaussian and Lorentzian component to vary sinusoidally with azimuth to accommodate PSF anisotropy. We also tried to set the ideal diffraction-limited PSF as one of the components. The guess PSF we found to reproduce the measured intensity in the artificial image best, denoted $K$, is given by
\begin{equation}
K(r,\phi)=\sum^{5}_{i=1}\left[1+A_i{\rm cos}(u_i\phi+v_i)\right]w_i\left[\frac{1}{2\pi\sigma^2_i}{\rm exp}\left(-\frac{r^2}{2\sigma^2_i}\right)\right].
\label{aiso5g}
\end{equation}
That is, the linear combination of five Gaussian functions, with weight $w_i$ and standard deviation $\sigma_i$, the amplitudes of which vary sinusoidally with azimuth, with amplitude $A_i$, period of $2\pi{}u_{i}$ radians (where $u_{i}\in\mathbb{Z}$) and phase $v_i$.

Modelling the guess PSF as the linear combination of four Gaussian or three Gaussian and a Lorentzian, as done in the cited works, still reproduces measured intensity in the artificial image reasonably well. The retrieved PSFs are however, negative at parts (i.e., unphysical) from the LMA converging to solutions where $|A_i|>1$. And introducing additional sinusoidal terms to the azimuth dependence of each Gaussian and Lorentzian function did not alleviate this problem. The linear combination of five Gaussians appeared necessary to reach a physical solution while reproducing the measured intensity in the artificial image in both the radial and azimuthal dimensions.

The guess PSF was applied to the artificial image by evaluating $K$ (Eqn. \ref{aiso5g}) at pixel scale intervals (0.504 arcsec) on a $251\times251$ grid, the centre element representing the origin ($r=0$), and taking the convolution of the artificial image with the result. On convergence, the value at each grid element represents the integral of the PSF over the element. The retrieved PSF therefore describes the pixel integrated PSF. This was done, instead of filling the grid with the pixel integrated value of $K$, for a practical reason. When correcting HMI observations for stray light via deconvolution with the PSF, it is the pixel integrated PSF that is required.

Care was taken to repeat the LMA optimization, varying the initial value of the free parameters, to reduce the likelihood that the solution lies in a local chi-square minimum. To accommodate the requirement that $u_{i}\in\mathbb{Z}$, we executed the LMA optimization with no constraint on the value of $u_{i}$, rounded the retrieved $u_{i}$ to the nearest integer and repeated the process with these parameters fixed.

The PSF derivation method described here implicitly assumes there is no interaction between solar radiation and the venusian atmosphere. We will qualify this statement, and detail the adjustments made to the artificial image and the mean transit image to account for the influence of the venusian atmosphere on the retrieved PSF in Sect. \ref{venusatm}. The method also ignores motion blurring from the lateral movement of the venusian disc relative to the solar disc. The displacement of the venusian disc within the exposure time of the instrument is, on average, about 0.015 pixels and can therefore be neglected without significant loss of accuracy.

In the following, we denote the PSF retrieved as described above, a preliminary estimate of the stray light behaviour of the instrument, by $\psfone$. The retrieved value of the parameters of $\psfone$ are listed in Table. \ref{aiso5gp}. The best fit value of the scale factor applied to the artificial image is 1.0029, and the radius of the disc of zeroes, 29.29 arcsec. Though only a slight departure from the initial values (unity and 29.5 arcsec), this correction to the amplitude of the artificial image and the size of the disc of zeroes effected a marked improvement in the chi-square statistic.

\begin{sidewaystable}
\caption{Parameter values (and associated formal regression error) of the guess PSF retrieved neglecting ($\psfone$) and accounting for the influence of the venusian atmosphere ($\psftwo$).}
\label{aiso5gp}
\centering
\begin{tabular}{lcccccc}
\hline\hline
PSF & Gaussian component & $w_i$ & $\sigma_i$ [Arcsec] & $A_i$ & $u_i$ & $v_i$ [Radians] \\
\hline
$\psfone$ & $i=1$ & $0.641\pm0.002$ & $0.470\pm0.001$ & $0.131\pm0.002$ & 1 & $-1.85\pm0.02$ \\
& $i=2$ & $0.211\pm0.002$ & $1.155\pm0.008$ & $0.371\pm0.006$ & 1 & $2.62\pm0.01$ \\
& $i=3$ & $0.066\pm0.002$ & $2.09\pm0.02$ & $0.54\pm0.01$ & 2 & $-2.34\pm0.01$ \\
& $i=4$ & $0.0467\pm0.0005$ & $4.42\pm0.02$ & $0.781\pm0.006$ & 1 & $1.255\pm0.004$ \\
& $i=5$ & $0.035\pm0.004$ & $25.77\pm0.04$ & $0.115\pm0.001$ & 1 & $2.58\pm0.01$ \\
\hline
$\psftwo$ & $i=1$ & $0.747\pm0.001$ & $0.417$ & $0.164\pm0.002$ & 1 & $-2.22\pm0.01$ \\
& $i=2$ & $0.126\pm0.003$ & $1.45\pm0.01$ & $0.48\pm0.01$ & 1 & $2.36\pm0.01$ \\
& $i=3$ & $0.049\pm0.003$ & $2.10\pm0.02$ & $0.74\pm0.04$ & 2 & $-2.36\pm0.01$ \\
& $i=4$ & $0.0428\pm0.0004$ & $4.66\pm0.02$ & $0.776\pm0.007$ & 1 & $1.194\pm0.006$ \\
& $i=5$ & $0.035\pm0.004$ & $26.16\pm0.05$ & $0.122\pm0.002$ & 1 & $2.63\pm0.01$ \\
\hline
\multicolumn{7}{p{.7\textwidth}}{\textbf{Notes.} The PSFs are given by the linear combination of five Gaussian functions (Eqn. \ref{aiso5g}), denoted by $i$, with weight $w_i$ and listed in ascending order by the standard deviation, $\sigma_i$. The amplitude of each Gaussian component modulates sinusoidally with azimuth, with amplitude $A_i$, period of $2\pi{}u_i$ radians (where $u_{i}\in\mathbb{Z}$) and phase $v_i$. There are no associated formal regression errors for $u_i$, and in the case of $\psftwo$, $\sigma_{i=1}$ as the value of these parameters were fixed in the LMA optimization (see text).} \\
\end{tabular}
\end{sidewaystable}

In Fig. \ref{aisofit1} we plot the intensity along different radii from the centre of the venusian disc; from the mean transit image (black curves) and reproduced in the artificial image by the convolution with $\psfone$ (red curves). Also plotted is the intensity reproduced in the artificial image by fixing the $A_i$ at zero (blue curves). In this instance, the variation with azimuth arises solely from limb darkening, which enters the process through the surface fit to the mean transit image. Evidently, limb darkening alone cannot account for all the observed variation with azimuth, confirming that the PSF of the instrument is significantly anisotropic. By allowing the amplitude of each Gaussian component in the guess PSF to vary sinusoidally with azimuth, we are able to reproduce most of the observed intensity azimuth dependence. The close overall agreement between observed and reproduced intensities in the radial dimension is illustrated in Fig. \ref{aisofit2}.

\begin{cfig}
\includegraphics[width=.93\textwidth]{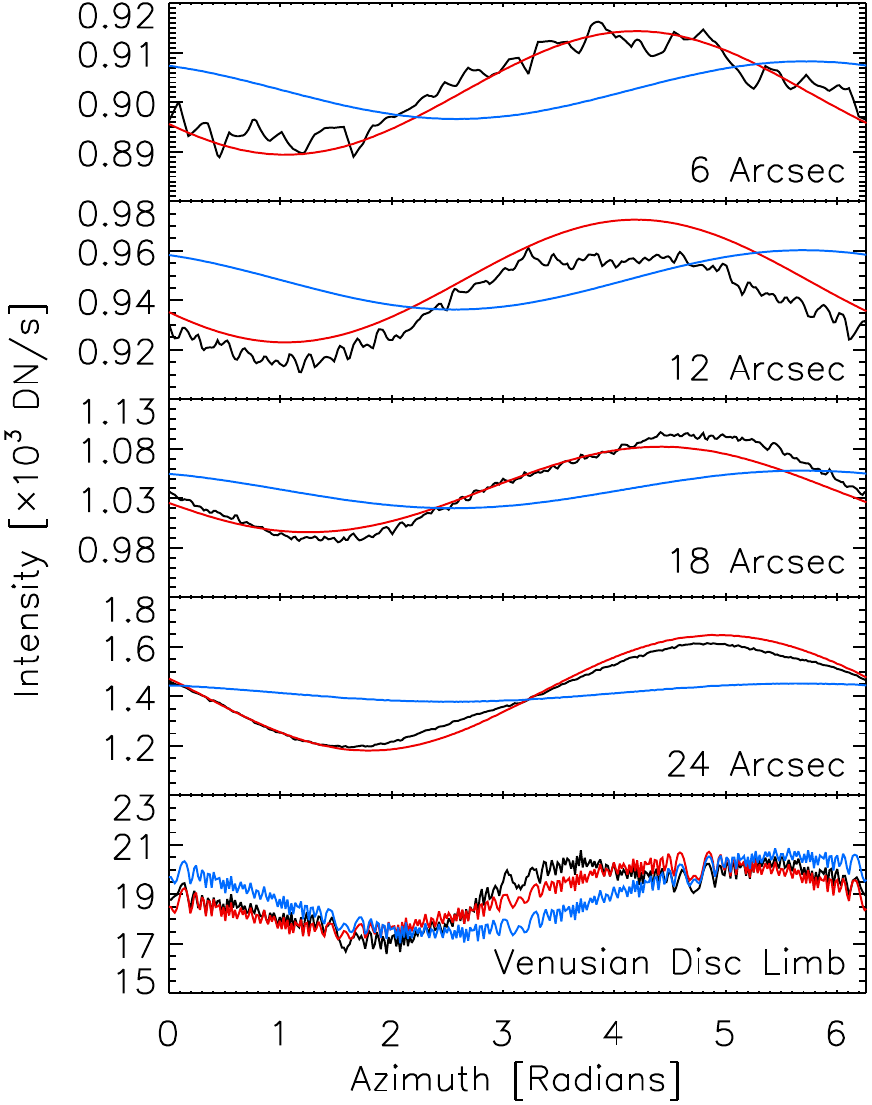}
\caption{Intensity on the venusian disc at distances of 6, 12 18 and 24 arcsec from the centre, and along its limb (taking the radius of the venusian disc retrieved along with $\psfone$ by the LMA optimization, 29.29 arcsec), as a function of azimuth. The black curves follow the values from the mean transit image and the red curves the values reproduced in the artificial image by the convolution with $\psfone$. The blue curves represent the intensity obtained in the artificial image by fixing $A_i$ at zero. The intensity fluctuations along the venusian limb arises from aliasing artefacts.}
\label{aisofit1}
\end{cfig}

\begin{cfig}
\includegraphics{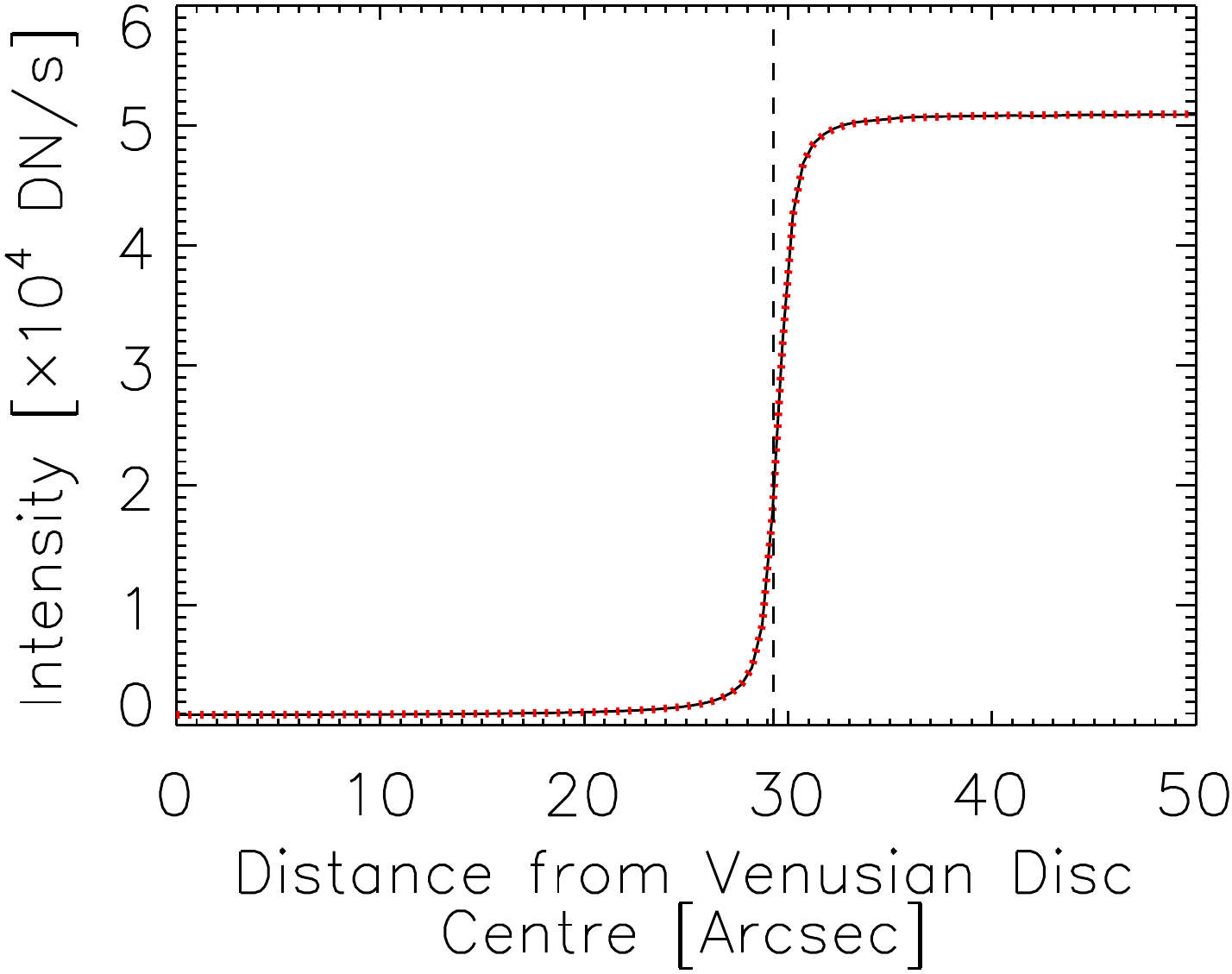}
\caption{Radial intensity as a function of distance from the centre of the venusian disc; from the mean transit image (black solid curve) and the convolution of the artificial image with $\psfone$ (red dotted curve). The dashed line marks the position of the Venus limb as returned along with $\psfone$ by the LMA optimization.}
\label{aisofit2}
\end{cfig}

Here we choose to describe the PSF as the linear combination of five Gaussian functions over more physically realistic models, such as the convolution of the diffraction-limited PSF with a Voigt function \citep{wedemeyerbohm08}. Apart from yielding a closer agreement between the PSF-blurred artificial image and the mean transit image, this functional form is more amenable to incorporating the complex azimuthal dependence. The retrieved parameter values of $\psfone$ (Table \ref{aiso5gp}) and the azimuth dependence of measured intensity on the venusian disc at different radii (Fig. \ref{aisofit1}) suggest that the overall amplitude and phase of the anisotropy of the PSF varies with radial distance.

The linear combination of Gaussian functions is not a physically realistic model of real PSFs for the following reasons:
\begin{itemize}
	\item It allows solutions with Strehl ratios exceeding unity, which is unphysical.
	\item The Fourier transform of the Gaussian function, and therefore the modulation transfer function (MTF) of such PSF models, is non-zero above the Nyquist limit. Correcting observations for stray light by the deconvolution with such a PSF can introduce aliasing artefacts from the enhancement of spatial frequencies above the Nyquist limit.
\end{itemize}
We will address these two potential issues in Sects. \ref{venusatm} and \ref{gc}, respectively.

The approach taken here to derive the PSF is broadly similar to that applied to images of Mercury in transit from Hinode/SOT/BFI by \cite{mathew09}. Specifically, by minimising the difference between observed intensity and that produced in a model of the aperture diffraction and stray light-free image (termed here the artificial image) by the convolution with a guess PSF. There are two significant departures.

Firstly, in this study, the artificial image is given by the surface fit to the mean transit image, with a disc of zeroes representing the venusian disc. In the cited work, the authors filled the mercurian disc in the recorded image with zeroes and took the result as the artificial image. As stated earlier in this subsection, having excluded the venusian disc and surroundings in the regression, the surface fit to the mean transit image is, in this excluded region, an approximation of the intensity had Venus been absent. For this we consider the approach taken here to yield a more realistic model of the instrumental scattered light-free image.

Secondly, as mentioned in the introduction, while the earlier effort assumed an isotropic form to the PSF, here we allowed the PSF to vary with azimuth. We were motivated by the observation that the stray light behaviour of the instrument is evidently anisotropic (Sect. \ref{dataselection} and Fig. \ref{aisofit1}).

\subsection{Image restoration method}
\label{imagerestoration}

To correct HMI observations for aperture diffraction and stray light, we utilised the implementation of the Richardson-Lucy algorithm, RLA \citep{richardson72,lucy74}, included in the IDL Astronomy User's Library, max\_likelihood.pro.

The RLA is an iterative method for restoring an image blurred by a known PSF, in our study, the guess PSF, $K$. Let $f_{k}$ denote the estimate of the restored image from the $k$-th iteration, $f_{k+1}$ is given by
\begin{equation}
f_{k+1}=f_{k}\circ\left(\left(f_{k}\ast{}K\right)\star{}K\right),
\label{rla}
\end{equation}
where the $\circ$, $\ast$ and $\star$ symbols represent the pixel-by-pixel product, convolution and correlation, respectively. The method has been shown, empirically, for data obeying Poisson statistics, to converge towards the maximum likelihood solution \citep{shepp82}. Following \cite{mathew09}, we employed a threshold for the chi-square between the original image and $f_{k}\ast{}K$ as the stopping rule. Here we set the threshold at $99.99\%$ confidence level.

In Figs. \ref{dim1} and \ref{dim2}a we show the result of restoring one of the 42 continuum filtergrams averaged to yield the mean transit image (recorded at 02:25:37 UTC, June 6, 2012), hereafter referred to as the test transit image, with $\psfone$. The image restoration sharpened the test transit image considerably and removed most of the signal on the venusian disc. The restoration however, also left a ringing artefact around the venusian disc; manifest as the bright halo in the grey scale plot (middle panel, Fig. \ref{dim1}) and the peak in the radial intensity profile (black solid curve, Fig. \ref{dim2}a). Restoring other observations taken nearby in time (within a few hours of the test transit image), we found similar artefacts in the boundary of active region faculae, and sunspot penumbra and umbra.

\begin{cfig}
\includegraphics[width=\textwidth]{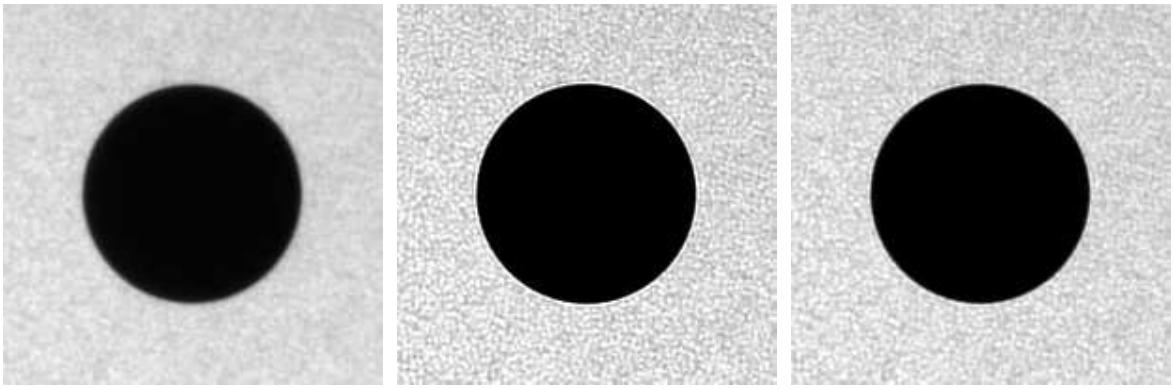}
\caption{$201\times201$ pixel inset, centred on Venus, of the test transit image; before (left) and after image restoration with $\psfone$ (middle), and with $\psftwo$ (right). The three grey scale plots are saturated at $6\times10^4\:\rm{DN/s}$, about $120\%$ of the mean photospheric level.}
\label{dim1}
\end{cfig}

\begin{cfig}
\includegraphics[width=\textwidth]{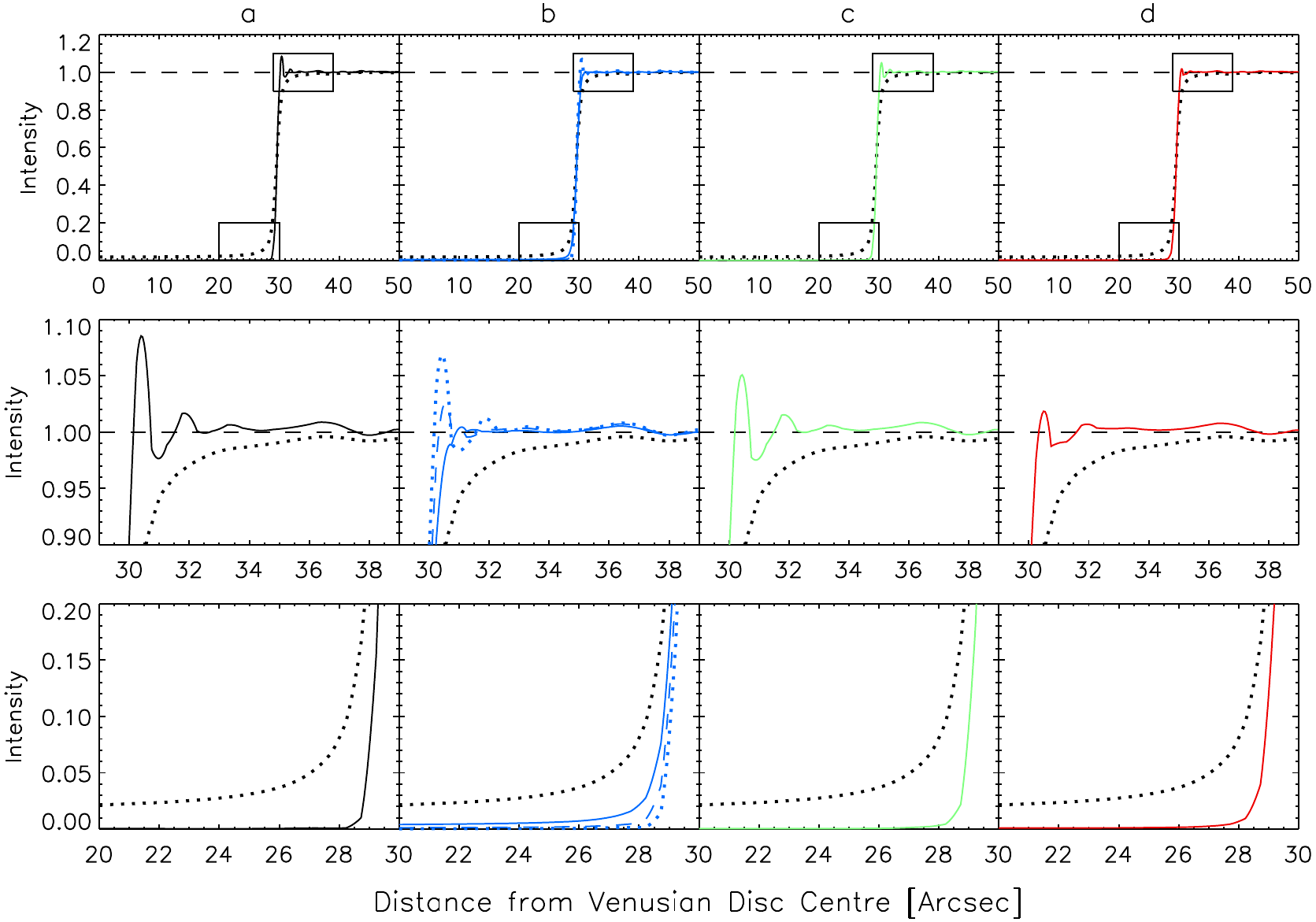}
\caption{Top: Radial intensity in the test transit image, corrected for limb darkening, as a function of distance from the centre of the venusian disc before (black dotted curves) and after image restoration with the various PSF estimates. a) With $\psfone$ (black solid curve). b) With the PSFs obtained by blurring the edge of the disc of zeroes in the artificial image with kernels representing Gaussian functions with standard deviations of 0.2, 0.3 and 0.4 arcsec (blue dotted, dashed and solid curves). c) With the PSF retrieved by subtracting the estimated aureole intensity from the mean transit image (green solid curve). d) With $\psftwo$ (red solid curve). Middle and bottom: Blow-up insets of the boxed areas. The radius of the venusian disc, a free parameter in the LMA optimization, is in all instances about 29.3 arcsec and not marked to avoid cluttering. The horizontal dashed lines follow unit intensity. The test transit image was corrected for limb darkening by normalizing it by the surface fit, computed as done for the mean transit image in Sect. \ref{psfderivation}.}
\label{dim2}
\end{cfig}

\cite{mathew09} in the similar study with images of Mercury in transit from Hinode/SOT/BFI noted similar ringing artefacts around the mercurian disc upon image restoration with the RLA. The authors attributed it to Gibb's phenomenon, ringing artefacts in the Fourier series representation of discontinuous signals. In the PSF derivation and image restoration process described here, discrete Fourier transforms, DFTs were executed in convolution and correlation computations. We found that repeating the derivation of $\psfone$ and the restoration of the test transit image without performing any DFTs in the convolution and correlation computations had negligible effect on the ringing artefact, ruling out Gibb's phenomenon as the main cause in this instance. In the following subsection we will demonstrate the ringing artefact found here to arise from us not taking the interaction between solar radiation and the venusian atmosphere into account in deriving $\psfone$.

\begin{cfig}
\includegraphics[width=\textwidth]{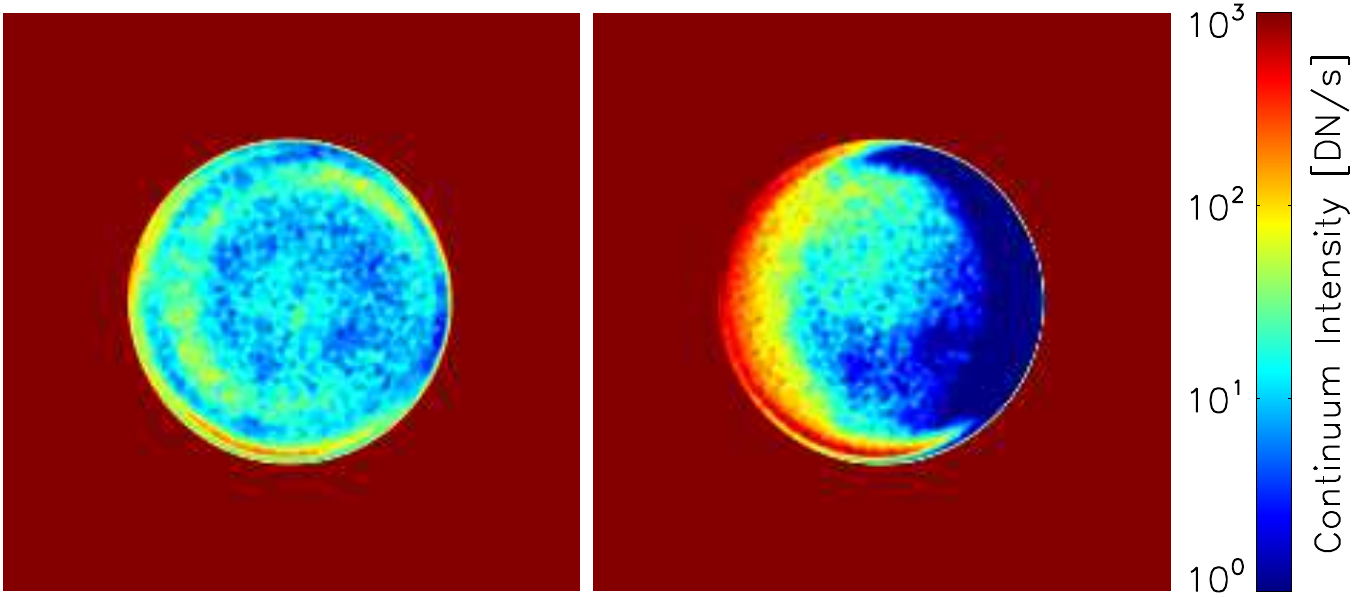}
\caption{$201\times201$ pixel inset, centred on Venus, of the test transit image after image restoration with $\psfone$ (left) and the same, except with $A_{i}$ set to zero (right).}
\label{dim4}
\end{cfig}

Figure \ref{dim4} is a colour scale plot of the venusian disc in the restored test transit image (left panel). The plot is saturated at about $2\%$ of the mean photospheric level to reveal the spatial distribution of residual intensity (instrumental scattered light not removed by the image restoration) on the venusian disc. Also shown is the result of restoring the test transit image with $\psfone$ excluding the anisotropy of the PSF by setting $A_{i}$ at zero (right panel). There is a gross, broadly east-west graduation of the residual intensity in the latter, not apparent in the former, where the residual intensity level is significantly more uniform across the venusian disc. This demonstrates the necessity to constrain the anisotropy of the PSF to properly correct HMI observations for instrumental scattered light.

\subsection{Interaction between solar radiation and the venusian atmosphere}
\label{venusatm}

As stated in Sect. \ref{psfderivation}, the PSF derivation method described so far builds on the assumption that there is no interaction between solar radiation and the venusian atmosphere.

In representing the venusian disc as a disc of zeroes in the artificial image, we have presumed that the body would, in the absence of aperture diffraction and stray light, be completely dark and exhibit a discrete edge. Diffusion and scattering of solar radiation in the venusian atmosphere can, however, render the edge of the venusian disc diffused.

The PSF is retrieved from matching the mean transit image to the convolution of the artificial image and the guess PSF. This is valid if all measured intensity came directly from the Sun. This is, however, not the case; there is a bright halo around the venusian disc when it is in transit (termed the aureole) from the refraction of solar radiation by the upper layers of the atmosphere towards the observer. 

\subsubsection{Diffusion and scattering of solar radiation in the venusian atmosphere}
\label{venusatm1}

To elucidate the influence of diffusion and scattering of solar radiation in the venusian atmosphere on the retrieved PSF we repeated the derivation, approximating the action of diffusion and scattering by blurring the edge of the disc of zeroes in the artificial image prior to the convolution with the guess PSF. We generated a copy of the artificial image that is unity everywhere outside the venusian disc and zero inside, convolved it with a Gaussian kernel, and scaled the original artificial image by the result. This procedure introduces Gaussian blur that is confined to near the edge of the disc of zeroes. We repeated the derivation of the PSF with different degrees of Gaussian blurring.

In Fig. \ref{dim3} (top panel) we display the PSFs retrieved after blurring the edge of the disc of zeroes with kernels representing Gaussian functions with standard deviations of 0.2, 0.3 and 0.4 arcsec (blue dotted, dashed and solid curves) along $\psfone$ (black dashed curve). In Fig. \ref{dim2}b we have the radial intensity profile of the test transit image before and after image restoration with these PSFs. The stronger the blurring, the narrower the core of the PSF and the weaker the ringing artefact. The narrowest Gaussian kernel (0.2 arcsec) returned a PSF that is still very similar to $\psfone$ while the broadest (0.4 arcsec) yielded a PSF that is unphysical, significantly narrower at the core than the ideal diffraction-limited PSF. As the Gaussian blurring is confined to near the edge of the disc of zeroes, the retrieved PSFs do not differ significantly from $\psfone$ beyond a few arcseconds from the centre of the PSF.

\begin{cfig}
\includegraphics[width=.7\textwidth]{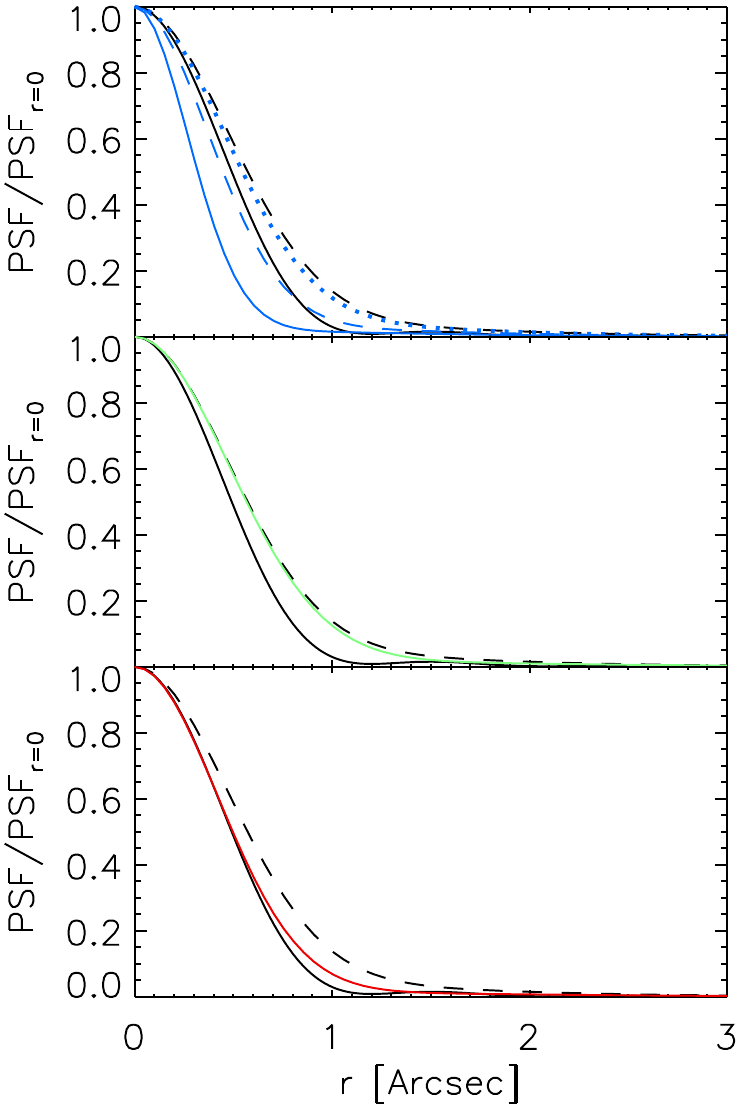}
\caption{Comparison between $\psfone$ (black dashed curves), the ideal diffraction-limited PSF (black solid curves) and the other retrieved PSFs. The blue curves (top) represent the PSFs retrieved after blurring the edge of the disc of zeroes in the artificial image by kernels representing Gaussian functions with standard deviations of 0.2 (dotted), 0.3 (dashed) and 0.4 arcsec (solid). The green curve (middle) corresponds to the PSF found by subtracting the estimated aureole intensity from the mean transit image. The red curve (bottom) corresponds to $\psftwo$. For illustration purposes, we set the value of $A_{i}$ at zero (i.e., ignoring the azimuthal dependence) and normalized each PSF to the level at $r=0$. As the retrieved PSFs represent estimates of the pixel integrated true PSF, the ideal diffraction-limited PSF was smoothed with a box function of pixel scale width to allow a direct comparison.}
\label{dim3}
\end{cfig}

\subsubsection{Refraction of solar radiation in the venusian atmosphere}
\label{venusatm2}

We estimated the contribution by the aureole to apparent intensity in the mean transit image. For this purpose we examined the $854\times854$ pixel crop, centred on the venusian disc, of a continuum filtergram taken shortly ($\sim10$ seconds) before the venusian disc moved completely into the solar disc (recorded at 22:25:33 UTC, June 5, 2012), hereafter referred to as the ingress image. The ingress image is expressed as a grey scale plot in Fig. \ref{aureole1}.

The aureole is only directly observable at ingress and egress (i.e., when the venusian disc is only partially within the solar disc), in the part of the venusian disc outside the solar disc. This is because the aureole is much dimmer than the photosphere and therefore difficult to distinguish from direct solar radiation. Generally, the intensity of the aureole increases with the proportion of the venusian disc sitting inside the solar disc \citep{tanga12}. Therefore, observations taken right before second contact (such as the ingress image) or right after third contact give the closest direct indication of the intensity of the aureole when the venusian disc is entirely within the solar disc. The intensity of the aureole also varies with azimuth. This is, at least in part, because it is modulated by the spatial distribution of photospheric intensity \citep{tanga12} and variation in the physical structure of the venusian atmosphere with latitude \citep{pasachoff11}.

Here we looked at the intensity of the aureole in the ingress image over the minor sector marked in Fig. \ref{aureole1} (blue lines), where it is relatively stable with azimuth.

\begin{cfig}
\includegraphics[width=.89\textwidth]{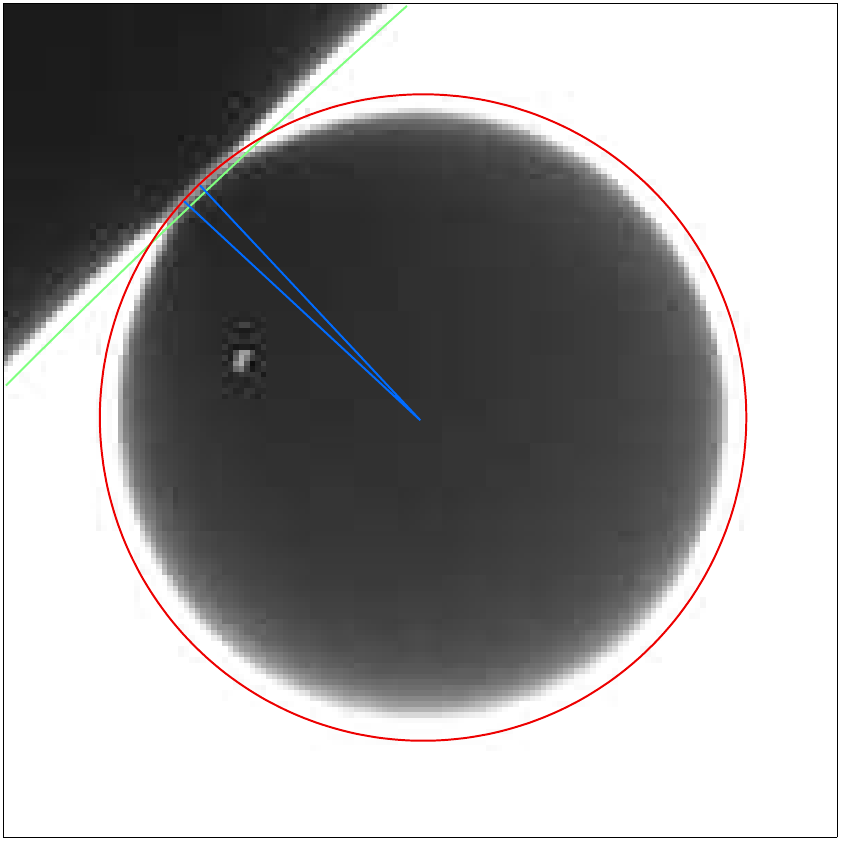}
\caption{$151\times151$ pixel inset, centred on Venus, of the ingress image. The edge of the venusian disc (as given by the point of inflexion on $\ir$) and of the solar disc are indicated by the red and green contours respectively. They do not coincide with the apparent boundaries due to the low grey scale saturation level. The grey scale is saturated at $3000\:\rm{DN/s}$ ($\sim5\%$ of the mean photospheric level at disc centre) to allow the aureole, the bright arc on the part of the venusian disc outside the solar disc, which is much dimmer than the solar disc, to be visible. The blue lines mark the minor sector within which the intensity of the aureole is relatively stable with azimuth. The bright feature on the northwest quadrant of the venusian disc is an artefact of cosmic ray hits on the CCD.}
\label{aureole1}
\end{cfig}

The radial intensity profile over the minor sector marked in Fig. \ref{aureole1} is plotted in Fig. \ref{aureole2} (circles, top panel). The peak near the edge of the venusian disc (dashed line) corresponds to the aureole while the slowly varying background is largely instrumental scattered light from the solar disc. We subtracted the polynomial fit to the background (red curve) from the radial intensity profile. To the background-subtracted radial intensity profile (circles, bottom panel) we fit the linear combination of two Gaussian functions (blue curve). We then scaled this fit by the quotient of the integrated photospheric intensity behind the venusian disc in the mean transit image and in the ingress image\footnote{The intensity of the photosphere behind the venusian disc in the mean transit image is given by the surface fit described in Sect. \ref{psfderivation}. For the ingress image, we binned the image pixels on the solar disc by $\mu$, excluding the venusian disc and surroundings, and took the bin-averaged intensity. The intensity behind the venusian disc was then estimated from the polynomial fit to these bin-averaged intensities given the $\mu$ of each image pixel within the venusian disc.}. The result (green curve) represents an estimate of the radial intensity profile of the aureole in the mean transit image.

\begin{cfig}
\includegraphics[width=.9\textwidth]{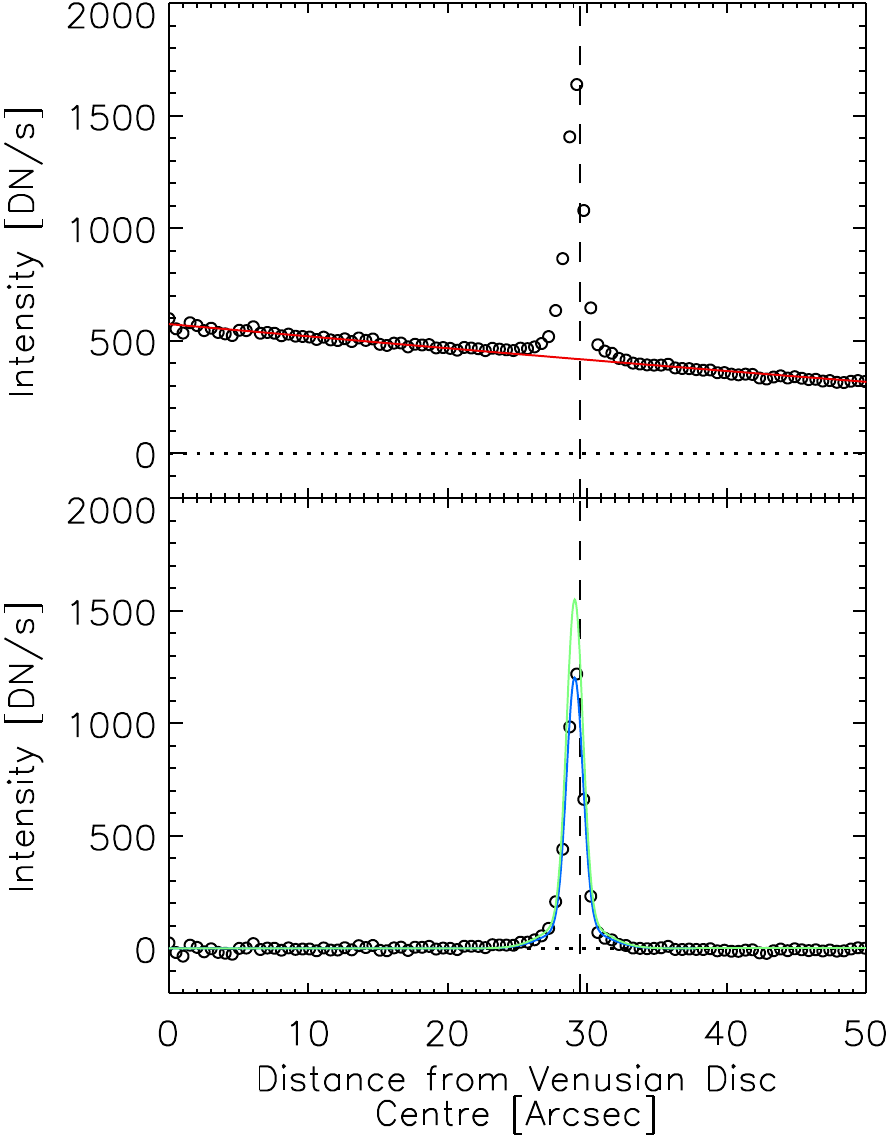}
\caption{Intensity in the ingress image, averaged over the minor sector marked in Fig. \ref{aureole1}, as a function of distance from the centre of the venusian disc (circles); before (top panel) and after (bottom panel) subtracting the polynomial fit to the slowly varying background (red curve). The blue curve corresponds to the sum-of-two-Gaussians fit to the background-subtracted series, while the green curve is the same after scaling by the quotient of the total photospheric intensity behind the venusian disc in the mean transit image and in the ingress image. The black dashed and dotted lines denote the position of the Venus limb (as given by the point of inflexion on $\ir$) and the zero intensity level, respectively.}
\label{aureole2}
\end{cfig}

In Fig. \ref{dim3} (middle panel) we compare the PSF retrieved after first subtracting the estimated radial intensity profile of the aureole from the mean transit image (green curve) with $\psfone$ (black dashed curve). In Fig. \ref{dim2}c we have the radial intensity profile in the test transit image, before and after image restoration with this PSF. The effect of removing the contribution by the aureole to observed intensity on the retrieved PSF and the ringing artefact in the restored test transit image is similar as that from introducing Gaussian blur to the edge of the disc of zeroes in the artificial image. The retrieved PSF is slightly narrower than $\psfone$ at the core. The ringing artefact in the restored test transit image is slightly weaker. As the aureole is concentrated near the edge of the venusian disc, removing it from the mean transit image made little difference to the retrieved PSF beyond a few arcseconds from the core.

In removing the contribution of the aureole from the mean transit image as described above, we have made two simplifying assumptions:
\begin{itemize}
	\item One, that the intensity of the aureole is directly proportional to the integrated photospheric intensity behind the venusian disc.
	\item Two, that the intensity of the aureole does not change with azimuth.
\end{itemize}
\cite{tanga12} recently published a model of aureole intensity, relating it to the spatial distribution of photospheric intensity and physical structure of the venusian atmosphere. This is, to our knowledge, the only model of its kind reported in the literature. Given the fact that the aureole is blurred by instrumental scattered light, the uncertainties over the structure of the venusian atmosphere, and in the interest of simplicity, we favoured the rather approximate approach taken here over a more rigorous computation based on the model of \cite{tanga12}. The estimated peak intensity of the aureole in the mean transit image is about 1500 DN/s (blue curve, Fig. \ref{aureole2}), much smaller than the photospheric level ($\sim5\times10^4 \:\rm{DN/s}$, Fig. \ref{aisofit2}). Taking into account this as well as the relatively minor effect of subtracting the radial intensity profile of the aureole from the mean transit image on the retrieved PSF, we surmise that the uncertainty introduced by the two assumptions listed is likely minimal.

By repeating the derivation of the PSF and restoration of the test transit image, first blurring the edge of the disc of zeroes in the artificial image or subtracting the estimated contribution by the aureole to the mean transit image, we have demonstrated that the interaction between solar radiation and the venusian atmosphere has a palpable impact on the width of the core of the retrieved PSF. Both adjustments yielded PSFs that were narrower at the core compared to $\psfone$ (Fig. \ref{dim3}). And the narrower the core of the PSF, the weaker the ringing artefact in the restored test transit image (Fig. \ref{dim2}). The width of the core of $\psfone$, derived with no consideration of the venusian atmosphere, is over-estimated and the over-sharpening this produces when used to restore HMI data shows up as ringing artefacts near where the signal is changing rapidly.

\subsubsection{Final estimate of the PSF}

We arrived at our final estimate of the PSF by making the following changes to the derivation procedure described in Sect. \ref{psfderivation}.
\begin{itemize}
	\item Firstly, we subtracted the estimate of the radial intensity profile of the aureole from the mean transit image.
	\item Secondly, we blurred the edge of the disc of zeroes in the artificial image with a Gaussian kernel, the width of which we allowed to be a free parameter in the LMA optimization.
	\item Lastly, we fixed the width of the narrowest Gaussian component in the guess PSF such that the full width at half maximum, FWHM of the component is similar to that of the pixel-integrated ideal diffraction-limited PSF. The pixel integration was achieved by smoothing the ideal diffraction-limited PSF with a box filter of HMI pixel scale width.
\end{itemize}
The parameters of the PSF so derived, hereafter referred to as $\psftwo$, are summarized in Table \ref{aiso5gp}. The adjustments to the PSF derivation procedure yielded a PSF that is significantly narrower at the core compared to $\psfone$, though not more than the ideal diffraction-limited PSF (bottom panel, Fig. \ref{dim3}). The agreement between the aureole-subtracted mean transit image, and the convolution of the Gaussian-blurred artificial image with $\psftwo$ is similar as in the $\psfone$ instance, illustrated in Figs. \ref{aisofit1} and \ref{aisofit2}, and therefore not plotted here.

The best fit value of the scale factor applied to the artificial image is 1.0028 and the radius of the disc of zeroes 29.33 arcsec. The retrieved standard deviation of the Gaussian kernel is 0.26 arcsec. The scale height of the venusian atmosphere, at 15.9 km or approximately 0.08 arcsec, is of similar order. The degree of Gaussian blurring introduced is, as far as one can infer from such a comparison, physically plausible.

The intention here is to recover a conservative estimate of the PSF, making use of the fact that the PSF of the instrument cannot be narrower at the core than the ideal diffraction-limited PSF. Also, it was necessary to fix the width of the narrowest Gaussian component as allowing both this and the standard deviation of the Gaussian kernel to be free parameters leads to a degeneracy of the LMA optimization\footnote{Specifically, the LMA converged to different solutions for the PSF, some of which are narrower at the core than the ideal diffraction-limited PSF, depending on the initial value of the free parameters.}. Though a conservative estimate, restoring the test transit image with $\psftwo$ still removed most of the intensity on the venusian disc while largely suppressing the ringing artefacts (Figs. \ref{dim1} and \ref{dim2}d).

As stated in Sect. \ref{psfderivation}, a potential hazard of modelling the PSF as the linear combination of Gaussian functions is that it allows solutions with Strehl ratios greater than unity. This functional form is only appropriate when the weight and width of the broader Gaussian components, representing the non-ideal contribution to the PSF (instrumental effects other than aperture diffraction) are sufficiently high to avoid this \citep{wedemeyerbohm08}. As evident in Fig. \ref{aisofit3}, this is indeed the case here for both $\psfone$ (dashed curve) and $\psftwo$ (red curve). The greater integral under both PSFs compared to the pixel-integrated ideal diffraction-limited PSF, all normalized to the level at $r=0$, indicates Strehl ratios of less than unity. (We cannot compute the Strehl ratio of $\psfone$ and $\psftwo$ directly as they describe the pixel-integrated PSF.)

\begin{cfig}
\includegraphics[width=\textwidth]{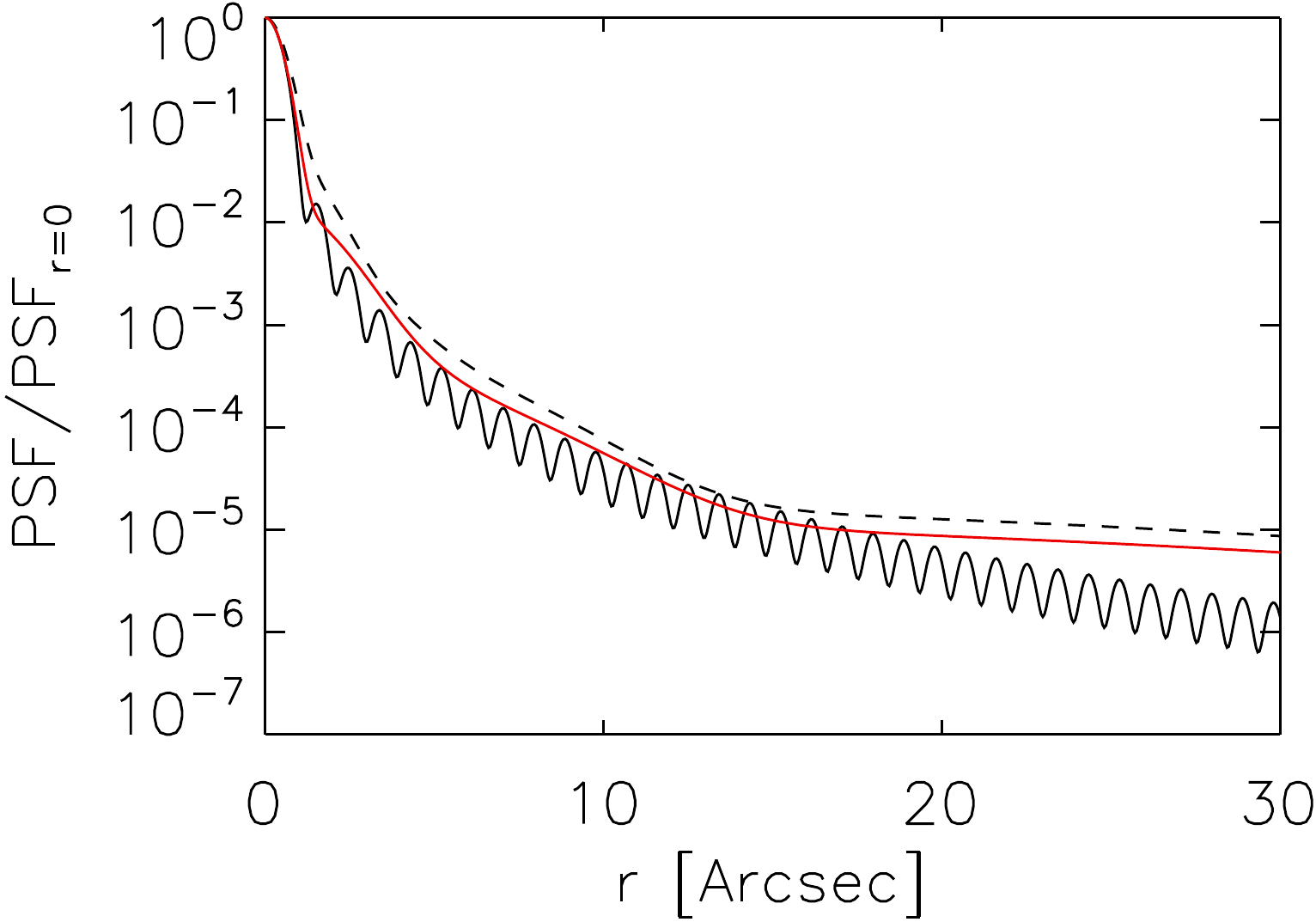}
\caption{As in Fig. \ref{dim3} (bottom panel), but with the PSF on a logarithmic scale and over an extended radial range.}
\label{aisofit3}
\end{cfig}

\subsection{Partial lunar eclipse and solar limb observations}
\label{psl}

As stated in the introduction, we consider the observations of Venus in transit the most appropriate available for recovering the PSF of the HMI, preferring them over data from partial lunar eclipses and of the solar limb.

It is challenging to constrain the PSF in both the radial and azimuthal dimensions with data from partial lunar eclipses due to the combination of the geometry, as well as jitter and defocus issues brought on by the lunar occultation itself.
\begin{itemize}
	\item The radius of curvature of the terminator, the edge of the lunar disc, is much greater than the width of the PSF. So at any given point along the terminator, the spatial distribution of signal smeared onto the lunar disc largely reflects the PSF in the direction of the centre of curvature. Therefore, a given partial lunar eclipse image only contains information about the PSF within a limited range of azimuths. Circumventing this limitation by looking at multiple images with the terminator at different orientations is not straightforward due to the variation of the PSF with position in the FOV.
	\item The terminator is uneven from lunar terrain. This makes it complicated to model the aperture diffraction and stray light-free image as we did here for the mean transit image with the artificial image (Sect. \ref{psfderivation}). A possible solution is to reduce the problem from 2D to 1D by looking at the radial intensity profile over segments of the terminator relatively free of lunar terrain features. This is, however, only appropriate when the terminator is near solar disc centre, where we can take the solar background to be uniform.
	\item The image stabilization system, ISS of the instrument was not always functional during the partial lunar eclipses due to the lunar disc blocking the diodes necessary for its operation, increasing the jitter.
	\item Having the lunar disc occult a significant proportion of the solar disc, and therefore greatly reducing the amount of impinging radiation, causes the front window of the instrument to cool, resulting in defocus \citep{schou12}.
	\item There are observations made when the lunar disc was just starting to cover the solar disc and not blocking the ISS diodes. These data do not suffer jitter and defocus problems, but we cannot resolve the lunar terrain issue by reducing the problem from 2D to 1D as described above, as the variation in the solar background from limb darkening is significant here.
\end{itemize}

Our approach in this study is to constraint the PSF by the spatial distribution of intensity about a closed bright and dark boundary as this allows us to recover the full azimuthal dependence (i.e., all directions). For this we can either employ observations of Venus in transit or of the solar limb. The PSF of HMI likely varies with position in the FOV, as shown for the part of the FOV transversed by Venus in Sect. \ref{dataselection}. The longer the boundary used to constrain the PSF, the greater the contribution by the variation of the PSF with position in the FOV to observed intensity fluctuation along and near the boundary, which introduces bias to the retrieved PSF. The venusian disc occupies only about $0.06\%$ of the FOV by area, and the solar disc, over $60\%$. The variation of the PSF over the part of the FOV occupied by the venusian disc is likely minimal, making these observations more suited for the purpose.

In view of the issues associated with deducing the PSF from partial lunar eclipse and solar limb data, we utilised the observations of Venus in transit though the interaction between solar radiation and the venusian atmosphere is challenging to account for, leaving us with only  a conservative estimate of the PSF (Sect. \ref{venusatm}).

\section{Application of the derived PSF to HMI observations}
\label{p2results}

\subsection{Granulation contrast}
\label{gc}

Restoring HMI data with $\psftwo$ is not exact. This is due to the approximate account of the influence of the venusian atmosphere in the derivation of $\psftwo$ (Sect. \ref{venusatm}) and from applying a single PSF to the entire FOV (so ignoring the variation of the PSF with position in the FOV, discussed in detail in Sect. \ref{psfdep1}). In this subsection we examine the effect of image restoration with $\psftwo$ on apparent granulation contrast, represented by the RMS intensity contrast of the quiet Sun\footnote{Intensity variation in the quiet Sun arises mainly from granulation.}. We compare the values deduced from HMI continuum observations and from synthetic intensity maps generated from a 3D MHD simulation. The purpose is to demonstrate that image restoration with $\psftwo$, with all its limitations, still yields reasonable estimates of the aperture diffraction and stray light-free intensity contrast.

The side CCD continued to observe in the continuum for about an hour after the end of the transit of Venus. For this analysis we employed a continuum filtergram from this period (recorded at 04:35:59 UTC, June 6, 2012), hereafter referred to as the test continuum filtergram. Of the various types of data available from HMI, the continuum filtergram represents the closest to a near instantaneous continuum capture (exposure time of $\sim0.135$ seconds). This implies minimal loss of apparent contrast from averaging in time. It is worth noting however, that the continuum bandpass ($-$344 m\AA{} from line centre), whilst close to the clean continuum, may be slightly affected by the far wing of the Fe I 6173 \AA{} line.

The intensity contrast of pixels corresponding to quiet Sun in the test continuum filtergram was computed largely following the method of \cite{yeo13}, who examined the intensity contrast of small-scale magnetic concentrations utilizing the 45-s continuum intensity and line depth data products from the front CCD. As in the cited work, the intensity contrast at a given image pixel is defined here as the normalized difference to the mean quiet-Sun intensity.

First, we identified magnetic activity present using the 45-s longitudinal magnetogram from the front CCD closest in time ($<1\:{\rm minute}$) to the test continuum filtergram\footnote{The 720-s longitudinal magnetogram data product from the side CCD, generated from the regular filtergram sequence, is evidently not available when this CCD is observing in the continuum.}. Let $\vbl/\mu$ denote the magnetogram signal, the mean line-of-sight magnetic flux density within a given image pixel, corrected (to first order) for foreshortening by the quotient with $\mu$. The magnetogram was resampled to register with the test continuum filtergram. Image pixels in the test continuum filtergram corresponding to points where $\bmu>10\:\rm{G}$ in the resampled magnetogram were taken to contain significant magnetic activity and masked, leaving quiet Sun.

The test continuum filtergram was corrected for limb darkening by normalizing it by a fifth order polynomial in $\mu$ fit to the quiet Sun pixels \citep[following][]{neckel94}. Let $\inorm$ denote the limb darkening corrected intensity. Next, we derived the mean $\inorm$ of the quiet Sun, $\iqs$ as a function of position on the solar disc. As similarly noted for the 45-s continuum and line depth data products by \cite{yeo13}, there are distortions in HMI filtergrams such that $\iqs$ is not at unity but varying with position on the solar disc. (This is not to be confused with the spatial distortion present in HMI data discussed in Sect. \ref{dataselection}.)

We sampled the solar disc at 16-pixel intervals in both the vertical and horizontal directions. At each sampled point, we retrieved the median intensity of all the quiet Sun pixels inside a $401\times401$ pixel window centred on the point. We then fit a bivariate polynomial surface to the values so obtained from the entire disc. This surface describes $\iqs$ as a function of position on the solar disc. The intensity contrast at a given image pixel is then given by the value of $\frac{\inorm}{\iqs}-1$ there.

Finally, we derived the RMS intensity contrast of the quiet Sun as a function of $\mu$. To this end we grouped the quiet Sun pixels by $\mu$ in bins with a width of 0.01 and took the RMS intensity contrast within each bin. The results are expressed in Fig. \ref{contrastqs1} (black curve). Also plotted are the values from first restoring the test continuum filtergram with $\psfone$ (blue curve), and with $\psftwo$ (red curve). Quiet Sun pixels near the limb ($\mu<0.2$) were excluded. Towards the limb, the spread in measured intensity contrast in HMI data is dominated by scatter from the combination of the diminishing signal-to-noise ratio and the limb darkening correction \citep{yeo13}.

The decline in granulation contrast with distance from disc centre, seen here for the test continuum filtergram, is a known, well reported phenomenon \citep[see][and references therein]{sanchezcuberes00,sanchezcuberes03}. Also within expectation, image restoration resulted in greater RMS contrasts, by a factor of about 2.6 near limb in the $\psfone$ instance, going up to 3.2 at disc centre, and going from 1.9 to 2.2 for $\psftwo$.

\begin{cfig}
\includegraphics[width=\textwidth]{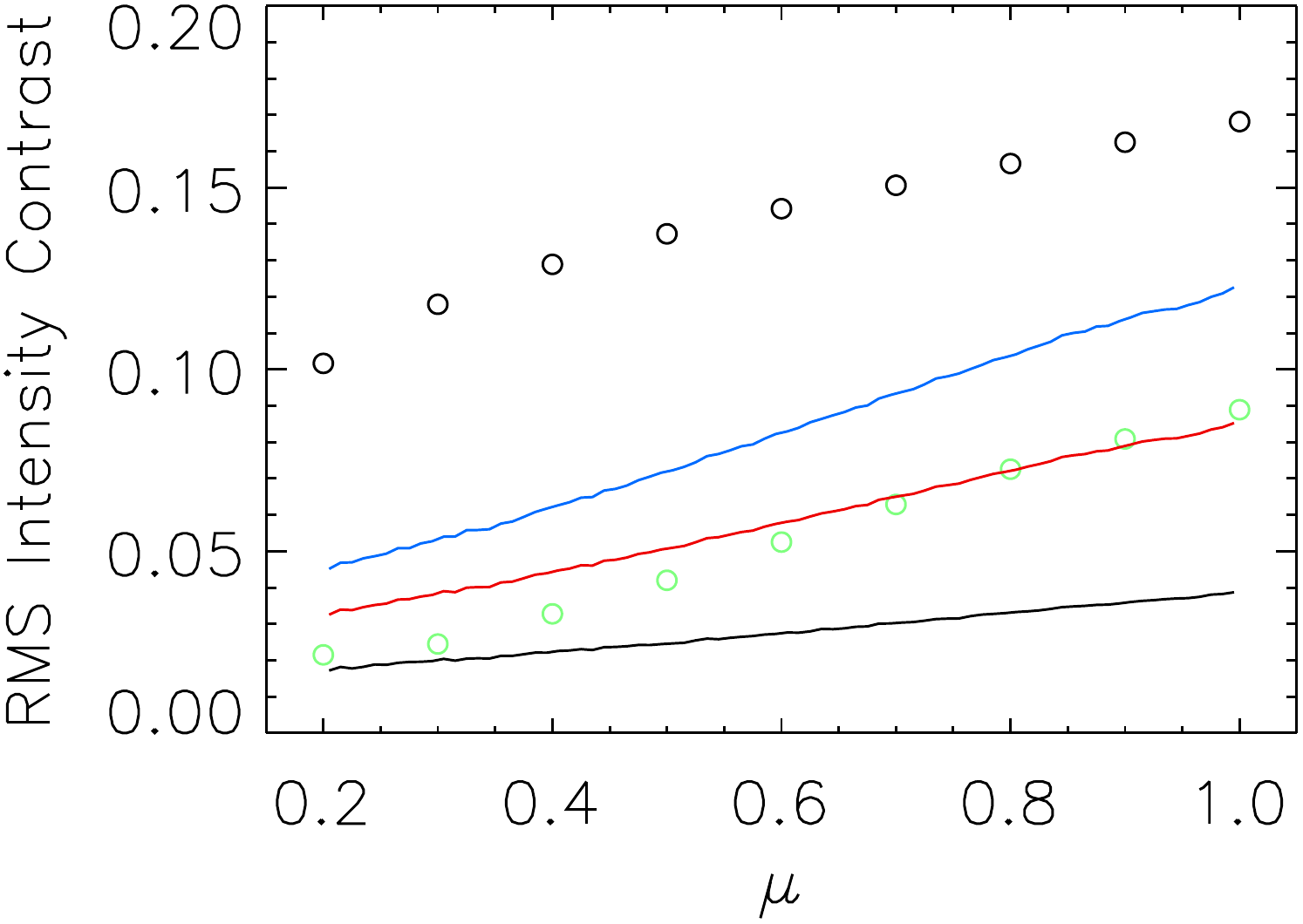}
\caption{RMS intensity contrast of the quiet Sun in the test continuum filtergram as a function of $\mu$, before (black curve), and after image restoration with $\psfone$ (blue curve) and with $\psftwo$ (red curve). The black circles represent the values from the synthetic intensity maps and the green circles the same, rescaled to reflect the proportion arising from spatial frequencies up to the cutoff spatial frequency of the restored (with $\psftwo$) test continuum filtergram (see text).}
\label{contrastqs1}
\end{cfig}

The apparent granulation contrast is not only significantly enhanced by image restoration but also rather sensitive to differences between $\psfone$ and $\psftwo$. This makes the RMS intensity contrast of the quiet Sun a suitable check of the goodness of $\psftwo$ for the restoration of HMI data. We compared the values obtained here with values from synthetic intensity maps, artificial images of the quiet Sun produced from a numerical simulation. To this end, synthetic Stokes spectra were generated by applying the LTE radiative transfer package SPINOR/STOPRO \citep{solanki87,frutiger00}, to snapshots of a 3D MHD simulation performed with the MURaM code \citep{vogler03,vogler05b}, as done in, for example, \cite{danilovic10,danilovic13}.

The simulation, set up as in \cite{danilovic13}, represents a layer encompassing the solar surface in the quiet Sun. The mean vertical magnetic flux density is $50\:{\rm G}$. The simulation ran over about 23 minutes solar time after reaching a statistically steady state. Synthetic Stokes profiles were computed for ten snapshots of the simulation output, recorded at intervals of approximately two minutes solar time. From each snapshot we produced nine synthetic intensity maps corresponding to $\mu$ of 0.2, 0.3, 0.4 and so on, up to 1.0, rotating the snapshot along one dimension. The computational domain of the simulation spans $6\times6\:{\rm Mm}$ in the horizontal, 1.4 Mm in depth, the top of the box lying about 0.5 Mm above the mean optical depth unity level, in a $288\times288\times100$ grid. This translates into a pixel scale of ($0.0287\mu$) arcsec and $0.0287$ arcsec, in the rotated and static direction, in the synthetic intensity maps.

The Stokes $I$ and $Q$ components of the synthetic spectra were convolved with a Gaussian function with a FWHM of 75 m\AA{} and sampled at $-$327 m\AA{} from the centre of the Fe I 6173 \AA{} line in order to yield synthetic intensity maps mimicking the polarization (Stokes $I+Q$) and bandpass of the test continuum filtergram. The FWHM and central wavelength of the continuum bandpass were estimated from the main lobe of the CCD centre filter transmission profile\footnote{The filter transmission profiles of the HMI varies slightly with position in the CCD. The central wavelength of the main lobe, at $-$327\AA{} from line centre, differs from the bandpass position of $-$344\AA{} stated earlier. Quoted bandpass positions for HMI are theoretical figures derived assuming the filter transmission profiles are delta functions.}.

The synthetic intensity maps were resampled in the foreshortened direction such that the pixel scale is similar along both dimensions. The RMS contrast of a given intensity map is given by the RMS value of $\frac{I}{\left\langle{I}\right\rangle}-1$ over all points, $\left\langle{I}\right\rangle$ denoting the mean intensity of the map. The mean RMS contrasts from the synthetic intensity maps at each $\mu$ level for which we simulated data are plotted along the measured values from the test continuum filtergram in Fig. \ref{contrastqs1} (black circles).

The RMS intensity contrast of the quiet Sun in the test continuum filtergram and the synthetic intensity maps cannot be compared directly due to the gross difference in the pixel scale (0.504 versus 0.0287 arcsec). Resampling the synthetic intensity maps to HMI's pixel scale is not feasible as the resampled synthetic intensity maps will extend only $16\times16$ pixels in the $\mu=1.0$ case, going down to $3\times16$ pixels for $\mu=0.2$. Simulations with considerably larger computational domains are necessary to yield synthetic intensity maps from which we can compute the RMS contrast at HMI's pixel scale with statistical confidence. What we did instead was to estimate, by comparing the power spectra of the synthetic intensity maps and the test continuum filtergram, the contribution to intensity variations in the synthetic intensity maps by spatial frequencies up to the resolution limit of the test continuum filtergram.

In Fig. \ref{analysisee} we plot the encircled energy of the power spectrum of the $361\times361$ pixel crop, centred on solar disc centre, of the test continuum filtergram. There are no sunspots present in this crop. We define the cutoff spatial frequency as the spatial frequency at which the encircled energy of the power spectrum reaches 0.99, taken here as an indication of the resolution limit. The cutoff spatial frequency is $0.75\:{\rm cycle/arcsec}$ for the original test continuum filtergram (black curve), and $0.79\:{\rm cycle/arcsec}$ and $0.78\:{\rm cycle/arcsec}$ for the iterations restored with $\psfone$ (blue curve) and with $\psftwo$ (red curve).

As mentioned in Sect. \ref{psfderivation}, image restoration with a PSF that is the linear combination of Gaussian functions can potentially introduce aliasing artefacts from the enhancement of spatial frequencies above the Nyquist limit ($0.99\:{\rm cycle/arcsec}$ for HMI). While image restoration with $\psfone$ and $\psftwo$ enhanced image contrast, indicated here by the rightward displacement of the encircled energy profile for the restored iterations of the test continuum filtergram, it made little difference to the resolution limit which is also significantly lower than the Nyquist limit. Even after image restoration, almost all energy is confined to spatial frequencies well below the Nyquist limit. Aliasing artefacts from the restoration, if present, are likely negligible.

\begin{cfig}
\includegraphics[width=\textwidth]{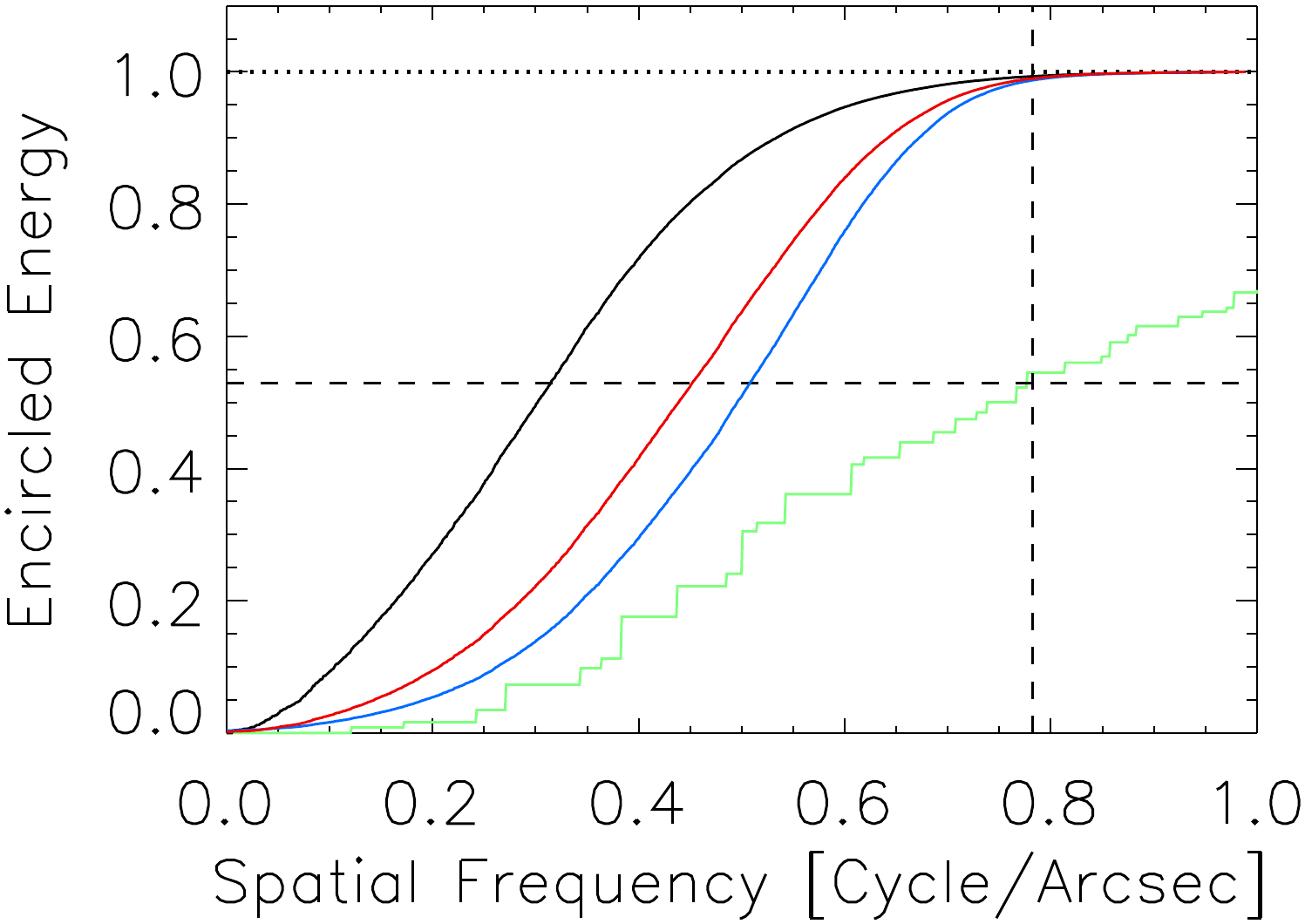}
\caption{The encircled energy of the power spectrum of the $361\times361$ pixel inset, centred on the centre of the solar disc, of the test continuum filtergram, before (black), and after image restoration with $\psfone$ (blue) and with $\psftwo$ (red). The green series gives the encircled energy of the mean power spectrum of the synthetic intensity maps corresponding to $\mu=1.0$. The vertical dashed line marks the cutoff spatial frequency (see text) of the restored (with $\psftwo$) test continuum filtergram and the horizontal dashed line the encircled energy of the mean power spectrum of the $\mu=1.0$ synthetic intensity maps at this spatial frequency. The dotted line denotes encircled energy of unity.}
\label{analysisee}
\end{cfig}

At each $\mu$ level for which we generated synthetic intensity maps, we computed the power spectrum of each intensity map and then the encircled energy of the mean power spectrum. Following that we estimated the encircled energy at the spatial cutoff frequency of the restored (with $\psftwo$) test continuum filtergram, illustrated in Fig. \ref{analysisee} for $\mu=1.0$. The encircled energy here gives the proportion of observed intensity variation in the synthetic intensity maps arising from spatial frequencies up to the spatial cutoff frequency of the test continuum filtergram. The product of this quantity with the RMS intensity contrast of the quiet Sun from the synthetic intensity maps (green circles, Fig. \ref{contrastqs1}) then represents an approximation of the RMS contrast if the spatial resolution of the synthetic intensity maps were similar to that of the test continuum filtergram. This treatment is very approximate, ignoring the fact that the spatial frequency response of the test continuum filtergram and synthetic intensity maps, up to the cutoff, are in all likelihood not similar.

In this analysis we had,
\begin{itemize}
 \item defined the cutoff spatial frequency as the level where the encircled energy of the power spectrum reaches 0.99, and
 \item used the cutoff spatial frequency of the copy of the test continuum filtergram restored with $\psftwo$ to find the factors by which to rescale the RMS intensity contrast of the synthetic intensity maps (as the comparison between this restored version of the test continuum filtergram and the synthetic intensity maps is of greatest interest).
\end{itemize}
As stated above, the restoration of the test continuum filtergram made little difference to the cutoff spatial frequency. Also, the encircled energy of the power spectrum of the synthetic intensity maps does not vary strongly with spatial frequency in the regime of the cutoff spatial frequency of the test continuum filtergram, as visibly evident for the $\mu=1.0$ example in Fig. \ref{analysisee}. Hence, tests showed that the level of the rescaled RMS contrast is not sensitive to small variations in the threshold encircled energy level chosen in the definition of the cutoff spatial frequency. The result is also not changed significantly if we employ the cutoff spatial frequency of the unrestored and restored with $\psfone$ versions of the test continuum filtergram to derive the rescaling factors instead.

The RMS intensity contrast of the quiet Sun in the restored (with $\psftwo$) test continuum filtergram (red curve, Fig. \ref{contrastqs1}) and in the synthetic intensity maps, rescaled as described above (green circles), are of gratifyingly similar magnitude, less than or close to 0.01 apart at most $\mu$, especially near disc centre. They do, however, differ in that the latter exhibits a steeper decline with distance from disc centre. The diverging trend with decreasing $\mu$ is likely, at least in part, from
\begin{itemize}
	\item the approximate way of accounting for the influence of the venusian atmosphere in the derivation of $\psftwo$ (Sect. \ref{venusatm}),
	\item applying a single PSF to the entire FOV, so ignoring the variation of the PSF with position in the FOV (discussed in detail in Sect. \ref{psfdep1}),
	\item the difference in the spatial frequency response of the test continuum filtergram and the synthetic intensity maps,
	\item sensor noise and its centre-to-limb variation (CLV), and
	\item Doppler shift of the spectral line from the motion of SDO and the rotation of the Sun, which may produce small, $\mu$-dependent effects on apparent intensity in the continuum bandpass through the line wing.
\end{itemize}
The observation that these two series are close, even with these factors present, confers confidence that image restoration with $\psftwo$, though not exact, returns a reasonable approximation of the true aperture diffraction and stray light-free intensity contrast.

A quantitative comparison of the RMS intensity contrast of the quiet Sun presented here for HMI and other measurements reported in the literature would require taking into account instrumental differences such as the spatial resolution and bandpass, which is beyond the scope of this study. Due to HMI's limited spatial resolution, the RMS contrast, even after image restoration with $\psftwo$, remains below the values returned from spaceborne and balloon-borne (i.e., similarly seeing-free) observatories at finer spatial resolutions, namely Hinode \citep{danilovic08,mathew09,wedemeyerbohm09} and SUNRISE \citep{hirzberger10}. (Note though, that the divergence is also due in part to the different bandpass of the various instruments.)

\subsection{Effect of image restoration on the Dopplergram, longitudinal magnetogram, continuum intensity and line depth data products}
\label{ssmcss}

In this subsection we discuss the effect of image restoration with $\psftwo$ on the Dopplergram, longitudinal magnetogram, continuum intensity and line depth data products. We examine the influence on the apparent continuum and line-core intensity, and magnetic field strength of small-scale magnetic concentrations, as well as sunspots and pores. We will also describe the result of image restoration on the apparent amount of magnetic flux on the solar surface and the line-of-sight velocity.

For this purpose we utilised a set of simultaneous (generated from the same sequence of filtergrams) 720-s Dopplergram, longitudinal magnetogram, continuum intensity and line depth images from the side CCD, taken shortly after this CCD resumed collection of the regular filtergram sequence, about an hour after Venus left the solar disc (at 05:35:32 UTC, June 6, 2012). Here we will refer to the result of subtracting the line depth image from the continuum intensity image, giving the intensity in the Fe I 6173 \AA{} line, as the line-core intensity image.

As mentioned in Sect. \ref{dataselection}, HMI data products cannot be corrected for stray light by the deconvolution with the PSF but instead we must correct either the Stokes parameters or the filtergrams used to compute the data products. The data set was restored for stray light by applying image restoration with $\psftwo$ to the corresponding 720-s Stokes parameters, and returning the result to the HMI data processing pipeline. A $200\times200$ pixel inset of the original and restored version of the data set, near disc centre ($\mu>0.93$), featuring active region NOAA 11494, is shown in Fig. \ref{spotinset}. The enhanced image contrast and visibility of small-scale structures is clearly evident.

The 720-s Milne-Eddington inversion data product includes the vector magnetogram. Since the inversion procedure employed to obtain this data product assumes a magnetic filling factor of unity everywhere, the process treats noise in the Stokes $Q$, $U$ and $V$ parameters as signal, creating pixel-averaged horizontal magnetic field strengths of $\sim100\:{\rm G}$ in the vector magnetogram even in the very quiet Sun. For ease of interpretation we confined ourselves to the longitudinal magnetogram data product here.

\begin{cfig}
\includegraphics[width=\textwidth]{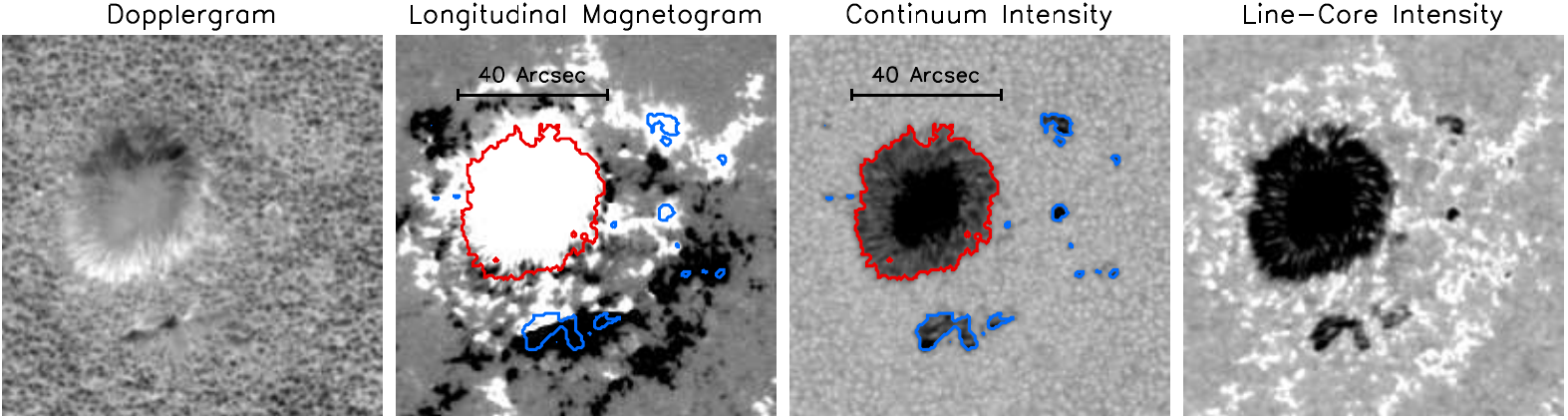}
\includegraphics[width=\textwidth]{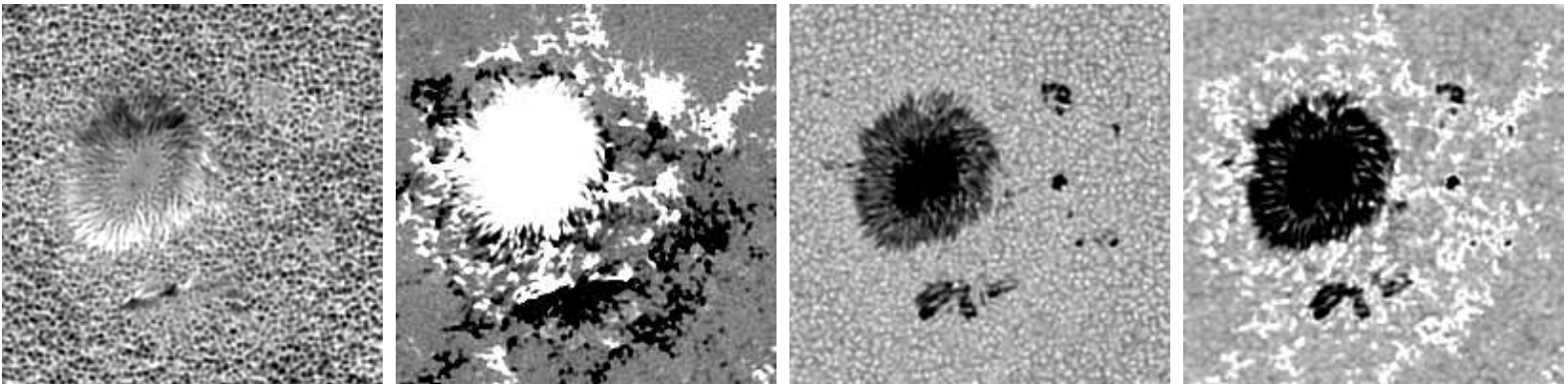}
\caption{$200\times200$ pixel ($101\times101$ arcsec) inset, near disc centre ($\mu>0.93$), encompassing active region NOAA 11494, of the simultaneously recorded 720-s data products examined, before (top) and after (bottom) image restoration with $\psftwo$. The grey scale is saturated at $-1200$ and $1200\:{\rm ms^{-1}}$ for the Dopplergrams, at $-100$ and $100\:{\rm G}$ for the longitudinal magnetograms, and at $0.6$ and $1.2$ for the continuum and line-core intensity images. The Dopplergram was corrected for the velocity of SDO relative to the Sun and for differential rotation (Sect. \ref{ssmcss4}). Both the continuum and line-core intensity images were normalized to the mean quiet-Sun level (Sect. \ref{ssmcss1}). The red and blue contours in the grey scale plot of the uncorrected longitudinal magnetogram and continuum intensity image follow $\inorm=\iqp$, the quiet Sun to penumbra boundary. The colour coding is to distinguish the big sunspot feature (red) from the smaller sunspots and pores (blue), treated separately in Fig. \ref{speg}.}
\label{spotinset}
\end{cfig}

\subsubsection{Intensity contrast and magnetogram signal of small-scale magnetic concentrations}
\label{ssmcss1}

Both the original and restored continuum and line-core intensity images were normalized by the fifth order polynomial in $\mu$ fit to the quiet Sun pixels. Then the intensity contrast at each image pixel was computed following the procedure applied to the test continuum filtergram in Sect. \ref{gc}. For the line-core intensity image, the normalization not only corrects for limb darkening, but also the centre-to-limb weakening of the Fe I 6173\AA{} line \citep{norton06,yeo13}.

Sunspots were identified by applying a continuum intensity threshold representing the quiet Sun-to-penumbra boundary, denoted $\iqp$. We took the threshold value for MDI continuum intensity images, taken at 6768 \AA{}, from \cite{ball12}, 0.89, and estimated the equivalent level at HMI's wavelength, 6173 \AA{}. Assuming sunspots to be perfect blackbodies and an effective temperature of $5800\:{\rm K}$ for the quiet Sun, the result is a threshold value of 0.88. This is a crude approximation, ignoring the difference in spatial resolution and variation in the continuum formation height with wavelength \citep{solanki98a,sutterlin99,norton06}. Pores were also isolated by the application of this threshold. In the following we count these features to the sunspots and do not mention them separately.

We selected the image pixels where $\mu>0.94$ (i.e., near disc centre), excluding sunspots (i.e., all points with $\inorm<\iqp$) and all points within three pixels of a sunspot. The selected points were binned by $\bmu$ such that we end up with 800 bins of equal population. We then took the mean $\bmu$, as well as the median continuum and line-core intensity contrast within each bin. The values for the uncorrected and restored copy of the data set are represented by the black and red curves respectively in Fig. \ref{contrastnf}. These profiles depict the intensity contrast of small-scale magnetic concentrations as a function of $\bmu$, which serves as an approximate proxy of the magnetic filling factor \citep{ortiz02,yeo13}, in the continuum and core of the Fe I 6173 \AA{} line. This is effectively a repeat of part of the analysis of \cite{yeo13}, except now on HMI observations corrected for aperture diffraction and stray light.

\begin{cfig}
\includegraphics[width=\textwidth]{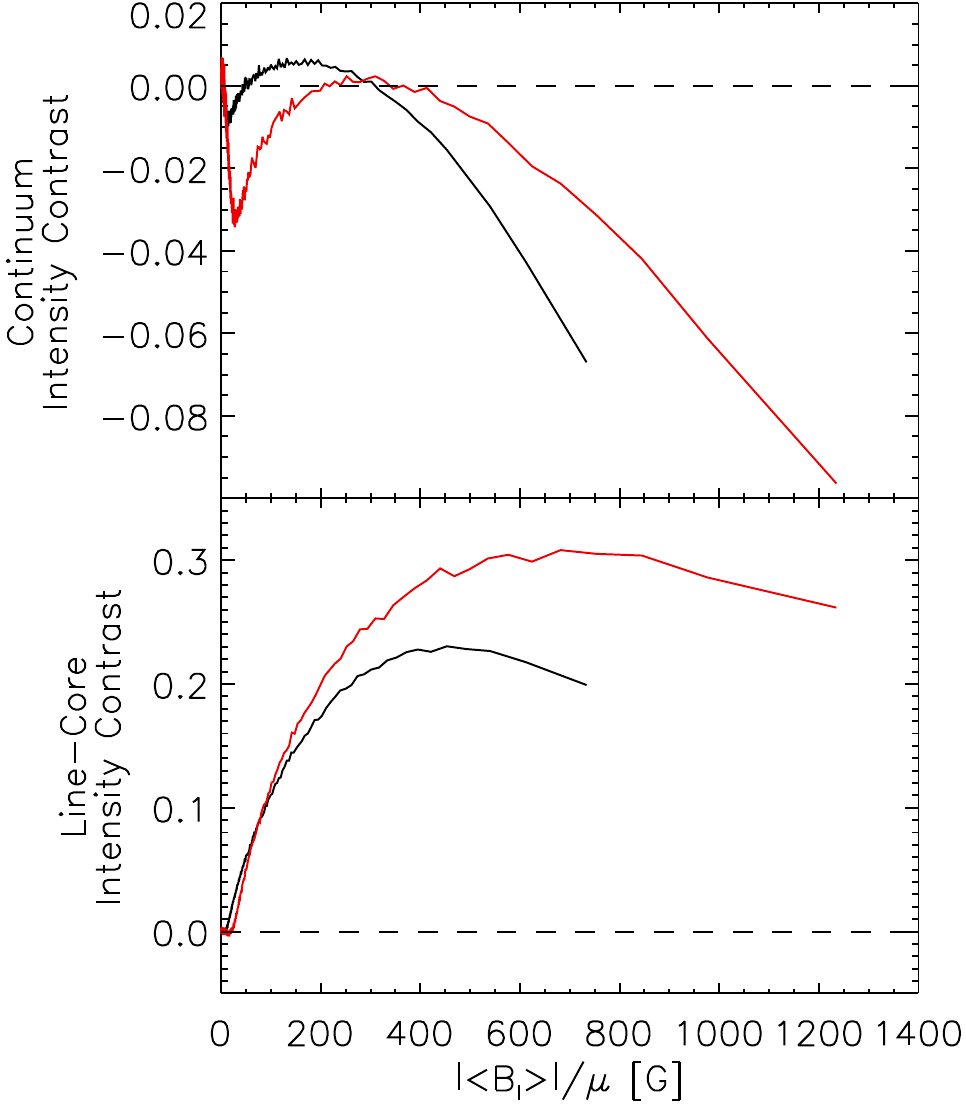}
\caption{The continuum (top) and line core (bottom) intensity contrast of small-scale magnetic concentrations, near disc centre ($\mu>0.94$), as a function of $\bmu$. The black and red curves correspond to the values from the original and restored (with $\psftwo$) data sets, respectively.}
\label{contrastnf}
\end{cfig}

Comparing the magnetogram with the continuum intensity image, we found the magnetogram signal associated with sunspots to extend beyond the continuum intensity boundary, given by the $\inorm=\iqp$ locus (for example, in Fig. \ref{spotinset}). This was similarly noted by \cite{yeo13}, who attributed it to the encroachment of the magnetic canopy of sunspots, and the smearing of polarized signal originating from sunspots onto its surrounds by instrumental scattered light. Hence, close to sunspots, the magnetogram signal is not entirely from local, non-sunspot magnetic features alone and would introduce a bias into the intensity contrast versus $\bmu$ profiles (Fig. \ref{contrastnf}) if left unaccounted for. In the cited work pixels contiguous to sunspots and with $\bmu$ above a certain threshold level were masked. Here, we excluded only all points within three pixels of each sunspot, observing that excluding pixels further than this distance made no appreciable difference to the resulting intensity contrast versus $\bmu$ profiles. This measure is sufficient here as, unlike in the earlier study which examined almost all disc positions ($\mu>0.1$), we are only looking at image pixels near disc centre ($\mu>0.94$). Near disc centre, the influence of magnetic canopies, which are largely horizontal, on the longitudinal magnetogram signal near sunspots is not as significant or extensive as at disc positions closer to the limb.

The continuum and line-core intensity contrast versus $\bmu$ profile of small-scale magnetic concentrations near disc centre presented here for the uncorrected data set (black curves, Fig. \ref{contrastnf}) is nearly identical to that by \cite{yeo13} (Figs. 9 and 10 in their paper), who employed similar data and method of derivation. The profiles from the restored data set (red curves) span a wider range, by a factor of about 1.3 in the continuum and line-core intensity contrast, and 1.7 in $\bmu$, but are qualitatively similar in form.

Image restoration produced an absolute increase in the continuum and line-core intensity contrast everywhere except around the peak of the continuum contrast versus $\bmu$ profile. The lower maximum in the profile from the restored data set, compared to the profile from the uncorrected data set ($2.6\times10^{-3}$ versus $5.7\times10^{-3}$), is likely from the enhanced contrast of dark intergranular lanes.

In Fig. \ref{contrastnfpeak} we show the uncorrected and restored continuum intensity and $\bmu$  along a 21-pixel cut across example magnetic features near disc centre ($\mu=0.97$). The troughs and peaks in the intensity curve (top panel) correspond to intergranular lanes and granules respectively. The magnetic features, the peaks in the $\bmu$ curve (bottom panel), sit inside the intergranular lanes. The $\bmu$ level at the core of these magnetic features lie in the regime of the peak of the continuum contrast versus $\bmu$ profile. The stray light correction boosted the magnetogram signal at the core of these magnetic features but also rendered them darker here, even from positive contrast to negative, as radiation originating from nearby granulation is removed from the intergranular lanes. The spatial resolution of HMI is insufficient to resolve many of the magnetic elements. Consequently, measured intensities contain contributions not only from magnetic features but also from the intergranular lanes that host them.

\begin{cfig}
\includegraphics[width=\textwidth]{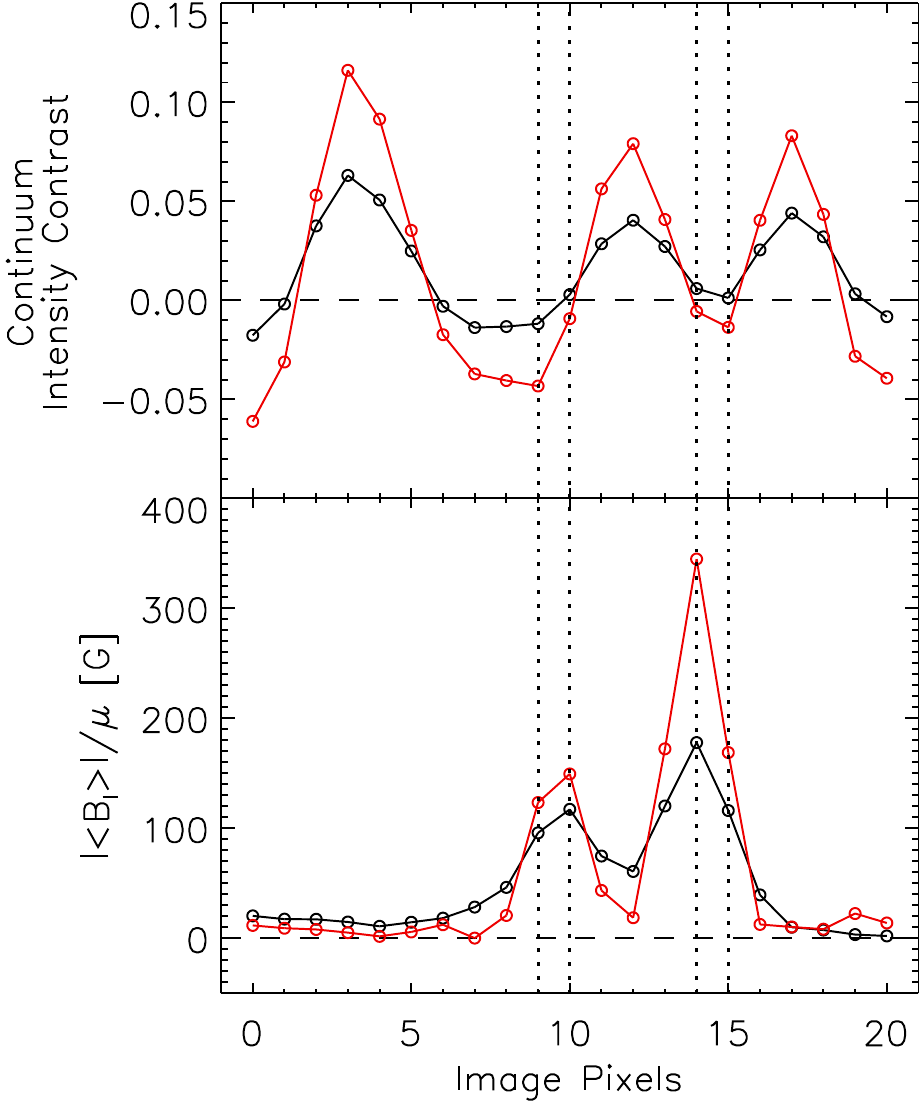}
\caption{The uncorrected (black circles) and restored (with $\psftwo$, red circles) continuum intensity contrast (top) and $\bmu$ (bottom) along a 21-pixel section across magnetic features near disc centre ($\mu=0.97$). The plotted points represent image pixel values and are connected by straight lines to aid the eye. The dotted lines highlight the image pixels inside the magnetic features where the stray light correction effected a decrease in continuum intensity contrast accompanied by an increase in $\bmu$. The dashed lines follow the zero level.}
\label{contrastnfpeak}
\end{cfig}

The intensity contrast of small-scale magnetic concentrations in both the continuum and spectral lines component of the solar spectrum, in particular the variation with position on the solar disc and magnetic field strength, is an important consideration in understanding the contribution by these features to variation in solar irradiance \citep{yeo13}. To the extent tested, image restoration with $\psftwo$ enhanced measured intensity contrast and $\bmu$ significantly but made little qualitative difference to the dependence of apparent contrast on $\bmu$. The analysis here was restricted to image pixels near disc centre ($\mu>0.94$). To extend the analysis to other disc positions, we would need to examine multiple full-disc images from different times featuring active regions at various disc positions as done by \cite{ortiz02} and \cite{yeo13}, beyond the scope of this paper.

\subsubsection{Intensity and magnetogram signal of sunspots and pores}
\label{ssmcss2}

In Fig. \ref{speg} we illustrate the change introduced by image restoration with $\psftwo$ on $\bl$, as well as the continuum and line-core intensity of the sunspots and pores defined by the $\inorm=\iqp$ contours in Fig. \ref{spotinset}. Signal enhancement is expressed as a function of the original level, separately for the big sunspot bounded by the red contours, and the smaller sunspots and pores bounded by the blue contours. We binned the image pixels by the uncorrected $\bl$ in intervals of $100\:{\rm G}$ and plotted the bin-averaged change in $\bl$ against the bin-averaged original $\bl$ (Fig. \ref{speg}a). This was repeated for the continuum and line-core intensity (Figs. \ref{speg}b and \ref{speg}c), taking a bin size of 0.05 in both instances.

\begin{cfig}
\includegraphics[width=\textwidth]{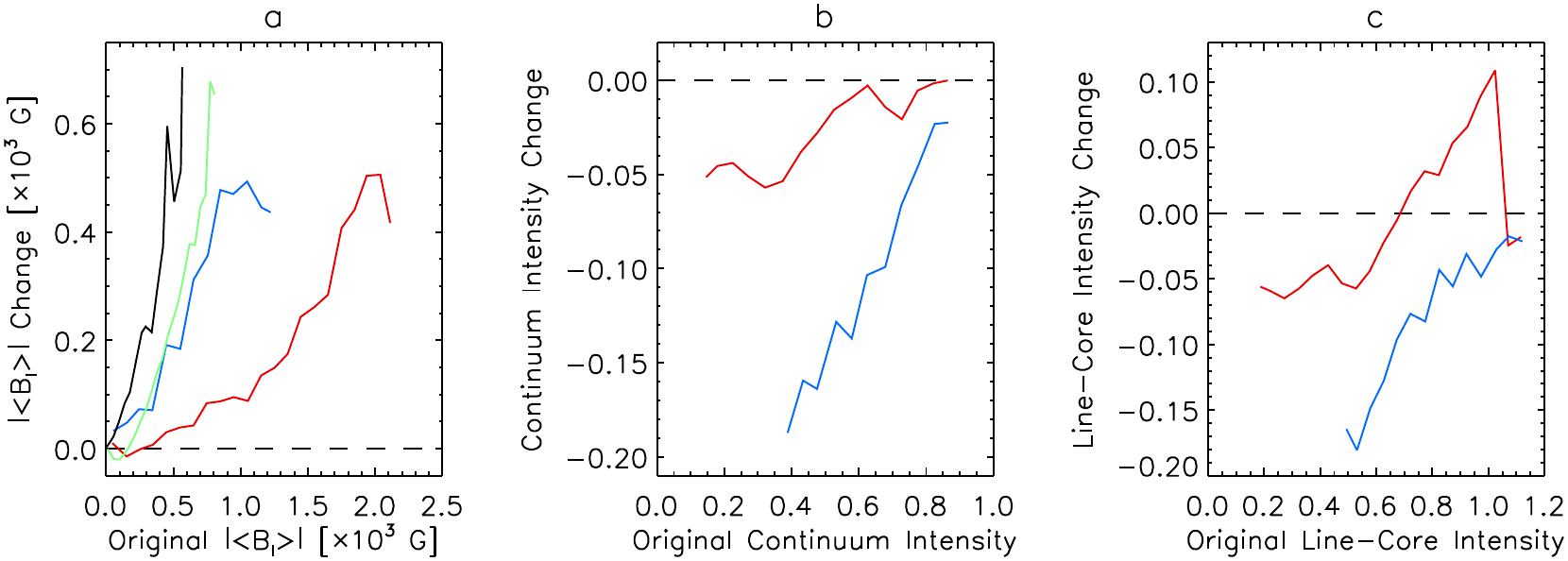}
\caption{The change in $\bl$ (left), as well as in the continuum (middle) and line-core intensity (right) introduced by image restoration with $\psftwo$, as a function of the original value. Both the continuum and line core intensities are normalized to the quiet-Sun level. The red and blue series follow the values derived from the sunspot and pore features encircled by the similarly colour-coded contours in Fig. \ref{spotinset}. The black and green curves (left) correspond to the change in $\bl$ in the quiet Sun field depicted in Fig. \ref{delta4}, and in a $201\times201$ pixel active region field near disc centre ($\mu>0.92$), respectively. The dashed lines mark the zero level.}
\label{speg}
\end{cfig}

Within expectation, the influence of image restoration on $\bl$ and intensity is highly correlated to the original values of these quantities. This comes largely from the fact that the darker regions, where $\bl$ is also typically higher, are more affected by stray light as scattered radiation forms a greater proportion of measured intensity, and therefore respond more strongly to image restoration. Also within expectation, the effect of restoration is more pronounced (greater absolute change) for smaller features, which are more susceptible to instrumental scattered light. An exception is the peak in the line-core intensity profile for the big sunspot feature (red curve, Fig. \ref{speg}c), which arose from the enhanced brightness of the bright filaments in the penumbra from the restoration, visible in Fig. \ref{spotinset}.

Given the variation in the response of sunspots to image restoration, it could have a profound effect on the apparent radiant and magnetic properties of these features. A full account of the effect of image restoration on sunspots, including the variation with size and disc position, would require examining a much larger sample of sunspots from multiple images taken at different times, which is outside the scope of this work (see \cite{mathew07} for such a study, based on MDI data).

\subsubsection{Amount of magnetic flux on the solar surface}
\label{ssmcss3}

We segmented the solar disc in the 720-s longitudinal magnetogram by $\mu$ (excluding points where $\mu<0.1$, $\sim1\%$ of the solar disc by area) into 50 annuli of equal area. Within each annulus, we computed the quantities listed below, plotted in Fig. \ref{delta3}. For each quantity, we derived the level in the uncorrected and restored magnetogram (black and red series, left axes), and the ratio of the restored and the uncorrected values (blue series, right axes), denoted by the $\Delta$ preffix.
\begin{itemize}
	\item The noise level, $\noise$ (Fig. \ref{delta3}a), given by the standard deviation of $\vbl$. The standard deviation was computed iteratively, points more than three standard deviations from the mean were excluded from succeeding iterations till convergence, to exclude magnetic activity. The variation of the noise level of HMI longitudinal magnetograms with position on the solar disc is dominated by a centre-to-limb increase \citep{liu12,yeo13}. It is therefore reasonable to represent the noise level within a given annulus by a single value of $\noise$. Image restoration increased the noise level, on average, by a factor of 1.6.
	\item The proportion of image pixels counted as containing significant magnetic activity, $\nmag$ (Fig. \ref{delta3}b), taken here as the points where $\bl>3\noise$.
	\item The mean $\bl$ of the image pixels counted as magnetic, $\mblmag$ (Fig. \ref{delta3}c).
	\item The product of $\nmag$ and $\mblmag$ (Fig. \ref{delta3}d). The quantity $\dflux$ represents the factor by which the apparent amount of line-of-sight magnetic flux changed from stray light removal.
\end{itemize}
Image restoration resulted in less image pixels being counted as magnetic (around $-10\%$ to $-25\%$, Fig. \ref{delta3}b), though the enhancement to the magnetogram signal (30$\%$ to 60$\%$, Fig. \ref{delta3}c) meant that there is an overall increase in the apparent amount of line-of-sight magnetic flux (10$\%$ to 40$\%$, Fig. \ref{delta3}d). Computing $\dflux$ taking all the annulus as a whole, the total amount of line-of-sight magnetic flux over the solar disc increased by a factor of 1.2.

\begin{cfig}
\includegraphics[width=\textwidth]{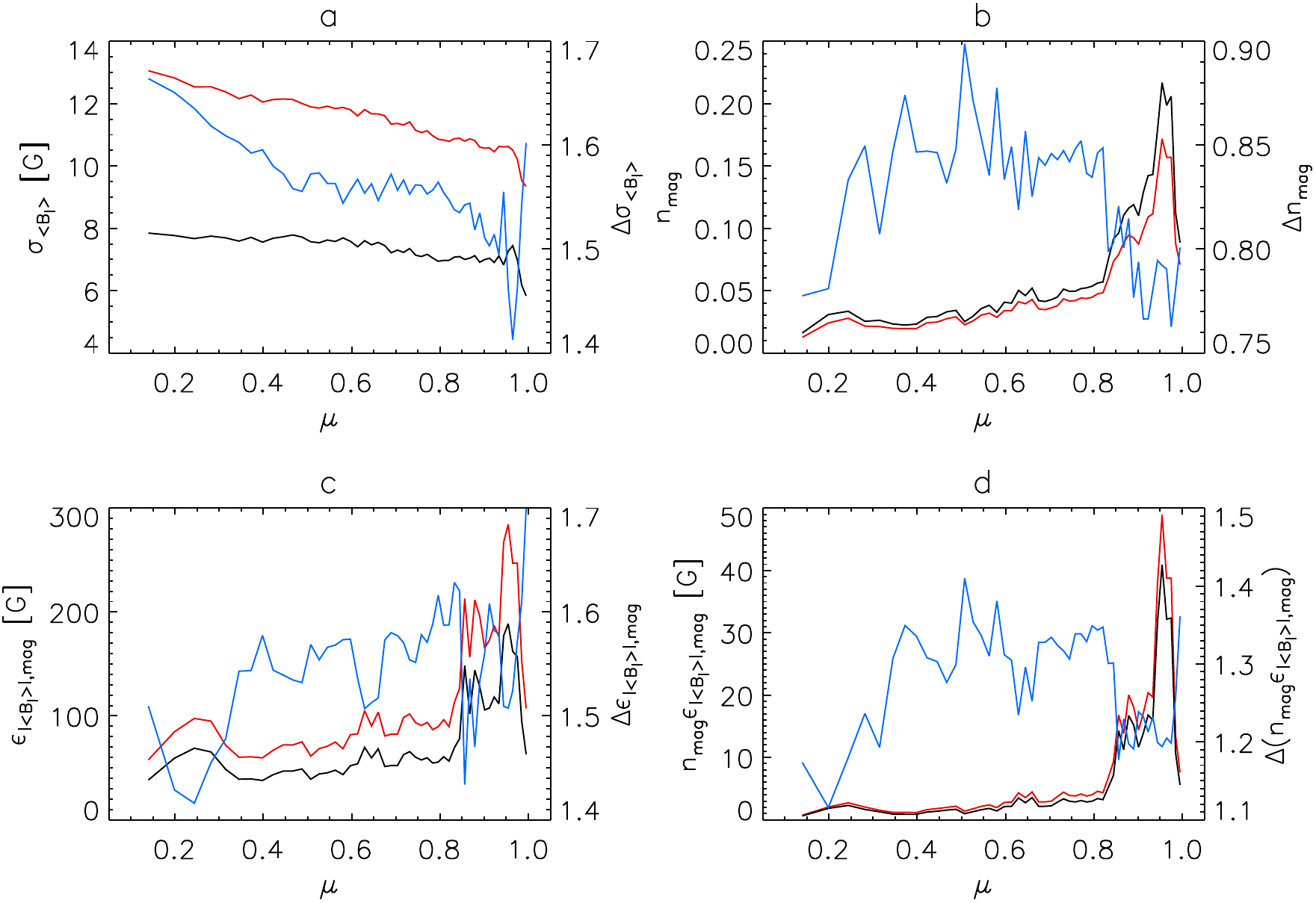}
\caption{Plotted as a function of $\mu$, a) the noise level of the 720-s longitudinal magnetogram, $\noise$, b) the proportion of image pixels counted as containing significant magnetic activity, $\nmag$, c) the mean $\bl$ of these points, $\mblmag$, d) and the product of $\nmag$ and $\mblmag$. Left axes: the uncorrected (black series) and restored with $\psftwo$ levels (red series). Right axes: the quotient of the restored and uncorrected values (blue series).}
\label{delta3}
\end{cfig}

In Fig. \ref{delta4}, we mark the location of image pixels counted as magnetic in the original and restored data, in a $201\times201$ pixel inset centred on the disc centre. Image restoration renders magnetic features spatially smaller as polarized radiation originating from these features, lost to the surrounding quiet Sun by aperture diffraction and stray light, is recovered (illustrated by the blue and red clusters). This change in the size of magnetic features likely depends on factors such as the surface area, and circumference to surface area ratio. The enhanced noise level also contributes to the smaller count in the restored data. Image restoration does recover some magnetic features smeared below the magnetogram signal threshold ($\bl=3\noise$) in the original data by instrumental scattered light (green clusters). Overall, less image pixels are counted as magnetic.

\begin{cfig}
\includegraphics[width=\textwidth]{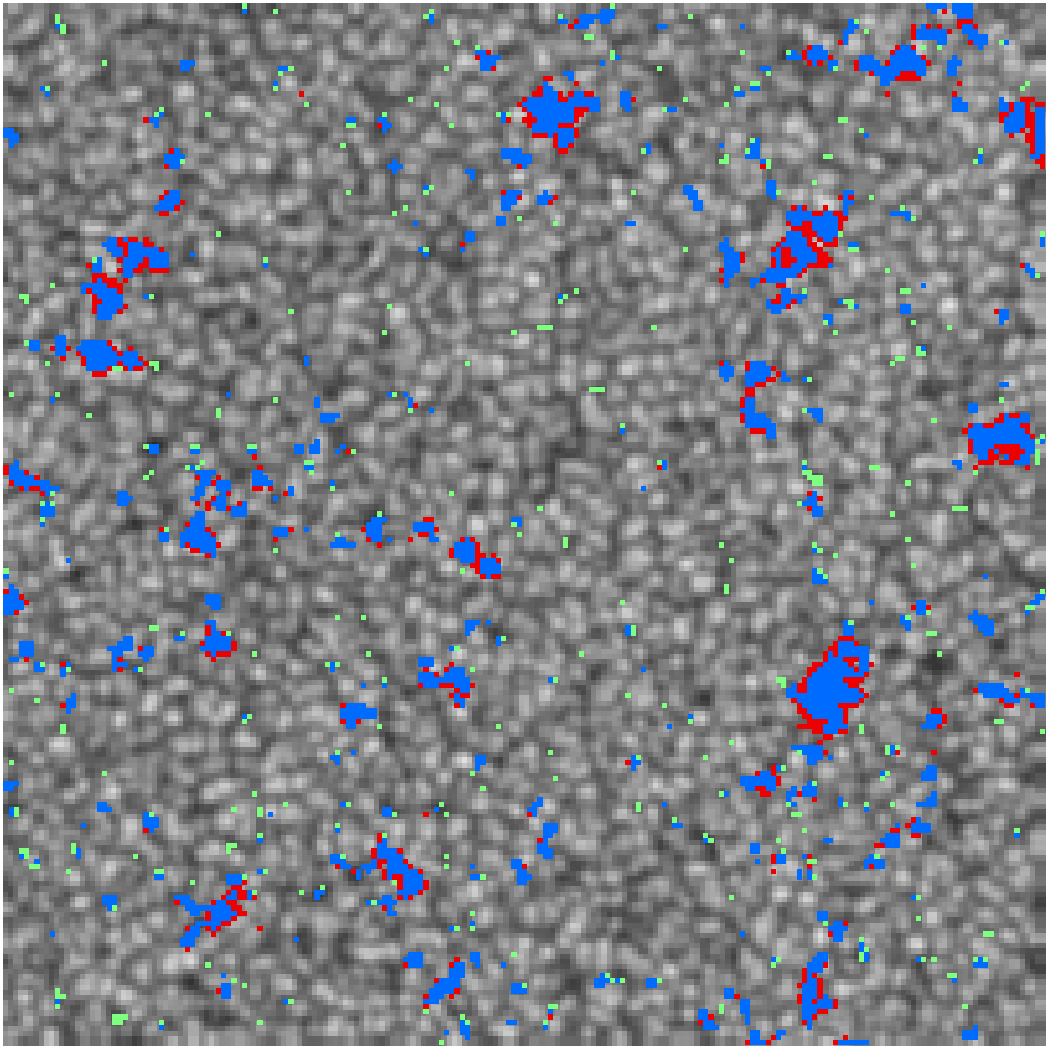}
\caption{$201\times201$ pixel inset, centred on disc centre, of the 720-s continuum intensity image, with the points displaying $\bl>3\noise$ in the 720-s longitudinal magnetogram highlighted. Blue corresponds to the points realizing this condition in both the uncorrected and restored magnetogram, red the points fulfilling it only in the uncorrected data and green the points satisfying it only in the restored data. The grey scale is saturated at 0.8 and 1.2 times the mean quiet-Sun level.}
\label{delta4}
\end{cfig}

The overall increase in $\noise$ towards the limb is partly related to the increase in low-level magnetogram signal fluctuations from the ubiquitous weak horizontal magnetic fields in the quiet Sun internetwork \citep{lites96,lites08,beck09}, which obtain a line-of-sight component near the limb, and magnetic features foreshortening towards the background noise regime when approaching the limb. A probable cause of the overall centre-to-limb increase in $\Delta\noise$ is the enhancement of these true signal contributions to apparent noise.

In Fig. \ref{speg}a we display the change in magnetogram signal as a function of the original signal, in the quiet Sun field illustrated in Fig. \ref{delta4} and in a $201\times201$ pixel crop of an active region near the disc centre ($\mu>0.92$), represented by the green and black curves, respectively. This was computed as done in Sect. \ref{ssmcss3} for the sunspots and pores in Fig. \ref{spotinset}, the results of which are also plotted for comparison (blue and red curves). The only difference is here we binned the image pixels by the uncorrected $\bl$ in intervals of $40\:{\rm G}$ instead of $100\:{\rm G}$. As done in Sect. \ref{ssmcss1}, we minimised the influence of pores present in the active region field by excluding the image pixels where $\inorm<\iqp$ and points up to three pixels away from them. (There are no image pixels where $\inorm<\iqp$ in the quiet Sun field.) The result from the quiet Sun and active region fields represent the effect of image restoration on the magnetogram signal of quiet Sun network, and active region faculae, respectively.

As noted for sunspots and pores (Sect. \ref{ssmcss2}), the enhancement of the magnetogram signal of network and faculae from image restoration is highly correlated to the original level. This is possibly from the restoration enhancing the signal in the core of magnetic features, where it is also typically stronger, while depressing the signal in the fringes, from the recovery of polarized radiation scattered from the core to the fringes and surrounding quiet Sun, as discussed above and visible for the magnetic features depicted in Fig. \ref{contrastnfpeak}.

The effect of image restoration on network and faculae is also more pronounced than in sunspots and pores, in particular for network. This is likely related to the smaller spatial scale of these features, which makes them more susceptible to stray light, and consequently they respond more acutely to restoration, than sunspots and pores. The stronger response of network compared to faculae is probably due to the fact that they appear in smaller clusters and the restoration of small-scale mixed polarities in the quiet Sun smeared out by instrumental scattered light.

As image restoration affects different magnetic features differently, the overall effect on the apparent amount of magnetic flux fluctuates with prevailing magnetic activity. This is the likely reason neither $\Delta\nmag$, $\Delta\mblmag$ nor $\dflux$ exhibit any obvious trend with $\mu$, modulated by the magnetic features present within each annulus. Importantly, the relatively acute effect of image restoration on network, and the fact that the solar disc is, by area, predominantly quiet Sun means most of the increase in the measured amount of magnetic flux comes from the enhancement of these features. This is consistent with the findings of \cite{krivova04}.

\subsubsection{Line-of-sight velocity}
\label{ssmcss4}

Here we are interested in the part of measured line-of-sight velocities in HMI Dopplergrams arising from convective motions on the solar surface. To this end, we corrected the 720-s Dopplergram for the contribution by the rotation of the Sun and the relative velocity of SDO to the Sun\footnote{Given by the radial, heliographic west and north velocity of the spacecraft relative to the Sun listed in the data header.}, $\vobs$, following the procedure of \cite{welsch13}. Oscillatory motions associated with $p-$modes are largely undetectable in the 720-s Dopplergram data product as it is a weighted combination of filtergram observations from a 1350-s period, much longer than the period of these oscillations ($\sim5$ minutes).

The rotation rate of the Sun varies with heliographic latitude, $\Phi$. Let $\vrot$ denote the velocity of the surface of the Sun from its rotation. We first determined the Stonyhurst latitude and longitude at each image pixel within the solar disc. Then, we derived the $\vrot$ at each latitude from the differential rotation profile by \cite{snodgrass83},
\begin{equation}
w\left(\Phi\right)=2.902-0.464\sin^{2}{\Phi}-0.328\sin^{4}{\Phi},
\label{snodgrass}
\end{equation}
which relates angular velocity, $w=\frac{|{\vrot}|}{R_{\rm S}\cos{\Phi}}$ ($R_{\rm S}$ being the radius of the Sun in metres) to $\Phi$. The contribution by $\vrot$ and $\vobs$ to measured velocity at a given image pixel is then given by the projection of $\vrot$ and $\vobs$ onto the line-of-sight to the point, taking the small-angle approximation. The projection of $\vobs$, in this instance, varied between 500 and $800\:\mps$ with position on the solar disc. The significant magnitude and variation of this with disc position arises from SDO's geosynchronous orbit about Earth and Earth's orbit about the Sun.

Let $\vl$ represent the signed Dopplergram signal, the mean line-of-sight component of the vector velocity over a given image pixel.

As done in Sect. \ref{ssmcss3}, we segmented the solar disc by $\mu$, into 50 equal annuli, excluding points where $\mu<0.1$. The image pixels where $\bmu>10{\rm G}$ in the 720-s longitudinal magnetogram were masked, leaving quiet Sun. Within each annulus, we binned the unmasked points by the uncorrected $\vl$ in intervals of $20\:\mps$, and retrieved the median original and restored $\vl$ within each bin, ignoring bins with less than 100 points. The factor enhancement of $\vl$ from image restoration, $\dvl$ is then given by the slope of the linear regression fit to the restored bin-median $\vl$ against the original, illustrated for the disc centremost interval of $\mu$ ($\mu>0.99$) in Fig. \ref{deltav}a. By performing this computation over small intervals of $\mu$, we avoid introducing scatter or bias in $\dvl$ from the $\mu$ dependence of the convective blueshift of the spectral line.

The enhancement of the Dopplergram signal in the quiet Sun from image restoration with $\psftwo$ is significant and exhibits an acute CLV; $\dvl$ increases monotonically with $\mu$, from about 1.4 near the limb to 2.1 at disc centre (Fig. \ref{deltav}b).

\begin{cfig}
\includegraphics[width=.87\textwidth]{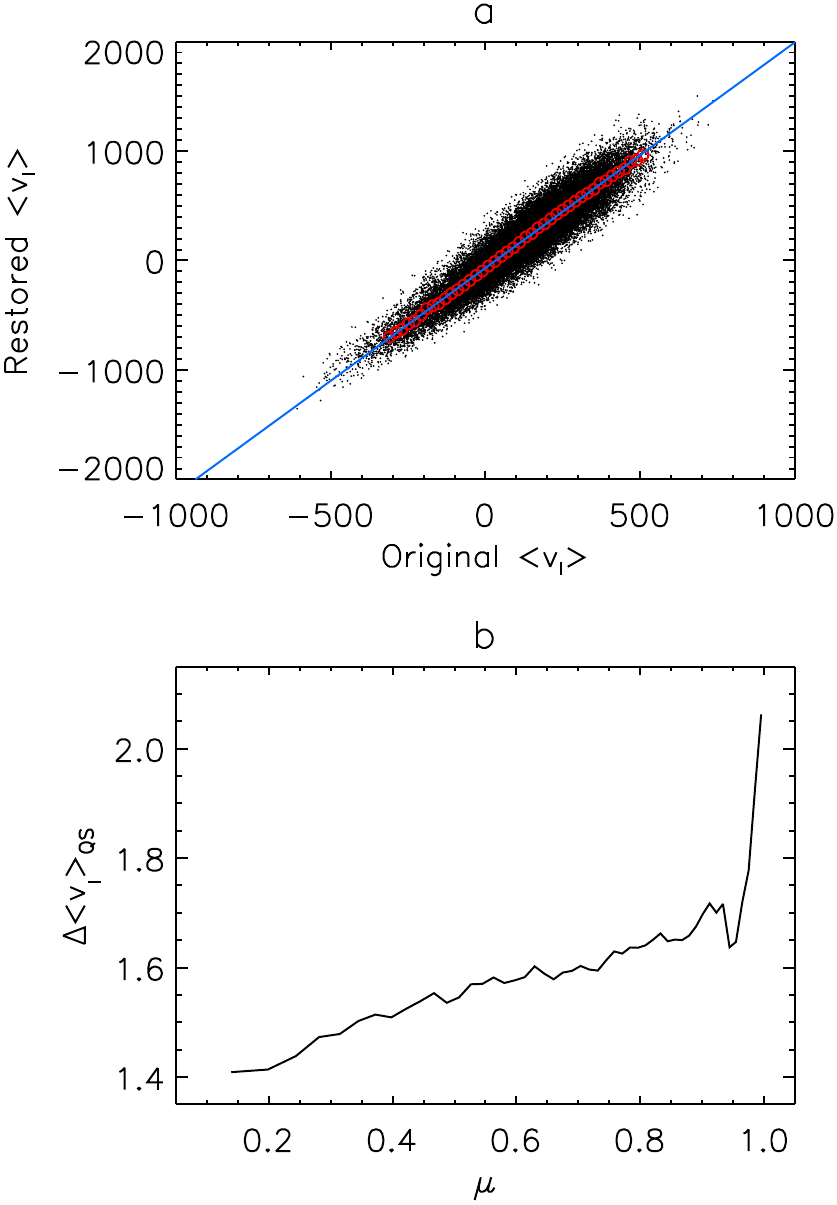}
\caption{a) Scatter plot of the Dopplergram signal, $\vl$ in the quiet Sun, near disc centre ($\mu>0.99$), in the restored (with $\psftwo$) and uncorrected versions of the 720-s Dopplergram. The red circles denote the bin-median (from binning the points by the uncorrected $\vl$ in intervals of $20\:\mps$) and the blue line the corresponding linear regression fit. b) The factor enhancement of $\vl$ in the quiet Sun from image restoration, $\dvl$ as a function of $\mu$.}
\label{deltav}
\end{cfig}

In looking only at the quiet Sun, we excluded phenomena localised in active regions (for example, Evershed flow in sunspots), leaving signal largely from granulation and supergranulation. Supergranulation flows are largely horizontal. The line-of-sight velocities of supergranulation flows are thus greatest near the limb and diminishes towards disc centre from foreshortening. In contrast, the apparent line-of-sight velocities of granulation flows diminish towards the limb. Approaching the limb, granulation is more and more difficult to resolve from foreshortening and the line-of-sight increasingly crossing into multiple granulation cells. The typical diameter of granulation and supergranulation cells is about $1\:{\rm Mm}$ and $30\:{\rm Mm}$, respectively. Granulation is therefore more affected by stray light and experiences greater signal enhancement from image restoration than supergranulation. The observed CLV of $\dvl$ is consistent with the converse CLV of the line-of-sight velocities of granulation and supergranulation flows, and the stronger effect of image restoration on granulation compared to supergranulation.

By correcting both the original and restored Dopplergram for differential rotation with Eqn. \ref{snodgrass}, we had implicitly assumed that this component of measured line-of-sight velocity is not significantly changed by stray light or its removal. Given the line-of-sight component of $\vrot$ varies gradually across the solar disc, this is true except very close to the limb. Therefore, the effect on this analysis is likely minute and confined to the annuli closest to the limb.

The pronounced Dopplergram signal enhancement effected by image restoration with $\psftwo$ could have an impact on the characterization of plasma flows in the solar surface with HMI data. In this study we will not attempt to examine the effects of image restoration on apparent Doppler shifts in active regions, $p-$mode oscillations (detectable in the 45-s Dopplergram data product of the front CCD) or the individual physical processes driving plasma motion on the solar surface.

\section{Variation of the PSF within the FOV, between the HMI CCDs and with time}
\label{psfdep}

Derived from observations of Venus in transit recorded on the side CCD, $\psftwo$ characterizes the stray light in the employed data, at the position in the FOV occupied by the venusian disc. Here we discuss the applicability of $\psftwo$ to the entire FOV, to data from the front CCD and, importantly, from other times.

\subsection{Variation of the PSF with FOV position}
\label{psfdep1}

Taking the test continuum filtergram (from Sect. \ref{gc}), we segmented the solar disc into eight equal sectors and computed the RMS intensity contrast of the quiet Sun within each sector. The level before and after image restoration with $\psftwo$ are illustrated in Fig. \ref{contrastqs2}a. There is scatter in the RMS contrast between the sectors which is more pronounced in the restored data. This enhanced divergence is, at least in part, caused by the fact that we restored the entire solar disc with a single PSF when the true PSF varies from sector to sector.

\begin{cfig}
\includegraphics[width=\textwidth]{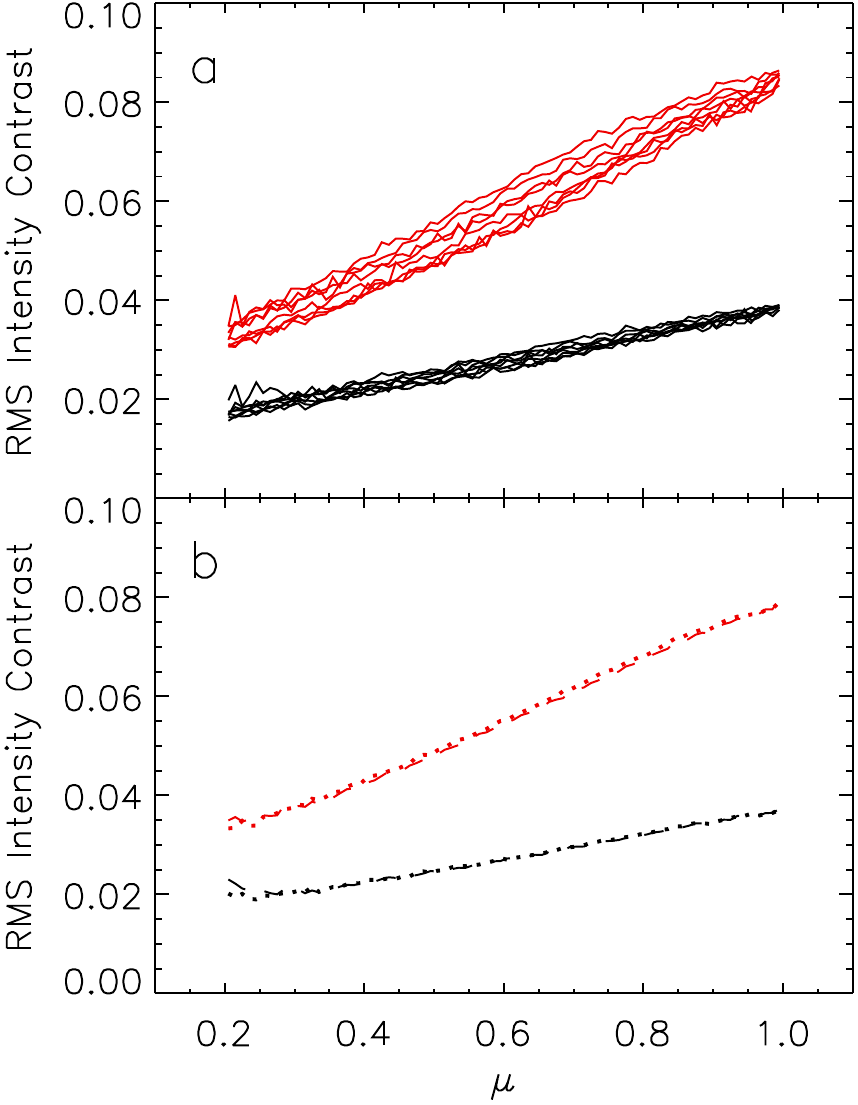}
\caption{RMS intensity contrast of the quiet Sun as a function of $\mu$, before (black lines) and after (red) image restoration with $\psftwo$. a) In the test continuum filtergram, the solar disc segmented into eight equal sectors. b) In the near-simultaneous filtergrams of similar bandpass and polarization from the side (dotted lines) and front (dashed) CCD.}
\label{contrastqs2}
\end{cfig}

The scatter in RMS contrast is, however, at least to the extent tested, relatively small in comparison to the absolute level. More importantly, the divergence between the sectors in the restored data (red curves) is small compared to the difference between the restored and the original data (black curves). This implies that the inhomogeneity introduced by applying a single PSF to the entire FOV is small in comparison to the contrast enhancement. Nonetheless, for sensitive measurements, care should be taken, where possible, to average measurements from different positions in the FOV after deconvolution with the PSF deduced here.

Next, we looked at the variation, over the solar disc, of the effect of image restoration with $\psftwo$ on the 720-s longitudinal magnetogram from Sect. \ref{ssmcss}. Sampling the solar disc at 16-pixel intervals in both the north-south and east-west directions, we centred a $401\times401$ pixel window over each sampled point and took the mean $\bl$ of all the solar disc pixels within the window, denoted $\mbl$. Let $\dbl$ represent the ratio of $\mbl$ in the restored and uncorrected data, representing the factor enhancement to $\mbl$ from the restoration.

\begin{cfig}
\includegraphics[width=.61\textwidth]{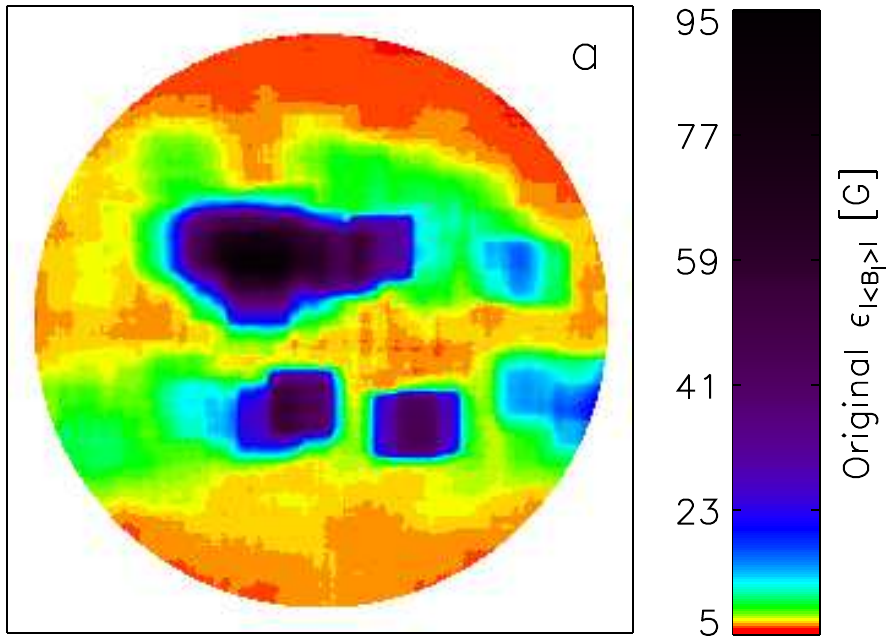}
\includegraphics[width=.61\textwidth]{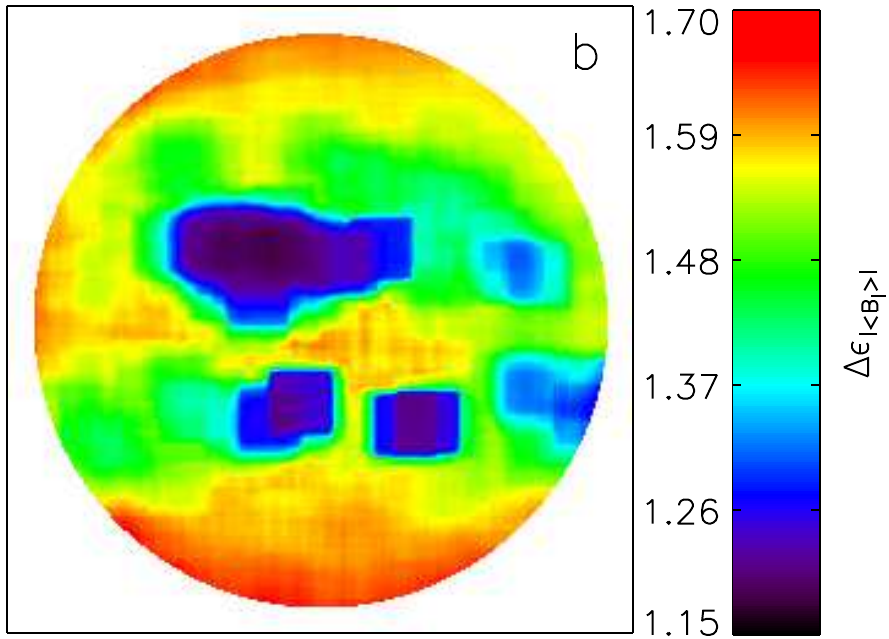}
\includegraphics[width=.61\textwidth]{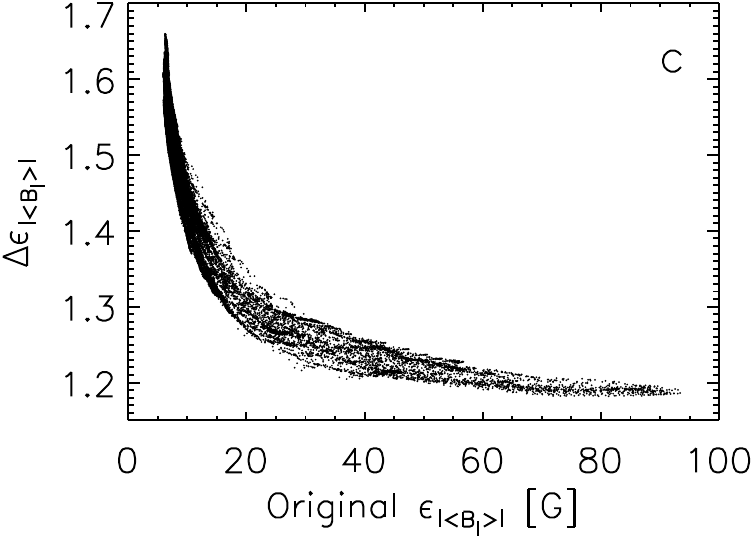}
\caption{a) Mean unsigned magnetogram signal, $\mbl$ of the original (not corrected for stray light) 720-s longitudinal magnetogram. a) Factor enhancement of $\mbl$ from image restoration with $\psftwo$, $\dbl$. c) Scatter plot of $\dbl$ against the original $\mbl$.}
\label{deltamag}
\end{cfig}

Expectedly, $\dbl$ varied with position on the solar disc (Fig. \ref{deltamag}b). This variation is driven by:
\begin{itemize}
	\item differences in the magnetic activity present in the sampling window (image restoration affects different magnetic features differently, Sect. \ref{ssmcss3}) and
	\item fluctuations in the effect of image restoration from the variation of the stray light behaviour of the instrument across the FOV.
\end{itemize}
The scatter plot of $\dbl$ versus the original $\mbl$ reveals an inverse relationship between the two quantities (Fig. \ref{deltamag}c). The cause of this correlation is that the restoration enhances the magnetogram signal in the quiet Sun more strongly than in active regions (Fig. \ref{speg}a). The relatively weak spread of the scatter plot suggests the variation in $\dbl$ with disc position is dominantly due to the inhomogeneous distribution of magnetic activity (Fig. \ref{deltamag}a) and the diverging effects of restoration on different magnetic features. This is further evidence that it is a reasonable approximation to restore the entire FOV of HMI for instrumental scattered light with a single PSF.

\subsection{Variation of the PSF between the HMI CCDs and with time}
\label{psfdep2}

In Fig. \ref{contrastqs2}b we show the RMS intensity contrast of the quiet Sun in two near-simultaneous filtergrams, one from each CCD, taken less than one minute apart. The two filtergrams were recorded about an hour after Venus left the solar disc (at around 05:36 UTC, June 6, 2012), shortly after the side CCD resumed collection of the regular filtergram sequence.

The two near-simultaneous filtergrams were taken in the same bandpass (-172 m\AA{} from line centre) and polarization (Stokes $I-V$). Any disparity in the RMS contrast between the two would arise mainly from differences in the performance of the two CCDs, including the stray light behaviour. The RMS contrast in the two filtergrams is very similar, even after image restoration with $\psftwo$ (red curves). Since we expect any disparity due to differences in the stray light behaviour of the two CCDs to be amplified by the application of the side CCD PSF to a front CCD filtergram, this agreement is encouraging.

Comparing the close similarity in the RMS intensity contrast of the quiet Sun in the two near-simultaneous filtergrams to the scatter between different FOV positions in the test continuum filtergram (top panel), it appears that the difference in the stray light behaviour of the two CCDs is much smaller than the variation with position in the FOV of the side CCD. The side and front CCDs are identical and share a common optical path \citep{schou12}, it is therefore within reason that the stray light behaviour, even the variation of the PSF with position in the FOV, is broadly similar.

The SDO satellite is in a geosynchronous orbit. Since the HMI commenced regular operation (May 1, 2010), the side and front CCDs recorded a continuum filtergram each during the daily pass through orbital noon and midnight. For the purpose of investigating the stability of the stray light behaviour of the instrument over time, we examined the solar aureole, the intensity observed outside the solar disc arising from instrumental scattered light, in these daily data\footnote{The solar aureole is not to be confused with the aureole about the venusian disc when it is in transit, discussed in Sect. \ref{venusatm}.}. We examined 2307 orbital noon and midnight continuum filtergrams from each CCD, spanning the 1157-day period of May 1, 2010 to June 30, 2013.

For each continuum filtergram, we averaged the solar aureole over all azimuths. From the resulting radial intensity profile, we determined the intensity of the solar aureole at distances of 1 and 10 arcsec from the edge of the solar disc. The derived intensities, normalized to the limb level, are expressed in Figs. \ref{analysistime}a and \ref{analysistime}b, respectively.

We excluded the points that are spurious or from continuum filtergrams with missing pixel values, leaving 2248 and 2288 points from the side and front CCDs respectively. To show up the broad trend with time we interpolated each time series at one day intervals and smoothed the result by means of binomial smoothing \citep{marchand83}. The time variation in the relative (to the limb level) intensity of the solar aureole at 1 and 10 arcsec from the limb reflects changes to the shape of the PSF near the core and in the wings respectively.

\begin{cfig}
\includegraphics[width=\textwidth]{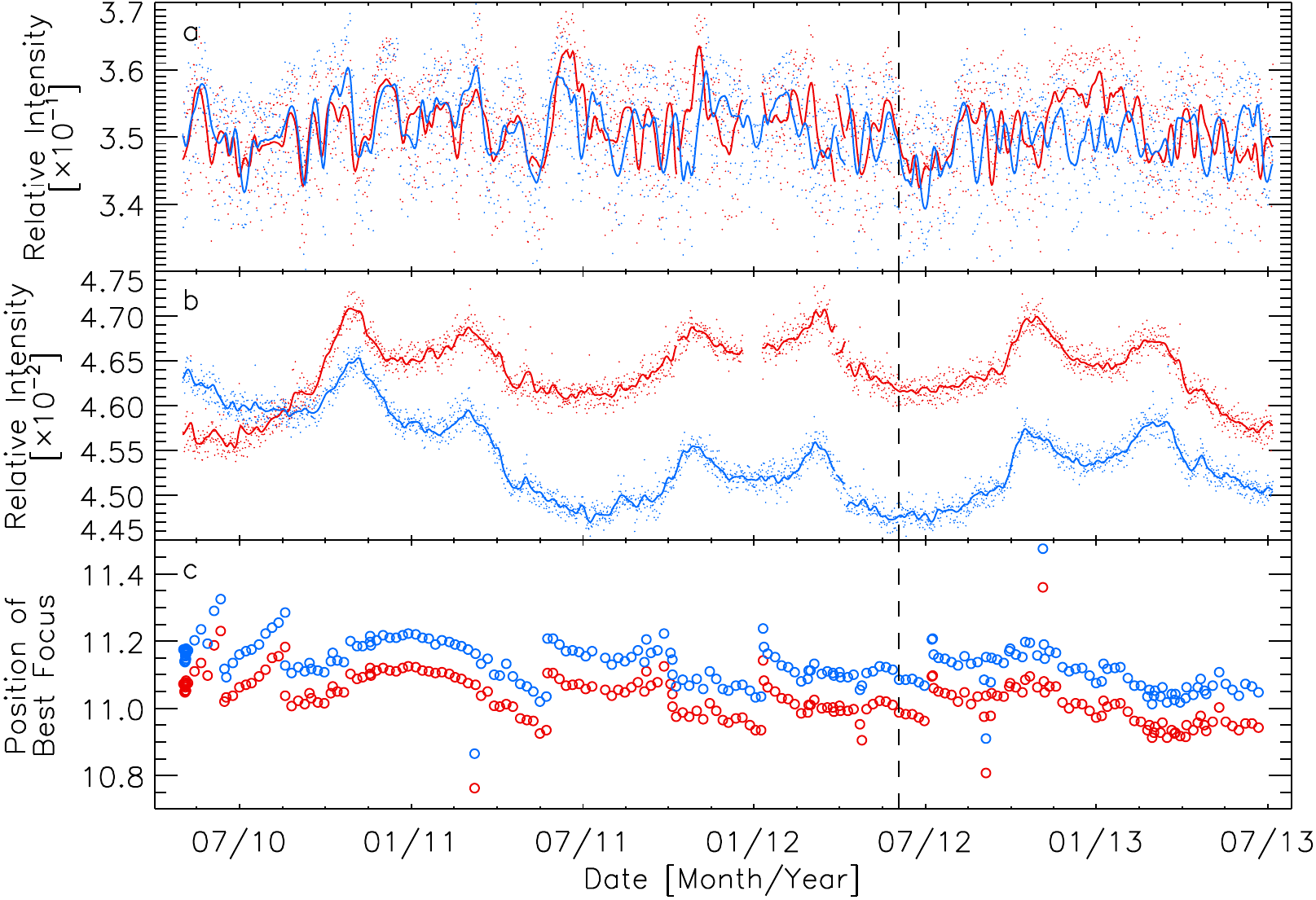}
\caption{Radial intensity of the solar aureole, at distances of a) 1 and b) 10 arcsec from the edge of the solar disc, in the orbital noon and midnight continuum filtergrams from the side (red) and front (blue) CCDs. The value from each filtergram is normalized to the level at the edge of the solar disc. The dots represent the measured values and the curves the smoothed time series. Segments of measured values spaced more than one day apart were treated as separate time series, giving the gaps in the curves. c) The position of best focus, in units of focus steps (see text), from the (approximately) weekly focus calibration of the side (red circles) and front CCDs (blue circles). The dashed line marks the time of the transit of Venus.}
\label{analysistime}
\end{cfig}

There is clear point-to-point fluctuation in the relative intensity of the solar aureole at 1 arcsec from the limb. This is due to the fact that near the limb, the intensity of the solar aureole decays rapidly with radial distance such that small variations in the width of the core of the PSF show up as large swings in the relative intensity at a fixed distance from and near the limb. Overall, the magnitude of the scatter, given by the standard deviation, is below $2\%$ of the mean level. Also, the side and front CCD series are relatively stable and similar, with no overt long term trends or divergence from one another. This suggests that the width of the core of the PSFs of the two CCDs are, to the extent tested, broadly similar and constant over the period examined.

For the relative intensity of the solar aureole at 10 arcsec from the limb, there is an approximately synchronous periodic variation between the two CCDs. There is also a gradual overall drift between May 2010 and July 2011, and between August 2012 and June 2013. The fluctuation of the two time series and the divergence between them is minute, less than $5\%$ of the overall level. The time variation of, and difference between the wings of the PSFs of the two CCDs, implied by these fluctuations, has likely little effect on image contrast. Even with the $\sim3\%$ offset between the front and side CCD time series around the time of the Venus transit, the RMS intensity contrast of the quiet Sun in the near-simultaneous filtergrams from the two CCDs (taken about an hour after the end of the transit) is practically identical (Fig. \ref{contrastqs2}b).

Going from 1 to 10 arcsec from the limb, the time dependence of the relative intensity of the solar aureole changes gradually from the trend seen at 1 arcsec to that seen at 10 arcsec. Beyond 10 arcsec the variation with time does not change significantly. The relative intensity of the solar aureole at 1 and 10 arcsec from the limb therefore constitute a reasonable representation of variation in the shape of the PSFs of the two CCDs with time.

From the near identity of the RMS intensity contrast of the quiet Sun in the near-simultaneous filtergrams from the two CCDs (Fig. \ref{contrastqs2}b) and the minute time variation of the PSFs of the two CCDs implied by the relative stability of the shape of the solar aureole in daily data (Figs. \ref{analysistime}a and \ref{analysistime}b), we assert that $\psftwo$ can be applied to observations from both CCDs for the period examined (May 1, 2010 to June 30, 2013) without introducing significant error.

This assertion is consistent with the state of the focus of the instrument over the period examined. The PSF of HMI, like any optical system, is strongly modulated by the focus. The instrument is maintained in focus by varying the heating of the front window and the position of two five-element optical wheels \citep{schou12}. By varying the elements of the two optical wheels placed in the optical path, they allow the focus to be adjusted in 16 uniform, discrete steps, each corresponding to about two-thirds of a depth of focus. In Fig. \ref{analysistime}c we plot the position of best focus, in units of the focus steps, from the regular (roughly weekly) focus calibration of the sensor, over the same period as the daily data. The focus of the two CCDs are remarkably similar and stable, differing from one another and varying over the period of interest by much less than a depth of focus. The contribution by focus to the variation of the PSF between the two CCDs and with time is most probably minute.

\section{Summary}
\label{summary}

In this paper we present an estimate of the PSF of the SDO/HMI instrument. The PSF was derived from observations of Venus in transit. We convolved a simple model of the venusian disc and solar background with a guess PSF iteratively, optimizing the agreement between the result of the convolution and observation. We modelled the PSF as the linear sum of five Gaussian functions, the amplitude of which we allowed to vary sinusoidally with azimuth. This azimuthal variation was necessary to reproduce the observations accurately. Recovering the anisotropy of the PSF was also shown to be important for the proper removal of stray light from HMI data by the deconvolution with the PSF. The interaction between solar radiation and the venusian atmosphere is complex and not straightforward to account for. The result is a conservative estimate of the PSF, similar in width to the ideal diffraction-limited PSF in the core but with more extended wings.

The PSF was derived with data from one of the two identical CCDs in the sensor. It therefore represents the stray light behaviour of that particular CCD, at the time of the transit of Venus, at the position in the FOV occupied by the venusian disc in the employed observations.

Comparing the apparent granulation contrast in different parts of a single image, we showed that although the variation in the stray light behaviour of the instrument with position in the FOV introduces uncertainty to measured contrast, amplified by restoring observations with a single PSF, the scatter is relatively minute and will likely have little quantitative influence if care is taken to average measurements from multiple FOV positions. This was confirmed by an examination of the uniformity, over the FOV, of the effect of image restoration on the 720-s longitudinal magnetogram data product.

The time variation of the shape of the solar aureole in daily data was taken as an indication of PSF changes over the period examined (May 1, 2010 to June 30, 2013). Based on the relatively weak time variation of the aureole, and the similarity of the aureole and granulation contrast in data from the two CCDs, we assert that the PSF derived here can be applied to observations from both CCDs over the period examined without introducing significant error.

Apparent granulation contrast, given here by the RMS intensity contrast of the quiet Sun, in HMI continuum observations restored by the deconvolution with the PSF, exhibit reasonable agreement with that in artificial images generated from a 3D MHD simulation at equal spatial sampling. This demonstrates that the PSF, though an approximation, returns a competent estimate of the aperture diffraction and stray light-free contrast. The restoration enhanced the RMS intensity contrast of the quiet Sun by a factor of about 1.9 near the limb ($\mu=0.2$), rising up to 2.2 at disc centre.

We also illustrated the effect of image restoration with the PSF on the 720-s Dopplergram and longitudinal magnetogram data products, and the apparent intensity of magnetic features.
\begin{itemize}
	\item For small-scale magnetic concentrations, image restoration enhanced the intensity contrast in the continuum and core of the Fe I 6173 \AA{} line by a factor of about 1.3, and the magnetogram signal by a factor of about 1.7.
	\item Magnetic features in the longitudinal magnetogram are rendered smaller, as polarized radiation smeared onto surrounding quiet Sun by instrumental scattered light is recovered.
	\item Image restoration increased the apparent amount of magnetic flux above the noise floor by a factor of about 1.2, mainly in the quiet Sun. This may be, in part, from the recovery of magnetic flux in opposite magnetic polarities lying close to one another partially cancelled out by stray light.
	\item The influence of image restoration on sunspots and pores varied strongly, as expected, within a given feature and between features of different sizes.
	\item Line-of-sight velocity due to plasma motions on the solar atmosphere increases by a factor of about 1.4 to 2.1. The variation comes from the restoration enhancing granulation flows more strongly than larger scale supergranulation flows. Given the spatial scale dependence, the effect on Doppler shifts from large scale motions such as the meridional flow and differential rotation is probably minute.
\end{itemize}
The pronounced effect of image restoration on the apparent radiant, magnetic and motional properties of solar surface phenomena could have a significant impact on the interpretation of HMI observations. For instance, the observation that restoring HMI longitudinal magnetograms renders magnetic features smaller while boosting the magnetogram signal, and the increase in the amount of magnetic flux is dominantly in the quiet Sun, can influence models of variation in solar irradiance based on HMI data. Specifically, with models that relate the contribution by small-scale magnetic concentrations to solar irradiance variations to the number density and magnetogram signal \citep[for example,][]{wenzler06,foukal11,ball12}.

\chapter[Reconstruction of total and spectral solar irradiance from 1974 to 2013 based on KPVT, SoHO/MDI and SDO/HMI observations \\ \textit{\footnotesize{(The contents of this chapter are identical to the submitted version of Yeo, K. L., Krivova, N. A., Solanki, S. K., Glassmeier, K. H., 2014, Reconstruction of total and spectral solar irradiance from 1974 to 2013 based on KPVT, SoHO/MDI and SDO/HMI observations, Astron. Astrophys., 570, A85.)}}]{Reconstruction of total and spectral solar irradiance from 1974 to 2013 based on KPVT, SoHO/MDI and SDO/HMI observations}
\label{paper3}

\begin{flushright}
{\it Yeo, K. L., Krivova, N. A., Solanki, S. K., Glassmeier, K. H.} \\
{\bf Astron. Astrophys., 570, A85 (2014)\footnoteD{The contents of this chapter are identical to the submitted version of Yeo, K. L., Krivova, N. A., Solanki, S. K., Glassmeier, K. H., 2014, Reconstruction of total and spectral solar irradiance from 1974 to 2013 based on KPVT, SoHO/MDI and SDO/HMI observations, Astron. Astrophys., 570, A85, reproduced with permission from Astronomy \& Astrophysics, \textcopyright{} ESO.}}
\end{flushright}

\section*{Abstract}

{\it Context.} Total and spectral solar irradiance are key parameters in the assessment of solar influence on changes in the Earth's climate.

\noindent
{\it Aims.} We present a SATIRE-S reconstruction of daily solar irradiance spanning 1974 to 2013, based on full-disc observations from the KPVT, SoHO/MDI and SDO/HMI.

\noindent
{\it Methods.} SATIRE-S ascribes variation in solar irradiance, on timescales greater than a day, to photospheric magnetism. The solar spectrum is reconstructed from the apparent surface coverage of bright magnetic features and sunspots in the daily data using the modelled intensity spectra of these magnetic structures. We cross-calibrated the various data sets, harmonizing the model input so to yield a single consistent time series as the output.

\noindent
{\it Results.} The model replicates $92\%$ ($R=0.957$) of the variability, and the secular decline between the 1996 and 2008 solar cycle minima in the PMOD TSI composite. The model also reproduced most of the variability in solar Lyman-$\alpha$ irradiance and the Mg II index. The ultraviolet solar irradiance measurements from the UARS and SORCE missions exhibit discrepant solar cycle variation, especially above 240 nm. As a result, the model while able to replicate the rotational variability in these records, aligned with certain observations better than others in terms of the long-term trends. The solar cycle variation in the ultraviolet in the reconstruction is confirmed by the close match to that in a SUSIM-based empirical model from a previous study. As with earlier similar investigations, the reconstruction cannot reproduce the long-term trends in SORCE/SIM spectrometry. We argue, from the apparent lack of solar cycle modulation in SIM SSI and the dissimilarity between the total flux recorded by the instrument and TSI, that unaccounted instrumental trends are present.

\noindent
{\it Conclusions.} The daily solar irradiance time series is consistent with observations from multiple sources, demonstrating its validity and utility for climate models. It also provides further evidence that photospheric magnetism is the prime driver of variation in solar irradiance on timescales greater than a day.

\section{Introduction}
\label{p3introduction}

Solar radiation is the principle source of energy entering the Earth system. As such, the variation in the Sun's radiative output, solar irradiance, is a prime candidate driver of externally driven changes to the Earth's climate. A significant body of publications citing correlations between solar variability and climate parameters exist in the literature, and numerous mechanisms have been mooted to explain these observations \citep[see e.g.,][for a review]{gray10}. The Earth's climate is believed to be modulated by variations in solar irradiance through effects related to its absorption at the Earth's surface, and in the atmosphere. An established mechanism is the interaction between ultraviolet irradiance and stratospheric ozone \citep[e.g.,][]{haigh94,solanki13}. These mechanisms are wavelength dependent, from the wavelength dependence of the variation in solar irradiance and its absorption in the Earth's surface and atmosphere \citep{haigh07}. Consequently, both total and spectral solar irradiance, TSI and SSI, are of interest in assessing the impact of variation in solar irradiance on the Earth's climate.

TSI has been monitored since 1978 through a succession of spaceborne radiometers. The measurements from individual radiometers diverge, mainly in terms of the absolute level and long-term trends, due to instrument-related factors such as the radiometric calibration, degradation with time and design. Accounting for these influences to combine the various data sets into a single time series is non-trivial and still a topic of debate \citep[e.g.,][]{scafetta09,krivova09b,frohlich12}. There are, at present, three published composites. Namely, the ACRIM \citep{willson03}, IRMB \citep{dewitte04b,mekaoui08} and PMOD \citep{frohlich00} composites. The long-term trends in these competing series do not agree. Notably, the successive solar minima in 1986, 1996 and 2008 exhibit conflicting levels and cycle-to-cycle variation in the three composites, encapsulating the difficulty in observing secular changes in TSI unambiguously.

As in the case of TSI, ultraviolet spectral irradiance (120 to 400 nm) has been measured from space on a regular basis since 1978. Again, the combination of the measurements from the succession of instruments into a single time series is an ongoing challenge \citep{deland08,deland12}. The difficulty in producing a composite record is, in this instance, compounded by the wavelength dependence of instrumental influences.

The series of GOME instruments \citep[the first of which was launched in 1996 onboard ERS-2,][]{weber98,munro06} and ENVISAT/SCIAMACHY \citep[launched in 2002,][]{skupin05a} made regular measurements of the solar spectrum in the 240 to 790 nm and 240 to 2380 nm range, respectively. However, these instruments are designed for atmospheric sounding measurements which do not require absolute radiometry rather than the monitoring of solar irradiance. As they lack the capability to track instrument degradation in-flight, long-term trends in solar irradiance cannot be recovered from the spectral observations. The spectral measurements from the Sun PhotoMeter, SPM on the SoHO/VIRGO experiment \citep{frohlich95,frohlich97}, spanning 1996 to the present, cover just three narrow (FWHM of 5 nm) bandpass, at 402, 500 and 862 nm. In the visible and infrared, continuous observation of spectral irradiance only started, in effect, with SORCE/SIM \citep{harder05a,harder05b}, which has been surveying the wavelength range of 200 to 2416 nm since 2003. SIM provides what is, at present, the only extended, radiometrically calibrated, record of SSI covering the ultraviolet to the infrared available.

The available body of solar irradiance observations is invaluable. However, given the limited period in time covered and uncertainties in the long-term variation, there is a need to augment observations with models relating solar variability to irradiance. Such models have been reported for almost as long as irradiance measurements were made from space \citep[e.g.,][]{wilson81,oster82}. The models that ascribe fluctuations in solar irradiance to the time evolution of magnetic concentrations in the photosphere \citep[from its influence on the temperature structure of the solar surface and atmosphere,][]{spruit83} have been particularly successful \citep{domingo09}. This is especially true of TSI. The recent TSI reconstructions by \cite{ball12} and \cite{chapman13}, based on such models, replicated $96\%$ and $95\%$ of the observed variation in SORCE/TIM radiometry \citep{kopp05a,kopp05b,kopp05c}, respectively. The model by \cite{ball12} also reproduced $92\%$ of the variation in the PMOD TSI composite.

In the case of SSI, gaping disagreements between models and observations persist, exemplified in the current debate on the long-term trends in the SORCE/SIM record \citep{harder09,lean12,unruh12,wehrli13,ermolli13}. The ultraviolet irradiance (200 to 400 nm) measured by SIM declined from 2004 to 2008 by a factor of two to six more than in other measurements and models \citep[sans][discussed below]{fontenla11}. The range of two to six reflects the spread between these other measurements and models. The pronounced drop in the ultraviolet, almost twice the decrease in TSI level over the same period, is accompanied and so compensated by an increase in visible irradiance (400 to 700 nm). Coming at a time where solar activity is declining, this elevation in the visible runs counter to models of solar irradiance, which point to visible solar irradiance varying in phase with the solar cycle instead.

The reconstruction of TSI and SSI by \cite{fontenla11}, based on PSPT observations \citep{coulter94,ermolli98} and the model atmospheres of \cite{fontenla09} with certain adjustments, is the only model so far to reproduce the long-term trends in SIM SSI, albeit qualitatively. However, the reconstruction failed to reproduce the solar cycle variation in TSI, which other models have generally been able to with reasonable success. Employing similar PSPT images and the model atmospheres of \cite{fontenla09} without modification, the analogous computation by \cite{ermolli13}, as with preceding models, reproduced observed rotational and cyclic variation in TSI but not the long-term behaviour of SIM SSI.

Apart from the models based on the regression of indices of solar activity to measured solar irradiance, commonly referred to as proxy models \citep[e.g.,][]{lean97,lean00,pagaran09,chapman13}, present day models of spectral solar irradiance have a similar architecture to one another \citep[reviewed in][]{ermolli13}. The proportion of the solar disc covered by magnetic features (such as faculae and sunspots) is deduced from full-disc observations. This information is converted to solar irradiance by means of the calculated intensity spectra of said features (derived by applying radiative transfer codes to semi-empirical model atmospheres). The Spectral And Total Irradiance REconstruction for the Satellite era, SATIRE-S \citep{fligge00,krivova03,krivova11a} is an established model of this type.

The SATIRE-S model has been applied to full-disc observations of intensity and magnetic flux from the Kitt Peak Vacuum Telescope \citep[KPVT,][]{livingston76,jones92} and the Michelson Doppler Imager onboard the Solar and Heliospheric Observatory \citep[SoHO/MDI,][]{scherrer95}. The solar irradiance reconstructions by \cite{krivova03,krivova06,krivova09a,krivova11b}, \cite{wenzler05b,wenzler06,wenzler09}, \cite{unruh08} and \cite{ball11,ball12,ball14}, spanning various periods between 1974 to 2009, achieved considerable success in replicating observed fluctuations in TSI, SSI measurements from the missions preceding SORCE and, at rotational timescales, SORCE SSI.

The reconstruction of solar irradiance with SATIRE-S has been curtailed by the deactivation of MDI in 2011. (The KPVT had ceased operation earlier in 2003.) In this study, we present a SATIRE-S reconstruction of total and spectral solar irradiance from 1974 to 2013. This work sought to update the preceding efforts based on KPVT and MDI data with similar observations from the Helioseismic and Magnetic Imager onboard the Solar Dynamics Observatory \citep[SDO/HMI,][]{schou12}.

The apparent surface coverage of the solar disc by magnetic features, the main input to the model, is instrument dependent. Concurrent observations from different instruments can diverge significantly from differences in spatial resolution, bandpass, calibration, stray light, noise and the like. The combination of the model output based on data from multiple instruments into a single time series constitutes one of the main challenges to such a study. Apart from the extension to the present time with HMI data, this study departs from the earlier efforts with the SATIRE-S model in how this combination is done. This improvement in the methodology is one of the main advances of this paper.

In the following, we briefly describe the SATIRE-S model (Sect. \ref{model}) and the data used in the reconstruction (Sect. \ref{p3data}). Thereafter, we detail the reconstruction process (Sect. \ref{analysis}), before a discussion of the result (Sect. \ref{p3discussion}) and summary statements (Sect. \ref{p3summary}).

\section{The SATIRE-S model}
\label{model}

SATIRE-S is currently the most precise version of the SATIRE model. The main assumption of the SATIRE model is that variations in solar irradiance, on timescales of days and longer, arise from solar surface magnetism alone. The different versions of the model differ by the data used to deduce the surface coverage of magnetic features \citep[see][]{krivova07,krivova11a,vieira11}. The SATIRE-S utilises continuum intensity images and longitudinal magnetograms. In the present iteration, the solar surface is modelled as being composed of quiet Sun, faculae, and sunspot umbrae and penumbrae.

Image pixels with continuum intensity below threshold levels representing the umbral (umbra to penumbra), and the penumbral (penumbra to granulation) boundary are classified as umbra and penumbra, respectively. Points with magnetogram signals exceeding a certain threshold and not already classed as umbra or penumbra (i.e., bright magnetic features), are denoted as faculae. (While bright magnetic features include both network and faculae, in this study we refer to them collectively as faculae for the sake of brevity.) The remaining image pixels are then taken to correspond to quiet Sun. Standalone facular pixels (i.e., not contiguous with any other) are reassigned as quiet Sun to minimize the misidentification of magnetogram noise fluctuations as bright magnetic features.

Let $\bl$ denote the longitudinal magnetogram signal (the pixel-averaged line-of-sight magnetic flux density) and $\mu$ the cosine of the heliocentric angle. At the spatial resolution of available full-disc magnetograms, the small-scale magnetic concentrations associated with bright magnetic features remain largely unresolved. This is approximately accounted for by scaling the filling factor of facular pixels (defined here as the effective proportion of the resolution element occupied) linearly with $\bmu$, from zero at 0 G, to unity at what is denoted $\bsat$, where after it saturates \citep[see][for details]{fligge00}. The quantity $\bsat$ is the sole free parameter in the model.

As small-scale magnetic concentrations are generally orientated roughly normally to the solar surface (due to magnetic buoyancy), the quantity $\bmu$ represents a first-order approximation of the pixel-averaged magnetic flux density. This approximation breaks down very close to the limb from the combined action of foreshortening and magnetogram noise. For this reason, image pixels near the limb ($\mu<0.1$, about $1\%$ of the solar disc by area) are ignored. Following \cite{ball11,ball12}, we counted facular pixels with $\bmu$ above an arbitrary but conservative cutoff value, denoted $\bcut$, as quiet Sun instead. These points correspond, especially towards the limb, mainly to the magnetic canopy of sunspots, rather than legitimate faculae \citep{yeo13}.

In this study, we used the same set of calculated intensity spectra, of umbra, penumbra, faculae and quiet Sun (at various values of $\mu$) as \cite{wenzler06,ball12} to convert the surface coverage of magnetic features to solar irradiance \citep[the derivation of these spectra is detailed in][]{unruh99}. The model output is the summation, over all the image pixels within the solar disc, of the intensity spectrum corresponding to each point as defined by the above analysis. The resulting spectrum, spanning the wavelength range of 115 to 160000 nm, and the integral represent the prevailing SSI and TSI at the sampled point in time. 

The appropriate value of the free parameter, $\bsat$ is recovered by comparing the reconstruction to measured TSI (see Sect. \ref{fixbsat}).

\section{Data selection and preparation}
\label{p3data}

\subsection{Daily full-disc continuum intensity images and longitudinal magnetograms}
\label{dailypairs}

In this study, we employed full-disc longitudinal magnetograms and continuum intensity images collected at the KPVT, and from the first-ever spaceborne magnetograph, SoHO/MDI \citep{scherrer95} and its successor instrument, SDO/HMI \citep{schou12}. The NASA/NSO 512-channel diode array magnetograph \citep{livingston76}, installed at the KPVT, started operation in 1974. In 1992, it was replaced with the NASA/NSO spectromagnetograph \citep{jones92}, itself retired in 2003. Here, we will refer to the two configurations of the KPVT, effectively two unique instruments for the purpose of this study, as $\kpone$ and $\kptwo$, respectively.

\begin{sidewaystable}
\centering
\caption{Summary description of the daily full-disc continuum intensity images and longitudinal magnetograms selected for this study.}
\centering
\begin{tabular}{lcccccc}
\hline\hline
 & & & Proportion of & & & \\
Instrument & No. of data days & Period [year.month.day] & period covered & Image size [pixel] & Pixel scale [$\as$] & Spectral line\\
\hline
$\kpone$ & 1371 & 1974.08.23 to 1993.04.04 & 0.20 & $2048\times2048$ & 1     & Fe I 8688 \AA{} \\
$\kptwo$ & 2055 & 1992.11.21 to 2003.09.21 & 0.52 & $1788\times1788$ & 1.14  & Fe I 8688 \AA{} \\
MDI      & 3941 & 1999.02.02 to 2010.12.24 & 0.91 & $1024\times1024$ & 1.98  & Ni I 6768 \AA{} \\
HMI      & 1128 & 2010.04.30 to 2013.05.31 & 1.00 & $4096\times4096$ & 0.504 & Fe I 6173 \AA{} \\
\hline
\end{tabular}
\label{datatable}
\end{sidewaystable}

To the extent permitted by available data (see Sects. \ref{datahmi} to \ref{datakpvt}), we selected, for each instrument, a continuum intensity image and a longitudinal magnetogram, recorded simultaneously or close in time, from each observation day. The number of selected daily continuum intensity images and longitudinal magnetograms from each instrument, and the period covered are summarized in Table \ref{datatable}. Also listed is the image size, pixel scale and spectral line surveyed.

\subsubsection{HMI}
\label{datahmi}

The HMI is a full-Stokes capable filtergram instrument. The instrument captures full-disc filtergrams continuously at 1.875s cadence, alternating between two CCDs, six positions across the Fe I 6173 \AA{} line and six polarizations. The filtergram data is combined to form continuum intensity images and longitudinal magnetograms at 45-s cadence. Generated from similar filtergram data, the two data products are exactly co-temporal.

For each day since the instrument commenced regular operation in April 30, 2010, up to 31 May, 2013, we took the average of seven consecutive 45-s continuum intensity images, resampled in space and time to co-register, and likewise the average of the corresponding 45-s longitudinal magnetograms. We will refer to the result as the 315-s continuum intensity image and longitudinal magnetogram. The averaging is to facilitate the segmentation of the solar disc by suppressing intensity and magnetogram signal fluctuations from noise and $p$-mode oscillations.

We took the first 315-s continuum intensity image and longitudinal magnetogram from each day, except for the period where we also have selected MDI observations (April 30 to December 24, 2010, see Sect. \ref{datamdi}). On these days we took the HMI data taken closest in time to the MDI longitudinal magnetogram.

Aided by the absence of atmospheric seeing, granulation and sunspot structures are starting to be resolved at HMI's spatial resolution ($\sim1\:\as$). To minimize the misclassification of darker non-sunspot and brighter sunspot features, the 315-s continuum intensity images were (after the correction for limb darkening, see Sect. \ref{equalsunspots}) convolved with a $7\times7$ pixel ($3.5\times3.5\:\as$) Gaussian kernel.

\subsubsection{MDI}
\label{datamdi}

The MDI returned observations from March 19, 1996 to April 11, 2011. The instrument recorded full-disc filtergrams at five positions, one near and four within the Ni I 6768 \AA{} line. A number of continuum intensity images and longitudinal magnetograms are produced each observation day. Unlike HMI, the two data products are generated from separate filtergrams and are therefore not co-temporal.

The SoHO spacecraft suffered two extended outages between June 1998 and February 1999 (usually referred to as the SoHO vacation). \cite{ball12} presented evidence that the response of MDI to magnetic flux might have changed over this period. Following the cited work, we excluded MDI data from before February 1999.

The flat field of MDI continuum intensity images is severe enough to impede the reliable identification of sunspots by the method employed in SATIRE-S (see Sect. \ref{equalsunspots}), and varied over the lifetime of the instrument \citep{krivova11b}. As in this earlier study, we corrected the intensity images for flat field by the division with the appropriate median filter, kindly provided by the MDI team. At least one median filter is produced per Carrington rotation period, but only up to Carrington period 2104 (November 26 to December 24, 2010). For this reason we did not consider data from after this Carrington period.

For each observation day between February 02, 1999 and December 24, 2010, we selected the level 1.5 continuum intensity image and level 1.8.2 5-min longitudinal magnetogram \citep{liu12} recorded closest in time and no more than 12 hours apart of one another. The continuum intensity images were rotated to the observation time of the corresponding longitudinal magnetogram and spatially resampled to co-register with the latter. We discarded the daily data from 18 days that were beset with gross instrumental artefacts, such as unusually strong noise fluctuations and smearing.

\subsubsection{KPVT}
\label{datakpvt}

Co-temporal full-disc continuum intensity images and longitudinal magnetograms, from spectropolarimetry of the Fe I 8688 \AA{} line, were collected at the KPVT on a daily basis with the $\kpone$ between February 1, 1974 and April 10, 1993, and with the $\kptwo$ between November 19, 1992 and September 21, 2003.

Some of these data carry artefacts from atmospheric seeing and instrumental effects, described in \cite{wenzler04,wenzler06}. In this study we consider just the 1757 $\kpone$ and 2055 $\kptwo$ daily continuum intensity image and longitudinal magnetogram identified by \citealt{wenzler06} (out of the 4665 and 2894 available) to be sufficiently free of these artefacts for the magnetic features of interest to be identified with confidence.

\begin{cfig}
\includegraphics[width=\textwidth]{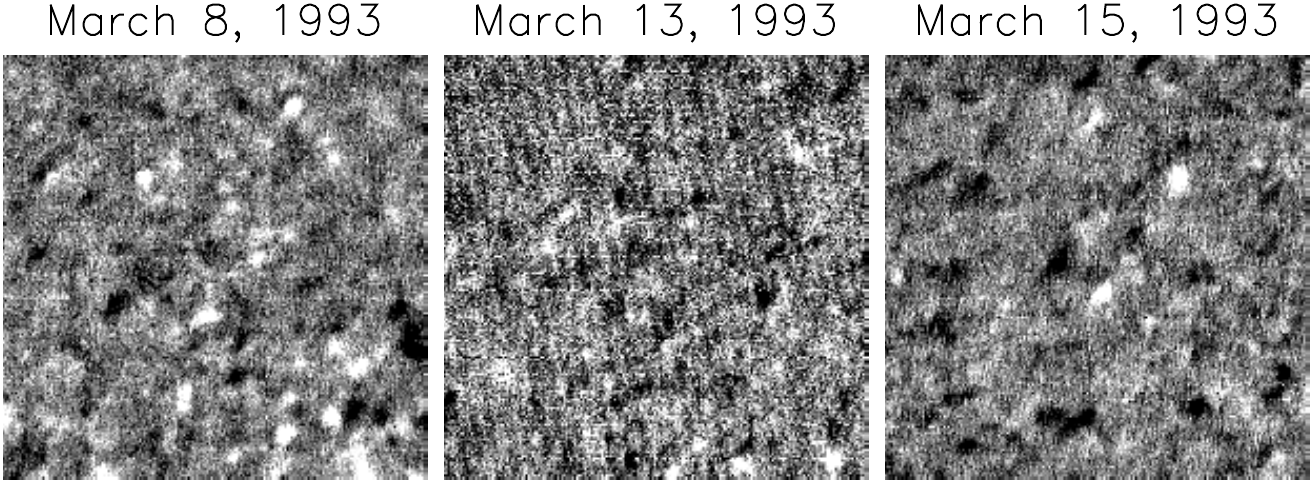}
\caption{$200\times200\:\as$ crop of one of the $\kpone$ magnetogram excluded from the analysis because of grid-like noise artefacts, described in the text (middle), contrasted against the similar inset of two magnetograms from nearby days that are markedly less affected by these instrumental artefacts (left and right). The grey scale is saturated at $\pm20\:{\rm G}$.}
\label{fotartefact}
\end{cfig}

On visual examination, we still found instrumental artefacts in some of the $\kpone$ magnetograms similar to what is depicted in Fig. \ref{fotartefact} (middle panel). The magnetogram signal is spurious along rows and columns of image pixels, producing the line and cross features in the grey scale plot. In 386 of the 1757 $\kpone$ magnetograms, large parts of the solar disc is pervaded with this grid-like noise pattern (concentrated between 1989 and 1992, where 268 of the 321 magnetograms are so affected). We excluded the data from these 386 days from the succeeding analysis.

\cite{arge02} compared the Carrington rotation synoptic charts of full-disc magnetograms from KPVT, Mount Wilson Observatory (MWO) and Wilcox Solar Observatory (WSO). The data sets spanned the period of 1974 to 2002. The authors noted that the total amount of unsigned magnetic flux in the KPVT charts, up to around 1990, appear slightly lower than the level in the MWO and WSO charts (Fig. 1 in their paper). This was attributed to the bias in the zero level of $\kpone$ magnetograms, which was stated to vary with time and with position in the solar disc (from centre to limb, and from east to west). The authors brought the magnetic flux levels in a subset of the KPVT synoptic charts from before 1990 to closer agreement with the MWO and WSO data by modifying the procedure by which $\kpone$ magnetograms are combined to form the synoptic charts. \cite{wenzler06} and \cite{ball12} attempted, by various approximations, to replicate the effect of this correction in $\kpone$ magnetograms from before 1990. \cite{wenzler06} multiplied the magnetogram signal by a factor of 1.242 while \cite{ball12} added 5.9 G to the absolute value.

Here, instead of trying to reproduce the effect of the procedure of \cite{arge02}, we examined the zero level bias of $\kpone$ magnetogram. In this study, we introduce the trimmed mean and trimmed standard deviation. These are the mean and standard deviation of a sample, computed iteratively, with data points more than three standard deviations from the mean omitted from succeeding iterations till no more points are removed.

\begin{cfig}
\includegraphics[width=\textwidth]{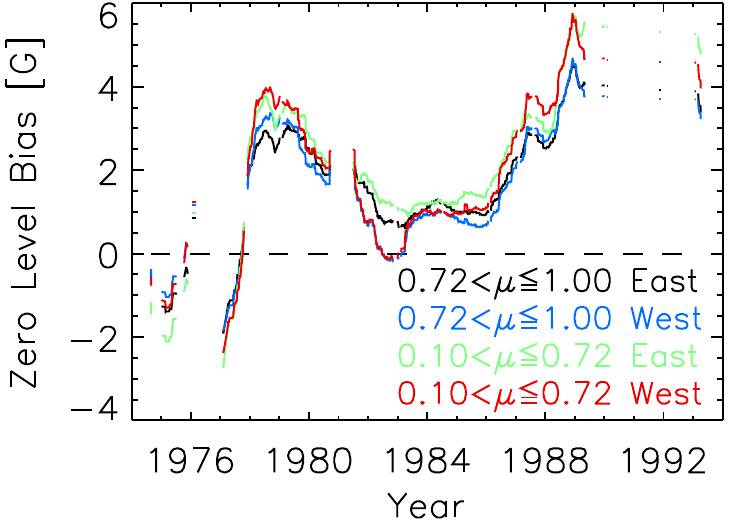}
\caption{Zero level bias of $\kpone$ magnetograms, around disc centre ($0.72<\mu\leq1.00$) and limb ($0.1<\mu\leq0.72$), and east and west of the central meridian. Here (and in all the subsequent time series plots in this paper), segments spaced more than 27 days apart are drawn separately, giving the gaps in the plots. The dashed line follows the null level.}
\label{fotzerobias}
\end{cfig}

We divided the solar disc along the central meridian and the $\mu=0.72$ locus, yielding four segments approximately equal in area. We computed, for each magnetogram, the trimmed mean of the signed magnetogram signal within each segment. The 361-day moving average of the time series of the trimmed mean within each segment (Fig. \ref{fotzerobias}) was taken as the zero level bias. The purpose of taking the moving average was to minimise error from the influence of magnetic activity on the trimmed mean in individual magnetograms. Consistent with the claims of \cite{arge02}, the zero level bias is non-zero, and it varied with time and between the four segments.

However, we found that subtracting the zero level bias from the magnetogram signal made negligible difference to the rest of the analysis, having on the whole, no appreciable effect on the apparent surface coverage and magnetic field strength of bright magnetic features. This is likely due to the fact that the zero level bias, though non-zero, is much weaker than noise fluctuations in the magnetogram signal \citep[the noise level of $\kpone$ magnetograms is around 8 G,][]{wenzler06}. In SATIRE-S, image pixels with magnetogram signals below the magnetogram signal threshold, which is based on the noise level, are counted as quiet Sun.

\cite{arge02} interpreted the observation that the magnetic flux level in the KPVT synoptic charts is slightly lower than in the MWO and WSO charts up to around 1990 (and broadly similar after) to indicate a problem with pre-1990 $\kpone$ data. As reported earlier in this section, most of the $\kpone$ magnetograms from between 1989 to 1992 are affected by the grid-like noise pattern depicted in Fig. \ref{fotartefact} (middle panel). The influence by these instrumental artefacts on the magnetogram signal might have contributed to the fact that the flux level in the KPVT charts from this period are no longer lower than in the MWO and WSO charts. (After this period, the KPVT charts are based on $\kptwo$ magnetograms.) We surmise that the disparity between the pre-1990 KPVT charts, and the MWO and WSO charts might be from systematic effects not accounted for in the analysis of \cite{arge02} and does not constitute any conclusive indication that a correction of pre-1990 $\kpone$ magnetograms is necessary.

We did not introduce the corrections to $\kpone$ magnetograms from before 1990 proposed by \cite{wenzler06} and \cite{ball12}, or subtract the zero level bias, determined here, from the magnetogram signal, leaving the $\kpone$ magnetograms as they are for the succeeding analysis.

\subsection{TSI measurements}
\label{datatsi}

As stated in Sect. \ref{model}, the appropriate value of the free parameter, $\bsat$ was recovered by comparing the reconstruction to TSI measurements (to be detailed in Sect. \ref{fixbsat}). For this purpose, we took the daily TSI measurements from four radiometers in current operation. They are, namely, ACRIMSAT/ACRIM3 \citep[version 11/13,][]{willson03}, the DIARAD \citep{crommelynck84,dewitte04a} and PMO6V \citep{brusa86} radiometers on SoHO/VIRGO \citep[level 2, version $6\_002\_1302$,][]{frohlich95,frohlich97}, and SORCE/TIM \citep[level 3, version 14,][]{kopp05a,kopp05b,kopp05c}.

There are two calibrations of the TSI measurements from DIARAD, one by IRMB and the other by PMOD/WRC. Hereafter, we will denote the two calibrations of the DIARAD record, both considered in the analysis, as $\dirmb$ and $\dpmod$, respectively.

The final reconstruction was also evaluated against the three published TSI composites (Sect. \ref{resulttsi}); ACRIM (version 11/13), IRMB (dated December 19, 2013, kindly provided by Steven Dewitte) and PMOD (version ${\rm d}41\_62\_1302$).

The absolute radiometric calibration of the various radiometers differ, in particular, TIM measurements are about $5\:\wms$ lower than that from preceding instruments. Tests conducted at the TSI radiometer facility, TRF with ground copies of ACRIM3, TIM and VIRGO revealed stray light effects in the ACRIM3 and VIRGO instruments \citep{kopp11}. Stray light correction introduced to the ACRIM3 record based on these tests brought it down to within $0.5\:\wms$ of TIM TSI \citep{kopp12}. The TSI measurements from Picard/PREMOS \citep{schmutz09,fehlmann12}, launched in 2011, the only TSI radiometer calibrated in vacuum at full TSI power levels before launch (also at the TRF), agree with TIM to within a similar margin. The results of these efforts have established the lower TSI level first registered by TIM as likely the more accurate.

Taking the 2008 solar minimum as the reference, we normalized all the TSI measurements and composites listed in this section, and the SATIRE-S reconstruction to the mean level in the TIM record over the period of November 2008 to January 2009\footnote{In this study, we took the position of solar cycle minima and maxima from the table published by NOAA at www.ngdc.noaa.gov/stp/space-weather/solar-data/solar-indices/sunspot-numbers/cycle-data/.}.

\section{Solar irradiance reconstruction}
\label{analysis}

\subsection{Harmonizing the model input from multiple instruments}
\label{crosscalibration}

The apparent surface coverage of magnetic features, the key input to the model, is modulated by the properties of the observing instrument. The influence of instrumental differences, left unaccounted for, can introduce inconsistencies between segments of the reconstruction based on data from different instruments. This is avoided here by treating the $\kpone$, $\kptwo$, MDI and HMI data sets in such a manner (described below in Sects. \ref{equalsunspots} and \ref{equalfaculae}), that we bring the apparent surface coverage of magnetic features in the various data sets, in the periods of overlap between instruments, to agreement.

\subsubsection{Umbra and penumbra}
\label{equalsunspots}

First, $\kptwo$, MDI and HMI continuum intensity images were corrected for limb darkening by the normalization to the fifth order polynomial in $\mu$ fit \citep[following][]{neckel94}.

In this study, we adopt the umbra and penumbra continuum intensity threshold for $\kptwo$ determined by \cite{wenzler06}, and for MDI by \cite{ball12}. \cite{wenzler06} set the $\kptwo$ penumbra threshold at 0.92, where the resulting sunspot area agrees with the sunspot area record by \cite{balmaceda09}. Assuming this value for $\kptwo$, \cite{ball12} found the penumbra threshold for MDI that brought the sunspot area in $\kptwo$ and MDI data into agreement over the period of overlap between the two to be 0.89. In both instances, the umbra threshold (0.70 for $\kptwo$ and 0.64 for MDI) was such that the umbra to sunspot area ratio is 0.2, a figure consistent with observations \citep{solanki03,wenzler05a}.

We set the umbra and penumbra threshold for HMI as done for MDI by \cite{ball12}. Comparing MDI and HMI continuum intensity images from the period of overlap, we fixed the HMI penumbra threshold at 0.87, the level that equalized the sunspot area in the two data sets. An umbra to sunspot area ratio of 0.2 was achieved with an umbra threshold of 0.59.

The difference in the umbra and penumbra threshold for the three data sets is primarily due to the fact that sunspot contrast is wavelength dependent.

Due to the 4-bit digitization, $\kpone$ continuum intensity images cannot be treated in a similar manner. For this data set, we employed the sunspot masks (indicating the image pixels occupied by sunspots) determined by \cite{wenzler06}. The sunspot masks were generated by binning the solar disc by $\mu$ and estimating from the intensity distribution within each bin the penumbra threshold for the bins closer to disc centre. This procedure overlooks pore-like structures that lack an umbral core. Following the cited work, we derived the umbra and penumbra area from the sunspot masks by
\begin{itemize}
	\item giving the image pixels marked as sunspots in the masks, umbra and penumbra filling factors of 0.2 and 0.8, respectively, and
	\item reclassifying the faculae pixels with the highest $\bmu$, up to 0.127 of the sunspot area, as penumbra.
\end{itemize}
The latter measure approximately recovers the pore-like structures omitted by the sunspot identification process, bringing the sunspot area in $\kpone$ into agreement with the sunspot area record by \cite{balmaceda09}.

\subsubsection{Faculae}
\label{equalfaculae}

While it is relatively straightforward to account for the effect of instrumental differences on apparent sunspot area, doing the same for faculae is considerably more subtle. The small-scale magnetic concentrations associated with bright magnetic features are largely unresolved even at HMI's spatial resolution ($\sim1\:\as$), leaving us to infer their presence and filling factor from the magnetogram signal. The magnetogram properties of the instruments considered in this study diverge significantly \citep{jones01,wenzler04,wenzler06,ball12,liu12}. Applying the same magnetogram signal threshold, $\bcut$ and $\bsat$ to all the data sets with no regards for these differences would create significant discrepancies between the segments of the reconstruction based on data from the different instruments.

The SATIRE-S reconstruction by \cite{ball12,ball14} combined the model output from $\kpone$, $\kptwo$ and MDI data into a single TSI/SSI time series. The authors accounted for instrumental differences by adjusting the parameters of the model to each data set (in a manner that does not introduce additional free parameters). This brought the model output from the various data sets into broad agreement. The small residual disagreement between the $\kpone$, $\kptwo$ and MDI segments of the reconstruction, caused by differences in the magnetogram response of the various instruments, was empirically corrected for by regression.

In this study, we took an alternative approach. We cross-calibrated the magnetograms and the faculae filling factor from the various data sets such that we can apply the same magnetogram signal threshold, $\bcut$ and $\bsat$ to all the data sets. Together with the harmonized umbra and penumbra areas (Sect. \ref{equalsunspots}), this yielded a consistent TSI/SSI time series without the need for any additional correction of the reconstructed spectra. 

For these modifications, we relied on the data from the various instruments taken close in time to one another. There are 11 days, between Jan 22 and April 4, 1993, with daily data from both $\kpone$ and $\kptwo$, recorded within about three hours of one another. We will term these observations the $\kpone$-$\kptwo$ co-temporal data. For the period of overlap between $\kptwo$ and MDI, and MDI and HMI, there are considerably more days with daily data from both instruments. For $\kptwo$ and MDI, 910 days between February 2, 1999 and September 21, 2003, and for MDI and HMI, 196 days between April 30 and December 24, 2010. This afforded us the option to restrict ourselves to just the observations taken closest in time to one another. We will denote the $\kptwo$ and MDI observations from 67 days, taken within one hour of one another, and MDI and HMI observations from 187 days, taken within five minutes of one another, as the $\kptwo$-MDI and MDI-HMI co-temporal data, respectively. It was in the interest of cross-calibrating the data sets as accurately as possible that we balanced examining just the daily data taken closest in time to one another and considering as many as possible, to the extent permitted by available data.

\begin{cfig}
\includegraphics[width=\textwidth]{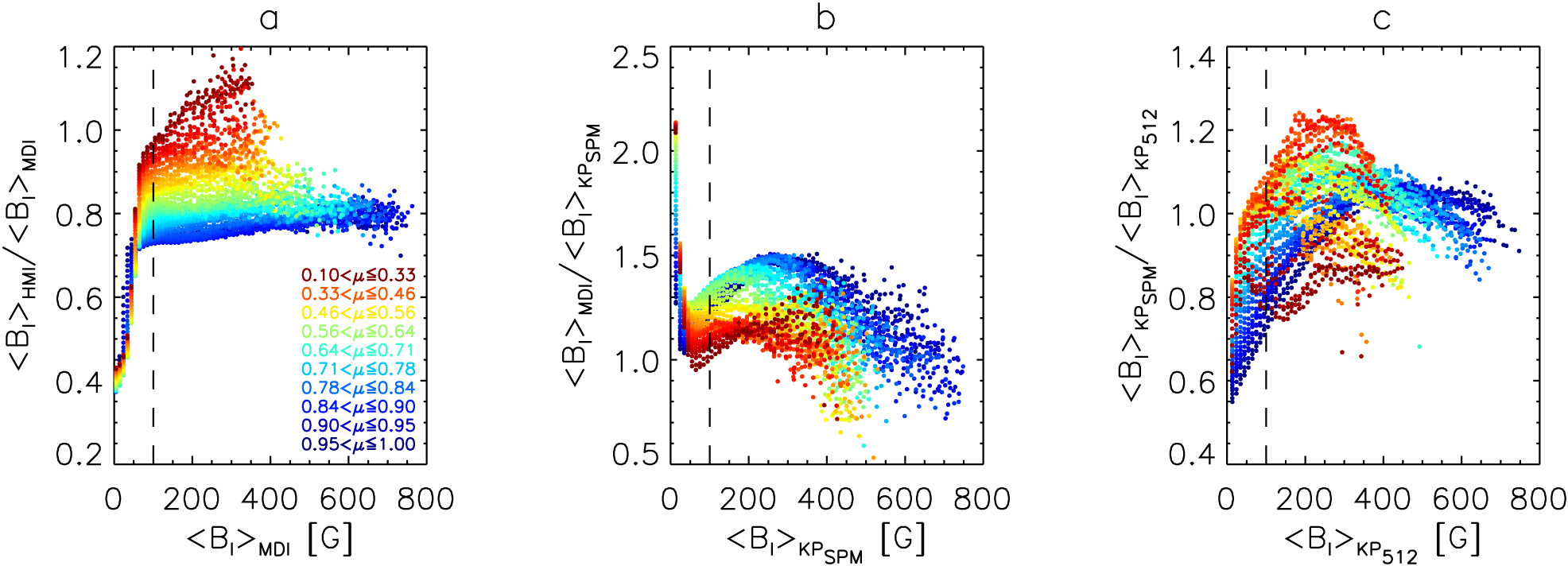}
\caption{a) Ratio of the HMI and MDI magnetogram signal, $\blhmi/\blmdi$ as a function of $\bl_{\rm MDI}$, within 68 partially overlapping intervals of $\mu$ (the boundary of ten of which are annotated). The dashed line, at $\blmdi=100\:{\rm G}$, marks the threshold below which the points are excluded from the derivation of the MDI magnetogram signal conversion function due to the increasing influence of magnetogram noise (see text). Also illustrated is the same from the comparison between b) $\kptwo$ and rescaled MDI magnetograms, and c) $\kpone$ and rescaled $\kptwo$ magnetograms.}
\label{mdihmihtgeqt}
\end{cfig}

\begin{cfig}
\includegraphics[width=\textwidth]{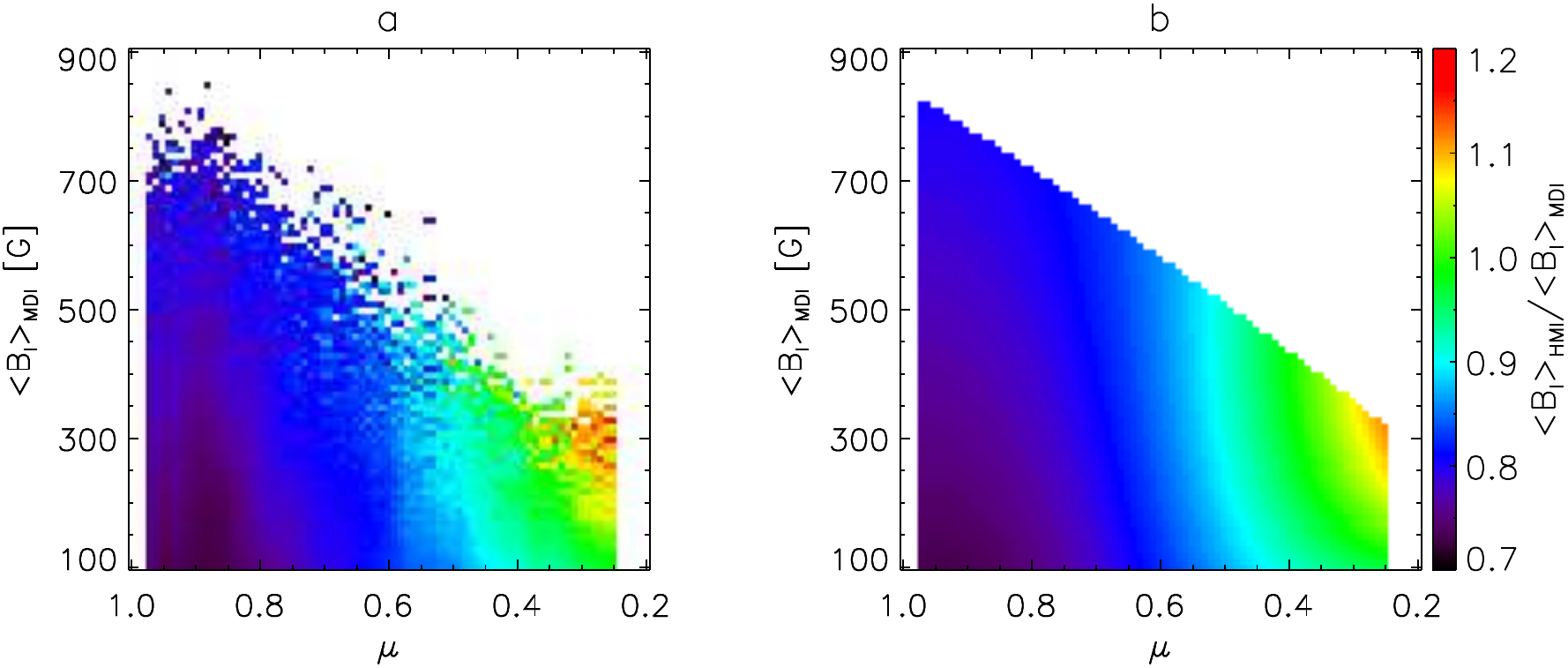}
\caption{a) The same as Fig. \ref{mdihmihtgeqt}a, except as a colour scale plot and excluding the points where $\blmdi<100\:{\rm G}$. b) The corresponding bivariate polynomial fit, the MDI magnetogram signal conversion function, over a similar range of $\mu$ and $\blmdi$.}
\label{mdihmihtgeqtfit}
\end{cfig}

Taking the MDI-HMI co-temporal data, we derived the ratio of the magnetogram signal in HMI and MDI magnetograms, $\blhmi/\blmdi$ as a function of the MDI magnetogram signal and $\mu$.
\begin{itemize}
	\item First, we resampled the HMI observations in space and time, to co-register with the corresponding MDI data. 
	\item Following that, we masked the image pixels counted as umbra or penumbra. Our interest here is the magnetogram signal of bright magnetic features. The influence of instrument-related factors on the magnetogram signal of these features, and of sunspots and pores differ \citep[e.g.,][]{yeo14a}, therefore the need to exclude the latter. Another consideration is the saturation of MDI  and HMI  magnetograms in sunspot umbrae. For MDI, this is due to limitations in the onboard algorithm used to generate the magnetograms \citep{liu07,liu12}, and for HMI, the Fe I 6173 \AA{} line shifting out of the spectral sampling range of the instrument (Sebastien Couvidat, personal communication).
	\item We divided the solar disc by $\mu$ into 68 partially overlapping intervals, each representing $\sim10\%$ of the solar disc by area. Within each $\mu$ interval, we compared the magnetogram signal in the MDI and HMI data by the following, based broadly on the histogram equalization method of \cite{jones01}.
	\item Taking each magnetogram, we ranked the unmasked image pixels by the signed magnetogram signal and took the mean of each successive ten points. We then matched the ranked and averaged MDI magnetogram signal to the corresponding values from the HMI magnetogram from the same day. Collating the MDI and HMI magnetogram signal pairs so derived from all the data days, we binned them by the MDI magnetogram signal in bins of 10 G width. Finally, we took the median MDI and HMI magnetogram signal within each bin, and the quotient of the two, $\blhmi/\blmdi$, expressed in Figs. \ref{mdihmihtgeqt}a and \ref{mdihmihtgeqtfit}a.
\end{itemize}
As evident in Figs. \ref{mdihmihtgeqt}a and \ref{mdihmihtgeqtfit}a, the relationship between the HMI and MDI magnetogram signal, represented by $\blhmi/\blmdi$, varies significantly with magnetogram signal and $\mu$. The marked decline in $\blhmi/\blmdi$ towards $\blmdi=0\:{\rm G}$ comes from the fact that MDI 5-min longitudinal magnetograms are significantly noisier than HMI 315-s longitudinal magnetograms \citep[see the noise level estimates by][]{ball12,liu12,yeo13}, and so does not reflect the true relationship between HMI and MDI magnetogram signal in this regime. Excluding the points where $\blmdi<100\:{\rm G}$ (a conservative threshold), we fit a bivariate polynomial (second order in $\blmdi$ and third order in $\mu$) to $\blhmi/\blmdi$, illustrated in  Fig. \ref{mdihmihtgeqtfit}b. This surface fit, hereafter referred to as the MDI magnetogram signal conversion function, gives the multiplicative factor for a given $\mu$ and $\blmdi$ that would convert the magnetogram signal in MDI magnetograms to the HMI equivalent.

We rescaled the magnetogram signal in the entire MDI data set to the HMI equivalent by applying the MDI magnetogram signal conversion function.

We repeated the entire process described above for the other data sets. That is, using the $\kptwo$-MDI co-temporal data to bring the magnetogram signal in the SPM data set to the rescaled MDI equivalent, and then the $\kpone$-$\kptwo$ co-temporal data to bring the magnetogram signal in the $\kpone$ data set to the rescaled $\kptwo$ equivalent. As between MDI and HMI, the relationship between $\kptwo$ and rescaled MDI, and $\kpone$ and rescaled $\kptwo$ magnetogram signal fluctuates with magnetogram signal and $\mu$ (Figs. \ref{mdihmihtgeqt}b and \ref{mdihmihtgeqt}c). Though there are only observations from 11 days in the $\kpone$-$\kptwo$ co-temporal data set, each magnetogram presents the order of $10^6$ point measurements of magnetic flux, providing sufficient statistics for the analysis.

\begin{cfig}
\includegraphics[width=\textwidth]{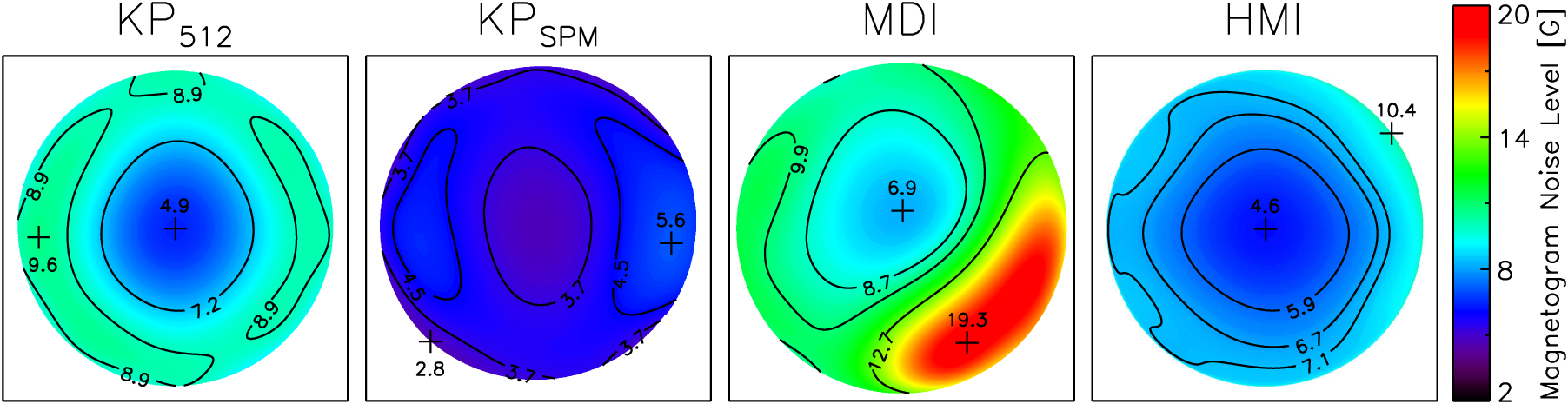}
\caption{Noise level of rescaled $\kpone$, $\kptwo$ and MDI magnetograms, and of HMI magnetograms, as a function of position in the field-of-view. The contours correspond to the first, second and third quartiles, and the crosses to the minimum and maximum points. The contour and point labels are in units of Gauss. We omit the second quartile in the $\kpone$ and $\kptwo$ plots to avoid cluttering. The $\kpone$, $\kptwo$ and HMI noise surfaces were resampled to MDI image size to allow a direct comparison.}
\label{magns}
\end{cfig}

The noise level of rescaled MDI magnetograms was estimated following the analysis of \cite{yeo13} with HMI 315-s longitudinal magnetograms. We selected the magnetogram from ten sunspot-free, low-activity days near the 2008 solar cycle minimum. For each magnetogram, we centred a $101\times101$ pixel window on each point on the solar disc and computed the trimmed standard deviation (defined in Sect. \ref{datakpvt}). Then, we took the median of the trimmed standard deviation from the ten magnetograms at each point on the solar disc. To these values we fit a polynomial surface, which represents the noise level of rescaled MDI magnetograms, plotted in Fig. \ref{magns}. Also illustrated are the noise surfaces of rescaled $\kpone$ and $\kptwo$ magnetograms, derived in a similar manner, and the noise surface of HMI magnetograms \citep[determined by][]{yeo13}. The noise level of rescaled MDI magnetograms is, at every (relative) position in the field-of-view, higher than the noise level of the other data sets.

\begin{cfig}
\includegraphics[width=\textwidth]{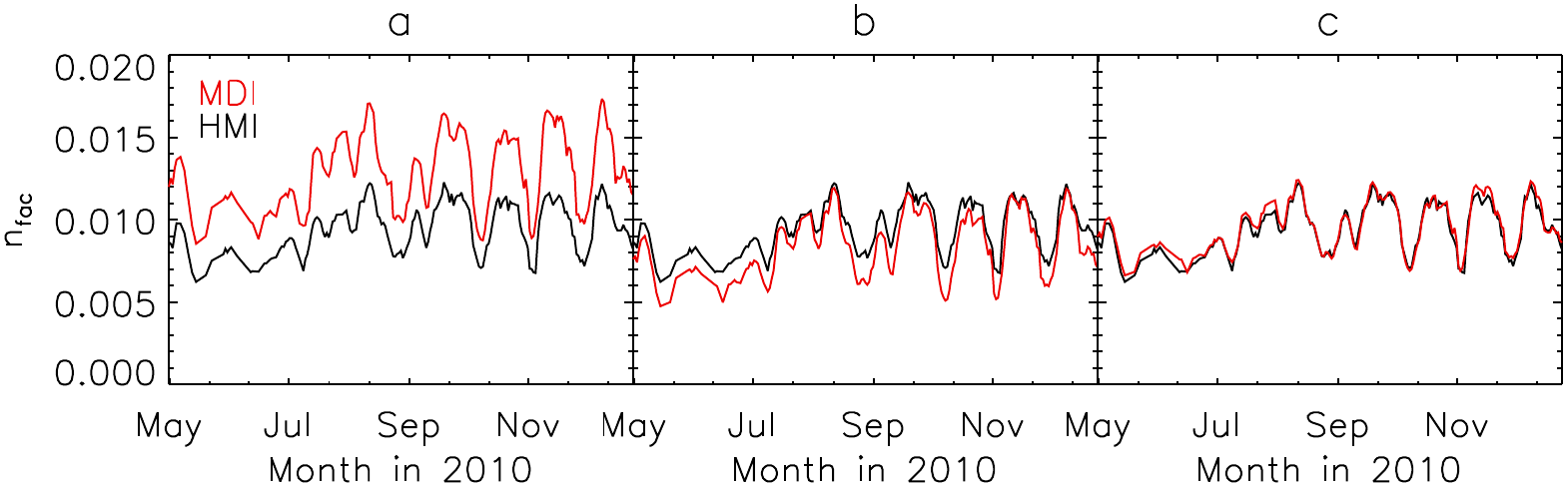}
\caption{a) Proportion of the solar disc covered by faculae, $\nfac$ in the MDI (red) and HMI (black) data sets over the period of overlap if we do not rescale the MDI magnetogram signal and faculae filling factor. b) The same, after rescaling the MDI magnetogram signal. c) The final result, incorporating the rescaled MDI magnetogram signal and faculae filling factor.}
\label{plotfacstats1}
\end{cfig}

The area covered by faculae was derived taking into account the fact that the MDI data set has (after the rescaling of the magnetogram signal) the highest magnetogram noise level (Fig. \ref{magns}) and the coarsest pixel scale of the data sets considered in this study (Table \ref{datatable}). To this end, we took the following measures.
\begin{itemize}
	\item We applied the same magnetogram signal threshold, given by three times the noise surface of rescaled MDI magnetograms, to all the data sets.
	\item In SATIRE-S, standalone facular pixels are reclassified as quiet Sun to minimise the inclusion of magnetogram noise fluctuations. This is at the expense of legitimate bright magnetic features occupying just a single image pixel. To ensure that we are discriminating against similar sized features in all the data sets, we excluded the bright magnetic features in the $\kpone$, $\kptwo$ and HMI data sets that would appear standalone if we resample the magnetograms to MDI pixel scale.
	\item Taking an arbitrary but conservative value of 600 G for $\bcut$, facular pixels with magnetogram signals above this level were also reclassified as quiet Sun \citep[following][]{ball11,ball12}. As stated in Sect. \ref{model}, this is to minimise the misclassification of the magnetic canopy of sunspots as faculae.
	\item The filling factor of each facular pixel was then determined by the linear relationship described in Sect. \ref{model}. We obtained a value of 230 G for $\bsat$ (see Sect. \ref{fixbsat}).
\end{itemize}
Let $\nfac$ represent the proportion of the solar disc covered by faculae, given by the sum over all facular pixels of the filling factor, normalised by the number of solar disc pixels. The $\nfac$ in the MDI and HMI data sets over the period of overlap is shown in Fig. \ref{plotfacstats1}. In Fig. \ref{plotfacstats1}a, we compare the $\nfac$ if we do not rescale the magnetogram signal in the MDI data set. As stated at the beginning of this section, applying the same model parameters to the MDI and HMI data sets without any consideration of the different magnetogram properties results in a gross divergence in the apparent faculae area. Rescaling the magnetogram signal of the MDI data set to the HMI equivalent removed most of this disparity (Fig. \ref{plotfacstats1}b). The residual difference, though small, is not negligible. The apparent facular area in the overlap between the $\kpone$, $\kptwo$ and MDI data sets exhibit a similar response to the analysis, not shown here to avoid repetition.

The magnetogram signal conversion functions are determined from the fit to observed magnetogram signal ratios (Fig. \ref{mdihmihtgeqtfit}), excluding the data below 100 G, where the apparent ratio is increasingly biased by magnetogram noise (Fig. \ref{mdihmihtgeqt}). The factors used to rescale the magnetogram signal below 100 G are given by the extrapolation of the fit to this regime. As a test, we repeated the analysis described in this section, fixing the $\bl<100\:{\rm G}$ segment of the conversion functions at the value at $\bl=100\:{\rm G}$ instead. This adjustment to the conversion function made no appreciable difference to $\nfac$, excluding the uncertainty in the conversion functions below 100 G as the main cause of the residual disparity in $\nfac$ between the various data sets.

The residual disparity in $\nfac$ is probably due to differences in spatial resolution. Specifically, in the point spread function and how it is sampled by the detector array. The effect of spatial resolution on apparent area is especially relevant to bright magnetic features, as they remain largely unresolved even at HMI's resolving power they are particularly susceptible to effects such as stray light and undersampling. \citep[As an example, the pronounced effect of stray light on the apparent surface area of bright magnetic features in HMI magnetograms was recently demonstrated by][]{yeo14a}. To correct for this directly, we would have to convolve the various data sets by suitable filters and resample the $\kpone$, $\kptwo$ and HMI data to MDI image size. These filters, different for each instrument, are not straightforward to determine \citep[see the discussions in][]{wedemeyerbohm08,yeo14a}. Here, we account for the residual effect of instrumental differences on $\nfac$ by the following steps instead.
\begin{itemize}
	\item Taking each magnetogram in the MDI-HMI co-temporal data, we divided the solar disc by $\mu$ in intervals of 0.01, and calculated the $\nfac$ within each segment.
	\item Interval by interval, we fit a second order polynomial to the scatter plot of the $\nfac$ from the MDI magnetograms versus the $\nfac$ from the corresponding HMI magnetograms. To improve the statistics, we included the values from adjacent $\mu$ intervals. The result is a series of functions that relate the apparent faculae area in MDI and HMI magnetograms at different distances from disc centre. 
	\item With these functions, we rescaled the faculae filling factor of the entire MDI data set to the HMI equivalent.
	\item This adjustment was then propagated to the $\kptwo$ and $\kpone$ data sets. We rescaled the faculae filling factor of the $\kptwo$ data set repeating the above analysis on the $\kptwo$-MDI co-temporal data, and finally of the $\kpone$ data set using the $\kpone$-$\kptwo$ co-temporal data.
\end{itemize}
This procedure brought the apparent faculae area in the $\kpone$, $\kptwo$, MDI and HMI data sets, in the periods of overlap, to agreement, illustrated for the overlap between MDI and HMI in Fig. \ref{plotfacstats1}c.

\subsection{The free parameter}
\label{fixbsat}

In the SATIRE-S model, the faculae filling factor of bright magnetic features is deduced from the magnetogram signal. It is given by the quotient of $\bmu$ and $\bsat$, up to $\bmu=\bsat$, beyond which it is fixed at unity \citep{fligge00}. The higher the value of $\bsat$, the sole free parameter in the model, the lower the apparent facular area, and vice versa. This is accentuated by the fact that the distribution of magnetogram signals is skewed towards lower magnetogram signal levels \citep{parnell09}. As a result, facular contribution to variation in reconstructed solar irradiance scales nearly inversely with $\bsat$. The appropriate value of $\bsat$ is obtained by optimizing the agreement between reconstructed and measured TSI. As mentioned in Sect. \ref{datatsi}, we considered the observations from ACRIM3, DIARAD, PMO6V and TIM for this purpose.

\begin{cfig}
\includegraphics[width=.83\textwidth]{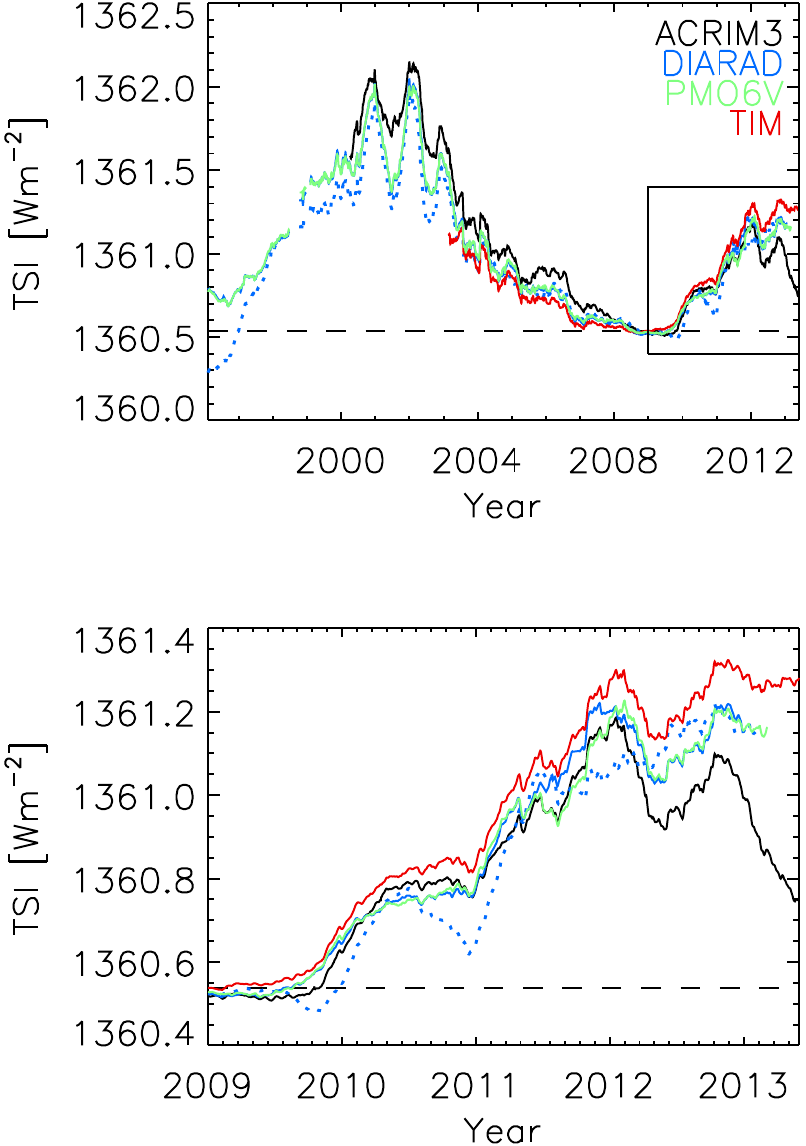}
\caption{181-day moving average of the daily TSI measurements from the ACRIM3 (black), DIARAD (blue), PMO6V (green) and TIM (red) radiometers. The dotted and solid blue curves correspond to $\dirmb$ and $\dpmod$, the calibration of the DIARAD record by IRMB and by PMOD/WRC. The $\dpmod$ series is almost completely hidden by the PMO6V series due to the close similarity. The dashed lines follow the mean TIM TSI level over the period of November 2008 to January 2009. The lower plot is a blow up of the boxed inset in the upper plot.}
\label{comparetsi}
\end{cfig}

There are discrepancies in the long-term trends in the TSI time series produced by ACRIM3, DIARAD, PMO6V and TIM (Fig. \ref{comparetsi}). The most notable is the conflicting secular variation between the 1996 and 2008 solar minima, between the $\dirmb$, and the $\dpmod$ and PMO6V records. This disparity is one of the reasons we considered multiple TSI records, all of which extend for at least about a decade for this part of the analysis. The motivation is to avoid introducing bias in $\bsat$ from relying on just a single or shorter TSI records \citep[such as, from Picard/PREMOS, launched only in 2011,][]{schmutz09}.

\begin{table}
\caption{To the nearest Gauss, the value of $\bsat$ that optimises the agreement between reconstructed and measured TSI. Also tabulated is the correlation ($R$ and $R^2$) and RMS difference ($k$) between each TSI record and the corresponding candidate reconstruction.}
\centering
\begin{tabular}{lcccc}
\hline\hline
TSI record & $\bsat$ [G] & $R$ & $R^2$ & $k$ [$\wms$] \\
\hline
ACRIM3                    & 204 & 0.963 & 0.928 & 0.155 \\
${\rm DIARAD}_{\rm IRMB}$ & 255 & 0.962 & 0.925 & 0.141 \\
${\rm DIARAD}_{\rm PMOD}$ & 230 & 0.969 & 0.940 & 0.131 \\
PMO6V                     & 230 & 0.979 & 0.959 & 0.107 \\
TIM                       & 220 & 0.960 & 0.921 & 0.108 \\
\hline
\end{tabular}
\label{fixbsattable}
\end{table}

Let $\eobs$ and $\esat$ denote observed and reconstructed TSI. In the earlier implementations of SATIRE-S by \cite{ball11,ball12}, $\bsat$ was fixed at the value that brought the slope of the linear fit to the scatter plot of $\eobs$ and $\esat$ to unity, taken as an indication of a close match in the overall trend. While a regression coefficient of unity would be a property of two TSI series that are similar, it is not by itself sufficient to indicate similarity or a linear relationship between the two. The divergence of the slope from unity is suitable as a measure of similarity only in the case where a linear relation between the two series in question can be assumed. Additionally, differences in the long-term trend between $\eobs$ and $\esat$ can possibly shift the slope towards, rather than away from unity. Here, we set $\bsat$ at the value that minimizes the root mean square, RMS difference between $\eobs$ and $\esat$, denoted $k$, instead. It can be shown that the RMS difference between $\eobs$ and $\esat$ is, apart from a factor, equivalent to the RMS of the normal distance of the scatter plot of $\eobs$ and $\esat$ to the $\eobs=\esat$ line. Quantifying the scatter about $\eobs=\esat$, this quantity measures the similarity between the two series independent of the underlying relationship.

We generated five candidate reconstructions by taking each of the five TSI records considered as the reference for recovering $\bsat$. The retrieved value of $\bsat$ in each case, determined to the nearest Gauss, is summarized in Table \ref{fixbsattable}. Recall, that facular contribution to variation in reconstructed solar irradiance scales nearly inversely with $\bsat$. The spread in the recovered value of $\bsat$ reflects the differences in the cyclical variation in the five TSI records (Fig. \ref{comparetsi}). The retrieved values of $\bsat$ in the $\dpmod$ and PMO6V cases (and therefore the corresponding candidate reconstructions) are exactly identical due to the close similarity in the long-term trend in these TSI records.

\begin{cfig}
\includegraphics[width=\textwidth]{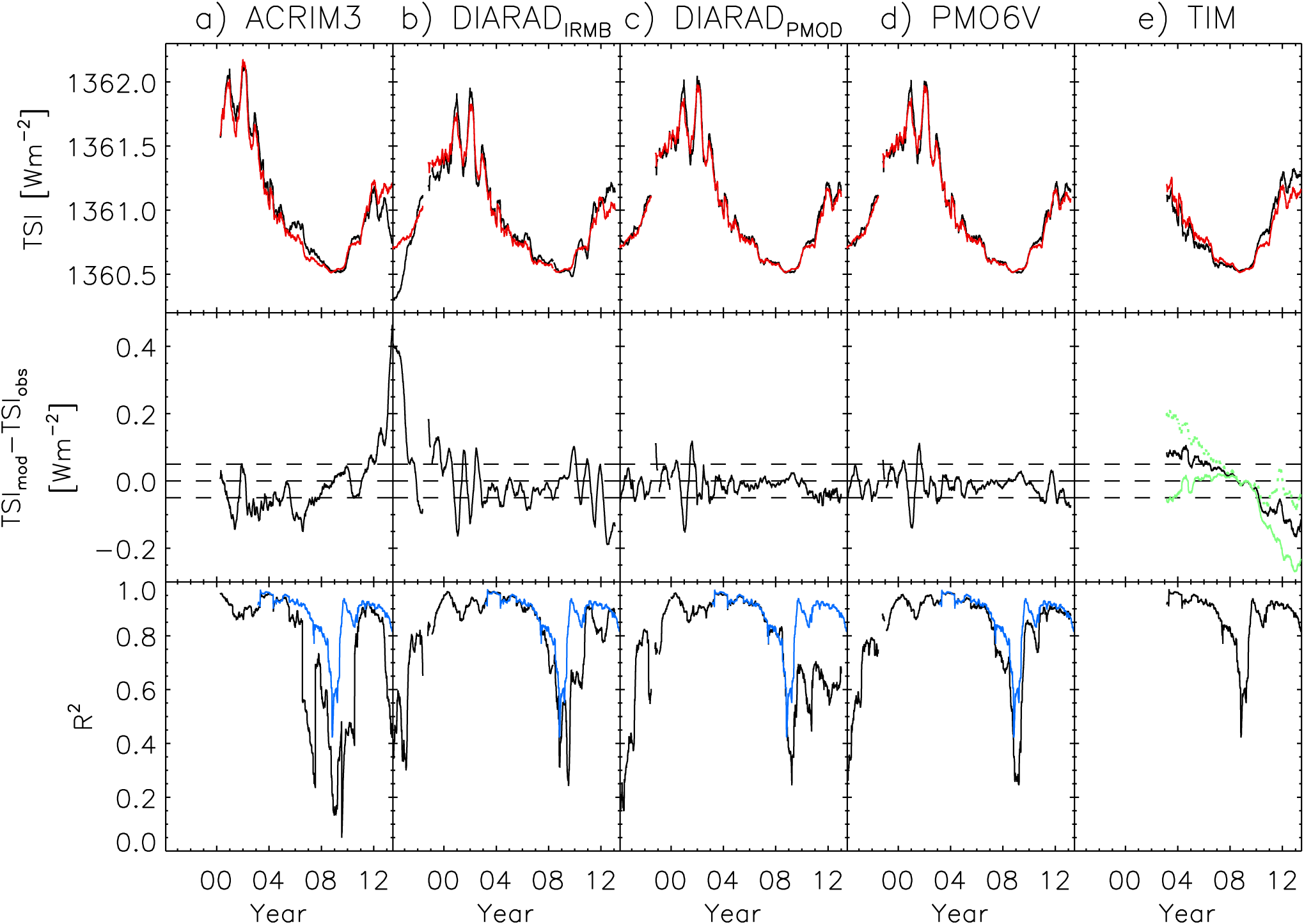}
\caption{Top: 181-day moving average of the five TSI records examined (black) and the TSI reconstructed by taking them as the reference for recovering $\bsat$ (red). Middle: The difference between the two, $\esat-\eobs$. The solid and dotted green curves in the TIM panel depict the same from taking just TIM measurements from before, and from after October 1, 2009 as the reference. The dashed lines mark the zero level and $\esat-\eobs=\pm0.05\:\wms$. Bottom: The $R^2$ between measured and modelled TSI within a 361-day window centred on each data day. The TIM series is plotted over the other series in blue for comparison.}
\label{fixbsatfig}
\end{cfig}

In Fig. \ref{fixbsatfig}, we plot the TSI from the candidate reconstructions, along the respective reference TSI record (top row), and the difference between the two (centre row). To elucidate the long-term fluctuations, we smoothed each time series with a 181-day boxcar filter.
\begin{itemize}
	\item ACRIM3 (Fig. \ref{fixbsatfig}a): The reconstruction is broadly lower than the TSI record, by $\lesssim0.1\:\wms$, up to 2007, after which it starts to drift (almost monotonically) higher, by $\sim0.25\:\wms$ between 2007 and 2012, and by the same margin in 2013 alone.
	\item $\dirmb$ (Fig. \ref{fixbsatfig}b): The reconstruction and the TSI record differ, for much of the period of comparison, by more than $\sim0.1\:\wms$. Notably, the level at the solar cycle minimum in 1996 is higher in the reconstruction by a gross margin of $\sim0.4\:\wms$.
	\item $\dpmod$ (Fig. \ref{fixbsatfig}c) and PMO6V (Fig. \ref{fixbsatfig}d): In both cases, modelled and measured TSI agree, to within $\sim0.05\:\wms$, over nearly the entire period of comparison.
	\item TIM (Fig. \ref{fixbsatfig}e): There is a continuous drift, of $\sim0.25\:\wms$ between 2003 and 2013, between model and measurement.
\end{itemize}
We derived the coefficient of determination, $R^2$ between reconstructed and observed TSI within a 361-day window centred on each data day (bottom row, Fig. \ref{fixbsatfig}), representing the short-term agreement. The agreement between modelled and measured TSI is relatively close, with $R^2$ generally above 0.8, except around solar cycle minima, where it dips as variation in solar irradiance diminishes and noise starts to play a role. In the ACRIM3 and $\dpmod$ cases, there are also other periods where the agreement deteriorated markedly. The closest short-term agreement is seen with the TIM data set.

In terms of the overall agreement, given by the correlation and the RMS difference, $k$ over the entire period of comparison, the closest alignment was found between the PMO6V record and the corresponding candidate reconstruction (Table \ref{fixbsattable}).

Only with the PMO6V record did reconstructed TSI exhibit consistent close agreement at both rotational and cyclical timescales (Fig. \ref{fixbsatfig}). For this reason, we retained the candidate reconstruction, derived taking the PMO6V record as the reference for recovering $\bsat$, for the succeeding analysis.

This is a departure from the preceding SATIRE-S reconstruction by \cite{ball12}. The authors determined $\bsat$ comparing the model output to TIM TSI over the period of 2003 to 2009. Their reconstruction exhibited close long-term agreement with the TIM record, while we noted an almost continuous drift between the TIM record and the corresponding candidate reconstruction (Fig. \ref{fixbsatfig}e). This disparity stems largely from the difference in the period of comparison (discussed below), although the fact that we had derived facular area by a significantly different analysis (Sect. \ref{equalfaculae}) may have contributed to a minor extent.

Here, the period of comparison extends to 2013. Taking, as in the earlier study, just TIM data up to October 31, 2009 (termed, for this discussion, the cutoff date) as the reference, we retrieved a higher value of 261 G for $\bsat$. This brought the pre-cutoff segment of the reconstruction down closer to the TIM record but at the same time amplified the post-cutoff divergence (green solid curve, Fig. \ref{fixbsatfig}e). Likewise, using only TIM measurements from after the cutoff date as the reference returned a lower $\bsat$ of 185 G, which raised the post-cutoff segment closer to TIM levels but also widened the pre-cutoff disparity (green dotted curve). As the solar cycle amplitude of modelled TSI is modulated by $\bsat$ and the cutoff date is incidentally near the 2008 solar cycle minimum, the slow, near-monotonic drift between the reconstruction and the TIM record found here is largely undetectable looking at data from before, or from after the cutoff date alone. No significance should be attached to this value of the cutoff date. Repeating this analysis, setting the cutoff date at earlier and later times around the 2008 minimum, we found similar trends in the recovered value of $\bsat$, and the pre-cutoff and post-cutoff segments of the reconstruction.

This observation, that disparity between model and measurement can possibly be hidden if we recover $\bsat$ comparing them over a span of time that sits largely within the ascending or descending phase of a solar cycle, is another reason for considering multiple, extended ($>10\:{\rm years}$) TSI records for this part of the analysis.

\subsection{Ultraviolet solar irradiance}
\label{uvfix}

\begin{cfig}
\includegraphics[width=\textwidth]{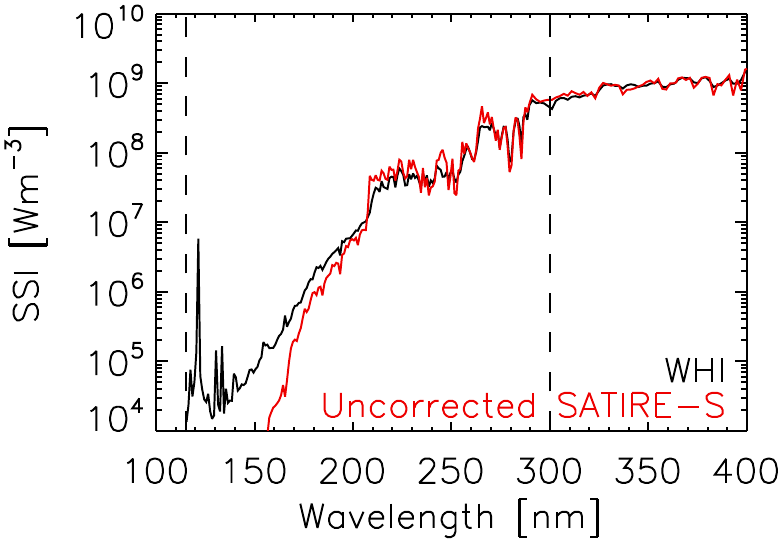}
\caption{The WHI reference solar spectrum for the period of April 10 to April 16, 2008, and the mean SATIRE-S spectrum over the same period prior to the correction of the 115 to 300 nm segment (see text), bounded by the dashed lines. The WHI reference solar spectrum is binned to the wavelength scale of the reconstruction.}
\label{refspec}
\end{cfig}

As detailed in \cite{unruh99}, the intensity spectra of umbra, penumbra, faculae and quiet Sun utilised in the reconstruction were synthesized with the ATLAS9 radiative transfer code \citep[which assumes local thermodynamic equilibrium, LTE,][]{kurucz93} representing spectral lines by opacity distribution functions, ODFs. Consequently, below $\sim300\:{\rm nm}$, modelled solar irradiance starts to diverge from observation and progressively so with decreasing wavelength \citep[as previously noted by][and illustrated in Fig. \ref{refspec}]{krivova06,unruh08}. The effects of the LTE and ODF simplifications emerge in the ultraviolet from the breakdown of the LTE approximation in the upper layers of the solar atmosphere, and estimating line blanketing using ODFs.

The limitations of the SATIRE-S model in the ultraviolet imposed by the LTE and ODF simplifications were previously accounted for as follows. \cite{krivova06} reported a close agreement between their SATIRE-S reconstruction of ultraviolet solar irradiance and the measurements from UARS/SUSIM \citep{brueckner93,floyd03} in the wavelength range of 220 to 240 nm (termed the reference interval). Taking the SUSIM record, the authors found, by regression, the relationship between each wavelength channel and the integrated flux in the reference interval. Solar irradiance over the wavelength range of 115 to 270 nm was then regenerated applying these relationships to the reconstruction of the integrated flux in the reference interval.

In this study, we accounted for the LTE and ODF simplifications by a modified analysis.
\begin{itemize}
	\item 180 to 300 nm: We offset this segment of the reconstruction (wavelength element by wavelength element) to the Whole Heliospheric Interval (WHI) reference solar spectra \citep[version 2,][]{woods09}. This somewhat less obtrusive approach was prompted by our observations that in this spectral range the model reproduces variations in SSI observations well (see Sect. \ref{resultuv}), though less so the absolute level. The intention here is to bring the absolute level of the reconstruction in this spectral range to a more realistic level, while retaining the variation returned by the model. We are prohibited from extending this approximation below 180 nm due to the gross disparity in the absolute level between modelled and measured solar irradiance there (Fig. \ref{refspec}).
	\item 115 to 180 nm: We regenerated this segment of the reconstruction by an analysis similar to that previously employed by \cite{krivova06}, described above. Except here we based the correction on the observations from the FUV (115 to 180 nm) instrument on SORCE/SOLSTICE \citep[level 3, version 12,][]{mcclintock05,snow05a} instead. This is in the interest of consistency with the correction introduced in the 180 to 300 nm segment; the wavelength range of 115 to 310 nm in the WHI reference solar spectra is provided by SOLSTICE spectrometry.
\end{itemize}

The WHI reference solar spectra represent the mean SSI over three periods within Carrington rotation 2068 (March 25 to March 29, March 30 to April 4, and April 10 to April 16, 2008). First, we binned the higher spectral resolution reference spectra to the wavelength scale of the reconstruction (given basically by the ATLAS9 code). Next, we took the difference between the reference spectra and the average reconstructed spectra over the same periods. Finally, we took the mean of the three difference spectra and offset the 180 to 300 nm segment of the SSI reconstruction by the result.

\begin{cfig}
\includegraphics[width=\textwidth]{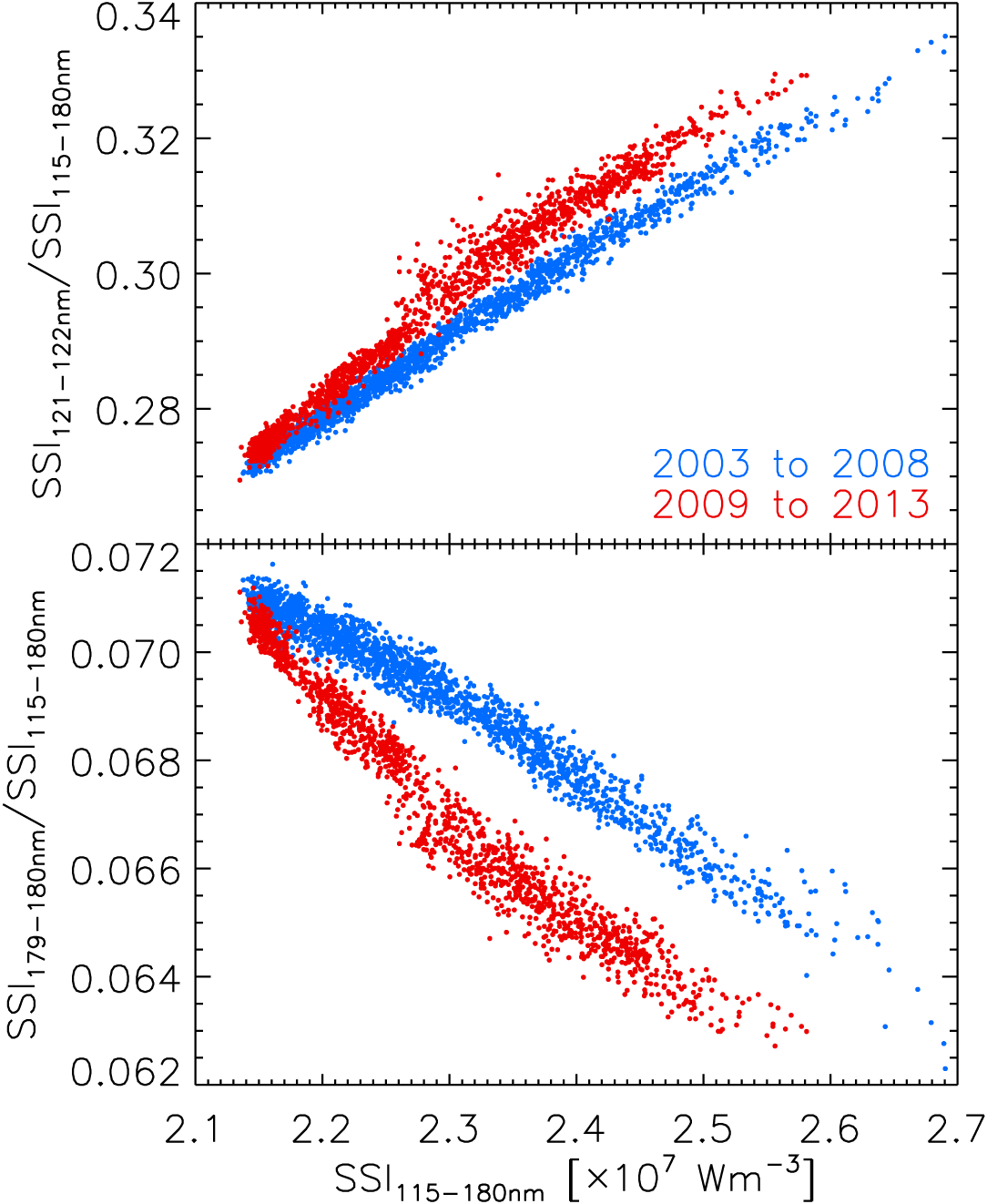}
\caption{Scatter plot of the ratio of SOLSTICE FUV SSI to the total flux registered by the instrument against the total flux, in the 121 to 122 nm (top), and 179 to 180 nm wavelength channels (bottom). The points from up to December 31, 2008 are drawn in blue, and from January 1, 2009 on in red.}
\label{solstice0}
\end{cfig}

The SOLSTICE FUV data set spans May 14, 2003 to July 15, 2013. We found a distinct disparity between the measurements from before 2009, and from 2009 onwards. This can be seen from the scatter plot of the ratio of measured flux in a given wavelength channel to the total flux registered by the instrument versus the total flux (two examples of which are illustrated in Fig. \ref{solstice0}). The two intervals of time, differently coloured, visibly do not overlap. This behaviour is seen for most of the wavelength channels and suggests that around the turn of 2008 to 2009, there was a shift in how the shape of SOLSTICE FUV spectra scaled with total flux. We will not speculate here if this shift is solar or instrumental in origin.

\begin{cfig}
\includegraphics[width=\textwidth]{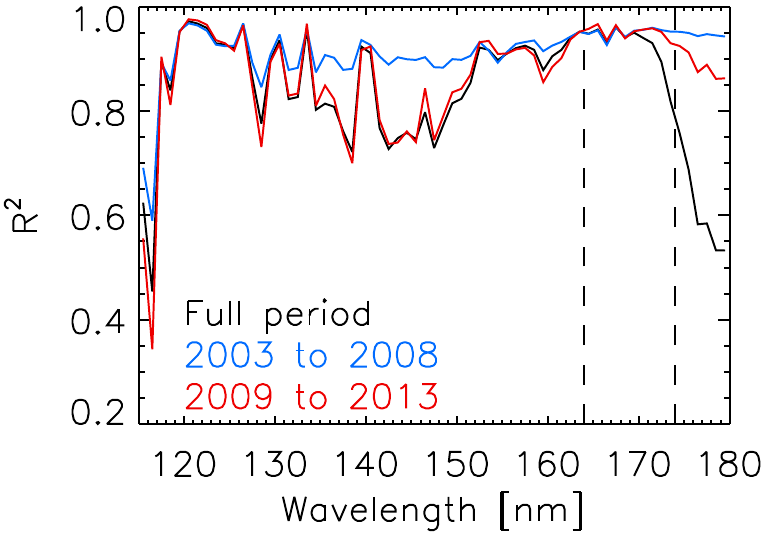}
\caption{The $R^2$ between the reconstruction (prior to the correction of the 115 to 300 nm segment) and SOLSTICE FUV, as a function of wavelength. The black curve represents the values obtained when considering the entire period of overlap between the two series, the blue and red curves data from before 2009 and from 2009 onwards alone. The dashed lines mark the reference interval, 164 to 174 nm (see text).}
\label{solstice1}
\end{cfig}

We also compared the reconstruction with the SOLSTICE FUV record, deriving the $R^2$ between the two as a function of wavelength. (The wavelength scale of the reconstruction and the SOLSTICE FUV record is similar, allowing us to make a channel-to-channel comparison.) Looking at data from before 2009 alone, measurement and model are highly correlated everywhere above $\sim120\:{\rm nm}$ (blue curve, Fig. \ref{solstice1}). Once we include data from 2009 onwards or examine them alone (black and red curves, respectively), the correlation at most wavelengths deteriorate significantly. Given the apparent shift in the property of SOLSTICE FUV observations around the turn of 2008 to 2009 (Fig. Fig. \ref{solstice0}), and the excellent consistency between measurement and model before this shift, we confined ourselves to the SOLSTICE FUV observations from up to December 31, 2008 for the following.

\begin{cfig}
\includegraphics[width=\textwidth]{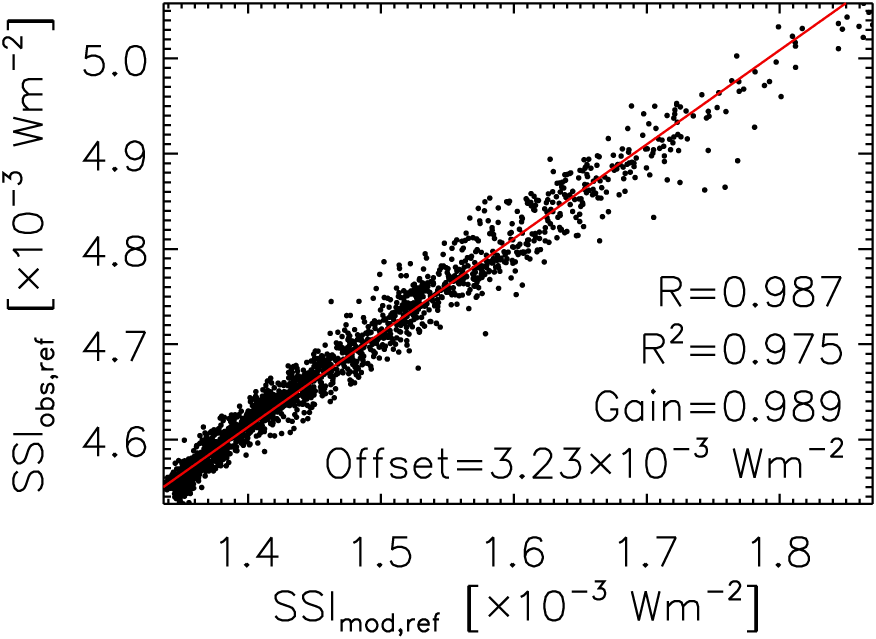}
\caption{Scatter plot of the integrated flux over the reference interval (164 to 174 nm) in the SOLSTICE FUV record, $\gobs$ and in the reconstruction, $\gsat$ (prior to the correction of the 115 to 300 nm segment). The red line is the straight line fit to the scatter plot. The $R$ and $R^2$ between the two series, and the gain and offset of the fit are indicated.}
\label{solstice2}
\end{cfig}

We took the 10 nm interval, 164 to 174 nm, as the reference interval. Let $\gobs$ and $\gsat$ denote the observed and reconstructed integrated flux in the reference interval. The linear regression of $\gobs$ to $\gsat$ is plotted along with the scatter plot of the two in Fig. \ref{solstice2}. The overall levels of $\gobs$ and $\gsat$ differ by a factor of about three. The variation with time however, is remarkably similar, as indicated by the proximity of the correlation coefficient, $R$ (0.987) and the gain of the fit (0.989) to unity. A regression slope that is close to unity, as discussed in Sect. \ref{fixbsat}, is inadequate by itself as a metric of similarity. However, together with a high $R$, it denotes that the two series are not only correlated, but the scale of the variation is closely similar. To a good approximation, $\gobs$ and $\gsat$ differ only by an offset of $3.23\times10^{-3}\:\wms$, the vertical intercept of the straight line fit.

Following, we fit a second order polynomial in $\gobs$ to the scatter plot of the measured solar irradiance in each SOLSTICE FUV channel versus $\gobs$. Finally, we applied the relationships defined by these fits to $\gsat+3.23\times10^{-3}\:\wms$ from the entire reconstruction to regenerated the 115 to 180 nm segment.

It must be emphasized that we are only introducing, into the reconstruction, an offset to the absolute level in the reference interval and the wavelength channel to reference interval relationships in SOLSTICE FUV data. The variation of the regenerated SSI with time is defined by the original reconstruction through $\gobs$, not by the SOLSTICE FUV record. By using the WHI reference solar spectra to correct the absolute level between 180 to 300nm, we are able to confine this regeneration to within 115 and 180 nm, a narrower wavelength range than in the analysis of \citealt{krivova06} (115 to 270 nm).

\subsection{Data gaps}
\label{datagaps}

Reconstructed solar irradiance from the $\kpone$, $\kptwo$, MDI and HMI data sets were collated into a single time series taking, on the days where the model output from more than one data set is available, the value from the succeeding instrument. That is, taking the model output from $\kptwo$ during the days with both $\kpone$ and $\kptwo$ data, from MDI where $\kptwo$ and MDI overlap, and from HMI where MDI and HMI overlap.

\begin{cfig}
\includegraphics[width=\textwidth]{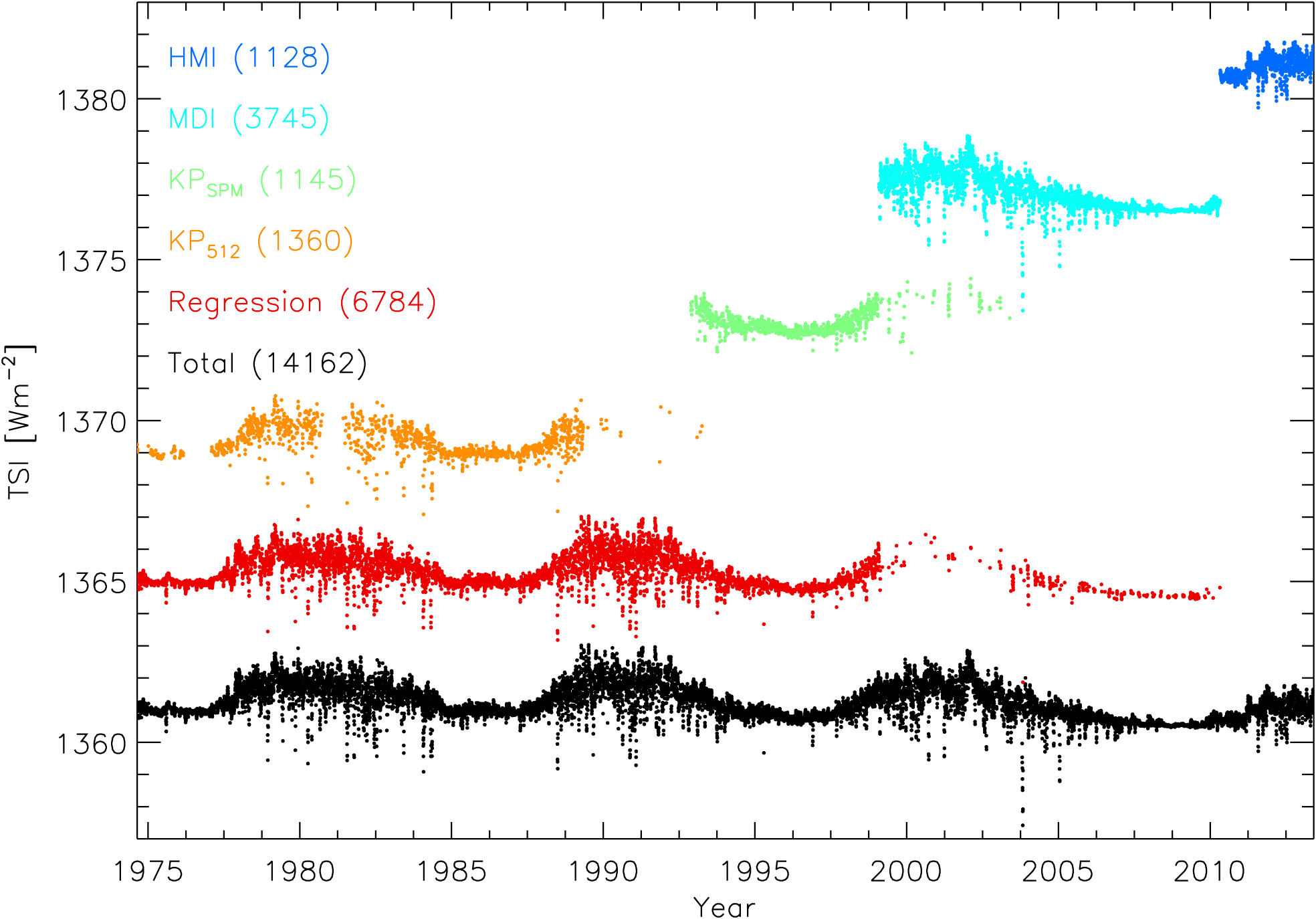}
\caption{Reconstructed daily total solar irradiance (black). The other series indicate the contribution to this time series by the SATIRE-S reconstruction based on the various data sets, and the regression series (see text), progressively offset to aid visibility. The values in parentheses indicate the number of daily values.}
\label{gapfill}
\end{cfig}

The reconstruction extends over the period of August 23, 1974 to May 31, 2013. Restricted by the availability of suitable magnetograms (Sect. \ref{dailypairs}), it covers just 7378 of the 14162 days within this period (Fig. \ref{gapfill}). The daily data cover, on average, just one of every five days in the $\kpone$ data set, one of every two for the $\kptwo$ data set and nine of every ten for the MDI data set (Table \ref{datatable}). The HMI data set is the only one that has no gaps. To yield an uninterrupted time series, we regressed indices of solar activity to the reconstruction, and used the relationships derived to estimate the solar spectra in the gaps from index data \cite[loosely following the method set out in][]{ball14}.

We employed the following solar activity index records; the Ottawa and Penticton adjusted 10.7 cm radio flux, F10.7 \citep{tapping87,tapping13}, the LASP Lyman-$\alpha$ composite \citep{woods00}, the IUP Mg II index composite \citep[version 4,][]{viereck99,skupin05b,skupin05c} and the projected sunspot area composite by \citealt{balmaceda09} (version 0613). Apart from the Mg II index composite by IUP, there is a competing composite by LASP \citep{viereck04,snow05b}. We selected the IUP record as the result of the regression agreed better with the SATIRE-S reconstruction.

For each wavelength element, we performed the (multiple) linear regression of each index, and each combination of two indices to the reconstruction. The indices-to-irradiance relationships so derived were then ranked by the correlation between the result of the regression and the reconstruction. The irradiance over the period of the reconstruction was estimated by applying, for each day, the highest ranked relationship for which the required index data is available. For this discussion, we will term this the regression series.

\begin{cfig}
\includegraphics[width=\textwidth]{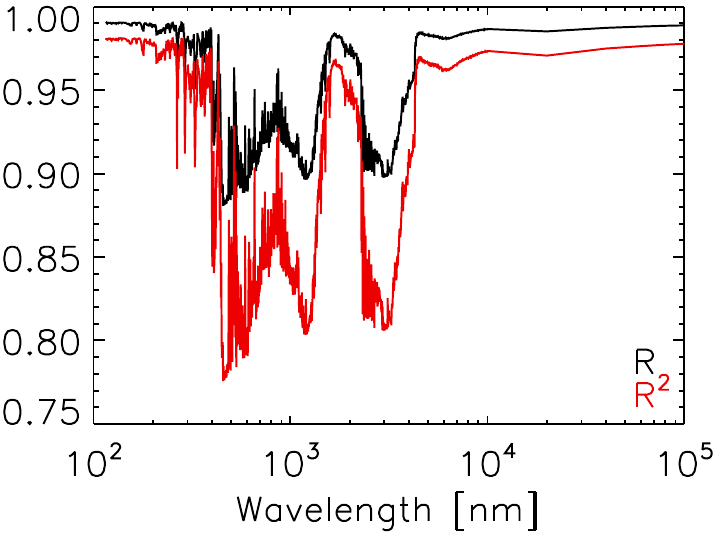}
\caption{The $R$ (black) and $R^2$ (red) between the SATIRE-S reconstruction and the regression series, as a function of wavelength.}
\label{satireversusindex}
\end{cfig}

Next, we offset each entry in the regression series by the average difference between the reconstruction and the regression series over the nearest 100 days where the reconstruction is available. This is to factor out any discrepancy in the long-term trend between the reconstruction and the regression series by adopting the long-term trend of the reconstruction in the regression series. The correlation between the SATIRE-S reconstruction and the regression series as a function of wavelength is depicted in Fig. \ref{satireversusindex}. Comparing the TSI from the two series, the $R$ ($R^2$) is 0.940 (0.883). Not unexpectedly, the regression series cannot reproduce all the variability in the SATIRE-S reconstruction. The agreement is, however, relatively good, sufficient for the intended purpose of the regression series. Finally, the gaps in the reconstruction were filled with values from the regression series.

It is worth noting that the indices-to-irradiance relationships were derived by regressing index records to the reconstruction, not to observations as in proxy models \citep[e.g.,][]{lean97,lean00,pagaran09,chapman13}. We are filling the gaps in the reconstruction by the regression of index data to the reconstruction, not a second independent model of solar irradiance.

\subsection{Error analysis}
\label{erroranalysis}

The reconstruction process is complex and involved multiple data sets. A rigorous determination of the uncertainty is far from straightforward, compounded by the following extenuating factors.
\begin{itemize}
	\item We did not examine the $\kpone$, $\kptwo$, MDI and HMI data sets for possible variation with time from changes in instrumental properties. This and uncertainty in the apparent surface coverage of magnetic features cannot be ascertained unambiguously from the data sets themselves.
	\item Magnetogram noise and finite resolution hinder us from identifying and including bright magnetic features that are weak and/or small \citep{krivova04}, and bright kilogauss magnetic concentrations have been observed even in the quiet Sun internetwork \citep{riethmuller13}. The amount of unresolved magnetic flux and its variation with time is not known and cannot be determined directly from the magnetogram data.
	\item The radiant behaviour of small-scale magnetic concentrations is not sufficiently understood to model its complexities fully. The LTE and ODF simplifications aside (Sect. \ref{uvfix}), the faculae intensity spectra utilised in the SATIRE-S reconstruction are based on a 1-D model atmosphere and applied to all bright magnetic features. This ignores 3-D effects on the apparent intensity contrast of small-scale magnetic concentrations \citep{solanki98b,holzreuter13}. This also overlooks the fact that their intensity contrast scales with magnetogram signal differently in the continuum and in spectral lines \citep{yeo13}, and quiet Sun network is, on a per unit magnetogram signal basis, brighter than active region faculae \citep{ortiz02,foukal11,kobel11}. It is our intention to refine this part of the model in the future, considering 3-D MHD simulations \citep[which are reaching the maturity required for quantitative comparison with observations,][]{afram11} and high spatial resolution observations from missions such as SUNRISE \citep{solanki10,barthol11}.
\end{itemize}
While a comprehensive error analysis is clearly onerous, a reasonable estimate of the reconstruction uncertainty can be obtained considering the uncertainty introduced by the steps taken to harmonize the model input from the various data sets (Sect. \ref{crosscalibration}), and in the free parameter, $\bsat$. This is similar to the approach taken by \cite{ball12,ball14}. The apparent uncertainty from these processes is influenced by, and would therefore largely encompass the error from other sources such as the indeterminate factors discussed above. We performed the following, for each wavelength element in the reconstruction and the TSI time series.

The uncertainty arising from the cross-calibration of the $\kpone$, $\kptwo$, MDI and HMI data sets is given by the RMS difference between the reconstruction based on the various data sets over the periods where they overlap. For the days with no SATIRE-S reconstruction, plugged with the regression series (Sect. \ref{datagaps}), we adopted the RMS difference between the SATIRE-S reconstruction and the regression series as the uncertainty.

As stated in Sect. \ref{fixbsat}, the amplitude of faculae contribution to variation in reconstructed solar irradiance scales nearly inversely with $\bsat$. The TSI records considered as the reference with which to recover $\bsat$ exhibit differences in terms of the amplitude of solar cycle variation (Fig. \ref{comparetsi}). Due to this discrepancies, we arrived at estimates of $\bsat$ ranging from 204 to 255 G (Table \ref{fixbsattable}). Eventually, we adopted 230 G, the value recovered with the PMO6V record as the reference (on account of the consistency between the resulting reconstruction and the record). We assume an uncertainty of $\pm30\:{\rm G}$ for $\bsat$.

The upper (lower) bound of the uncertainty range of the reconstruction is then given by the reconstruction generated with $\bsat$ set at 200 G (260 G), plus (minus) the cross-calibration error.

\section{Discussion}
\label{p3discussion}

\subsection{Comparison with observations}

In this section, we evaluate the reconstruction against
\begin{itemize}
	\item the ACRIM, IRMB and PMOD composite records of TSI (Sect. \ref{resulttsi}),
	\item the LASP Lyman-$\alpha$ composite, the Mg II index composites by IUP and by LASP (Sect. \ref{resultci}), and
	\item SSI observations from the UARS and SORCE missions (Sects. \ref{resultuv} and \ref{antiphase}).
\end{itemize}

\subsubsection{TSI composites}
\label{resulttsi}

\begin{cfig}
\includegraphics[width=\textwidth]{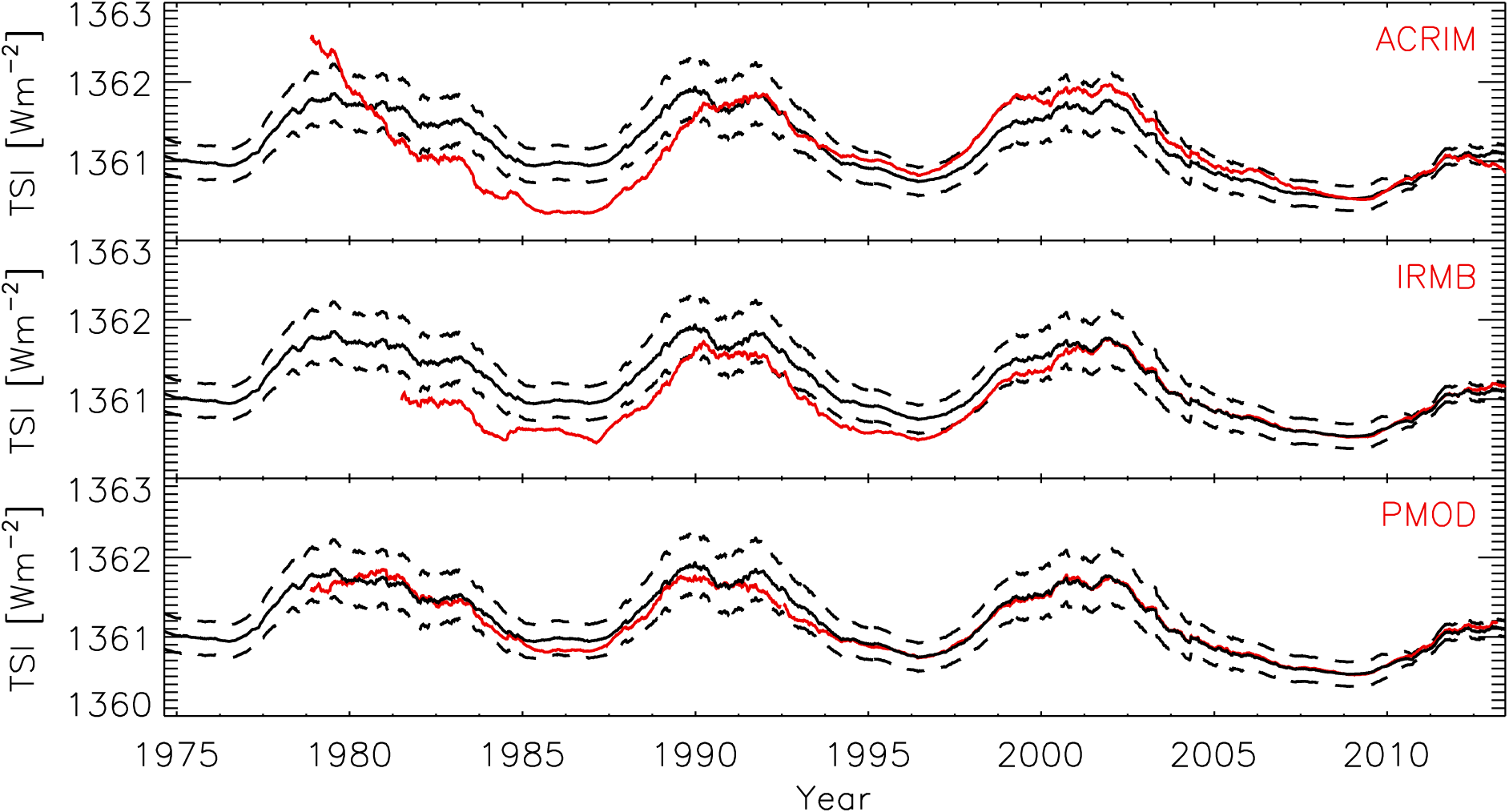}
\caption{361-day moving average of the TSI reconstruction (black), and the ACRIM, IRMB and PMOD composites (red). The dashed lines denote the uncertainty range of the reconstruction.}
\label{comparesatcps2}
\end{cfig}

\begin{table}
\caption{The difference between the TSI level at the solar cycle minima of 1976, 1986 and 1996, and the level at the 2008 minimum. We considered the mean level over the 3-month period centred on each minima.}
\centering
\begin{tabular}{lccc}
\hline\hline
Series & 1976 [$\wms$] & 1986 [$\wms$] & 1996 [$\wms$] \\
\hline
ACRIM    &         & $-0.205$ & $0.341$  \\
IRMB     &         & $-0.084$ & $-0.048$ \\
PMOD     &         & $0.273$  & $0.188$  \\
SATIRE-S & $0.423$ & $0.399$  & $0.236$  \\
\hline
\end{tabular}
\label{tsitable}
\end{table}

While broadly consistent with one another, the ACRIM, IRMB and PMOD composites do exhibit divergent decadal trends (Fig. \ref{comparesatcps2}). This is readily apparent in the conflicting cycle-to-cycle variation of the solar cycle minimum level (both in strength and direction, Table \ref{tsitable}), the focal point of the ensuing debate \citep{scafetta09,krivova09b,frohlich12}.

Preceding efforts with the SATIRE-S model, based on KPVT and/or MDI data, have found the greatest success in replicating the variability of the PMOD composite \citep{wenzler09,krivova11b,ball12}. Here, having updated the reconstruction method and extended the model to the present with HMI data, a re-evaluation is necessary.

\begin{table}
\caption{The $R$, $R^2$ and $k$ (RMS difference) between the reconstruction, and the ACRIM, IRMB and PMOD composites. Here and in the rest of the paper, these quantities are calculated excluding the part of the reconstruction provided by the regression series.}
\centering
\begin{tabular}{lccc}
\hline\hline
Series & $R$ & $R^2$ & $k$ [$\wms$] \\
\hline
ACRIM & $0.864$ & $0.747$ & $0.301$ \\
IRMB  & $0.897$ & $0.805$ & $0.251$ \\
PMOD  & $0.957$ & $0.916$ & $0.149$ \\
\hline
\end{tabular}
\label{tsitable2}
\end{table}

More than with the ACRIM or IRMB composites, the reconstruction replicated the variability of the PMOD composite (Table \ref{tsitable2}). The reconstruction and the PMOD composite are not only highly correlated to one another ($R=0.957$ and $R^2=0.916$) but as indicated by the weak RMS difference ($k=0.149\:\wms$) the absolute variation is similar.

\begin{table}
\caption{The same as Table \ref{tsitable2}, except between the reconstruction and the TSI measurements from ACRIM3, TIM and VIRGO. (This is not to be confused with Table \ref{fixbsattable}, where we compared the same TSI records with the corresponding candidate reconstruction, see Sect. \ref{fixbsat}.)}
\centering
\begin{tabular}{lccc}
\hline\hline
Series & $R$ & $R^2$ & $k$ [$\wms$] \\
\hline
ACRIM3   & $0.965$ & $0.931$ & $0.175$ \\
$\dirmb$ & $0.958$ & $0.917$ & $0.159$ \\
$\dpmod$ & $0.969$ & $0.940$ & $0.131$ \\
PMO6V    & $0.979$ & $0.959$ & $0.107$ \\
TIM      & $0.961$ & $0.924$ & $0.109$ \\
\hline
\end{tabular}
\label{tsitable3}
\end{table}

In Table \ref{tsitable3}, we also listed the correlation and RMS difference between the reconstruction and the TSI measurements from ACRIM3, TIM and VIRGO. Obviously, the closest alignment was found with the PMO6V record ($R=0.979$, $R^2=0.959$ and $k=0.107\:\wms$). This excellent agreement is all the more significant considering the fact that the PMO6V record, which extends 1996 to the present, encompasses an entire solar cycle minimum to minimum.

\begin{cfig}
\includegraphics[width=\textwidth]{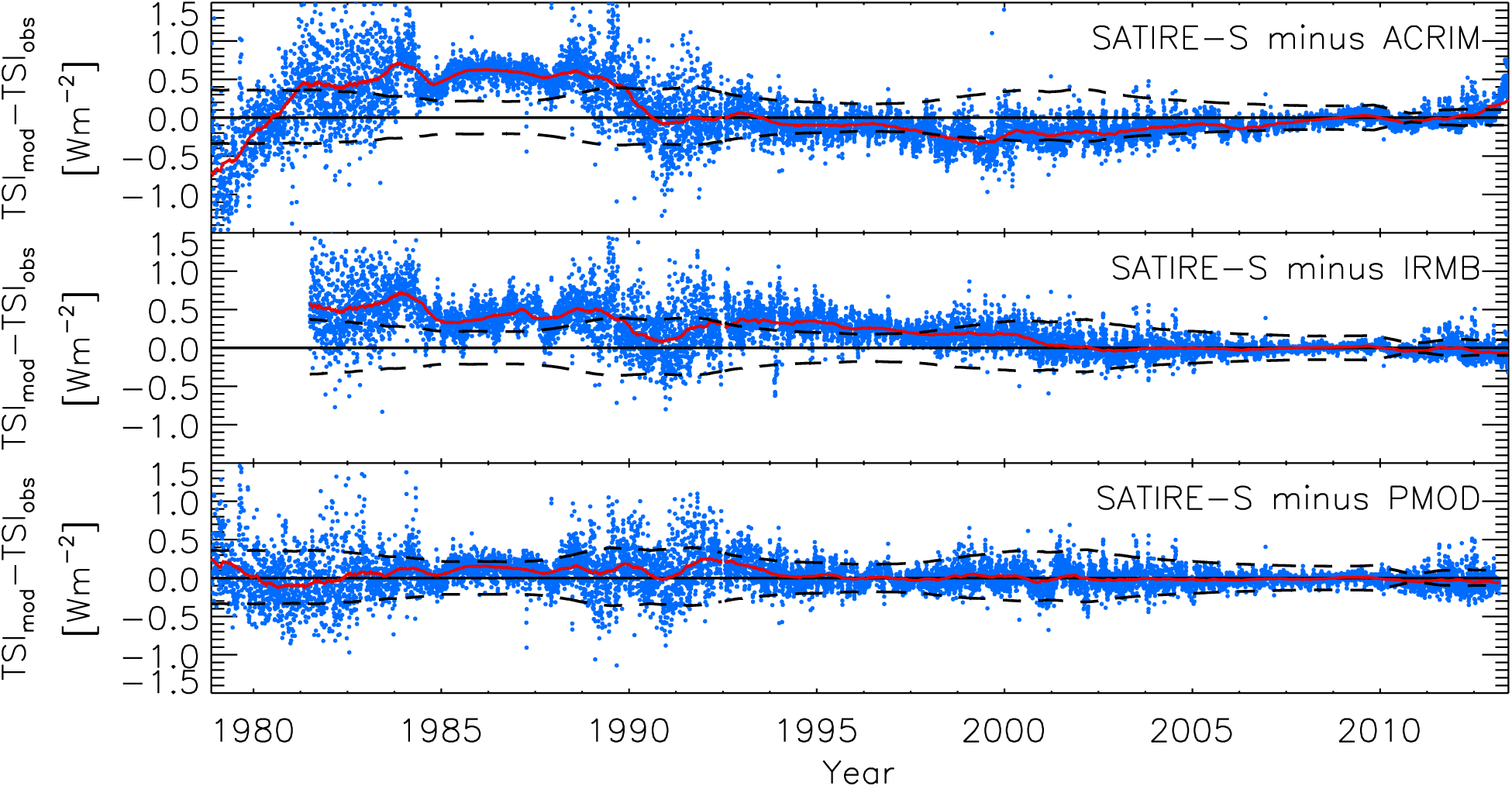}
\caption{The difference between the TSI reconstruction, and the ACRIM, IRMB and PMOD composites (blue dots), and the corresponding 361-day moving average (red curves). The dashed and solid black lines represent the reconstruction uncertainty and the zero level, respectively.}
\label{comparesatcps1}
\end{cfig}

In terms of the long-term trends, it is evident that the reconstruction is most consistent with the PMOD composite (Figs. \ref{comparesatcps2} and \ref{comparesatcps1}). The cyclic variation of the PMOD composite is replicated to well within the uncertainty limits of the reconstruction (bottom panel, Fig. \ref{comparesatcps2}), including the secular decline between the solar cycle minima of 1986, 1996 and 2008 (Table \ref{tsitable}).

\begin{cfig}
\includegraphics[width=\textwidth]{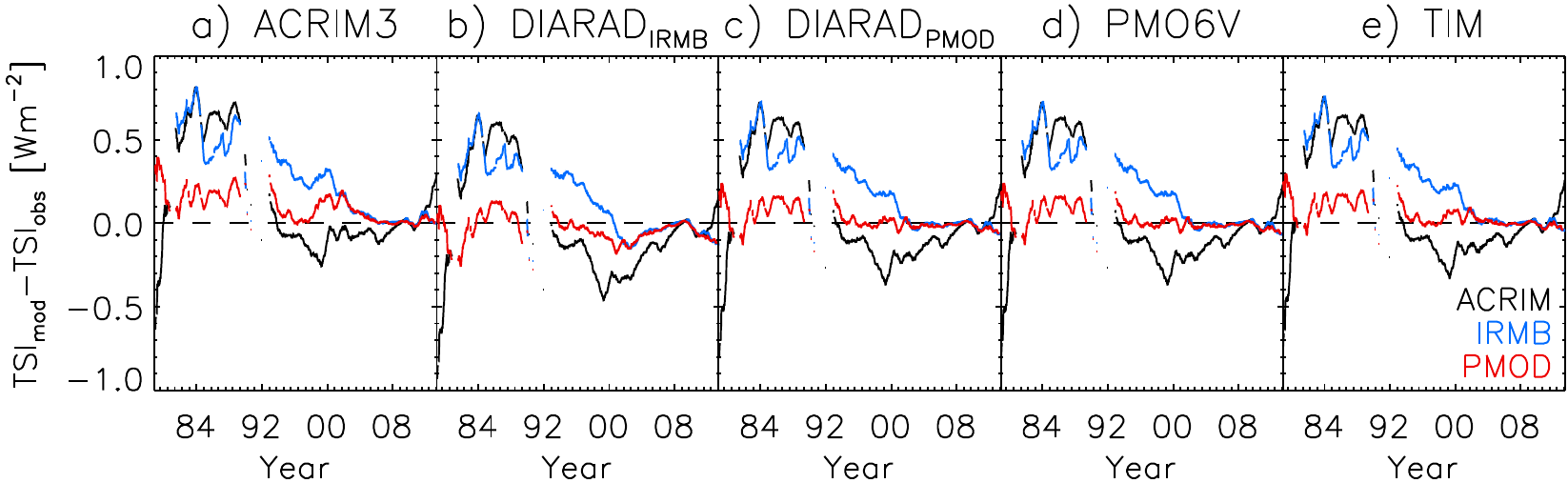}
\caption{361-day moving average of the difference between the five candidate TSI reconstructions described in Sect. \ref{fixbsat} (denoted by the TSI record referenced to determine $\bsat$), and the ACRIM (black), IRMB (blue) and PMOD composites (red). The dashed line represents the zero level.}
\label{comparesatcps0}
\end{cfig}

The fact that we had relied on the PMO6V record to fix the free parameter, $\bsat$ is not the reason the reconstruction is more aligned towards the PMOD composite (which uses the PMOD/WRC calibration of VIRGO radiometry in solar cycles 23 and 24) than the ACRIM and IRMB composites. In Fig. \ref{comparesatcps0} we express the smoothed difference between the five candidate reconstructions examined in Sect. \ref{fixbsat} and the three composites. There is no qualitative difference in how the candidate reconstructions compare with the three composites. For solar cycles 23 and 24, the ACRIM composite is based on the ACRIM2 and ACRIM3 records. For the IRMB composite, measurements from ACRIM2, ERBS, TIM and VIRGO (IRMB calibration). Using the ACRIM3 record as the reference to determine $\bsat$ does not swing the comparison towards the ACRIM composite (Fig. \ref{comparesatcps0}a). Neither does employing the $\dirmb$ and TIM records particularly favour the IRMB composite (Figs. \ref{comparesatcps0}b and \ref{comparesatcps0}e), nor the $\dpmod$ and PMO6V records the PMOD composite (Figs. \ref{comparesatcps0}c and \ref{comparesatcps0}d).

We noted earlier (in Sect. \ref{fixbsat}) that the candidate reconstructions generated taking the ACRIM3, $\dirmb$ and TIM records as the reference to recover $\bsat$ differ significantly from the respective reference in terms of the long-term variations (Fig. \ref{fixbsatfig}). $\bsat$ modulates the solar cycle amplitude of the reconstruction through its effect on apparent facular area. In using a given TSI record to determine $\bsat$, we are not adopting its long-term trend but rather just taking it as a reference to scale the overall amplitude of the reconstruction. The long-term trend, including the secular decline between the solar cycle minima of 1986, 1996 and 2008 in the reconstruction is inherent to it and not an outcome of using the PMO6V record to fix $\bsat$.

The observation here that we are able to replicate the secular change between the 1996 and 2008 solar cycle minima in the PMOD composite (i.e., PMOD/WRC calibration of VIRGO radiometry) using a model based on photospheric magnetism alone runs contrary to the claims of \cite{frohlich09,frohlich12,frohlich13}. The author noted that proxy models using chromospheric indices such as the Mg II index to represent bright magnetic features cannot fully replicate this secular decline and attributed the difference to a possible cooling/dimming of the photosphere. \cite{foukal11} proposed that the differing decadal trends exhibited by TSI and chromospheric indices might be due to the non-linear relationship between the two \citep{solanki04}. Another probable contributing factor is the long-term uncertainty in index data. We provide evidence for these in Sect. \ref{resultci}.

The long-term trend in the PMOD composite is well reproduced almost everywhere. A notable exception is the offset between the reconstruction and the PMOD composite between 1984 and 1994; the reconstruction is broadly higher by 0.1 to $0.3\:\wms$ (Fig. \ref{comparesatcps1}). Consequently, the reconstruction registers a stronger secular decline between the 1986 and 1996 solar cycle minima than the PMOD composite (Table \ref{tsitable}). To put this apparent disparity in context though, it is minute compared to the spread between the three composites, between the reconstruction and the other two composites, and between recent measurements of TSI (Fig. \ref{comparetsi}).

The offset between the reconstruction and the PMOD composite between 1984 and 1994 is unlikely to be related to the cross-calibration of the $\kpone$ and $\kptwo$ data sets (Sect. \ref{crosscalibration}). Going back in time, the reconstruction and the PMOD composite started to differ in 1994, while the period of overlap between $\kpone$ and $\kptwo$ data sets is November 1992 to April 1993. Also, any drift between model and measurement from how we combined the model output based on the $\kpone$ data set to the rest of the reconstruction would most likely amplify going back in time.  The apparent discrepancy could possibly be from unaccounted instrumental variation in the $\kpone$ and $\kptwo$ data sets, or in the PMOD composite, especially as the period of 1984 to 1994 encompasses the ACRIM gap \citep{scafetta09,krivova09b}.

\subsubsection{Solar Lyman-$\alpha$ irradiance and Mg II index composites}
\label{resultci}

Next, we compare the reconstruction with the LASP Lyman-$\alpha$ composite, and the competing Mg II index composites by IUP and by LASP (introduced in Sect. \ref{datagaps}).

In the wavelength range of 115 to 290 nm, the wavelength sampling of the reconstruction is 1 nm. We took the reconstructed solar irradiance in the 121 to 122 nm wavelength element, ${\rm SSI_{mod,121-122nm}}$, as the solar Lyman-$\alpha$ irradiance. The `Mg II index' of the reconstruction is given by
\begin{equation}
\frac{2\times{\rm SSI_{mod,279-281nm}}}{{\rm SSI_{mod,276-277nm}}+{\rm SSI_{mod,283-284nm}}},
\end{equation}
crudely following the definition by \cite{heath86}. The Lyman-$\alpha$ and Mg II index series so taken from the reconstruction are obviously not exactly equivalent to that from measurement, which are computed from the examination of higher spectral resolution line profiles. However, as we are only interested in comparing the relative variation, these approximations are still appropriate. The correlation between measurement and model is excellent. The reconstruction reproduced over $94\%$ of the variability in the LASP Lyman-$\alpha$ composite ($R=0.970$ and $R^2=0.942$), and over $96\%$ for the IUP Mg II index composite ($R=0.981$ and $R^2=0.963$). The agreement with the LASP Mg II index composite is poorer by a significant margin, but still very good ($R=0.948$ and $R^2=0.899$).

A direct quantitative comparison is not feasible given the approximate manner by which we estimated the solar Lyman-$\alpha$ irradiance and Mg II index from the reconstruction. Also, the absolute level of the LASP Mg II index composite is nearly twice that of the IUP composite\footnote{The Mg II index produced from measurements from different spectrometers differ in absolute terms since the wavelength sampling is usually not identical. There is therefore a degree of arbitrariness in how the Mg II index records generated from the various instruments are rescaled when combining them together to form the composite time series.}. To compare the long-term trends, we did the following. Take the Lyman-$\alpha$ instance, we regressed the LASP Lyman-$\alpha$ composite, detrended by subtracting the 361-moving average, to the similarly detrended reconstruction series. We rescaled the original LASP Lyman-$\alpha$ composite by the regression coefficient, in so doing matching the rotational variability to that of the reconstruction series. Finally, we offset the LASP Lyman-$\alpha$ composite to the reconstruction series at the 2008 solar cycle minimum. The IUP and LASP Mg II index composites were similarly rescaled and offset to the Mg II index generated from the reconstruction.

\begin{cfig}
\includegraphics[width=\textwidth]{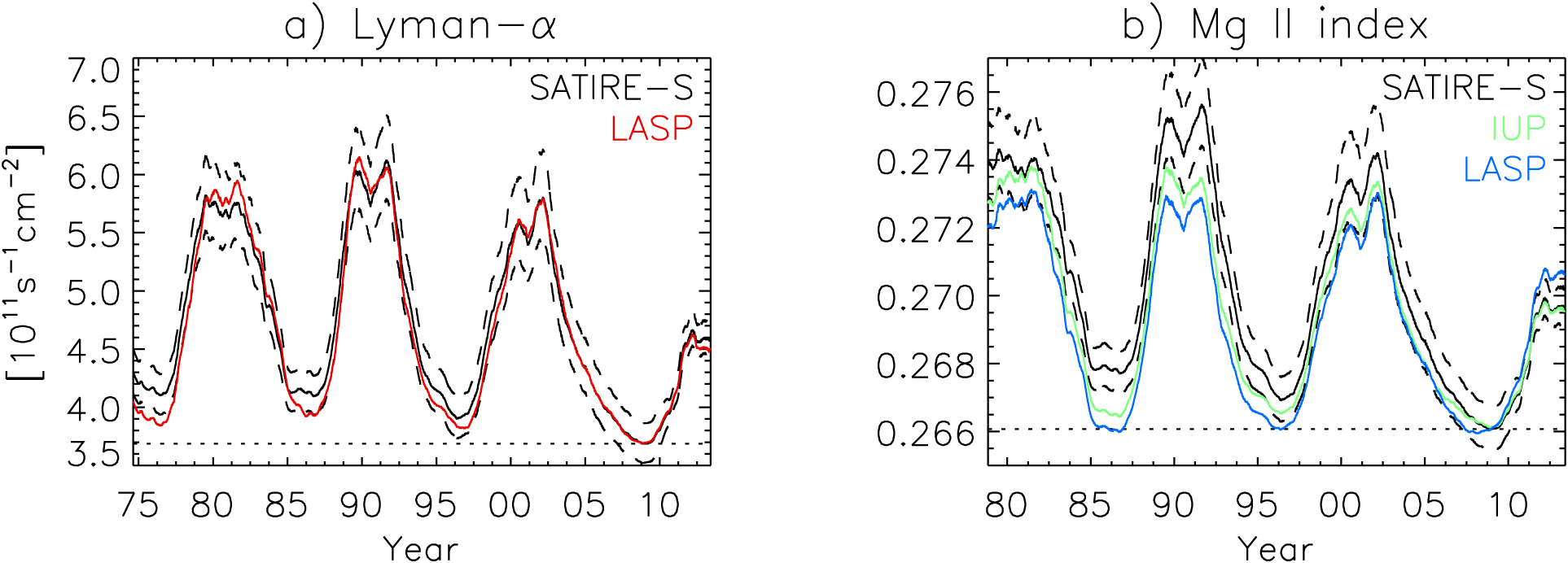}
\caption{361-day moving average of the a) solar Lyman-$\alpha$ irradiance and the b) Mg II index. The various time series were matched, in terms of the rotational variability and the level at the 2008 solar cycle minimum (dotted lines), to the time series generated from the SATIRE-S reconstruction to allow a direct comparison of the long-term trends (see text). The dashed lines indicate the lower and upper uncertainty of the SATIRE-S Lyman-$\alpha$ and Mg II index series.}
\label{comparesatssi}
\end{cfig}

The long-term trend in the LASP Lyman-$\alpha$ composite is reproduced in the reconstruction series to, at most places, well within the limits of uncertainty (Fig. \ref{comparesatssi}a). A notable exception is the noticeably higher levels in the reconstruction series at the solar cycle minima of 1976, 1986 and 1996. Even with this discrepancy, the secular trend in the solar cycle minima level in the LASP Lyman-$\alpha$ composite is largely replicated in the reconstruction. The lower solar cycle minima levels in the LASP Lyman-$\alpha$ composite compared to reconstructed solar irradiance at 121 to 122 nm is likely from the non-linear relationship between the two.

In the case of the Mg II index, the two composites and the Mg II index based on the reconstruction exhibit differences in the long-term trend greater than can be accounted within the reconstruction uncertainty (Fig. \ref{comparesatssi}b). The IUP composite does, however, register a similar, albeit weaker, secular decline between the 1996 and 2008 solar cycle minima as in the reconstruction series. The Mg II index, the line core-to-wing ratio of the Mg II h and k doublet at 280 nm, is relatively robust to instrument degradation as that largely cancels out in the division. The discrepancy between the two composites, of similar order as the difference to the reconstruction series, highlight the uncertainty still present (likely from the variation in instrument degradation with wavelength within the line profile). 

The non-linear relationship between chromospheric indices and solar irradiance, and the long-term uncertainty in index data has a significant impact on how the decadal trends in chromospheric indices and the reconstruction compare to one another (Fig. \ref{comparesatssi}). As noted in Sect. \ref{resulttsi}, these are the probable reasons why proxy models employing chromospheric indices to represent faculae brightening cannot replicate the secular decline in VIRGO TSI radiometry.

\subsubsection{UARS and SORCE ultraviolet solar irradiance}
\label{resultuv}

Compared to TSI, greater uncertainty persists over the measurement and modelling of ultraviolet solar irradiance. Measurements and models are broadly consistent at rotational timescales but diverge significantly in terms of the absolute level and long-term trends \citep{deland08,deland12,lean12,unruh12,ermolli13,solanki13}. This is due in part to the challenge in making reliable measurements (complicated by the wavelength dependence of instrumental influences), and the varying approaches taken by models of solar irradiance to account for non-LTE effects on reconstructed spectra \citep[see][and Sect. \ref{uvfix}]{fontenla99,krivova06,shapiro10}.

In this section, we compare the reconstruction with the daily ultraviolet spectral measurements from the SOLSTICE \citep[as archived on lasp.colorado.edu/lisird/, covering 119 to 420 nm,][]{rottman01} and SUSIM \citep[version 22, 115 to 410 nm,][]{brueckner93,floyd03} experiments onboard the UARS mission, and SOLSTICE \citep[level 3, version 12, 114 to 310 nm,][]{mcclintock05,snow05a} and SIM \citep[level 3, version 19, 240 to 2416 nm,][]{harder05a,harder05b} onboard SORCE. The UARS/SOLSTICE, SUSIM and SIM data sets contain glitches which manifest themselves as null or spurious measurements. We omitted the daily spectra with null measurements but otherwise employed these records as they are.

\begin{table*}
\caption{The correlation ($R$ and $R^2$) and RMS difference ($k$ and $k_{\rm rot}$, for the complete and detrended series, respectively) between the reconstruction and ultraviolet solar irradiance observations from the UARS and SORCE missions, over the indicated spectral intervals.}
\centering
\begin{tabular}{llcccc}
\hline\hline
Spectral interval & Series & $R$ & $R^2$ & $k$ [$\wms$] & $k_{\rm rot}$ [$\wms$] \\
\hline
120 to 180 nm & UARS/SOLSTICE & 0.979 & 0.959 & $3.85\times10^{-4}$ & $3.10\times10^{-4}$ \\
& UARS/SUSIM & 0.971 & 0.943 & $4.41\times10^{-4}$ & $3.86\times10^{-4}$ \\
& SORCE/SOLSTICE & 0.979 & 0.959 & $3.04\times10^{-3}$ & $1.98\times10^{-4}$ \\
\hline
180 to 240 nm & UARS/SOLSTICE & 0.521 & 0.271 & $2.51\times10^{-2}$ & $1.83\times10^{-2}$ \\
& UARS/SUSIM & 0.943 & 0.890 & $6.97\times10^{-3}$ & $3.99\times10^{-3}$ \\
& SORCE/SOLSTICE & 0.789 & 0.623 & $3.91\times10^{-2}$ & $2.83\times10^{-3}$ \\
& \cite{morrill11} & 0.976 & 0.952 & $5.48\times10^{-3}$ & $3.40\times10^{-3}$ \\
\hline
240 to 300 nm & SORCE/SOLSTICE & 0.432 & 0.187 & 0.102 & $2.05\times10^{-2}$ \\
& SORCE/SIM & 0.818 & 0.670 & 0.140 & $1.02\times10^{-2}$ \\
& \cite{morrill11} & 0.964 & 0.929 & $2.41\times10^{-2}$ & $1.07\times10^{-2}$ \\
\hline
\end{tabular}
\label{ssitable}
\end{table*}
		
\begin{cfig}
\includegraphics[width=\textwidth]{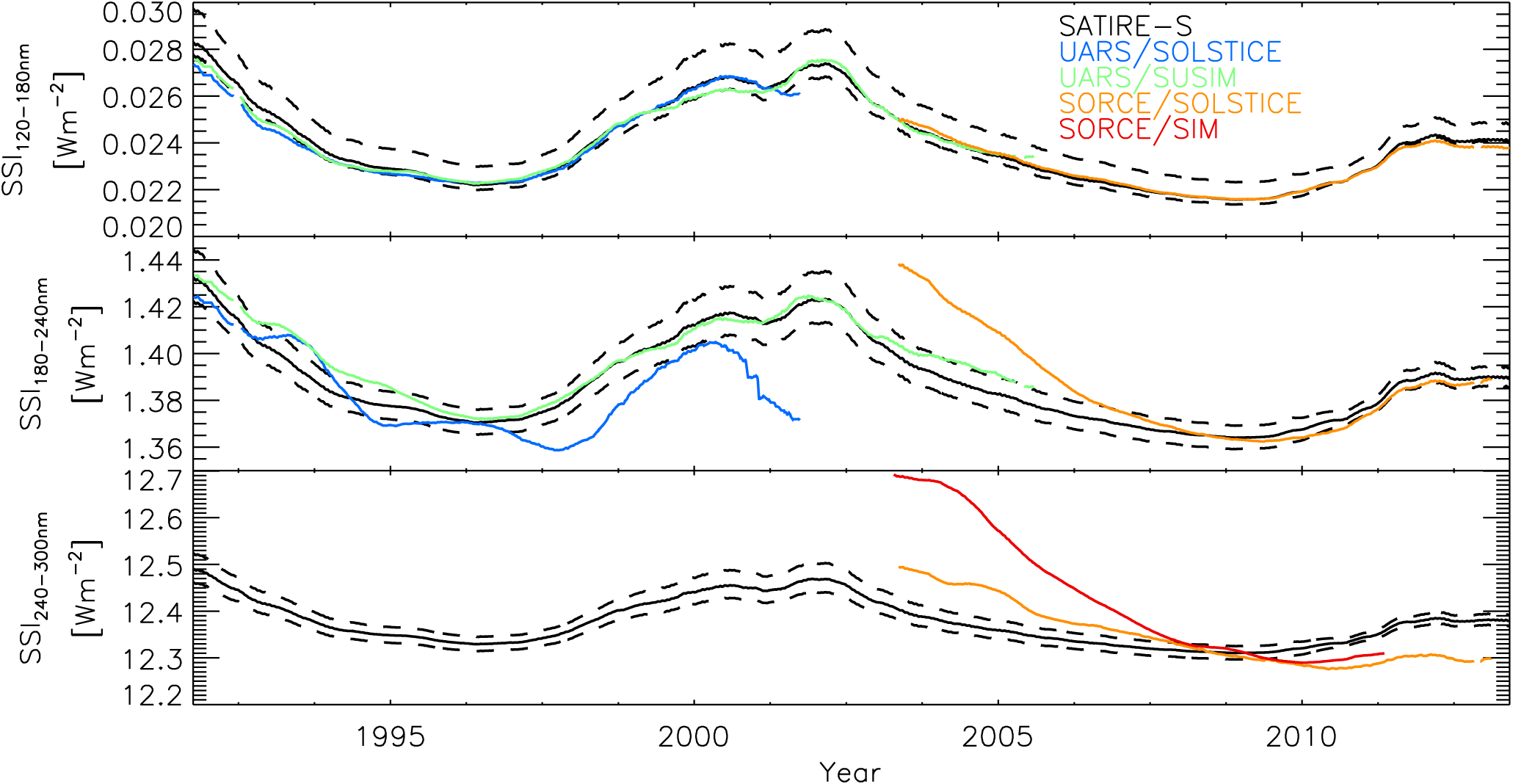}
\caption{361-day moving average of the integrated flux between 120 to 180 nm (top), 180 to 240 nm (middle) and 240 to 300 nm (bottom), in the reconstruction and the spectral measurements from the UARS and SORCE missions. The UARS and SORCE time series are offset to the level of the reconstruction at the 1996 and 2008 solar cycle minima, respectively. The dashed lines mark the uncertainty range of the reconstruction.}
\label{sus}
\end{cfig}

\begin{cfig}
\includegraphics[width=\textwidth]{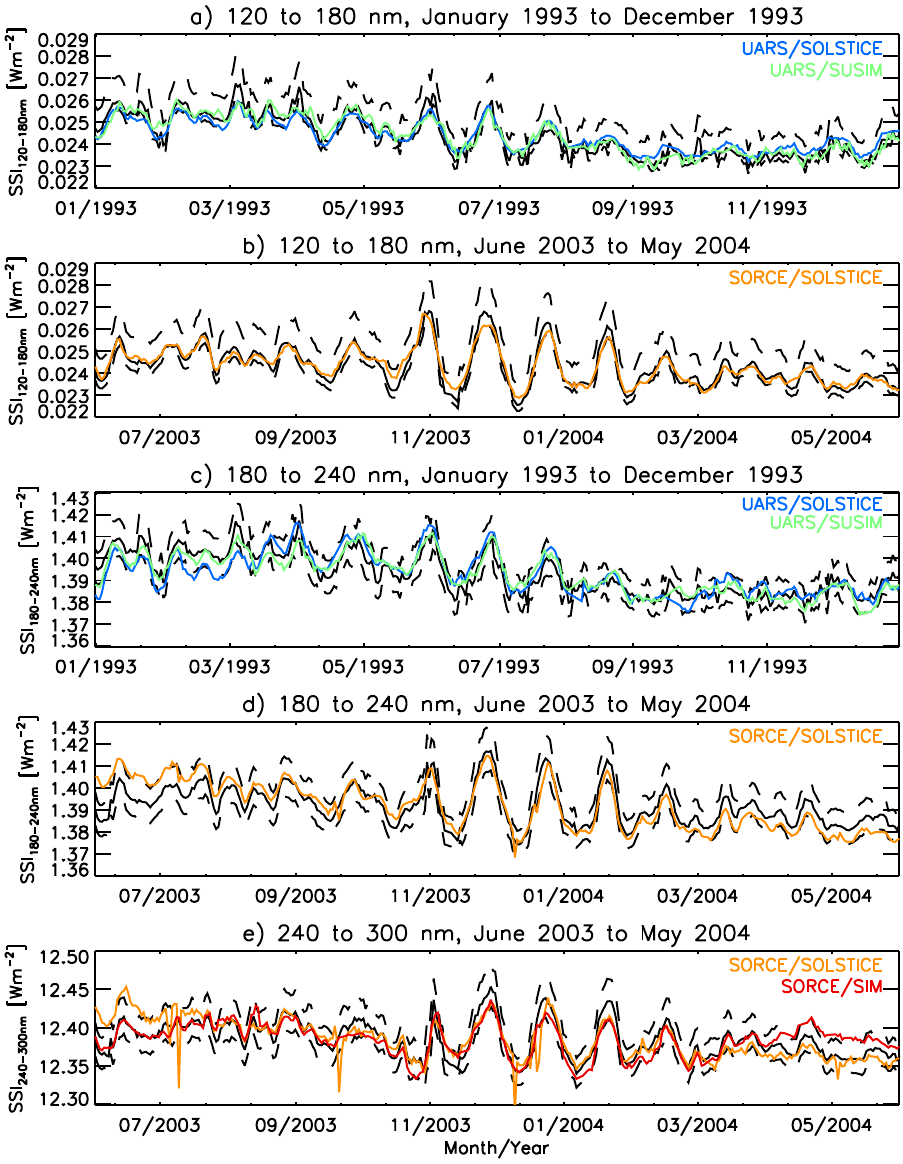}
\caption{Integrated flux between 120 to 180 nm, 180 to 240 nm and 240 to 300 nm, in the reconstruction (black curves) and in the spectral measurements from the UARS and SORCE missions, over the annotated periods. The UARS and SORCE time series are offset to the level of the reconstruction. The dashed lines represent the reconstruction uncertainty.}
\label{susr2}
\end{cfig}

For this discussion, we examine the integrated flux between 120 and 180 nm, 180 and 240 nm, and 240 and 300 nm. We excluded UARS/SOLSTICE and SUSIM observations from the comparison of 240 to 300 nm flux; approaching 300 nm, solar cycle variation is increasingly obscured by long-term stability issues \citep{deland04}. The integrated flux time series from the UARS and SORCE data sets were offset to the reconstruction at the 1996 and 2008 solar cycle minima, respectively. Given the spread in observed solar cycle variation, it is within expectation that in terms of the overall agreement (given by the correlation and RMS difference, Table \ref{ssitable}) and long-term variation (Fig. \ref{sus}), the reconstruction is aligned towards certain records more than others.
\begin{itemize}
	\item 120 to 180 nm (top panel, Fig. \ref{sus}): The overall agreement between the reconstruction, and the measurements from SUSIM and the two SOLSTICE experiments is excellent ($R>0.97$, $R^2>0.94$ and $k<3\times10^{-3}\:\wms$). The long-term variation in the three records is replicated to largely within the uncertainty of the reconstruction.
	\item 180 to 240 nm (middle panel, Fig. \ref{sus}): The reconstruction is a close match to the SUSIM record ($R=0.943$, $R^2=0.890$ and $k=6.97\times10^{-3}\:\wms$) but much less so with the observations from the two SOLSTICE experiments. There is a periodic modulation in the UARS/SOLSTICE record that is not present in the reconstruction or any of the other records. The SORCE/SOLSTICE record declined, between 2003 and 2008, by about twice as much as the reconstruction over the same period, but exhibited a similar overall trend after.
	\item 240 to 300 nm (bottom panel, Fig. \ref{sus}): The overall agreement between the reconstruction and the two SORCE records is poor, primarily from the gross divergence in the solar cycle variation. Between 2003 and 2008, the overall level in the SORCE/SOLSTICE and SIM records decreased by a factor of about two and four, respectively, more than in the reconstruction over the same period. Both records continued to decline after the solar cycle minimum of 2008, thus going from varying in phase to being in anti-phase with the solar cycle. This behaviour is not replicated in the reconstruction.
\end{itemize}
To compare the rotational variability, we detrended both measurement and model by the 361-day moving average. The corresponding RMS difference, denoted $k_{\rm rot}$, is listed in Table \ref{ssitable}. The rotational variability of all the records examined is, at most places, reproduced to within the reconstruction uncertainty (Fig. \ref{susr2}), encapsulated in the weak $k_{\rm rot}$ values. This suggests that the discrepancy between the reconstruction and certain measurements, detailed above, is likely a result of residual instrumental influences in said observations.

Ultraviolet solar irradiance below 242 nm, and between 242 and 310 nm is responsible for the production and destruction of ozone in the stratosphere, respectively. While the long-term variability of ultraviolet solar irradiance below 240 nm is relatively well constrained, above 240 nm the amplitude of solar cycle variation in measurements and models differ from one another by up to a factor of six \citep[see][Table \ref{ssitable} and Fig. \ref{sus}]{ermolli13}. Due to this spread, their application to climate models has led to qualitatively different results for the variation in mesospheric ozone \citep{haigh10,merkel11,ball14}.

As previously noted for SATIRE-S by \cite{krivova06,ermolli13,ball14}, between 240 and 400 nm, the amplitude of solar cycle variation in the SATIRE-S reconstruction is around twice that in the NRLSSI model \citep{lean97,lean00}, illustrated in Fig. \ref{morrill2}, and multiple times weaker than indicated by SIM SSI (Fig. \ref{comparesatsimssr}, discussed in the following section).

\begin{cfig}
\includegraphics[width=\textwidth]{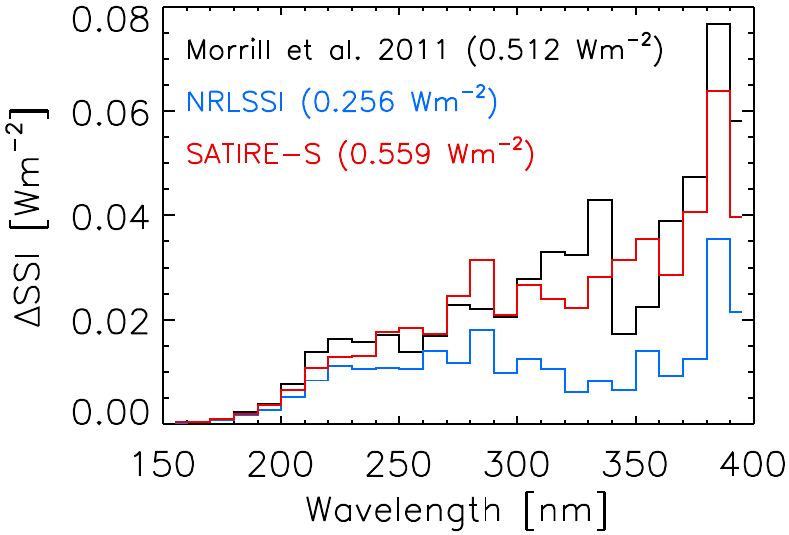}
\caption{The change in SSI, between the solar cycle maximum of 2000 and the minimum of 2008, in the \citealt{morrill11} (black), NRLSSI (blue) and SATIRE-S (red) models, as a function of wavelength. We considered the mean spectra over the 3-month period centred on the stated activity maximum/minimum, binned by wavelength in intervals of 10 nm. The integral between 240 and 400 nm is given in parentheses.}
\label{morrill2}
\end{cfig}

\begin{cfig}
\includegraphics[width=\textwidth]{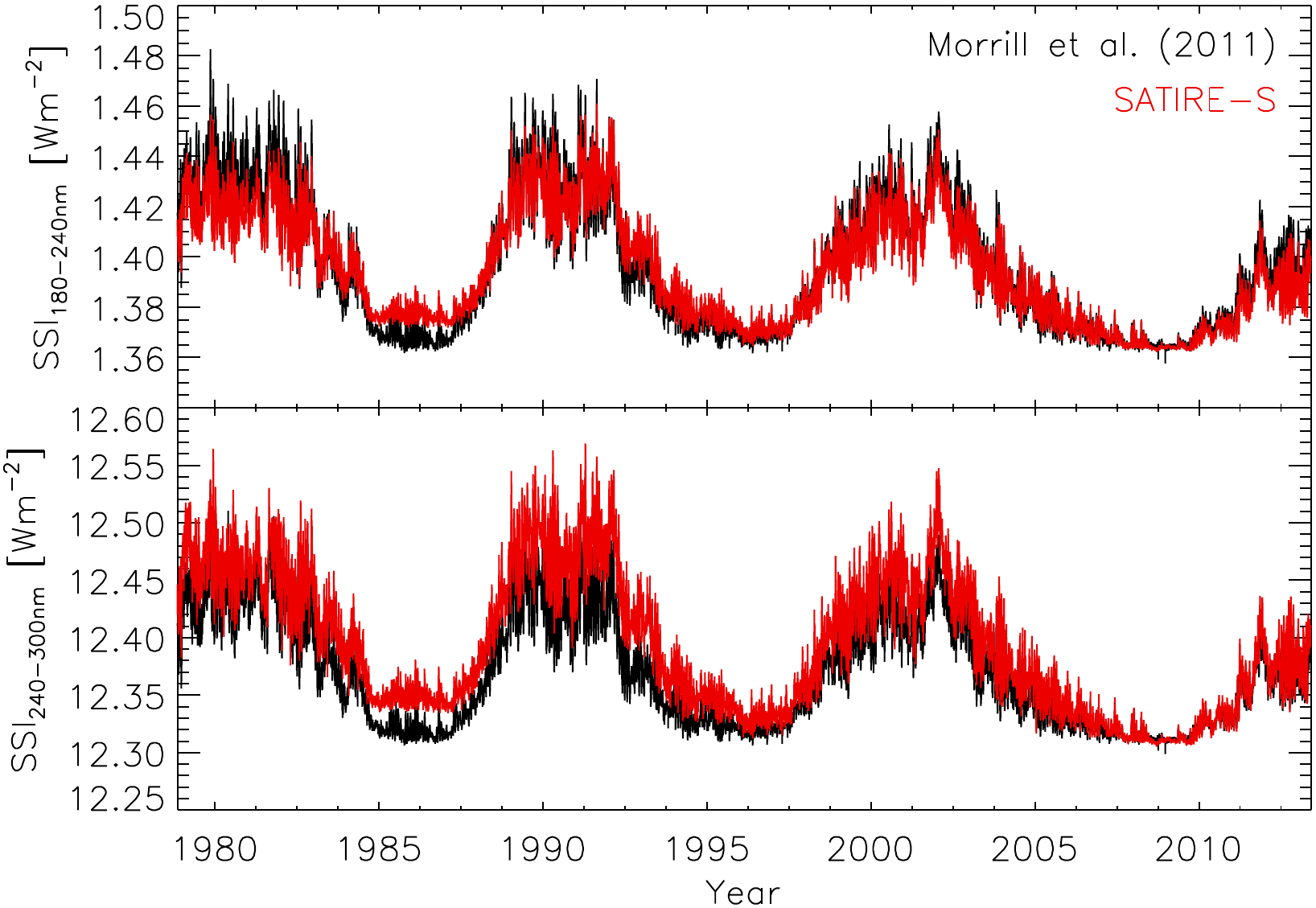}
\caption{Integrated flux between 180 and 240 nm (top), and 240 and 300 nm (bottom), in the \citealt{morrill11} (black) and the SATIRE-S reconstructions (red).}
\label{morrill}
\end{cfig}

We contrasted the SATIRE-S reconstruction against the empirical model of ultraviolet solar irradiance by \citealt{morrill11} (courtesy of Jeff Morrill). This model, covering 150 to 400 nm, is based on the regression of the Mg II index to SUSIM SSI. It therefore represents an approximation of SUSIM-like SSI with the long-term degradation corrected to that of the Mg II index. The \cite{morrill11} reconstruction, covering 1978 to 2013, is a close match to the SATIRE-S reconstruction (Table \ref{ssitable} and Fig. \ref{morrill}). The differences in the level at solar cycle minima reflect the similar divergence between the IUP Mg II index composite and the Mg II index taken from the SATIRE-S reconstruction, discussed in Sect. \ref{resultci}. Even with this discrepancy in the decadal trend, the amplitude of solar cycle variation is similar (Fig. \ref{morrill2}).

Between 120 and 400 nm, NRLSSI solar irradiance is given by the regression of the Mg II index to the rotational variability of the UARS/SOLSTICE record. The observation here that the solar cycle variation in the SATIRE-S reconstruction and the SUSIM-based model of \cite{morrill11} is similar lends credence to the variability reproduced in both models. It also supports the suggestion, stated in \cite{ermolli13}, that solar cycle variability in the ultraviolet in the NRLSSI might be underestimated from extending the result of the regression to rotational variability to longer timescales.

\subsubsection{SORCE/SIM SSI}
\label{antiphase}

\begin{cfig}
\includegraphics[width=\textwidth]{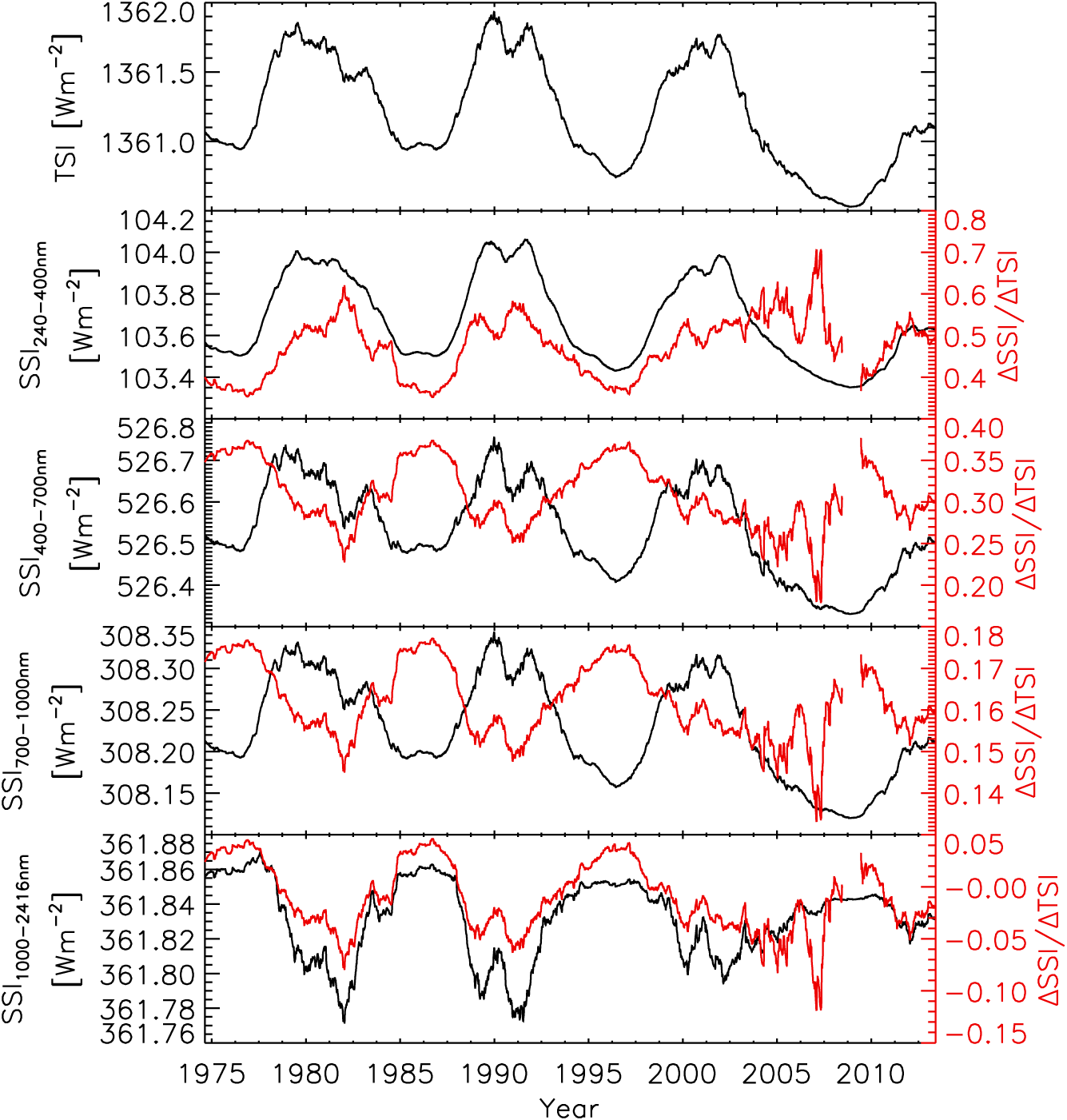}
\caption{Left: 361-day moving average of the TSI reconstruction and the integrated flux over the annotated spectral intervals in the SSI reconstruction (black). Right: Ratio of the variation (with respect to the 2008 solar cycle minimum) in SSI and in TSI, $\Delta{\rm SSI}/\Delta{\rm TSI}$, as given by the smoothed TSI and SSI time series, drawn in red. As $\Delta{\rm SSI}/\Delta{\rm TSI}$ is sensitive to noise where $\Delta{\rm TSI}$ is weak, the values from near the 2008 minimum are ignored, therefore the gap in the plots there.}
\label{uvvisnirir}
\end{cfig}

\begin{cfig}
\includegraphics[width=\textwidth]{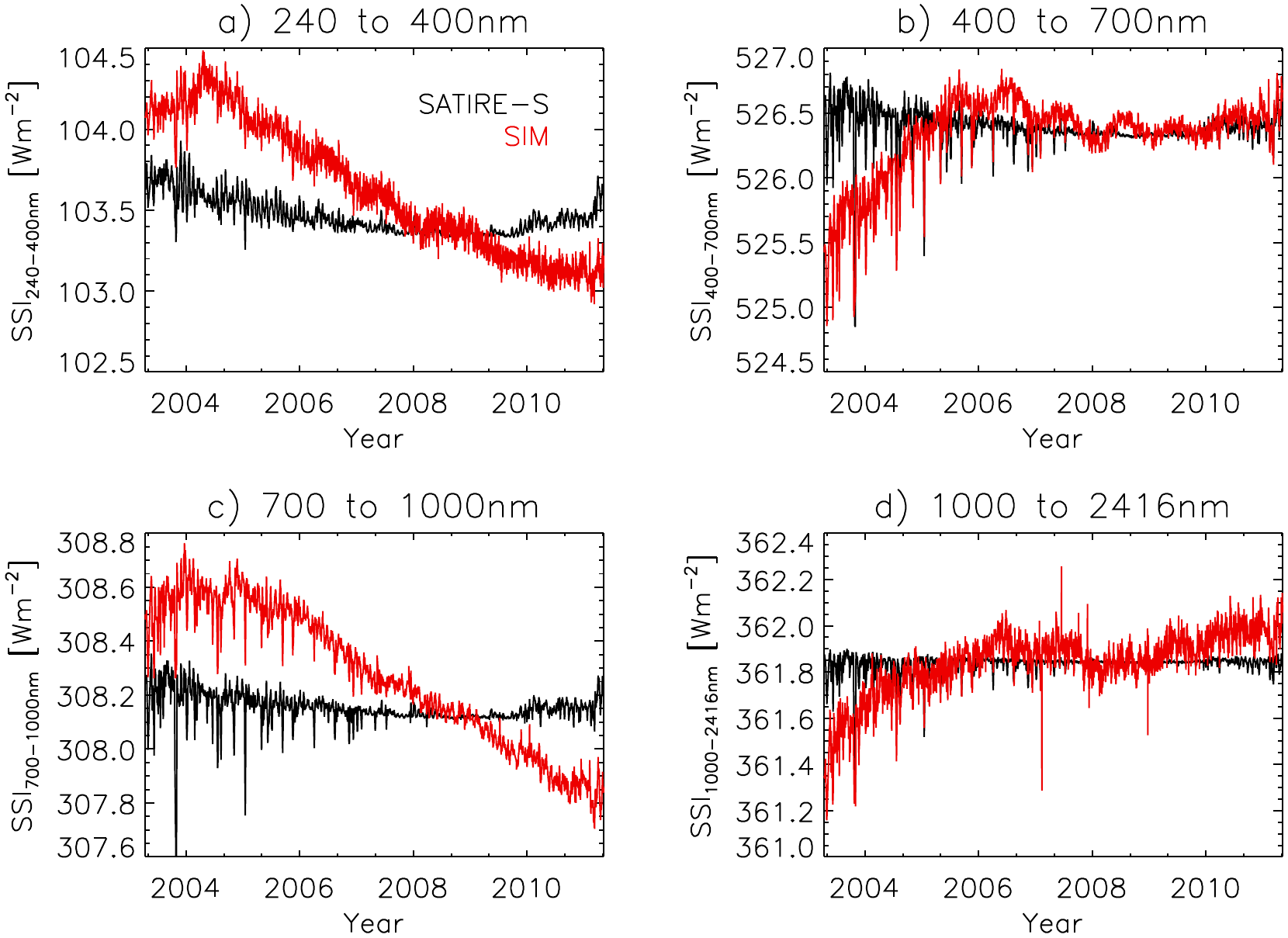}
\caption{Integrated flux over the annotated spectral intervals in the reconstruction (black) and in SIM SSI (red). The SIM time series are offset to the level of the reconstruction at the 2008 solar cycle minimum.}
\label{comparesatsim}
\end{cfig}

The recent release of SORCE/SIM SSI (level 3, version 19, dated November 19, 2013) spans the wavelength range of 240 to 2416 nm. As stated in the introduction, this is, at present, the only extended and continuous (in time) radiometrically calibrated SSI record covering the ultraviolet to infrared available. For this discussion, we examine the integrated flux in the reconstruction and the SIM data set over the ultraviolet (240 to 400 nm), visible (400 to 700 nm), near-infrared (700 to 1000 nm) and shortwave-infrared (1000 to 2416 nm).

In terms of the overall level, the reconstruction is, up to the near-infrared, varying broadly in phase with TSI (Fig. \ref{uvvisnirir}). In the shortwave-infrared, the radiation deficit due to sunspots become dominant due to the low (negative at parts) facular intensity contrast here \citep[as noted earlier for SATIRE-S by][]{unruh08,krivova11b,ball11}.

Also depicted in Fig. \ref{uvvisnirir} is the ratio of the variation (with respect to the level at the 2008 solar cycle minimum) within each spectral interval and in TSI, $\Delta{\rm SSI}/\Delta{\rm TSI}$. The value of $\Delta{\rm SSI}/\Delta{\rm TSI}$ around solar cycle maxima gives the relative contribution by the spectral interval to solar cycle variation in TSI (approximately, ultraviolet: $55\%$, visible: $30\%$, near-intrared: $16\%$ and shortwave-infrared: $-5\%$). $\Delta{\rm SSI}/\Delta{\rm TSI}$ at solar cycle minima represents the same for the secular variation in the TSI level at solar cycle minima (ultraviolet: $40\%$, visible: $35\%$, near-infrared: $17\%$ and shortwave-infrared: $5\%$). The variation in $\Delta{\rm SSI}/\Delta{\rm TSI}$ between solar cycle maxima and minima, including the change of sign in the shortwave-infrared, arises mainly from the fact that sunspots and pores are largely absent around activity minima (see Sect. \ref{facsptvar}).

The SIM record indicates a steady decline in ultraviolet and near-infrared flux since 2004 (Fig. \ref{comparesatsim}a and \ref{comparesatsim}c). Counter to expectation (that solar irradiance display solar cycle variation), there was no reversal in the overall trend after the 2008 solar cycle minimum. Between 2003 and 2006, SIM recorded a pronounced, almost monotonic increase in visible and shortwave-infrared flux (Fig. \ref{comparesatsim}b and \ref{comparesatsim}d), during a period solar activity is diminishing. Thereafter, the variation is much weaker and largely in phase with the solar cycle, declining till 2008 before ascending again. There is no obvious indication of any consistent trend (neither in phase nor in anti-phase) between the integrated flux within each spectral interval and the solar cycle.

\begin{table}
\caption{The correlation ($R$ and $R^2$) and scatter ($k$ and $k_{\rm rot}$) between the integrated flux over the indicated spectral intervals in the SIM record and in the SSI reconstruction.}
\centering
\begin{tabular}{lcccc}
\hline\hline
Spectral interval & $R$ & $R^2$ & $k$ [$\wms$] & $k_{\rm rot}$ [$\wms$] \\
\hline
240 to 400 nm   & 0.675 & 0.456 & 0.367 & $6.61\times10{-2}$ \\
400 to 700 nm   & 0.088 & 0.008 & 0.312 & $8.35\times10{-2}$ \\
700 to 1000 nm  & 0.380 & 0.145 & 0.258 & $3.39\times10{-2}$ \\
1000 to 2416 nm & 0.427 & 0.182 & 0.124 & $5.47\times10{-2}$ \\
240 to 2416 nm   & 0.411 & 0.169 & 0.583 & 0.179 \\
\hline
\end{tabular}
\label{simtable}
\end{table}

\begin{table}
\caption{The correlation ($R$ and $R^2$) between the total flux registered by SIM and the integrated flux in the SSI reconstruction over a similar wavelength range (i.e., 240 to 2416 nm), and the five TSI records discussed in Sect. \ref{fixbsat} and the TSI reconstruction. We only considered the data from the days where the reconstruction, SIM SSI and all the TSI records are available.}
\centering
\begin{tabular}{llcc}
\hline\hline
SSI Series & TSI Series & $R$ & $R^2$ \\
\hline
SIM ${\rm SSI_{240-2416nm}}$ & ACRIM3            & 0.481 & 0.232 \\
                             & $\dirmb$          & 0.474 & 0.225 \\
                             & $\dpmod$          & 0.418 & 0.175 \\
                             & PMO6V             & 0.428 & 0.183 \\
                             & TIM               & 0.328 & 0.108 \\
                             & SATIRE-S TSI      & 0.412 & 0.170 \\
\hline
SATIRE-S ${\rm SSI_{240-2416nm}}$ & ACRIM3       & 0.940 & 0.883 \\
                                  & $\dirmb$     & 0.963 & 0.927 \\
                                  & $\dpmod$     & 0.965 & 0.931 \\
                                  & PMO6V        & 0.971 & 0.942 \\
                                  & TIM          & 0.966 & 0.933 \\
                                  & SATIRE-S TSI & 0.999 & 0.999 \\
\hline
\end{tabular}
\label{simtable2}
\end{table}

The apparent non-correlation between the SIM record and the solar cycle is in conflict with what we found in the reconstruction (Fig. \ref{uvvisnirir}). It also does not corroborate with most other observations of ultraviolet solar irradiance (Fig. \ref{sus}) and VIRGO SPM photometry \citep{wehrli13}, which demonstrate clear solar cycle variation. The SATIRE-S model describes the effect of photospheric magnetism on the solar spectrum. As the amount of resolved magnetic flux and sunspots exhibit solar cycle variation, obviously so does reconstructed solar irradiance. The amplitude of the overall trend in the SIM record is also grossly stronger than what is reproduced in the reconstruction (Fig. \ref{comparesatsim}). Expectedly, the overall agreement between the reconstruction and SIM SSI, as given by the correlation and the RMS difference, is very poor (Table \ref{simtable}).

\begin{cfig}
\includegraphics[width=\textwidth]{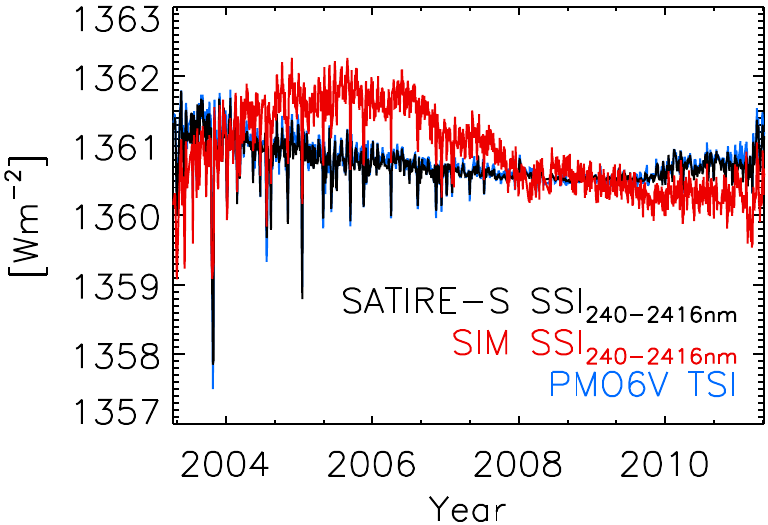}
\caption{The total flux recorded by SIM (red), the integrated flux in the reconstruction over a similar wavelength range (black) and PMO6V TSI (blue). The SIM and SATIRE-S time series were offset to the PMO6V series at the 2008 solar cycle minimum. The PMO6V series is largely hidden by the SATIRE-S series due to the close similarity.}
\label{intsim}
\end{cfig}

In Fig. \ref{intsim}, we compare the total flux recorded by SIM (red), the integrated flux in the reconstruction over the same wavelength range (black) and PMO6V TSI (blue). With the bulk of the energy in solar radiation confined within the spectral range surveyed by SIM ($>97\%$), the integrated flux over this spectral range should at least resemble TSI (as similarly argued by \citealt{lean12}). The total flux registered by SIM, as with the integrated flux over each of the spectral intervals examined earlier (Fig. \ref{comparesatsim}), exhibits no clear consistent relation to TSI or the solar cycle. For the reconstruction, the integrated flux over the spectral range of SIM reproduces at least $88\%$ ($R>0.94$ and $R^2>0.88$) of the variability in ACRIM3, TIM and VIRGO TSI radiometry, and effectively all of the variability in the TSI from the reconstruction (Table \ref{simtable2}). The total flux registered by SIM reproduces, in the best case, about 23\% of the variability in the ACRIM3 record ($R=0.481$ and $R^2=0.232$).

\begin{cfig}
\includegraphics[width=\textwidth]{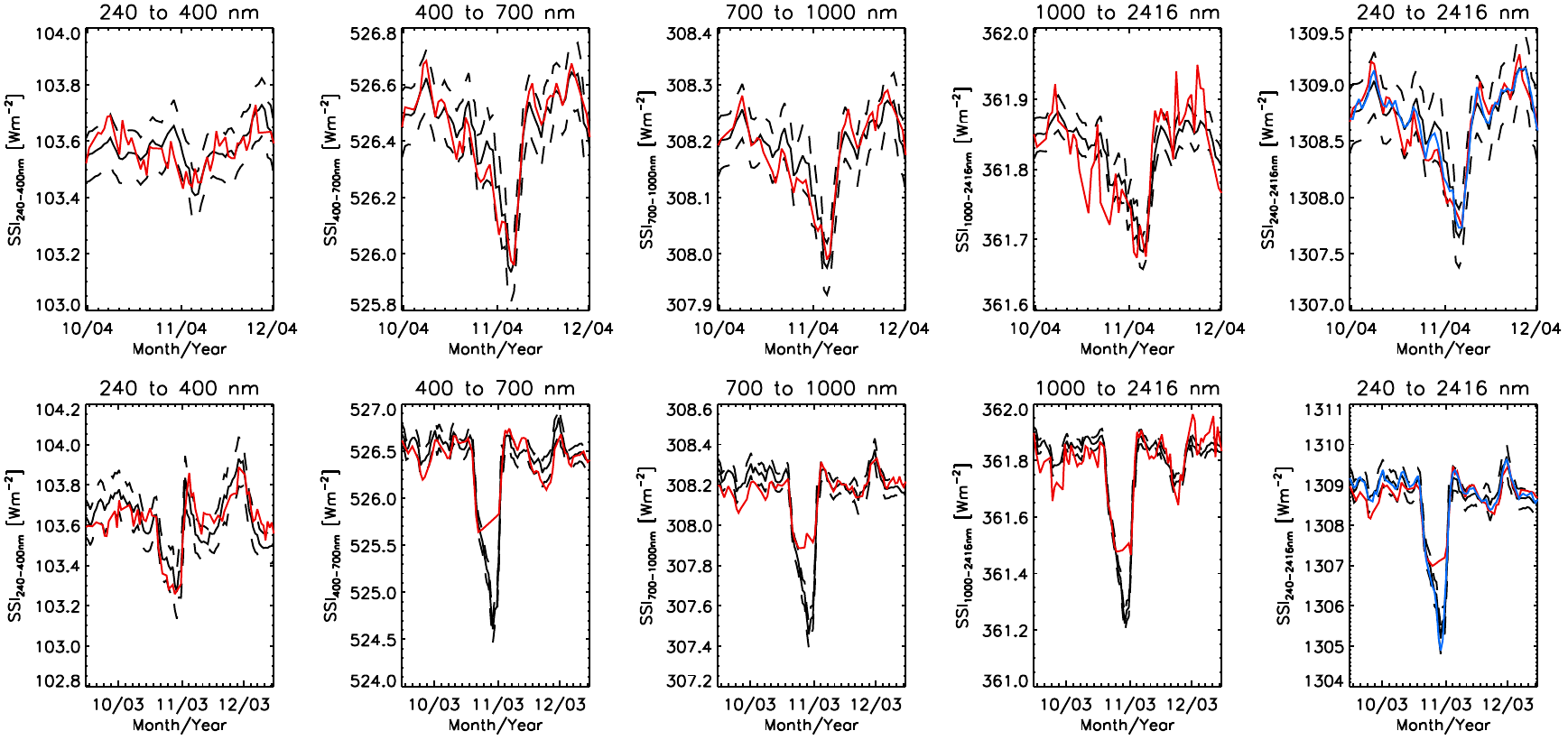}
\caption{The integrated flux over the annotated spectral intervals, in the reconstruction (black) and in the SIM record (red), between October 1 and November 30, 2004 (top), and September 15 and December 15, 2003 (bottom). The maximum sunspot number over the depicted periods is 99 (October 24, 2004) and 167 (October 29, 2003), respectively. PMO6V TSI is plotted along the integrated flux over 240 to 2416 nm (blue). The SIM and PMO6V time series are offset to the level of the reconstruction. The dashed lines denote the uncertainty range of the reconstruction.}
\label{ssrotational}
\end{cfig}

As with the ultraviolet solar irradiance records discussed in Sect. \ref{resultuv}, the reconstruction replicates the rotational variability in SIM SSI, largely to within uncertainty, in all the spectral intervals examined except the shortwave-infrared (the $k_{\rm rot}$ between measurement and model is tabulated in Table \ref{simtable}). That said, even with the more pronounced scatter, there is still a broad agreement between model and measurement in terms of the rotational variation in the shortwave-infrared (top row, Fig. \ref{ssrotational}).

During periods of significant sunspot activity, the dip corresponding to the passage of sunspots across the solar disc is in some instances significantly deeper in the reconstruction than in SIM SSI, an example of which is illustrated in Fig. \ref{ssrotational} (bottom row). If we examine the integrated flux in the reconstruction over the spectral range of SIM, the dip is much closer to that registered in concurrent TSI measurements. It is unlikely that the discrepancy between the total flux recorded by SIM and TSI measurements at these times can be accounted for by sunspot contrast below 240 nm and above 2416 nm. The weaker dips registered by SIM as compared to the reconstruction during periods of heightened sunspot activity is likely instrumental in origin.

Earlier in this section, we noted that there is no obvious solar cycle modulation in the SIM record (Fig. \ref{comparesatsim}), and the total flux recorded by the instrument fails to reproduce much of the observed variability in TSI (Fig. \ref{intsim} and Table \ref{simtable2}). In comparison, the reconstruction exhibits clear solar cycle variation (Fig. \ref{uvvisnirir}) and the integrated flux over the spectral range of SIM replicates most of the variability in ACRIM3, TIM and VIRGO TSI radiometry. These observations, together with the broad consistency between the reconstruction and the SIM record at rotational timescales implies that the overall discord between measurement and model (Fig. \ref{comparesatsim} and Table \ref{simtable}) is probably the result of residual instrumental trends in the SIM data set.

\subsection{Facular and sunspot contribution to variation in TSI}
\label{facsptvar}

From the intermediate products of the SATIRE-S reconstruction process, it is straightforward to compute the (magnetically) quiet solar spectrum (reconstructed solar spectrum assuming a solar disc that is entirely quiet Sun) and the variation in solar irradiance with respect to this base level from faculae, and from sunspots. Of course, this analysis excludes the part of the reconstruction provided by the regression series (Sect. \ref{datagaps}).

\begin{cfig}
\includegraphics[width=\textwidth]{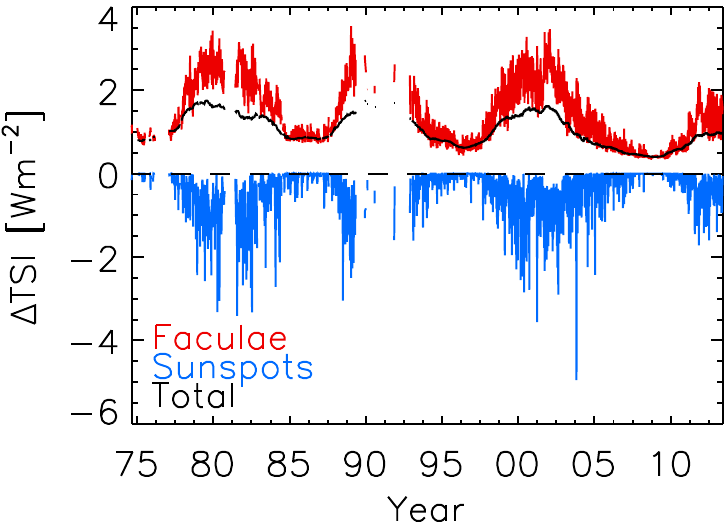}
\caption{TSI excess from faculae (red), and deficit from sunspots (blue), with respect to the calculated quiet Sun level ($1360.122\:\wms$, denoted by the dashed line) in the reconstruction. The sum, giving the net variation in TSI, is drawn in black, smoothed by means of a 361-day boxcar filter so as not to obscure the other series.}
\label{facspt}
\end{cfig}

The variation in TSI from faculae (denoted $\dtsifac$), and from sunspots ($\dtsispt$), with respect to the calculated quiet Sun level, $1360.122\:\wms$ (the integral under the quiet solar spectrum) is illustrated in Fig. \ref{facspt}. Between the solar cycle minima of 1986, 1996 and 2008, $\dtsifac$ (red) diminished, minimum-to-minimum, by over $0.2\:\wms$, while $\dtsispt$ (blue) declined, in absolute terms, by around $0.02\:\wms$. (We do not have any figures for the 1976 minimum as this period in the reconstruction is mainly covered by the regression series.) The secular variation of the solar cycle minima level in the TSI reconstruction (Table \ref{tsitable}) is dominantly from the variation in the intensity excess from faculae.

\begin{cfig}
\includegraphics[width=\textwidth]{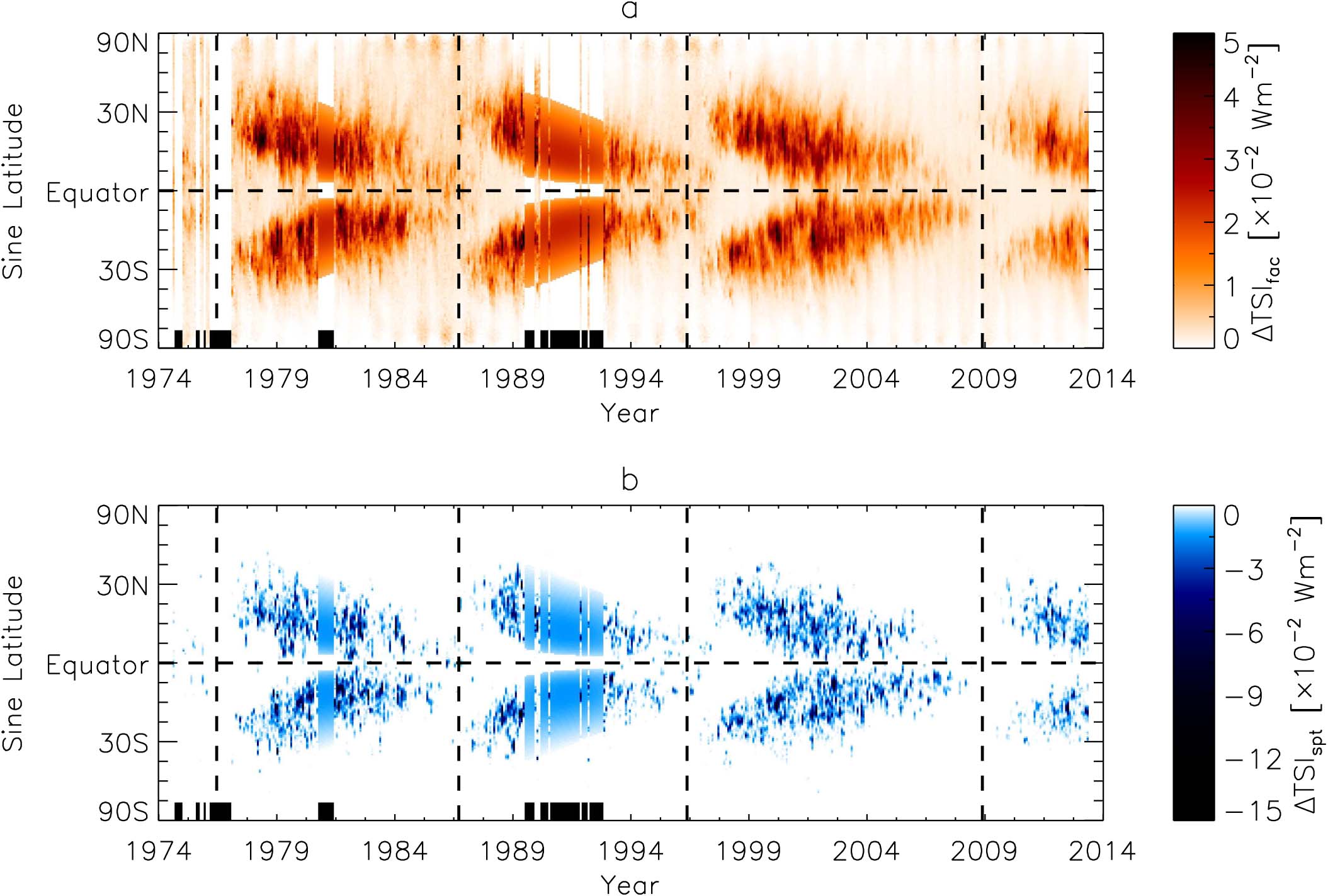}
\caption{Variation in reconstructed TSI with respect to the calculated quiet Sun level due to a) faculae ($\dtsifac$) and b) sunspots ($\dtsispt$), as a function of time and latitude (the monthly average in sine latitude intervals of 0.01). The black bars along the horizontal axes mark the months with no values from the lack of suitable magnetogram data. The gaps around the maxima of solar cycles 21 and 22 are filled by the interpolation. The vertical dashed lines denote solar cycle minima and the horizontal dashed lines the equator.}
\label{cpsbutterfly}
\end{cfig}

We histogrammed $\dtsifac$ and $\dtsispt$ from each data day by sine latitude (bin width of 0.01) and concatenated the monthly average histogram from the entire reconstruction, expressed in Fig. \ref{cpsbutterfly}. This is an update and extension of the similar figure by \citealt{wenzler05a}, generated with the SATIRE-S model using similar $\kpone$ and $\kptwo$ data.

Obviously, since $\dtsifac$ traces bright magnetic features (concentrated in active regions) and $\dtsispt$, sunspots and pores, the figure demonstrates Sp{\"o}rer's law, bearing resemblance to butterfly diagrams of sunspot area/position and magnetic flux \citep[e.g., Figs. 4 and 14 in][]{hathaway10}. A similar diagram to the $\dtsispt$ butterfly diagram (Fig. \ref{cpsbutterfly}b), based on the photometric sunspot index, was recently presented by \citealt{frohlich13} (Fig. 5 in their paper). The vertical stripes in the $\dtsifac$ butterfly diagram (Fig. \ref{cpsbutterfly}a) correspond to an annual modulation associated with Earth's inclined orbit about the Sun. The northern and southern hemispheres of the Sun, and therefore the photospheric magnetism present in each, come alternately into greater view. 

\begin{cfig}
\includegraphics[width=\textwidth]{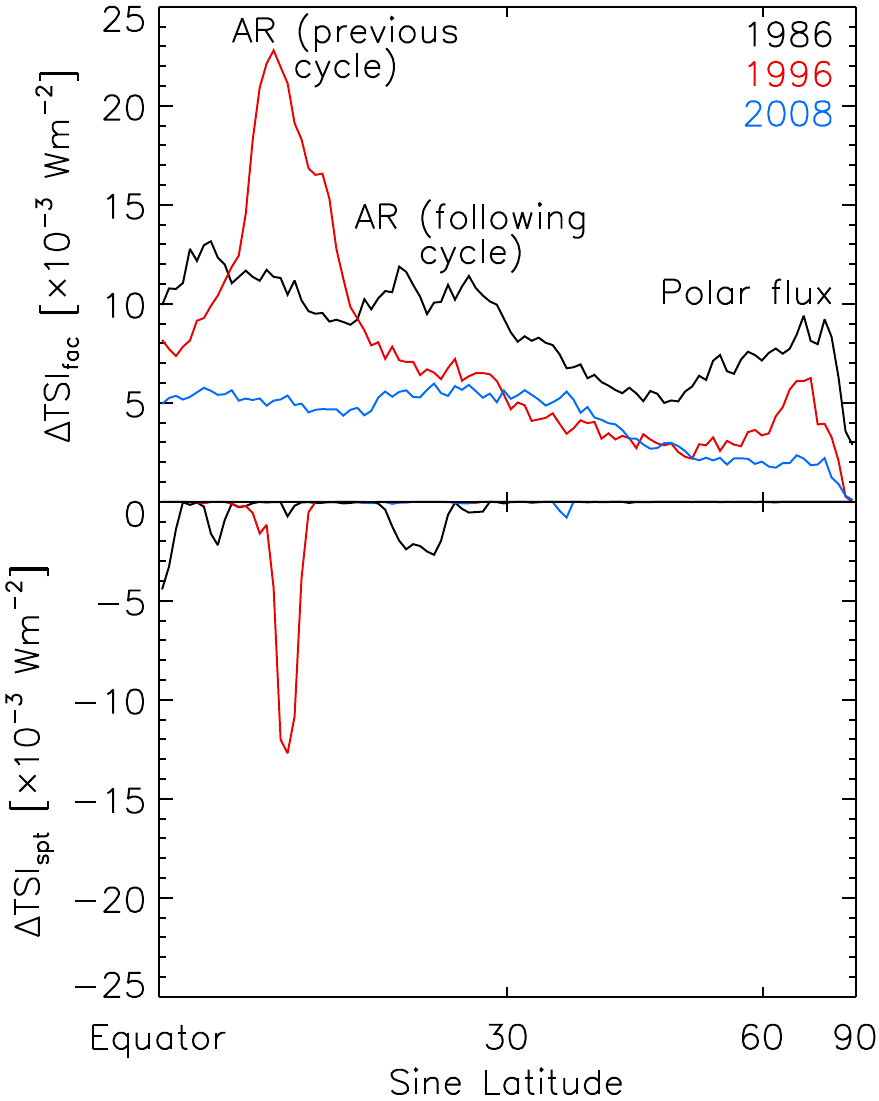}
\caption{Top: Mean variation in reconstructed TSI, with respect to the calculated quiet Sun level, from faculae ($\dtsifac$, top) and from sunspots ($\dtsispt$, bottom) around the solar cycle minima in 1986 (black), 1996 (red) and 2008 (blue), as a function of distance from the equator in sine latitude. AR is short for active regions.}
\label{secular}
\end{cfig}

We took the average of the daily $\dtsifac$ histograms over the three-month period centred on each of the last three solar cycle minima. Then, we summed the values from the southern and northern hemispheres, giving the mean intensity excess from faculae around each minimum as a function of distance from the equator (top panel, Fig. \ref{secular}).

Looking at the mean $\dtsifac$ histogram from the 1986 minimum (black series), there are three broad peaks. The maximum near the equator is from active regions associated with the preceding cycle, and that at middle latitude active regions associated with the succeeding cycle. The elevation towards high latitude corresponds to magnetic flux transported polewards over the previous cycle by meridional circulation. (What these peaks represent is visibly apparent in the $\dtsifac$ butterfly diagram, Fig. \ref{cpsbutterfly}a.) Comparing the mean $\dtsifac$ histogram from the three solar cycle minima, the secular trend in total $\dtsifac$ (the main component of the corresponding secular variation in reconstructed TSI) is influenced by the prevailing magnetic activity in the three latitude regions.

We repeated this analysis with the $\dtsispt$ histograms (bottom panel, Fig. \ref{secular}). The dips in the mean $\dtsispt$ histograms are located at similar latitudes as the low and middle latitude peaks in the the mean $\dtsifac$ histograms, associated with the same active regions. They are much weaker in terms of the magnitude and minimum-to-minimum variation, showing, as in Fig. \ref{facspt}, that the secular trend of the solar cycle minima level in the TSI reconstruction is primarily faculae-driven.

\section{Summary}
\label{p3summary}

In this paper, we present a SATIRE-S model reconstruction of daily total and spectral solar irradiance. The reconstruction, spanning the period of 1974 to 2013, is based on full-disc observations of continuum intensity and line-of-sight magnetic field from the KPVT, SoHO/MDI and SDO/HMI. Gaps in the time series, from the limited availability of suitable magnetogram data, were plugged by the regression of indices of solar activity to the reconstruction.

This work extends the preceding reconstruction based on similar observations from the KPVT and MDI by \cite{ball12,ball14}, which covered the period of 1974 to 2009. Apart from the extension to the present time with HMI data, we updated the reconstruction method.

The most significant change to the reconstruction method is the procedure by which the model input from the various data sets were combined such that they yield a consistent time series as the model output. The approach taken in previous studies tailored the parameters of the model to each data set (in such a manner that it does not introduce additional free parameters) and any residual discrepancy between the reconstructed spectra generated from the various data sets were accounted for by regression.

In this study, we compared the magnetogram signal and the computed faculae filling factor in the various data sets and brought them into agreement in the periods of overlap between them by regression. This allowed us to apply the same model parameters (apart from the umbra and penumbra intensity thresholds, which are wavelength dependent) to all the data sets. The updated procedure yielded a consistent TSI/SSI time series without the need for any additional correction of the reconstructed spectra.

The model has a single free parameter which modulates the contribution by bright magnetic features to variability in reconstructed solar irradiance. This free parameter is recovered by the comparison with TSI measurements. We considered the observations from four TSI radiometers in current operation, ACRIM3, DIARAD (both the IRMB and PMOD/WRC calibrations of the instrument output) and PMO6V on VIRGO, and TIM. Each of these TSI record resulted in a somewhat different value for the free parameter (and hence reconstructed solar irradiance time series). Based on the particularly close consistency between the PMO6V record and the TSI reconstruction generated using it as the reference to retrieve the free parameter, we adopted this candidate reconstruction for further study.

The agreement between the reconstruction and the PMO6V record, which extends 1996 to 2013, is excellent. The correlation coefficient, $R$ is 0.979 ($R^2=0.959$), implying that over solar cycles 23 and 24, at least $96\%$ of TSI variability is due to the time evolution of solar surface magnetism. The reconstruction also reproduces the secular decline between the 1996 and 2008 solar cycle minima in the PMO6V record. This counters the suggestion by \cite{frohlich09} that a dimming of the solar photosphere, rather than photospheric magnetism, might be responsible for the observed secular decline.

We also evaluated the reconstruction against the ACRIM, IRMB and PMOD composite records of TSI, the LASP Lyman-$\alpha$ composite, the Mg II index composites by IUP and by LASP, and SSI measurements from the UARS and SORCE missions.

As with previous efforts with SATIRE-S, we found the closest match with the PMOD composite ($R=0.957$ and $R^2=0.916$ over the span of the series, 1978 to 2013). The long-term trend in the PMOD composite is well reproduced in the reconstruction almost everywhere except between 1984 and 1994, where the reconstruction is broadly higher by 0.1 to $0.3\:\wms$. This discrepancy is, however, minute compared to the spread between TSI composites, or between the TSI radiometers in current operation. We attribute this apparent disparity to possible unresolved instrumental trends in the KPVT data set or in the PMOD composite, especially as the period in question encompasses the ACRIM gap.

The solar Lyman-$\alpha$ irradiance and Mg II index taken from the reconstruction exhibit excellent correlation with the LASP Lyman-$\alpha$ composite ($R=0.970$ and $R^2=0.942$) and the IUP Mg II index composite ($R=0.981$ and $R^2=0.963$), respectively. The secular trend in the solar cycle minima level in the LASP Lyman-$\alpha$ composite is also, to largely within uncertainty, reproduced. There are weak but palpable discrepancies, in terms of the amplitude of solar cycle variation, between the IUP and LASP Mg II index composites, and the reconstruction series. The implied long-term uncertainty of the two Mg II index composites could have contributed to the observation that proxy models of solar irradiance based on the Mg II index cannot replicate the secular decline between the 1996 and 2008 solar cycle minima in VIRGO TSI radiometry.

We contrasted the reconstruction against the ultraviolet (120 to 300 nm) solar irradiance measurements from the SOLSTICE and SUSIM experiments onboard UARS, and SOLSTICE and SIM onboard SORCE. As observations of ultraviolet solar irradiance exhibit divergent long-term trends, especially above 240 nm, the agreement between measurement and model varied. The reconstruction reproduces at least $89\%$ of the variability between 120 and 240 nm in SUSIM SSI ($R>0.94$ and $R^2>0.89$), and $95\%$ for 120 to 180 nm observations from the two SOLSTICE instruments ($R>0.97$ and $R^2>0.95$). It was less successful with measurements above 180 nm from the two SOLSTICE experiments and SIM SSI. At rotational timescales however, the reconstruction exhibits reasonable agreement with all the records examined, implying that the disparity between the model and some of these measurements at longer timescales arises from unresolved instrumental trends in the latter. The validity of the reconstruction in the ultraviolet is further supported by the excellent alignment with the SUSIM-based proxy model of ultraviolet solar irradiance by \citealt{morrill11} ($R>0.96$ and $R^2>0.92$). Importantly, the solar cycle variation between 240 and 400 nm in the SATIRE-S and the \cite{morrill11} reconstructions is similar (this spectral interval is crucial for climate models but the long-term trend is very poorly constrained by measurements and preceding models).

As stated in the introduction, SIM is the first and, at present, only instrument purpose-built to return radiometrically calibrated spectral measurements spanning the ultraviolet to the infrared at a regular basis. Also stated, the record indicates long-term trends that are difficult to reconcile with other observations and models of solar irradiance. The reconstruction reproduces the rotational variability in the SIM record but as with previous similar studies, failed to reproduce the long-term trends. We argue, from the lack of constancy in how SIM SSI relate to the solar cycle, and the disparity between the total flux recorded by the instrument and TSI (the wavelength range surveyed by SIM accounts for over $97\%$ of the energy in solar radiation), that unaccounted instrumental trends are present in the data set. In contrast, the reconstruction exhibits clear solar cycle variation at all wavelengths and the integrated flux over the wavelength range of SIM replicates at least $88\%$ of the variability in ACRIM3, TIM and VIRGO TSI radiometry ($R>0.94$ and $R^2>0.88$). The present quandary between SIM observations, and other measurements and models of solar irradiance emphasizes the need for continual monitoring of the solar spectrum over a wide spectral range.

The SATIRE-S model describes the variation in the solar spectrum from bright magnetic features (classed as `faculae'), and sunspots and pores (classed as `sunspots'). We examined the intensity excess from faculae, and the intensity deficit from sunspots, at the solar cycle minimum of 1986, 1996 and 2008 (there was insufficient data around the 1976 minima) as a function of latitude. This allowed us to visualize the contribution by low and middle latitude magnetic activity (mostly from active regions), and polar magnetic flux to the observed secular variation.

The results of this work significantly strengthen support for the hypothesis that variation in solar irradiance on timescales greater than a day is driven by photospheric magnetic activity \citep{foukal86,fligge00,solanki02,preminger02,krivova03}. The reconstruction is consistent with observations from multiple sources, confirming its reliability and utility for climate modelling. The SATIRE-S daily total and spectral solar irradiance time series is available for download at www.mps.mpg.de/projects/sun-climate/data.html.

\chapter[Solar cycle variation in solar irradiance \\ \textit{\footnotesize{(The contents of this chapter are identical to the submitted version of Yeo, K. L., Krivova, N. A., Solanki, S. K., 2014, Solar cycle variation in solar irradiance, Space Sci. Rev., online.)}}]{Solar cycle variation in solar irradiance}
\label{paper4}

\begin{flushright}
{\it Yeo, K. L., Krivova, N. A., Solanki, S. K.} \\
{\bf Space Sci. Rev., online (2014)\footnoteE{The contents of this chapter are identical to the submitted version of Yeo, K. L., Krivova, N. A., Solanki, S. K., 2014, Solar cycle variation in solar irradiance, Space Sci. Rev., online, reproduced with permission from Space Science Reviews, \textcopyright{} Springer Science+Business Media.}}
\end{flushright}

\section*{Abstract}

The correlation between solar irradiance and the 11-year solar activity cycle is evident in the body of measurements made from space, which extend over the past four decades. Models relating variation in solar irradiance to photospheric magnetism have made significant progress in explaining most of the apparent trends in these observations. There are however, persistent discrepancies between different measurements and models, in terms of the absolute radiometry, secular variation and the spectral dependence of the solar cycle variability. We present an overview of solar irradiance measurements and models, and discuss the key challenges in reconciling the divergence between the two.

\section{Introduction}

The 11-year solar activity cycle, the observational manifest of the solar dynamo, is apparent in indices of solar surface magnetism such as the sunspot area and number, 10.7 cm radio flux, and in the topic of this paper, solar irradiance. The observational and modelling aspects of the solar cycle are reviewed in \cite{hathaway10} and \cite{charbonneau10}, respectively. Solar irradiance is described in terms of what is referred to as the total and spectral solar irradiance, TSI and SSI. They are defined, respectively, as the aggregate and spectrally resolved solar radiative flux (i.e., power per unit area, and power per unit area and wavelength) above the Earth's atmosphere, normalized to one AU from the Sun. Factoring out the Earth's atmosphere and the variation in the Earth-Sun distance, TSI and SSI characterize the radiant behaviour of the Earth-facing hemisphere of the Sun.

Although the variation of the radiative output of the Sun with solar activity has long been suspected \citep{abbot23,smith75,eddy76}, it was not observed directly till satellite measurements, free from the effects of atmospheric seeing, became available. TSI and SSI (at least in the ultraviolet) have been monitored regularly from space through a succession of satellite missions, starting with Nimbus-7 in 1978 \citep{hickey80,willson88,frohlich06,deland08,kopp12}. A correlation between variations in TSI and the passage of active regions across the solar disc was soon apparent \citep{willson81,hudson82,oster82,foukal86}, leading to the development of models relating the variation in solar irradiance to the occurrence of bright and dark magnetic structures on the solar surface. While not the only mechanism mooted, models that ascribe variation in solar irradiance at timescales greater than a day to solar surface magnetism have been particularly successful in reproducing observations \citep{domingo09}. At timescales shorter than a day, excluded from the discussion here, intensity fluctuations from acoustic oscillations, convection and flares begin to dominate \citep{hudson88,woods06,seleznyov11}.

The measurement and modeling of the variation in solar irradiance over solar cycle timescales, a minute proportion of the overall level (about $0.1\%$ in the case of TSI), is a considerable achievement. Though significant progress has been made over the past four decades, considerable discrepancies remain between different measurements and models, chiefly in terms of the absolute radiometry, secular variation and the spectral dependence of the cyclic variability \citep[see the recent reviews by][]{ermolli13,solanki13}. In the following, we present a brief overview of the current state of solar irradiance observations (Sect. \ref{measurements}) and models (Sect. \ref{models}). We then discuss the key issues in reconciling measurements and models (Sect. \ref{discussionssr}), before giving a summary (Sect. \ref{summaryssr}).

\section{Measurements}
\label{measurements}

\subsection{Total solar irradiance, TSI}
\label{tsimeasurements}

\begin{cfig}
\includegraphics[width=\textwidth,trim=0cm .4cm 0cm 0cm,clip=true]{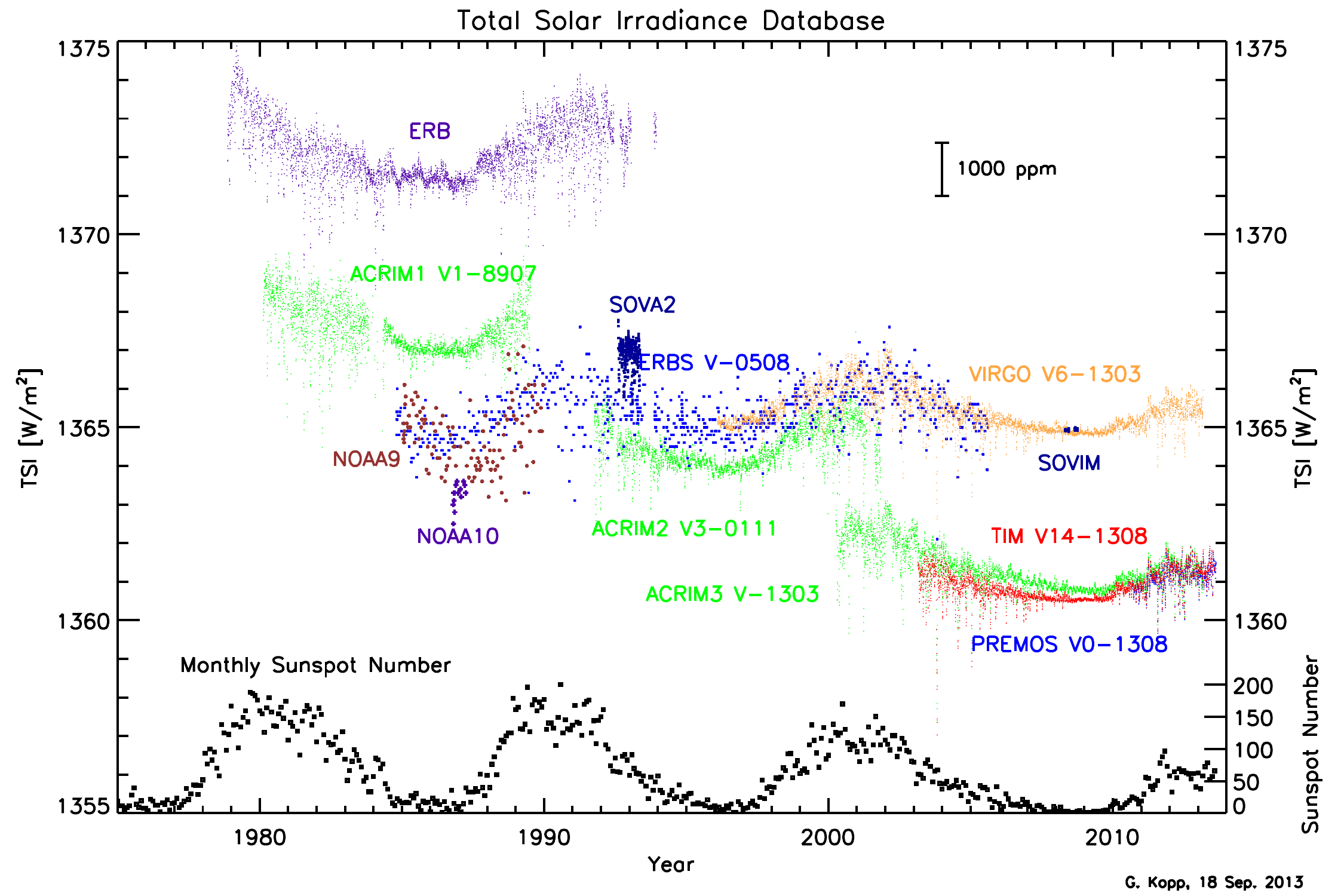}
\caption{The measurements from the succession of TSI radiometers sent into orbit since 1978 (colour coded) and the monthly mean of the sunspot number (lower right axis, black). Each TSI record is annotated by the name of the instrument name, and where applicable, the version number, and/or the year and month of the version. Courtesy of G. Kopp (http://spot.colorado.edu/~koppg/TSI/).}
\label{tsi}
\end{cfig}

The measurements from the succession of TSI radiometers sent into space since 1978, collectively representing a nearly uninterrupted record, exhibit clear solar cycle modulation. This is illustrated in Fig. \ref{tsi} by the comparison with the monthly sunspot number. All these instruments are based on active cavity electrical substitution radiometry \citep{butler08,frohlich10}. Succinctly, TSI is measured by allowing solar radiation into a heated absorptive cavity intermittently and adjusting the heating power as necessary to maintain thermal equilibrium. While these observations are sufficiently accurate to trace solar cycle variability (only about $0.1\%$ of the overall level), the measurements from the various instruments are offset from one another by a greater margin, reflecting the uncertainty in the absolute radiometry.

With the early instruments, specifically, Nimbus-7/ERB \citep{hickey80,hoyt92}, SMM/ACRIM1 \citep{willson79} and ERBS/ERBE\footnote{Nimbus-7/ERB denotes the Earth Radiation Budget instrument onboard Nimbus-7, SMM/ACRIM1 the Active Cavity Radiometer Irradiance Monitor onboard the Solar Maximum Mission, and ERBS/ERBE the Earth Radiation Budget Experiment onboard the similarly named satellite.} \citep{lee87}, the spread in absolute radiometry arose mainly from the uncertainty in the aperture area. As the determination of the aperture area improved, so the absolute radiometry from the succeeding missions converged. That is, up till the Total Irradiance Monitor, TIM \citep{kopp05a,kopp05b,kopp05c} onboard the SOlar Radiation and Climate Experiment, SORCE \citep[launched in 2003,][]{rottman05}.

The measurements from TIM were about $5\:\wms$ lower than the concurrent observations from ACRIMSAT/ACRIM3 \citep{willson03} and SoHO/VIRGO\footnote{ACRIMSAT/ACRIM3 refers to the ACRIM radiometer onboard the ACRIM SATellite, and SoHO/VIRGO the Variability of IRradiance and Gravity Oscillations experiment onboard the Solar and Heliospheric Observatory.} \citep{frohlich95,frohlich97}. Tests conducted at the TSI Radiometer Facility, TRF \citep{kopp07} with ground copies of ACRIM3, TIM and VIRGO revealed unaccounted stray light effects in ACRIM3 and VIRGO \citep{kopp11,kopp12,fehlmann12}. Correction for scattered light subsequently introduced to the ACRIM3 record based on these tests brought the absolute radiometry down to significantly closer agreement with TIM (within $0.5\:\wms$). The similar proximity between the measurements from the PREcision MOnitor Sensor, PREMOS onboard Picard \citep[launched in 2010,][]{schmutz09,fehlmann12} with TIM radiometry provided further evidence that the lower absolute level first registered by TIM is likely the more reliable. (PREMOS is the first TSI radiometer to be calibrated in vacuum at full solar power levels prior to flight, also at the TRF.)

\begin{cfig}
\includegraphics[width=\textwidth]{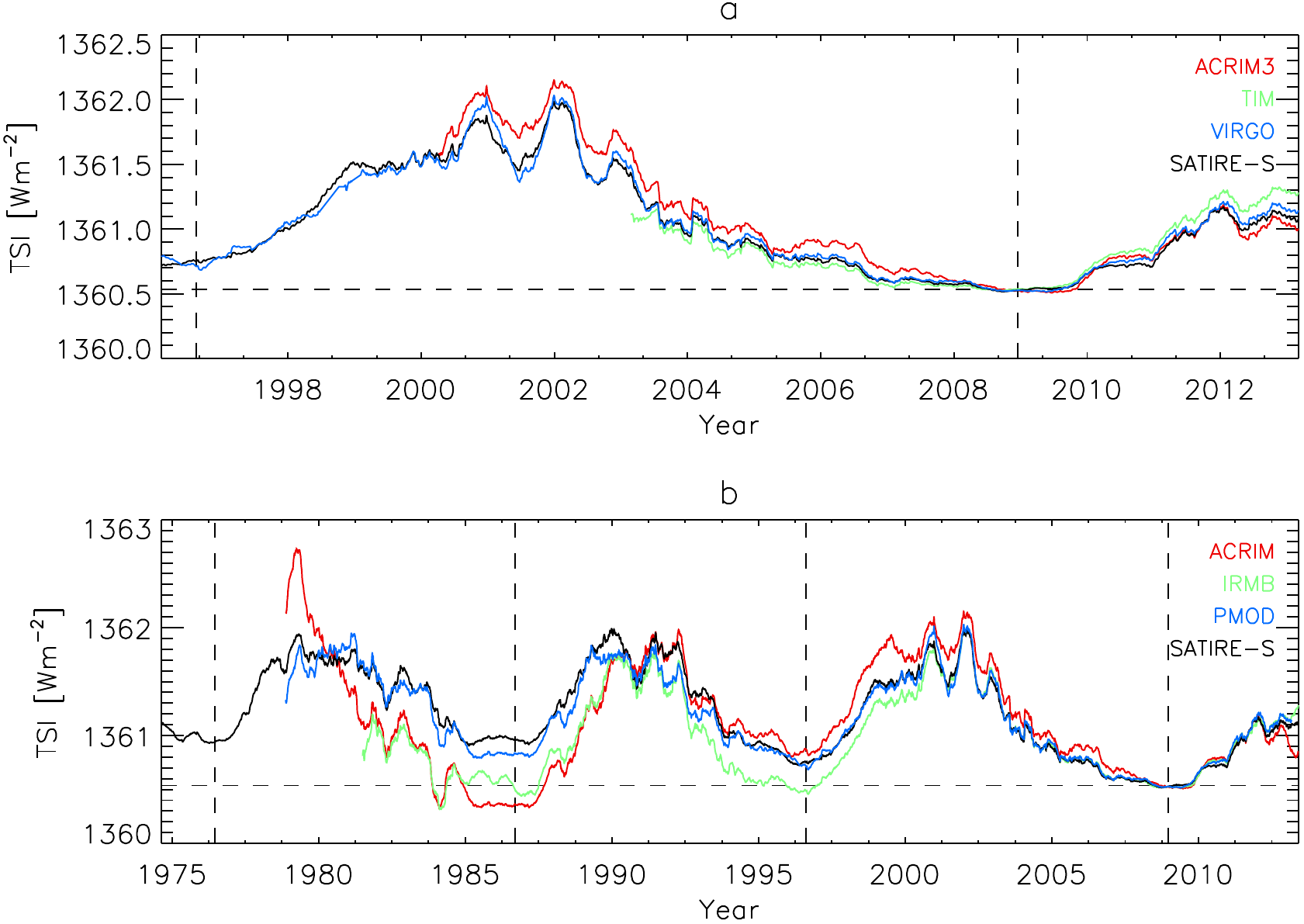}
\caption{a) TSI measurements from ACRIMSAT/ACRIM3 (version 11/13), SORCE/TIM (level 3, version 14) and SoHO/VIRGO (level 2, version $6\_002\_1302$). b) The ACRIM (version 11/13), IRMB (version dated December 19, 2013, provided by S. Dewitte) and PMOD (version ${\rm d}41\_62\_1302$) composite records of TSI. Also plotted is the SATIRE-S reconstruction of TSI. The vertical dashed lines mark the position of solar cycle minima. All the time series were normalized to TIM at the 2008 solar cycle minimum (horizontal dashed line) and smoothed with a 181-day boxcar filter.}
\label{comparetsissr}
\end{cfig}

Due to the limited lifetime of TSI radiometers, there is no single mission that covered the entire period of observation, and apart from VIRGO (1996 to present), there is no other record that extends more than a complete solar cycle minimum-to-minimum. The composition of the observations from the various missions into a single time series, obviously essential, is non-trivial due to ageing/exposure degradation, calibration uncertainty and other instrumental issues. The ACRIM, PREMOS, TIM and VIRGO instruments are designed with redundant cavities to allow in-flight degradation tracking. Even with this capability and the best efforts of the respective instrument teams, significant uncertainty persists over the long-term stability \citep[conservatively, $0.2\:\wms$ or 20 ppm per year,][]{solanki13}, reflected in the discrepant amplitude of solar cycle variation in the ACRIM3, TIM and VIRGO records (Fig. \ref{comparetsissr}a). Accounting for changes in instrument sensitivity, which are often particularly severe during early operation and can see discrete shifts (such as that suffered by ERB and VIRGO), has proven to be particularly precarious \citep{hoyt92,lee95,dewitte04a,frohlich06}. It is worth mentioning here that the observations from the various radiometers do largely agree at solar rotational timescales (roughly one month), where apparent variability is much less affected by the instrumental influences discussed above.

There are, at present, three composite records of TSI, published by the ACRIM science team \citep{willson03}, IRMB \citep{dewitte04b,mekaoui08} and PMOD/WRC\footnote{IRMB is short for L'Institut Royal M\'{e}t\'{e}orologique de Belgique, and PMOD/WRC Physikalisch-Meteorologisches Observatorium Davos/World Radiation Center. The composite by IRMB is also variously referred to as the RMIB or SARR composite.} \citep{frohlich00,frohlich06}. These competing time series, while broadly similar, do differ in terms of the amplitude of solar cycle variation \citep[discussed in detail by][]{frohlich06,frohlich12}, most readily apparent in the conflicting secular trend of the solar cycle minima level (Fig. \ref{comparetsissr}b). The TSI reconstructions presented by \cite{wenzler09,ball12,yeo14b}, based on the SATIRE-S model \citep{fligge00,krivova03,krivova11a}, have found the greatest success in replicating the solar cycle variation in the PMOD composite (see Sect. \ref{modelsatires} for more details).

\subsection{Spectral solar irradiance, SSI}
\label{ssimeasurements}

\begin{cfig}
\includegraphics[width=\textwidth,natwidth=3408,natheight=1388]{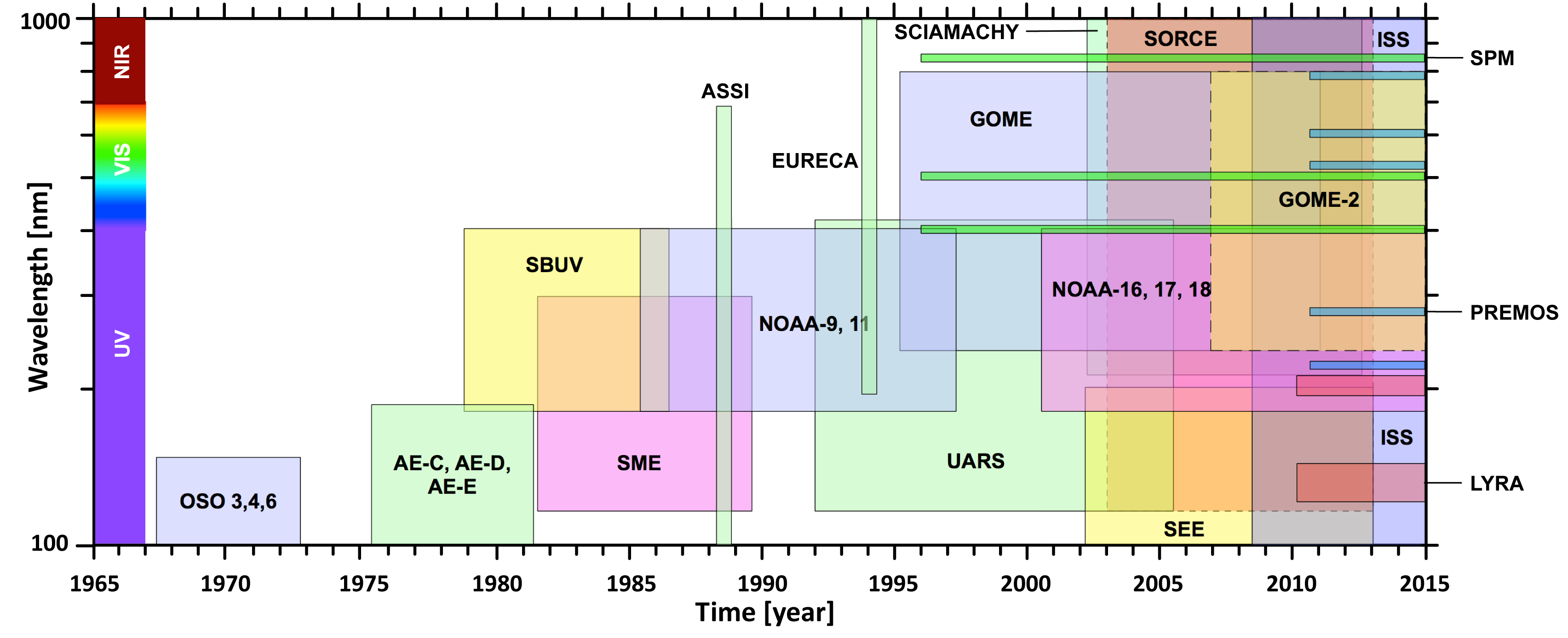}
\caption{Timeline and spectral range of the space missions making observations of the solar spectrum (above 100 nm). Taken from \cite{ermolli13}.}
\label{f03}
\end{cfig}

The solar spectrum has been probed through a miscellany of spaceborne instruments over the past five decades (Fig. \ref{f03}), differing in the regularity of measurements and the spectral range surveyed. As with TSI, ultraviolet solar irradiance (120 to 400 nm) has been monitored, almost without interruption, from space since 1978.

\begin{cfig}
\includegraphics[width=\textwidth]{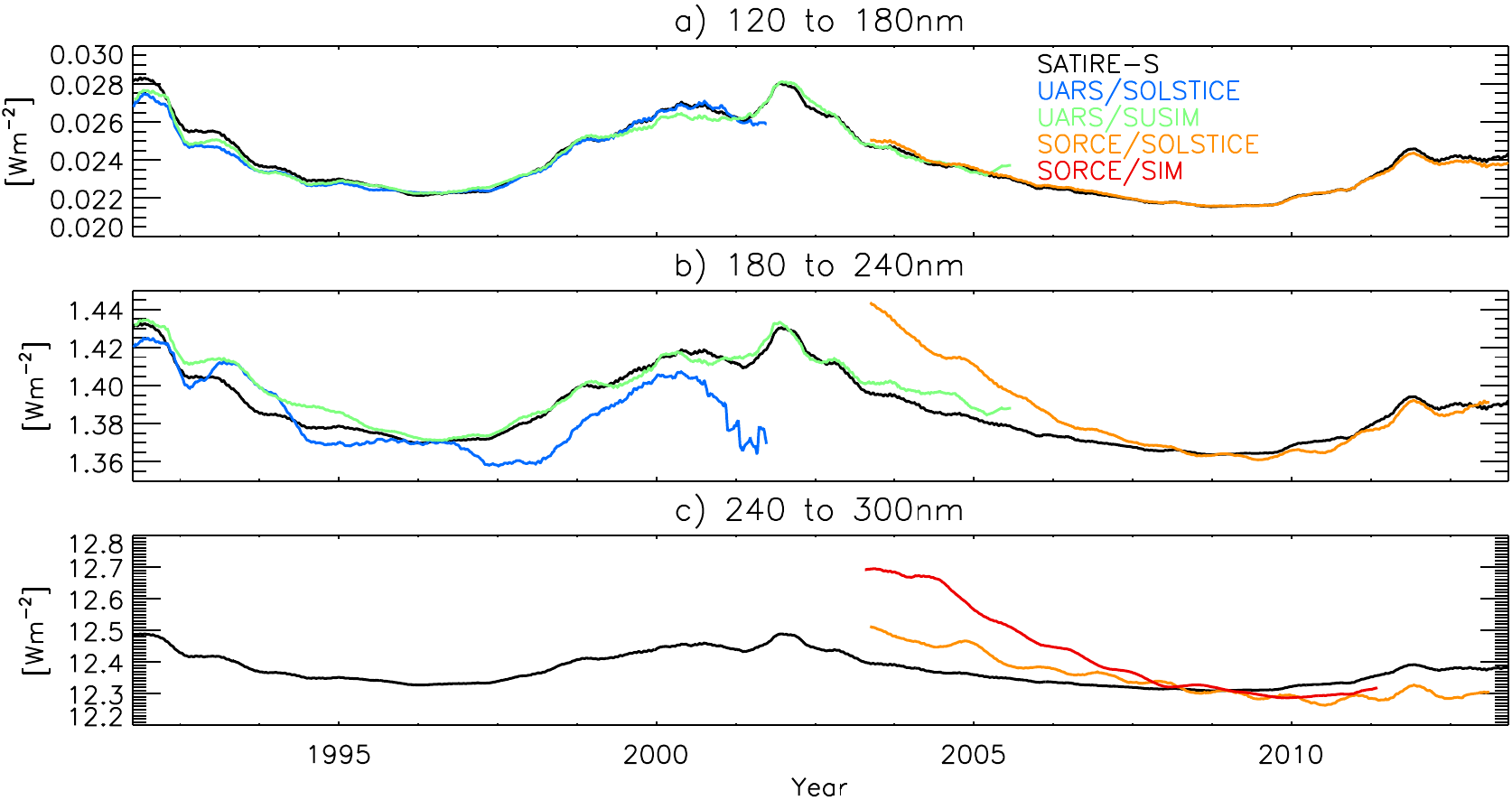}
\caption{Integrated solar irradiance between a) 120 and 180 nm, b) 180 and 240 nm (centre), and c) 240 and 300 nm (bottom), in the SATIRE-S reconstruction of SSI (black). Also drawn are the measurements from the UARS and SORCE missions, offset to SATIRE-S at the 1996 and 2008 solar cycle minima, respectively. All the time series were smoothed by taking the 181-day moving average. Adapted from \cite{yeo14b}.}
\label{satireuarssorcessr}
\end{cfig}

The key features of the body of ultraviolet solar irradiance measurements is similar to that of TSI, discussed in Sect. \ref{tsimeasurements}. The observations from the various spectrometers display similar rotational variability but diverge in terms of the absolute radiometry and the amplitude of solar cycle variation, especially at wavelengths above 240 nm \citep[see][and Fig. \ref{satireuarssorcessr}]{deland12,unruh12,ermolli13,yeo14b}. In particular, the measurements from the SIM \citep{harder05b,harder05a} and SOLSTICE \citep{mcclintock05,snow05a} experiments onboard SORCE exhibit much stronger solar cycle variation than registered by the SOLSTICE \citep{rottman01} and SUSIM \citep{brueckner93,floyd03} instruments onboard the precessor mission, UARS\footnote{SIM denotes the Spectral Irradiance Monitor, SOLSTICE the SOLar STellar Irradiance Comparison Experiment, SUSIM the Solar Ultraviolet Spectral Irradiance Monitor and UARS the Upper Atmosphere Research Satellite.} \citep[by a factor of three to ten, depending on wavelength, see Fig. 7 in][]{deland12}. This disparity, while broadly within the uncertainty in the long-term stability of said instruments, is much greater than encountered between pre-SORCE instruments \citep{deland12,ermolli13}. The uncertainty in the long-term stability of ultraviolet solar irradiance observations, on the order of 0.1 to $1\%$ per year \citep[varying with wavelength and between instruments,][]{snow05a,merkel11,deland12}, is of similar magnitude as the variation over the solar cycle. It is also grossly greater than the uncertainty afflicting TSI measurements. Again, due to the limited lifetime of spaceborne instrumentation, there is no record that extends beyond a complete solar cycle minimum-to-minimum, with the exception of the observations from NOAA-9 SBUV/2\footnote{The second generation Solar Backscatter UltraViolet spectrometer onboard the ninth National Oceanic and Atmospheric Administration satellite.} \citep{deland98}.

The solar cycle modulation in, and disparity between the pre-SORCE measurements is illustrated in the time series plot by \citealt{deland08} (Fig. 2 in their paper), which is qualitatively analogous to Fig. \ref{tsi}. The authors presented the first published effort to compose the ultraviolet solar irradiance observations from these instruments into a single time series. The result, spanning the period of 1978 to 2005, still contains overt instrumental trends for which the appropriate correction is not known. The challenge in the account of instrumental influences is substantially greater than with TSI, exacerbated by the wavelength dependence of instrumental effects, and differences in the design, operation and calibration.

The series of GOME instruments \citep[the first of which was onboard ERS-2, launched in 1996,][]{weber98,munro06} and ENVISAT/SCIAMACHY\footnote{GOME is short for Global Ozone Monitoring Experiment. ERS-2 refers to the second European Remote Sensing satellite, and ENVISAT/SCIAMACHY the SCanning Imaging Absorption spectroMeter for Atmospheric CHartographY onboard the ENVIronmental SATellite.} \citep[launched in 2002,][]{skupin05a} made regular measurements of the solar spectrum in the 240 to 790 nm and 240 to 2380 nm wavelength range, respectively. However, these instruments, purpose built for atmospheric sounding rather than solar irradiance monitoring, lack the capability for in-flight degradation tracking. This renders the observations unsuitable for tracing the solar cycle variation of the solar spectrum. The long-term stability of the narrowband (FWHM of 5 nm) photometry at 402, 500 and 862 nm from the Sun PhotoMeter, SPM on the SoHO/VIRGO experiment \citep{frohlich95,frohlich97} is similarly problematic \citep[though considerable progress has been made, see][]{wehrli13}.

\begin{cfig}
\includegraphics[width=\textwidth]{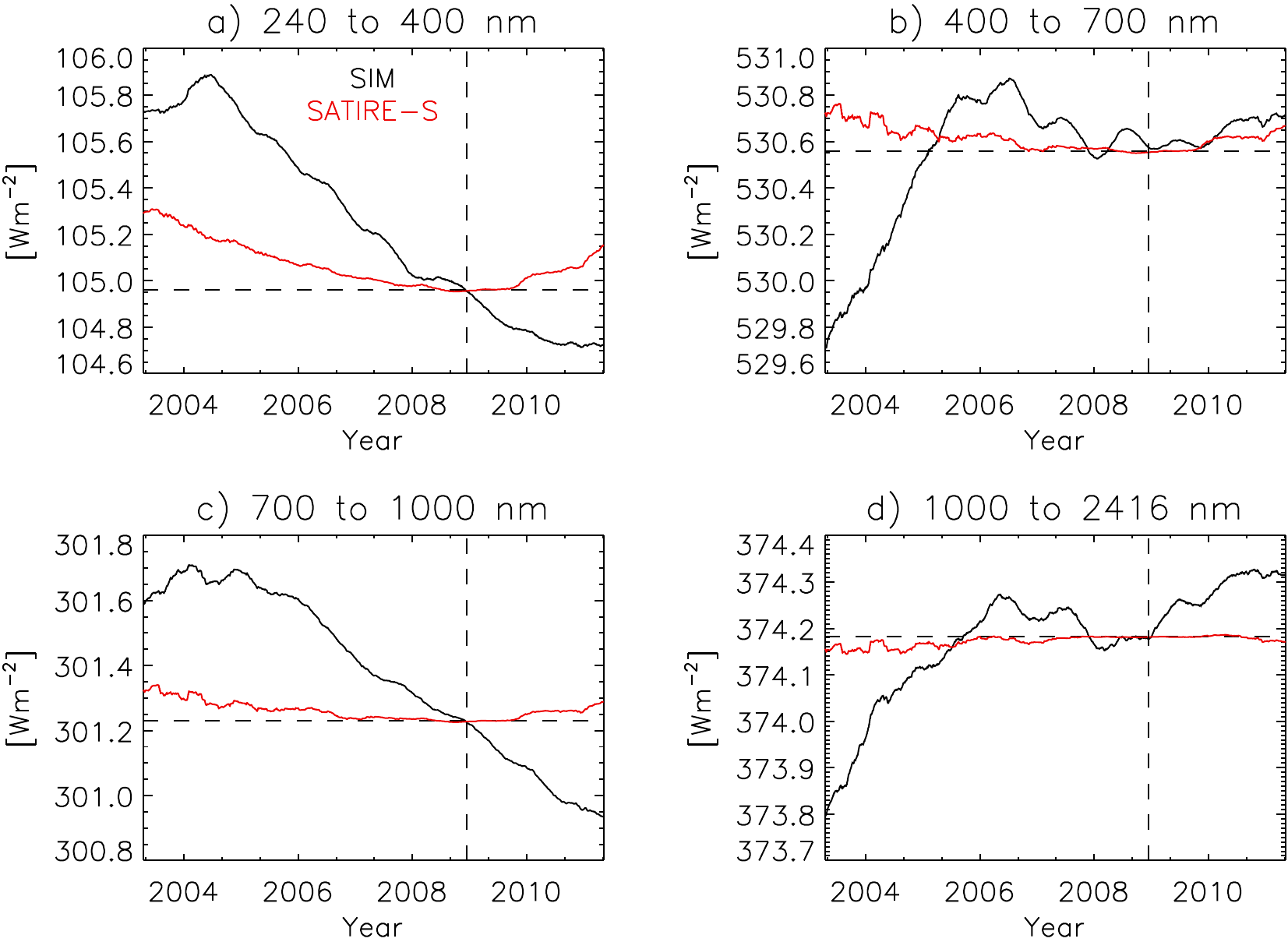}
\caption{181-day moving average of the integrated solar irradiance over the spectral intervals specified above each panel, in the SORCE/SIM record (red) and the SATIRE-S reconstruction (black). The SATIRE-S time series were offset to the corresponding SIM time series at the 2008 solar cycle minimum, the position and level at which is indicated by the dashed lines. Adapted from \cite{yeo14b}.}
\label{comparesatsimssr}
\end{cfig}

\begin{cfig}
\includegraphics[width=\textwidth]{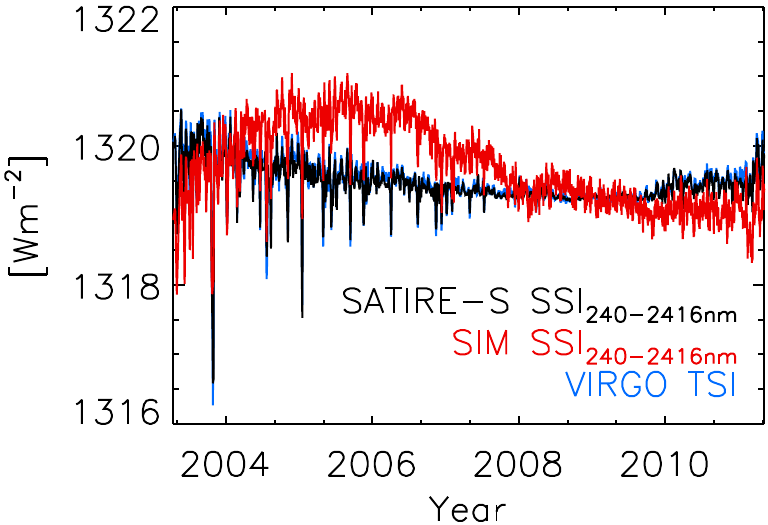}
\caption{Total flux registered by SIM (red). Also drawn is the integrated flux in the SATIRE-S reconstruction over a similar wavelength range (black) and VIRGO TSI (blue), both offset to match the level of the SIM time series at the 2008 solar cycle minimum. The VIRGO time series is largely hidden by the SATIRE-S time series due to the close similarity. Adapted from \cite{yeo14b}.}
\label{comparesatsimssr2}
\end{cfig}

Regular monitoring of the solar spectrum beyond the ultraviolet only started, in effect, with SIM. This instrument provides the only radiometrically calibrated, continuous/extended (in time) record of the solar spectrum spanning the ultraviolet to infrared (the latest release, version 19, dated November 19, 2013, covers 240 to 2416 nm) presently available, illustrated in Figs. \ref{comparesatsimssr} and \ref{comparesatsimssr2}.

\begin{cfig}
\includegraphics[width=\textwidth]{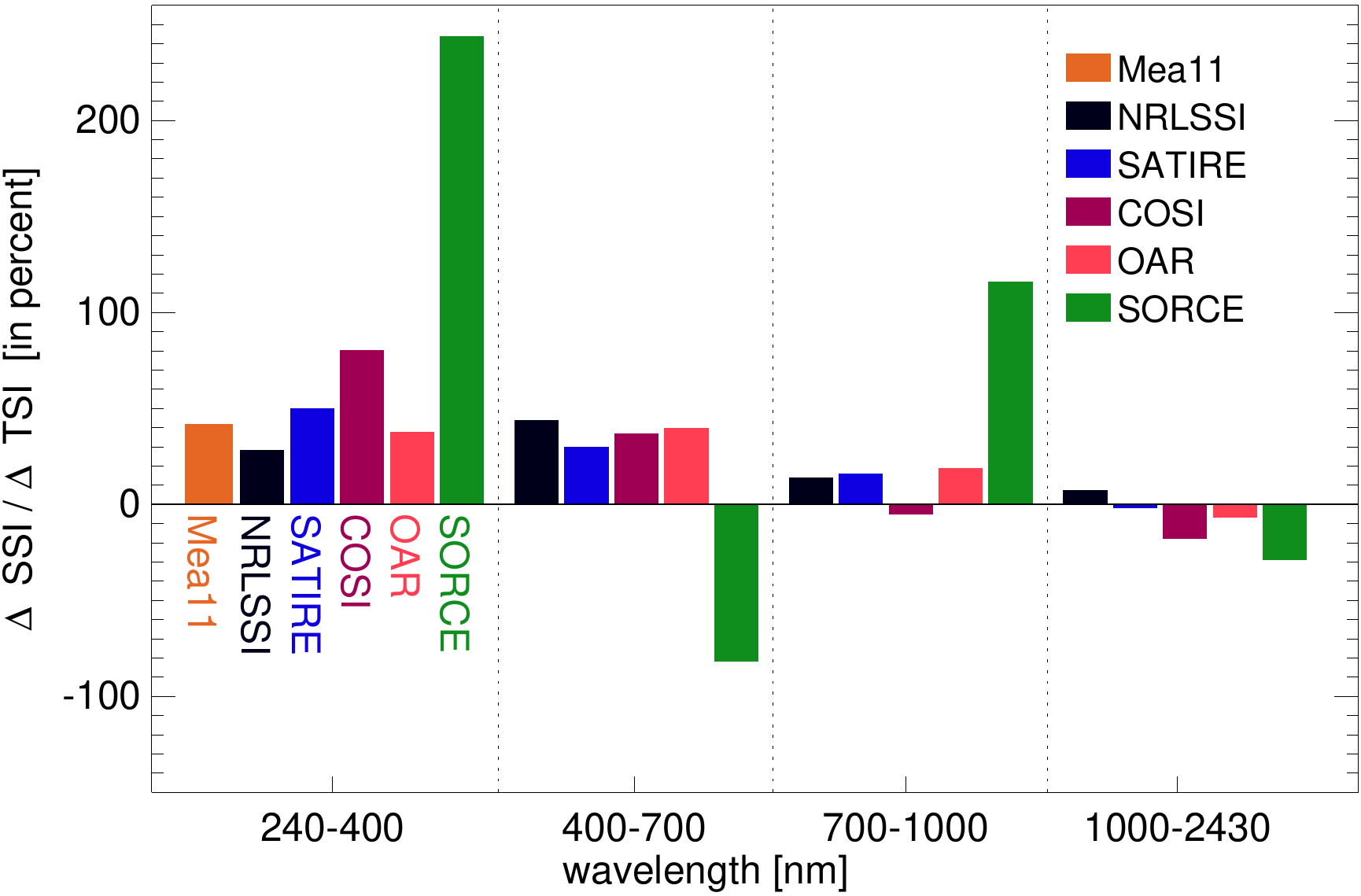}
\caption{Ratio of the integrated variation in the same spectral intervals examined in Fig. \ref{comparesatsimssr}, ${\rm \Delta{}SSI}$ and the corresponding variation in TSI, ${\rm \Delta{}TSI}$ between solar cycle maximum and minimum (2002 and 2008) in the models discussed in Sect. \ref{models3}. For the model by \cite{morrill11}, denoted Mea11, we took ${\rm \Delta{}TSI}$ from the SATIRE-S model. Also depicted is the same for SORCE measurements between 2004 and 2008, with ${\rm \Delta{}SSI}$ and ${\rm \Delta{}TSI}$ from the SIM and TIM records, respectively. Adapted from \cite{ermolli13}.}
\label{barchart}
\end{cfig}

Between 2004 and 2008 (which is within the declining phase of solar cycle 23), SIM registered a decline in ultraviolet flux (integrated solar irradiance between 240 and 400nm, Fig. \ref{comparesatsimssr}a) that is, in absolute terms, almost double the drop in TSI over the same period (Fig. \ref{barchart}). Up to 2006, this pronounced downward trend was accompanied by a comparable increase in the visible (400 to 700 nm, Fig. \ref{comparesatsimssr}b), in apparent anti-phase with the solar cycle. The decline in the ultraviolet is multiple times greater than projections from other measurements (i.e., pre-SORCE ultraviolet solar irradiance) and models of solar irradiance, summarized in Fig. \ref{barchart} \citep{harder09,ball11,deland12,lean12,unruh12,ermolli13,yeo14b}. (We exclude, for now, the solar irradiance reconstruction presented by \citealt{fontenla11}, where steps were taken to bring the model into qualitative agreement with SIM SSI, to be discussed in Sect. \ref{modelsrpm}.) The increase in the visible prior to 2006 is in conflict with VIRGO SPM photometry (see next paragraph) and present-day models, all of which see visible solar irradiance varying in phase with the solar cycle. The variation in the infrared (700 to 2416 nm, Figs. \ref{comparesatsimssr}c and \ref{comparesatsimssr}d) over this period is also significantly stronger than in model reconstructions. The investigations of \cite{ball11,deland12,lean12,unruh12,yeo14b} did, however, note similar rotational variability in SIM SSI and in these other measurements and models.

Looking at the entire SIM record (the current version extends from 2003 to 2011), there is no constancy in how the overall trend in SIM SSI (at a given wavelength) relate to the solar cycle, neither in phase nor in anti-phase (Fig. \ref{comparesatsimssr}). Apart from the solar cycle modulation evident in pre-SORCE ultraviolet solar irradiance measurements, this also runs counter to the positive correlation between SPM visible (500 nm) photometry and TSI over the similar period of 2002 to 2012 reported by \cite{wehrli13}. There being no other extended record of SSI covering a similar spectral range, we cannot rule out completely that segments of the solar spectrum may vary in a non-cyclic manner as apparent in SIM SSI but it is almost irrefutable that the integral of the solar spectrum over all wavelengths, TSI, does exhibit variation that is clearly in phase with the solar cycle. The spectral range surveyed by SIM accounts for more than $97\%$ of the power in solar radiation; the total flux recorded by SIM should closely replicate the variability in TSI, but that is evidently not the case (Fig. \ref{comparesatsimssr2}). In contrast, the reconstruction of the integrated solar irradiance over the spectral range of SIM from SSI models replicates most of the variability in measured TSI \citep[see][and Fig. \ref{comparesatsimssr2}]{ball11,lean12,yeo14b}.

The discrepancies between SIM SSI, and other measurements and models, discussed above, were taken in various studies to indicate that there are unaccounted instrumental effects in the SIM record \citep{ball11,deland12,lean12,unruh12,yeo14b}. This is favoured over alternative interpretations such as the apparent trends between 2004 and 2008 implying a change in the physics of the Sun during this period compared to earlier times, or that there are gaping insufficiencies in our understanding/modeling of the physical processes driving variations in solar irradiance. As we will argue in Sect. \ref{models1}, the decline in visible flux between 2003 and 2006 (Fig. \ref{comparesatsimssr}a) is not consistent with our current understanding of solar surface magnetism and its effect on solar irradiance.

Evidently, considerable uncertainty afflicts the direct observation of the variation in SSI over solar cycle timescales. The situation is set to improve, with the continuing efforts to calibrate SCIAMACHY and SORCE spectrometry, and spectral measurements expected from ISS/SOLSPEC \citep[intermittent in time but radiometrically calibrated,][]{thuillier09} and the upcoming JPSS\footnote{ISS/SOLSPEC denotes the SOLar SPECtrum experiment onboard the International Space Station, and JPSS the Joint Polar Satellite System.} mission (set to be launched in 2016), which will carry an improved version of the SIM instrument \citep{richard11}. For more extensive reviews of the measurement of solar irradiance, we refer the reader to \cite{deland08,deland12,domingo09,kopp12,frohlich12,ermolli13,solanki13}.

\section{Models}
\label{models}

The collection of satellite measurements of solar irradiance, while evidently core to our understanding of the solar cycle variation in the radiative output of the Sun, cover a limited period in time and suffer significant uncertainty. Models of solar irradiance serve both to complement these observations and to advance our understanding of the physical processes driving the apparent variability.

\subsection{Solar surface magnetism}
\label{models1}

\begin{cfig}
\includegraphics[width=.96\textwidth]{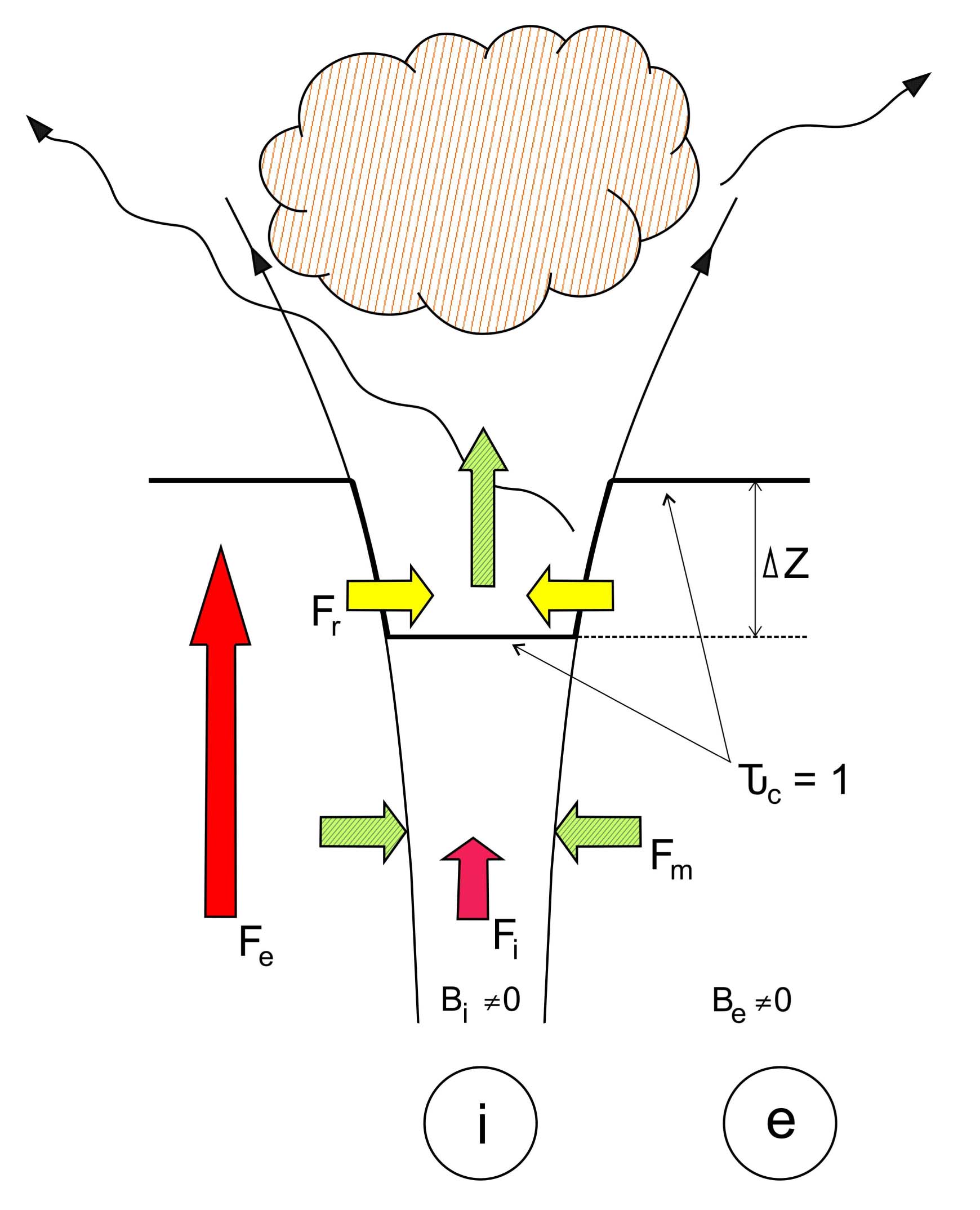}
\caption{Schematic of a thin magnetic flux tube. The thickened line follow the optical depth unity ($\tau_{c}=1$) surface, depressed by $\Delta{}Z$ within the flux tube. The red arrows represent convective and radiative energy transport in the interior and exterior of the flux tube (denoted by the i and e subscripts, respectively). The yellow arrows correspond to radiative heating through the side walls of the depression, and the gree arrows mechanical energy transfers. The cloud symbolizes the heated chromospheric layers of the magnetic structure. Reproduced from \cite{solanki13}, which followed \cite{zwaan78}.}
\label{fluxtubemodel}
\end{cfig}

A feature of the 11-year solar activity cycle is the cyclical emergence and evolution of kilogauss strength magnetic concentrations in the photosphere \citep{solanki93b}, the main properties of which can be explained by describing them as magnetic flux tubes \citep[see Fig. \ref{fluxtubemodel} and][]{spruit76,spruit83,solanki93a}. They range in physical extent, from on the order of $10^{1}$ to $10^{5}$ km in cross section. The lower end corresponds to the small-scale magnetic elements which make up active region faculae, quiet Sun network and internetwork \citep{lagg10,riethmuller13}, and the upper end, sunspots and pores.

Solar irradiance is modulated by photospheric magnetic activity from its effect on the thermal/radiant property of the solar surface and atmosphere. The influence of magnetic concentrations on the local temperature structure of the solar surface and atmosphere varies strongly with the size of the magnetic feature, as described below.

\begin{cfig}
\includegraphics[width=\textwidth]{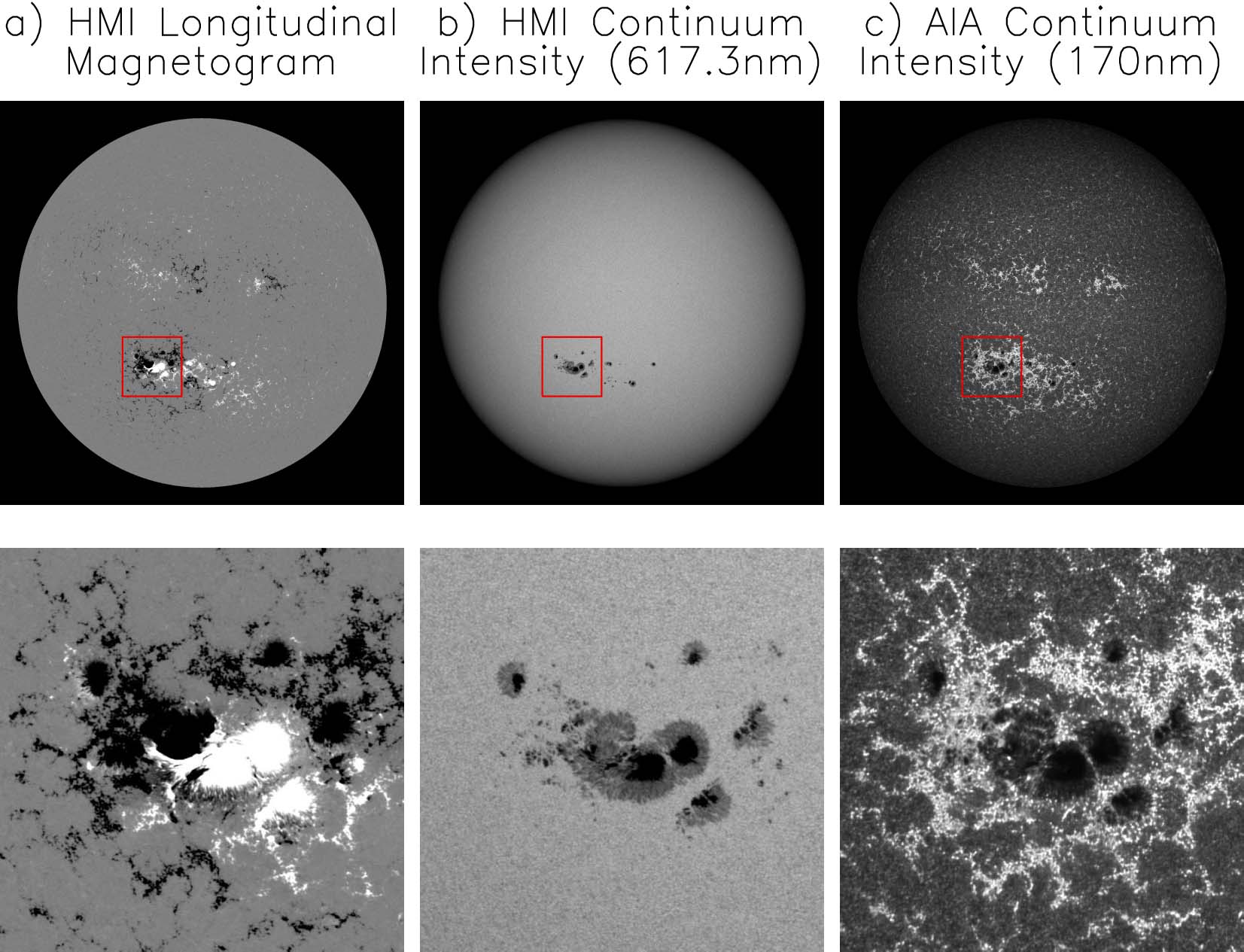}
\caption{Top: Concurrent observations, from July 11, 2012, of the a) line-of-sight magnetic field, and the continuum intensity at b) 617.3 nm and c) 170 nm from the Helioseismic and Magnetic Imager, HMI \citep{schou12} and the Atmospheric Imaging Assembly, AIA \citep{lemen12} onboard the Solar Dynamics Observatory. Bottom: Blow up of the boxed region, featuring active region NOAA 11520. The grey scale is saturated at $\pm30\:{\rm G}$, and at about $60\%$ and $120\%$, and $20\%$ and $300\%$ of the mean quiet Sun level at disc centre, respectively.}
\label{cma17}
\end{cfig}

As a consequence of pressure balance, the interior of magnetic concentrations is evacuated. The lower density creates a depression in the optical depth unity surface and magnetic buoyancy (the result of which is flux tubes are largely vertical). The intensity contrast in the continuum is influenced by the competing effects of magnetic suppression of convection and radiative heating from surrounding granulation through the side walls of the depression. Sunspots and pores are dark from the magnetic suppression of convection within these features \citep[see the reviews by][and Fig. \ref{cma17}]{solanki03,rempel11}. For small-scale magnetic concentrations, this is overcome by the lateral heating, rendering them bright \citep{spruit76,spruit81,grossmann94}, especially away from the disc centre as the side walls come into greater view \citep{spruit76,keller04,carlsson04,steiner05}. Heated by mechanical and resistive dissipations \citep{musielak03,moll12}, and radiation from deeper layers \citep{knoelker91}, the temperature gradient within small-scale magnetic elements is steeper, so enhancing their intensity within spectral lines and in the ultraviolet, formed at greater heights than the visible continuum \citep[see][and Fig. \ref{cma17}c]{frazier71,mitchell91,morrill01,ermolli07,yeo13,rietmuller10}.

Models describing the variation in solar irradiance, at timescales greater than a day, by the intensity deficit and excess facilitated by photospheric magnetism have achieved substantial success in reproducing measured solar irradiance \citep[see][and Sects. \ref{models2} and \ref{models3}]{domingo09}. While other plausible mechanisms have been proposed \citep{wolff87,kuhn88,cossette13}, related to physical processes in the solar interior, direct evidence is not straightforward to obtain and consequently still largely lacking.

\begin{cfig}
\includegraphics[width=\textwidth]{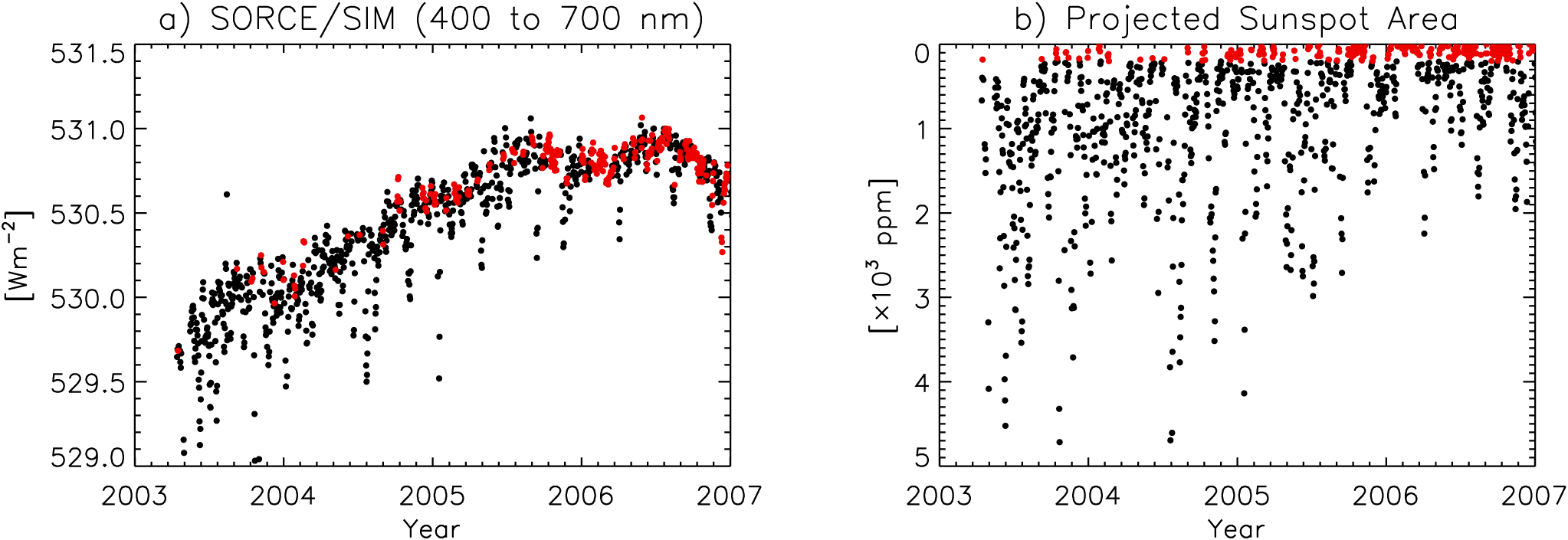}
\caption{a) Integrated solar irradiance in the visible in SORCE/SIM spectroscopy, and b) the concurrent projected sunspot area \citep[from the composite record by][version 0613]{balmaceda09}. Points corresponding to days where the projected sunspot area is less than 200 ppm are highlighted in red. The vertical axis of the projected sunspot area plot is inverted to aid interpretation.}
\label{comparesimsa}
\end{cfig}

The increase in visible solar irradiance registered by SIM between 2003 and 2006 (Fig. \ref{comparesatsimssr}b, discussed in Sect. \ref{ssimeasurements}) came at a time solar activity is declining. For the solar cycle variation in photospheric magnetism to be compatible with this trend in visible solar irradiance, small-scale magnetic concentrations would have to be dark in the visible. The intensity deficit from sunspots and pores, while declining over this period, is not driving the upward trend in SIM visible solar irradiance. This is demonstrated in Fig. \ref{comparesimsa} by the variation of SIM visible flux over the days with minimal sunspot activity.

Indeed, studies examining the intensity contrast of small-scale magnetic concentrations in the visible continuum as a function of magnetogram signal have noted negative contrasts near disc centre at both low and high magnetogram signal levels \citep[summarized in Table 2 of][]{yeo13}. However, the results of recent investigations suggest that these apparent negative contrasts are related to the fact that small-scale magnetic elements congregate mainly within dark intergranular lanes \citep{schnerr11,kobel11}, and observational effects related to the limited spatial resolution and telescope diffraction \citep{rohrbein11}, than any indication that small-scale magnetic concentrations are, near disc centre, dark in the visible continuum. Taken together with the intensity enhancement towards the limb and within spectral lines \citep{topka97,ortiz02,yeo13}, it is highly unlikely that small-scale magnetic concentrations (at least overall) might be dark in the visible. The upward trend in early SIM visible solar irradiance is not consistent with our current understanding of solar surface magnetism and its effect on solar irradiance.

\subsection{Model architectures}
\label{models2}

There are two broad categories of solar irradiance models, commonly referred to as proxy and semi-empirical models, distinguished by the modelling approach.

As stated in the introduction, the measurement of solar irradiance from space quickly revealed an apparent correlation between TSI and the passage of active regions across the solar disc. This was followed by the development of proxy models, aimed at reconstructing solar irradiance by the multivariate regression of indices of solar activity to measured solar irradiance. The index data serve as proxies of the effects of bright and dark magnetic structures on the radiative output of the Sun.

Sunspot darkening is usually represented by sunspot area or the PSI \citep[photometric sunspot index,][]{hudson82,frohlich94}, and facular brightening by chromospheric indices such as the Ca II K \citep{keil98}, Mg II \citep{heath86} or F10.7 \citep[10.7 cm radio flux,][]{tapping87} indices. (In this context the term sunspots includes pores, and faculae, quiet Sun network.) The Ca II K and Mg II indices are given by the ratio of the disc-integrated flux in the line core (of the Ca II K line, and the Mg II h and k doublet, respectively) to that at nearby reference wavelengths. The line core to `continuum' ratio is preferred over absolute flux as it is more robust to instrument degradation.

The TSI reconstructions by the group at the San Fernando Observatory, SFO \citep{chapman96,chapman12,chapman13,preminger02} employ sunspot and faculae indices derived from full-disc photometric images obtained at the observatory. These models proved to be particularly successful among proxy models. By employing full-disc imagery instead of Sun-as-a-star measures such as the indices listed above, they include the centre-to-limb variation of sunspot and faculae contrast, albeit only at the photometric bandpass. The latest iteration, based on visible red and Ca II K observations, reproduced most of the variation in TIM radiometry \citep[$R^2=0.95$,][]{chapman13}.

Since the proxy model approach relies on the availability of reliable solar irradiance measurements, it is not straightforward to reconstruct SSI by this method due to the long-term stability issues in available measurements and the relative paucity of observations outside the ultraviolet (Sect. \ref{measurements}). Making use of the fact that the effects of instrument degradation on apparent variability is relatively benign at shorter timescales, long-term stability issues are sometimes circumvented by fitting index data to measured rotational variability \citep[either by detrending the data or confining the regression to rotational periods,][]{lean97,pagaran09,thuillier12} and then assuming the indices-to-irradiance relationships so derived to all timescales. As we will discussed in Sect. \ref{discussionproxy}, the assumption that the underlying relationship between indices of solar activity and solar irradiance is similar at all timescales is not likely valid.

The next level of sophistication in the modelling of solar irradiance is realized by semi-empirical models. In these models, the solar disc is segmented by surface feature type, termed `components'. The filling factor (proportion of the solar disc or a given area covered) and time evolution of each component is deduced from indices of solar activity or suitable full-disc observations. This information is converted to solar irradiance employing the intensity spectra of the various components, calculated using spectral synthesis codes with semi-empirical model atmospheres of said feature types \citep{fontenla99,fontenla09,unruh99,shapiro10}. The reconstruction of the solar spectrum is given by the filling factor weighted-sum of the component intensity spectra. The semi-empirical model atmospheres describe the temperature and density stratification of the solar atmosphere within each component, constrained and validated by observations (therefore the term `semi-empirical').

The semi-empirical approach has the advantage that it yields SSI independent of the availability of reliable measurements. Additionally, for the models that rely on full-disc imagery for the filling factor of the solar surface components (i.e., the exact disc position of magnetic features is a known), the centre-to-limb variation of the radiant behaviour of each component can be taken into account by generating and applying the corresponding intensity spectrum at varying heliocentric angles.

At present, there are five models reported in the literature capable of returning the full solar spectrum, reviewed in \cite{ermolli13}. They are,
\begin{itemize}
	\item NRLSSI \citep[Naval Research Laboratory Solar Spectral Irradiance,][]{lean97,lean00},
	\item SATIRE-S \citep[Spectral And Total Irradiance REconstruction for the Satellite era,][]{unruh99,fligge00,krivova03,krivova06,krivova11a,yeo14b},
	\item SRPM \citep[Solar Radiation Physical modelling,][]{fontenla99,fontenla04,fontenla06,fontenla09,fontenla11},
	\item OAR \citep[Observatorio Astronomico di Roma,][]{penza03,ermolli11,ermolli13} and
	\item COSI \citep[COde for Solar Irradiance,][]{haberreiter08,shapiro10,shapiro11,shapiro13}.
\end{itemize}
Apart from the NRLSSI, these models adopt the semi-empirical approach. In the following, we discuss the recent results obtained with the SATIRE-S model by \citealt{yeo14b} (Sect. \ref{modelsatires}) before giving an overview of the other models listed (Sect. \ref{modelsrpm}). For a broader review of models of solar irradiance, we refer the reader to \cite{domingo09,ermolli13,solanki13}.

\subsection{SSI models capable of returning the full solar spectrum}
\label{models3}
 
\subsubsection{SATIRE-S}
\label{modelsatires}

The SATIRE-S model \citep{fligge00,krivova03,krivova11a} relies on full-disc observations of magnetic field and intensity to segment the solar disc into quiet Sun, faculae, sunspot umbra and sunspot penumbra. It has been applied to longitudinal magnetograms and continuum intensity images collected at the KPVT \citep[in operation from 1974 to 2003,][]{livingston76,jones92}, as well as from SoHO/MDI \citep[1996 to 2011,][]{scherrer95} and SDO/HMI\footnote{In full, the Kitt Peak Vacuum Telescope (KPVT), the Michelson Doppler Imager onboard the Solar and Heliospheric Observatory (SoHO/MDI)), and the Helioseismic and Magnetic Imager onboard the Solar Dynamics Observatory (SDO/HMI).} \citep[2010 to the present,][]{schou12} to reconstruct total and spectral solar irradiance over various periods between 1974 and 2013 \citep{krivova03,wenzler06,ball12,ball14,yeo14b}. In the latest iteration \citep{yeo14b}, KPVT, MDI and HMI magnetograms were cross-calibrated in such a way that the model input from all the data sets combine to yield a single consistent TSI/SSI time series spanning the entire period of 1974 to 2013 as the output. Apart from the NRLSSI, which extends back to 1950, this is the only other daily reconstruction of the full solar spectrum (from present-day models) to cover multiple solar cycles.

At present, the model employs the intensity spectra of quiet Sun, faculae, umbra and penumbra from \cite{unruh99}, generated with the ATLAS9 spectral synthesis code \citep{kurucz93}. As ATLAS9 assumes local thermodynamic equilibrium (LTE), it fails in the ultraviolet or below approximately $300\:{\rm nm}$. (Solar ultraviolet radiation is formed in the upper photosphere and lower chromosphere, where the plasma is increasingly collisionless.) This is accounted for by offsetting/rescaling the 115 to 300 nm segment of the reconstructed spectra to the Whole Heliospheric Interval (WHI) reference solar spectra by \cite{woods09} and SORCE/SOLSTICE spectrometry \citep[detailed in][]{yeo14b}. SATIRE-S is the only semi-empirical model to include non-LTE effects through such an approach. The SRPM, OAR and COSI models make use of various non-LTE spectral synthesis codes, which differ from one another in the method non-LTE effects are approximated \citep[see][]{fontenla99,uitenbroek02,shapiro10}.

Recall, due to uncertainties in the amplitude of solar cycle variation, the three published TSI composites exhibit conflicting decadal trends (Sect. \ref{tsimeasurements}). The TSI from the reconstruction is a significantly closer match to the PMOD composite than to the ACRIM and IRMB time series, replicating most of the variability ($R^2=0.92$), including the solar cycle variation (Fig. \ref{comparetsissr}b). Reconstructed TSI also exhibits excellent agreement with the measurements from individual instruments such as ACRIM3, TIM and VIRGO. In particular, the record produced by the PMO6V\footnote{VIRGO TSI is actually given by the combination of the measurements from two onboard radiometers, DIARAD and PMO6V. The solar cycle variation in the DIARAD and PMO6V records is nearly identical but they do differ at rotational timescales resulting in very different results for the correlation on comparison with other measurements or models.} radiometer on VIRGO ($R^2=0.96$). The secular decline between the 1996 and 2008 solar cycle minima in VIRGO radiometry is reproduced to within $0.05\:\wms$ (Fig. \ref{comparetsissr}a). This agreement between SATIRE-S TSI and VIRGO radiometry, which extends all of solar cycle 23 and on to the present, is significant. It implies that at least $96\%$ of the variability in solar irradiance, over this period, including the secular variation between the 1996 and 2008 solar cycle minima can be explained by solar surface magnetism alone.

\begin{cfig}
\includegraphics[width=\textwidth]{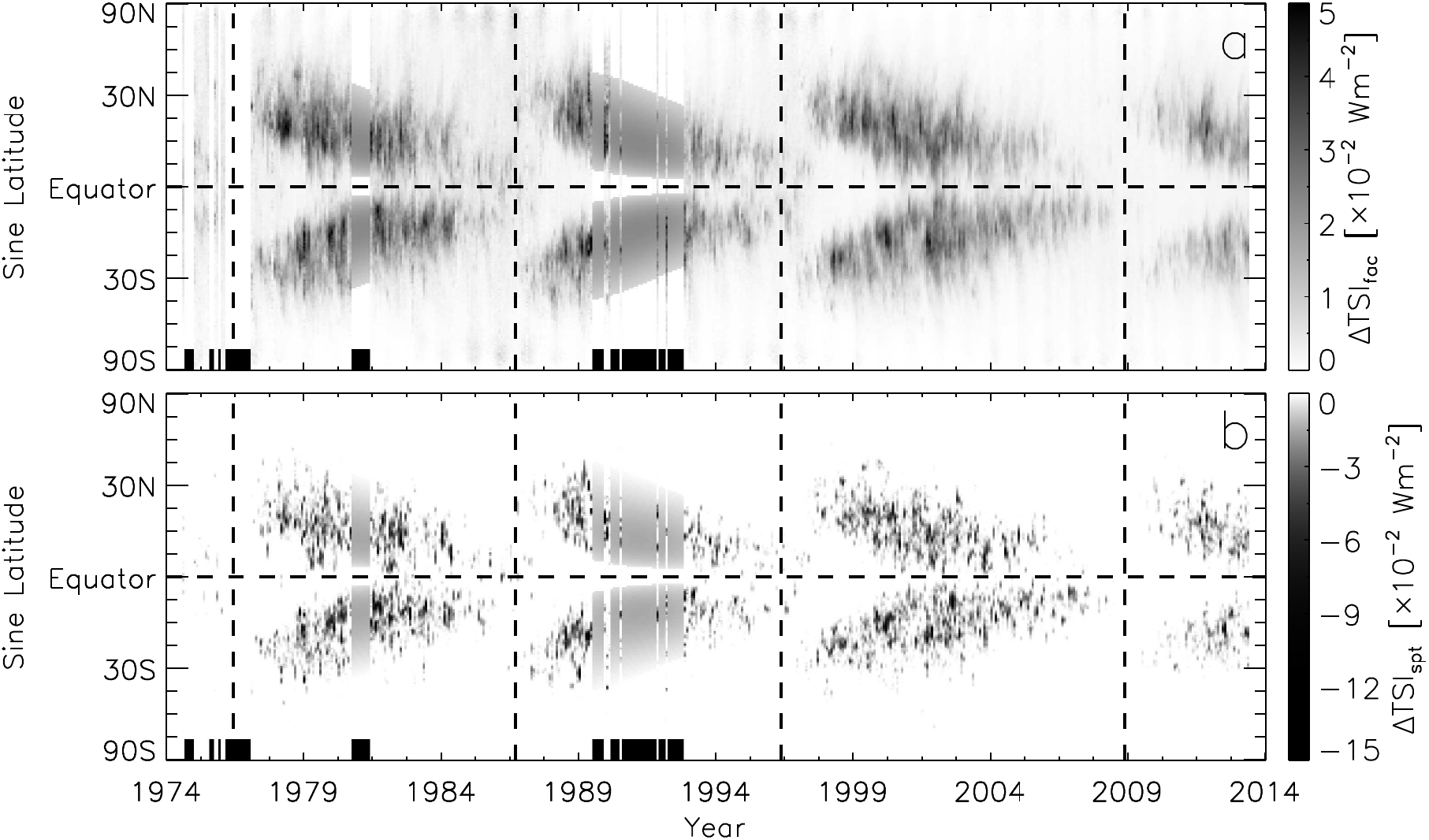}
\caption{Variation in TSI from a) faculae brightening, $\dtsifac$ and b) sunspot darkening, $\dtsispt$ in the SATIRE-S model, as a function of time and latitude (the monthly average in sine latitude intervals of 0.01). The black bars along the horizontal axes mark the months with no values from the lack of suitable magnetogram data. The gaps around the maxima of solar cycles 21 and 22 are filled by the interpolation. The horizontal and vertical dashed lines denote the equator and position of solar cycle minima, respectively. Adapted from \cite{yeo14b}.}
\label{cpsbutterflyssr}
\end{cfig}

The bolometric facular brightening and sunspot darkening, $\dtsifac$ and $\dtsispt$ (with respect to the TSI level of the magnetically quiet Sun, which emerges in the computation), binned and averaged by month and sine latitude, is expressed in Fig. \ref{cpsbutterflyssr}. Since solar surface magnetism is concentrated in active regions, it follows then that the latitudinal distribution of the associated intensity excess/deficit demonstrate Sp{\"o}rer's law, resembling butterfly diagrams of sunspot area/position and magnetic flux \citep[for example, Figs. 4 and 14 in][]{hathaway10}. A diagram similar to Fig. \ref{cpsbutterflyssr}b, based on the PSI, was recently presented by \cite{frohlich13}.

\begin{cfig}
\includegraphics[width=\textwidth]{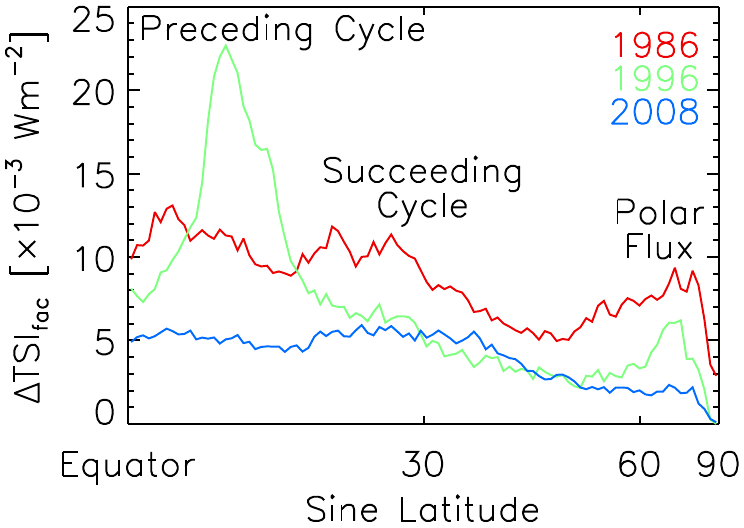}
\caption{Latitudinal distribution of facular brightening ($\dtsifac$) in SATIRE-S at the 1986, 1996 and 2008 solar cycle minima, taken from Fig. \ref{cpsbutterflyssr}a. Facular brightening is influenced by active regions in the low and mid-latitudes (associated with the preceding and succeeding solar cycles, respectively), and polar flux at high latitudes. Adapted from \cite{yeo14b}.}
\label{secularssr}
\end{cfig}

Since sunspots are largely absent around solar cycle minima, the minimum-to-minimum variation in SATIRE-S TSI is dominantly from the change in facular brightening. In Fig. \ref{secularssr}, we illustrate the latitudinal distribution of faculae brightening at the last three solar cycle minima, obtained from the butterfly diagram (Fig. \ref{cpsbutterflyssr}a). Around the 1986 solar cycle minimum, facular brightening is elevated close to the equator, around mid-latitudes, and towards high latitudes (red curve). The broad peak near the equator and at mid-latitudes correspond to active regions associated with the preceding cycle and the succeeding cycle, respectively. The increase towards high latitudes relate to magnetic elements transported polewards by meridional circulation over the course of the previous cycle (i.e., polar flux). Comparing the latitudinal distribution at the three solar cycle minima depicted, the minimum-to-minimum trend in faculae brightening (and therefore TSI) is modulated by the prevailing magnetic activity in the three latitude regions. The low flat profile of the blue curve, which corresponds to the 2008 solar cycle minimum, reflects the near-complete absence of any form of activity during this period, which contributed significantly to the secular decline between the 1996 and 2008 solar cycle minima.

\begin{cfig}
\includegraphics[width=\textwidth]{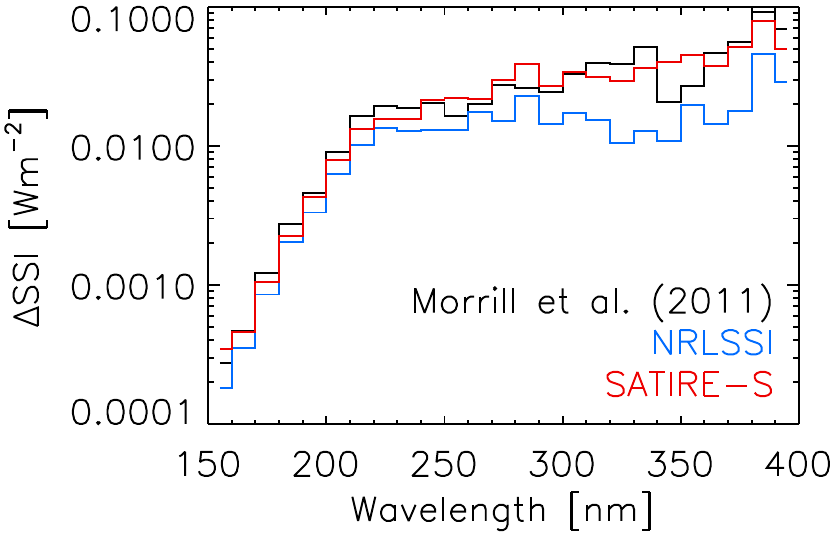}
\caption{The change in solar irradiance (${\rm \Delta{}SSI}$), integrated over 10 nm intervals, between 2000 and 2008 (solar cycle maximum and minimum), in the \cite{morrill11}, NRLSSI and SATIRE-S models. Adapted from \cite{yeo14b}.}
\label{comparesatmorrillssr}
\end{cfig}

As discussed in Sect. \ref{ssimeasurements}, ultraviolet solar irradiance (120 to 400 nm) measurements exhibit similar rotational variability but diverge in terms of the amplitude of solar cycle variation, especially above 240 nm, and between SORCE and pre-SORCE missions. Consequently, while the SATIRE-S reconstruction reproduces the rotational variability in ultraviolet solar irradiance observations from the UARS and SORCE missions, in terms of the variation over the solar cycle, it replicates certain records better than others (Fig. \ref{satireuarssorcessr}). It reproduced the solar cycle variation in SUSIM spectrometry between 115 and 240 nm, and in UARS/SOLSTICE and SORCE/SOLSTICE measurements between 115 and 180 nm. The reconstruction is also a very close match to the empirical model by \cite{morrill11}. (This model, based on matching the Mg II index to SUSIM SSI, represents an estimation of SUSIM-like spectrometry over the wavelength range of 150 to 400 nm.) Notably, the amplitude of solar cycle variation is similar, even above 240 nm (Figs. \ref{barchart} and \ref{comparesatmorrillssr}). The reconstruction also replicated the rotational variability in SIM SSI over the entire wavelength range of the record (240 to 2416 nm) but not the overall trend (Fig. \ref{comparesatsimssr}). It is worth noting here that there is no model reported thus far that is able to replicate the solar cycle variation in SIM SSI and TSI simultaneously.

\begin{cfig}
\includegraphics[width=\textwidth]{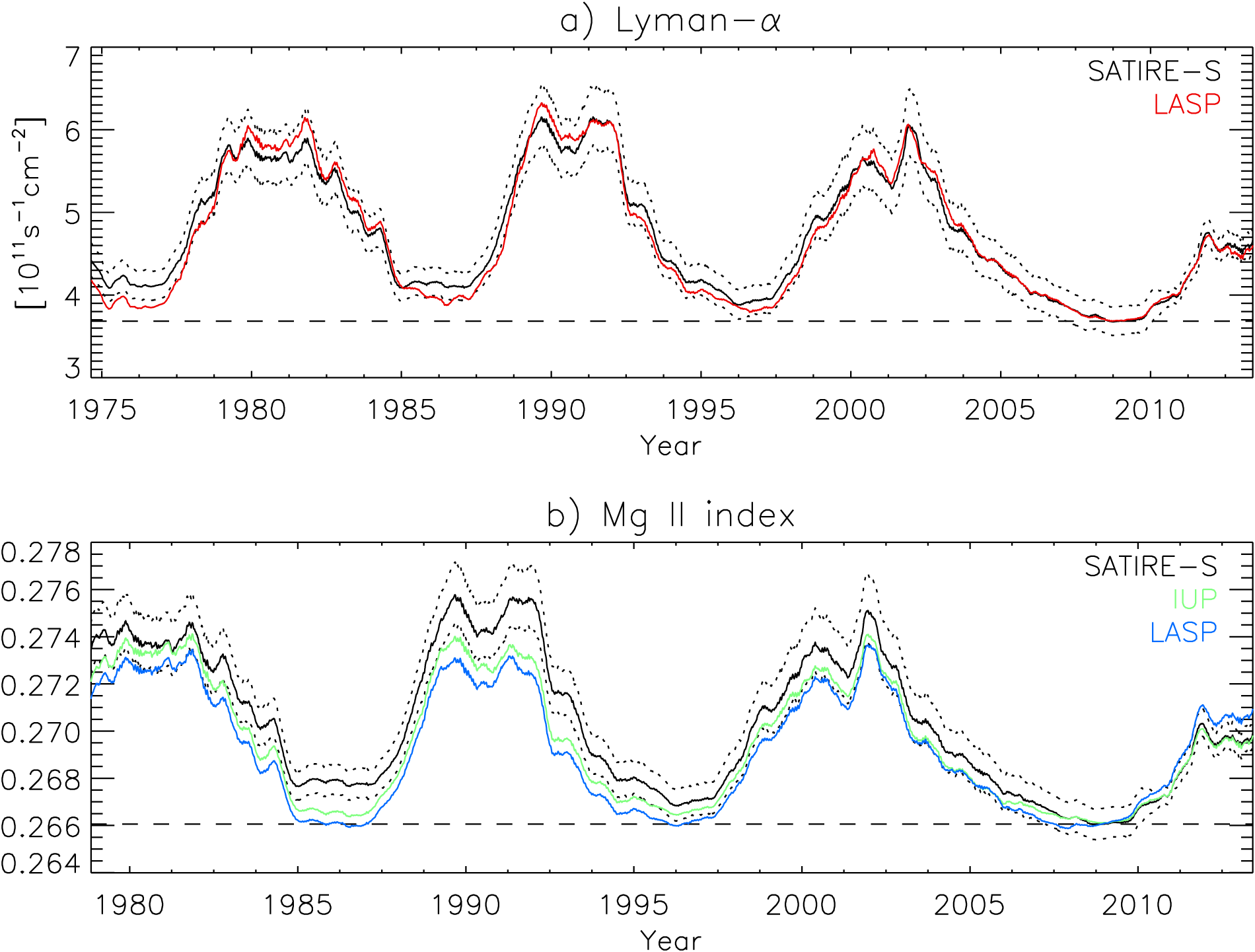}
\caption{a) The Lyman-$\alpha$ irradiance and b) Mg II index based on the SATIRE-S reconstruction of SSI (black solid lines). The reconstruction uncertainty is denoted by the dotted lines. Also illustrated, the LASP Lyman-$\alpha$ composite (red), and the IUP (green) and LASP (blue) Mg II index composites, regressed to the rotational variability and offset to the 2008 solar cycle minimum level (dashed lines) in the respective SATIRE-S time series. All the time series were smoothed with a 181-day boxcar filter. Adapted from \cite{yeo14b}.}
\label{comparesatssissr}
\end{cfig}

The Lyman-$\alpha$ irradiance of the reconstruction reproduces most of the variability ($R^2=0.94$), including the solar cycle trend (Fig. \ref{comparesatssissr}a), in the LASP Lyman-$\alpha$ composite \citep{woods00}. The Mg II index taken from the reconstruction is also highly correlated to the competing Mg II index composites by LASP \citep{viereck04,snow05b}, and by IUP \citep{viereck99,skupin05c,skupin05b}, in particular the latter ($R^2=0.96$). It was, however, less successful in replicating the decadal trend (Fig. \ref{comparesatssissr}b). That said, it did reproduce, to within model uncertainty, the secular decline between the 1996 and 2008 solar cycle minima in the IUP Mg II index composite.

\subsubsection{Other present-day models}
\label{modelsrpm}

The present-day models capable of returning the full solar spectrum, with the exception of the SRPM, are broadly consistent with one another. There are some differences, most notably in terms of the spectral dependence of the cyclic variability (Fig. \ref{barchart}). In this section, we will give a brief description of the NRLSSI, SRPM, OAR and COSI models, and highlight the key discrepancies between these models and SATIRE-S.

The NRLSSI \citep{lean97,lean00} describes the effect of sunspot darkening and faculae brightening on a model spectrum of the quiet Sun. The time evolution of sunspot darkening is given by the PSI calculated from sunspot area records, and facular brightening by the Ca II K, Mg II and F10.7 indices. In the ultraviolet (120 to 400 nm), the variation in solar irradiance is inferred from the multivariate regression of the index data to the rotational variability in UARS/SOLSTICE SSI (by first detrending both index and solar irradiance data). Above 400 nm, it is given by the sunspot and faculae contrast (as a function of wavelength) model from \cite{solanki98a}, modulated in time by the index data.

The models discussed here, apart from the SRPM, all see reconstructed SSI varying in phase with the solar cycle in the ultraviolet and visible (Fig. \ref{barchart}). In the infrared, facular contrast is weak (and depending on the model, negative at certain wavelengths), allowing sunspot darkening to dominate, such that the overall level at activity maximum can be lower than at minimum (illustrated for SATIRE-S in Fig. \ref{comparesatsimssr}d). Depending on the computed sunspot/faculae contrast adopted by the various models, they differ in the wavelength range and strength of this effect \citep[see Fig. 7 in][]{ermolli13}. This effect is relatively weak in the NRLSSI such that the integrated flux over the shortwave-infrared (1000 to 2416 nm) still varies in phase with the solar cycle, contrary to the other models (Fig. \ref{barchart}). The variation over the solar cycle between 240 and 400 nm is also weaker, almost half of that in SATIRE-S, attributed to confining the regression to rotational variability \citep{ermolli13}. The consistency between SATIRE-S and the SUSIM-based model of \cite{morrill11} gives further support to the amplitude of solar cycle variation in the ultraviolet exhibited by these two models (Fig. \ref{comparesatmorrillssr}). While SATIRE-S replicates the secular decline between the 1996 and 2008 solar cycle minima in VIRGO TSI radiometry (Fig. \ref{comparetsissr}a), the NRLSSI does not as this minimum-to-minimum variation is absent in the Mg II index composite employed (the one released by LASP, Fig. \ref{comparesatssissr}b).

The SRPM denotes the set of data and tools for semi-empirical modeling of solar irradiance, including a non-LTE spectral synthesis code, developed by \cite{fontenla99,fontenla04,fontenla06,fontenla09,fontenla11}. Currently, the package features semi-empirical model atmospheres for nine solar surface components, namely, quiet Sun internetwork, quiet Sun network lane, enhanced network, plage, bright plage, sunspot umbra and sunspot penumbra \citep[presented in][]{fontenla09}, and dark quiet Sun internetwork and hot facula \cite[introduced in][but not incorporated into the solar irradiance reconstruction reported by the authors]{fontenla11}. Earlier SRPM reconstructions (or indeed, all other models) cannot replicate the strong decline in the ultraviolet or the accompanying increase in the visible, in apparent anti-phase with the solar cycle, registered by SIM between 2004 and 2008 (Sect. \ref{ssimeasurements}). \cite{fontenla11} took this to indicate that the temperature gradient inside small-scale magnetic concentrations might be steeper than previous estimations and proceeded to adjust the temperature stratification of the model atmospheres introduced in \citealt{fontenla09} \citep[see Fig. 1 in][]{fontenla11}. The solar irradiance reconstruction presented, extending 2000 to 2009, is based on these modified model atmospheres and filling factors for the various components derived from full-disc images (in the visible red and Ca II K) acquired with the Precision Solar Photometric Telescope, PSPT at OAR \citep{coulter94,ermolli98}.

The modifications to the \cite{fontenla09} model atmospheres did result in reconstructed visible solar irradiance varying in anti-phase with the solar cycle. However, the solar cycle variation in the ultraviolet is still weaker than what was registered by SORCE/SOLSTICE. The model thus failed to reproduce the solar cycle variation in measured TSI, which all the other present-day models are able to do with reasonable success. All these shortcomings were taken by the authors to imply that the number of solar surface components is insufficient.

The full-disc images recorded with the PSPT at OAR were also employed in the series of proxy and semi-empirical reconstructions of solar irradiance reported by the OAR team \citep{penza03,domingo09,ermolli11}. Work on a new semi-empirical model is in progress \citep{ermolli13}. This latest effort considers the seven solar surface components defined in \cite{fontenla09}. The filling factors were obtained from PSPT observations spanning the period of 1997 to 2012 following \cite{ermolli10}. The intensity spectra corresponding to each component were calculated with the \cite{fontenla09} model atmospheres \citep[without the more recent modifications introduced by][]{fontenla11} using the non-LTE spectral synthesis code RH\footnote{Based on the work of, and abbreviated after \cite{rybicki91,rybicki92}.} \citep{uitenbroek02}. The computation is therefore, apart from employing the \cite{fontenla09} model atmospheres as is, broadly analogous to the SRPM reconstruction presented by \citealt{fontenla11}.

In the ultraviolet and visible, the OAR reconstruction is consistent with NRLSSI, SATIRE-S and COSI, exhibiting a similar disparity with SIM SSI (Fig. \ref{barchart}). Also in line with these other models, the reconstruction replicates most of the variability in TSI observations (specifically, TIM radiometry and the PMOD composite). In apparent contradiction to the conclusion of \cite{fontenla11}, the agreement with measured TSI suggests that the seven solar surface components described in \cite{fontenla09} are sufficient for the semi-empirical modeling of (at least) TSI. This also supports the conclusions of \cite{ball11,deland12,lean12,unruh12,yeo14b}, that the discrepancy between SIM SSI and models arise from unaccounted instrumental effects in SIM observations, than any indication that a significant rethink in how solar irradiance is modelled in current models is necessary (discussed in Sect. \ref{ssimeasurements}).

The non-LTE spectral synthesis code COSI \citep{haberreiter08,shapiro10} has been utilised to generate intensity spectra of solar surface components for semi-empirical modeling of solar irradiance \citep{haberreiter05}. The current implementation utilizes intensity spectra generated with the model atmospheres by \cite{fontenla99}. These were applied to sunspot number, $^{10}{\rm Be}$ and neutron monitor data to reconstruct solar irradiance back to the Maunder minimum period and over the Holocene \citep[at 1-year and 22-year cadence, respectively,][]{shapiro11}, and to SRPM PSPT-based filling factors covering a two week period in 2010 \citep{shapiro13}.

Contrary to the NRLSSI, SATIRE-S and OAR models, the \cite{shapiro11} reconstruction varied in anti-phase with the solar cycle in the near-infrared (700 to 1000 nm, Fig. \ref{barchart}), accompanied and compensated by an enhanced variability in the ultraviolet. This is attributed to the use of a single model atmosphere for sunspots, and for plage \citep[so not distinguishing between sunspot umbra and penumbra, and plage and bright plage,][]{ermolli13}. In response, \cite{shapiro13} reduced sunspot and plage contrast in such a manner that brought the reconstruction presented into alignment with SORCE/SOLSTICE measurements in the Herzberg continuum (190 to 222 nm). A more comprehensive approach is in development.

\section{Reconciling measurements and models}
\label{discussionssr}

\subsection{Proxy models}
\label{discussionproxy}

Solar irradiance observations, in particular SSI, suffer non-trivial uncertainty in the long-term stability (Sect. \ref{measurements}). As mentioned in Sect. \ref{models2}, in the regression of indices of solar activity to observations, certain proxy models confine the fitting to rotational variability to circumvent bias from instrumental trends, implicitly assuming that the relationship between the two is similar at all timescales.

Rotational variability in solar surface magnetism is dominated by active regions, and at longer timescales, by the magnetic flux distributed in the quiet Sun. The response of chromospheric indices to magnetic flux in active regions and quiet Sun is, evidently, not the same \citep{tapping87,solanki04,ermolli10,foukal11}. The weak solar cycle variation in the ultraviolet in the NRLSSI highlights the limitation of applying the relationship between chromospheric indices and solar irradiance at rotational timescales to longer timescales. 

The NRLSSI also does not replicate the secular decline between the 1996 and 2008 solar cycle minima in VIRGO TSI radiometry as it is absent in the LASP Mg II index composite. \cite{frohlich09,frohlich12,frohlich13} attributed the discrepant decadal trend in the LASP Mg II index composite and VIRGO TSI to a possible cooling/dimming of the photosphere between the two solar cycle minima. This disparity is more likely related to the non-linear relationship between chromospheric indices and solar irradiance, discussed in the previous paragraph \citep{foukal11}, and the uncertainty in the long-term stability of Mg II index data. While the LASP Mg II index composite is effectively level between the 1996 and 2008 solar cycle minima, the competing composite by IUP does exhibit a secular decline (Fig. \ref{comparesatssissr}b). This minimum-to-minimum drop in the IUP composite is replicated in the Mg II index produced from the SATIRE-S reconstruction, the TSI from which reproduces the secular decline in VIRGO TSI (Sect. \ref{modelsatires}). The discrepancy between the IUP and LASP composites demonstrates the sensitivity of the apparent decadal trend, even in a quantity as robust to instrumental effects as the Mg II index, to the long-term stability.

The rigorous reconstruction of solar irradiance through proxy models would require a greater understanding of the relationship between indices of solar activity and solar irradiance, and of the long-term stability in index data (still largely unknown).

\subsection{Semi-empirical models}
\label{discussionsemiempirical}

Present-day semi-empirical models (reviewed in Sect. \ref{models3}) all employ one-dimensional or plane-parallel model atmospheres. Various studies have pointed out that the intensity spectra synthesized from one-dimensional representations of the spatially inhomogeneous solar atmosphere do not necessarily reflect the true average property \citep{uitenbroek11,holzreuter13}.

The intensity contrast of network and facular magnetic features varies with distance from disc centre and magnetic field strength. In SATIRE-S, the magnetic field strength dependence is included by scaling the faculae filling factor of non-sunspot image pixels above the magnetogram noise floor by the magnetogram signal \citep[up to a saturation level, the free parameter in the model,][]{fligge00}. For the SRPM, OAR and COSI models, which employ full-disc Ca II K images as a proxy of magnetic field strength, after identifying sunspots, the rest of the solar disc is segmented into multiple components. These measures are not only empirical but also do not account (at least not fully) for the observation that the continuum and line core intensity contrast of small-scale magnetic concentrations scale differently with magnetogram signal \citep{yeo13}. As set out by \cite{unruh09}, three-dimensional model atmospheres would allow the possibility to relate the appropriate calculated intensity spectra to the magnetogram signal or Ca II K intensity directly.

At present, the centre-to-limb variation of the intensity of solar surface components is accounted in semi-empirical models by generating the intensity spectra at varying limb angles (by rotating the respective model atmospheres). In the continuum, the intensity contrast of small-scale magnetic elements increases with distance from disc centre before declining again close to the limb. The converse is observed within spectral lines, the difference coming primarily from the differing interaction between the line-of-sight and magnetic flux tubes at the continuum and spectral lines formation heights \citep{yeo13}. In employing one-dimensional model atmospheres, we do not capture these effects.

Three-dimensional model atmospheres based on observations \citep{socasnavarro11} and magnetohydrodynamics (MHD) simulations \citep{vogler05}, while growing in sophistication and realism, cannot as yet reproduce observations at all heights \citep{afram11}. A limiting factor is our understanding of the effects of spatial resolution (that is, the point spread function and how it is sampled by the imaging array) on observations \citep{danilovic08,danilovic13,rohrbein11}. This is especially severe for the small-scale magnetic elements which make up network and faculae, largely unresolved in current observations. The increasing availability of atmospheric seeing free observations from space and balloon-borne missions, in particular high spatial resolution imagery such as those from SUNRISE \citep{solanki10,barthol11}, will provide stringent constraints on model atmospheres. Space and balloon-borne telescopes have the advantage that the point spread function can be well constrained \citep{mathew07,mathew09,wedemeyerbohm08,yeo14a}, rendering them particularly useful for this purpose.

Another source of uncertainty is the treatment of non-LTE effects, which are highly complex and not fully understood. As a matter of necessity, the SATIRE-S, SRPM, OAR and COSI models take varying approximations to include non-LTE effects. While inexact, the one-dimensional model atmospheres and non-LTE approximations employed in present-day semi-empirical models are a practical necessity. As these simplifications are tested against and so constrained by observations, the output from current models are still reasonably realistic, demonstrated by the broad consistency between reconstructed solar irradiance and measurements (Sect. \ref{models3}).

\section{Summary}
\label{summaryssr}

The TSI observations from the succession of radiometers sent into space since 1978 readily reveal solar cycle modulation. The records from the various instruments however, differ in terms of the absolute level and the apparent amplitude of solar cycle variation, chiefly from the difficulty in accounting for instrument degradation. With the collaborative efforts of various teams, the absolute radiometry of present-day instruments is converging. The uncertainty in the long-term stability is evident in the conflicting decadal trends exhibited by the three published TSI composites.

Like TSI, ultraviolet (120 to 400 nm) solar irradiance has been monitored from space, almost without interruption, since 1978. Spectrometry is obviously a more complicated measurement, the uncertainty in the absolute radiometry and the amplitude of solar cycle variation is more severe than with TSI. Compounded by the wavelength dependence of instrumental influences, this translates into uncertainty in the spectral dependence of the cyclical variability. The problem is particularly acute above 240 nm, and between measurements from the SORCE satellite and preceding missions.

The SIM instrument onboard SORCE provides what is still the only extended (2003 to 2011) and radiometrically calibrated record of SSI spanning the ultraviolet to the infrared (240 to 2416 nm) available. The measurements from the first few years of operation (2003 to 2008) saw ultraviolet solar irradiance declining almost twice as rapidly as TSI, and visible solar irradiance ascending, in apparent anti-phase with the solar cycle. These trends conflict with the projections from other measurements and models \citep[except for][]{fontenla11}. Looking at the full period, the overall trend shows no obvious solar cycle modulation. The total flux recorded by the instrument, which surveys a wavelength range responsible for more than $97\%$ of the power in solar radiative flux, also fails to replicate the solar cycle variation evident in TSI.

Satellite monitoring of solar irradiance has been accompanied by the development of models aimed at recreating the observed variability. Solar irradiance is modulated by photospheric magnetism from its effect on the thermal structure and consequently the radiant behavior of the solar surface and atmosphere. Models of solar irradiance based on the assumption that variations at timescales greater than a day are driven by solar surface magnetism have achieved considerable success in replicating observations.

There are two broad categories of solar irradiance models, termed proxy and semi-empirical. Proxy models are based on the regression of indices of solar activity to solar irradiance observations. Semi-empirical models employ the intensity spectra of various types of solar surface features (calculated with spectral synthesis codes from semi-empirical model atmospheres) to reconstruct the solar spectrum from the apparent surface coverage of these feature types (derived from index data or full-disc observations).

We discussed the present-day models capable of returning the full solar spectrum; NRLSSI, SATIRE-S, SRPM, OAR and COSI. Apart from the NRLSSI, these models adopt the semi-empirical approach.

In the regression of index data to solar irradiance observations, certain proxy models including the NRLSSI (in the ultraviolet) restrict the fitting to the rotational variability to factor out any bias from the uncertainty in the long-term stability of the solar irradiance measurements employed. In doing so, these models implicitly assume that the relationship between chromospheric indices (utilized in these models as a proxy of faculae brightening) and solar irradiance at rotational timescales is applicable to longer timescales. Likely a consequence of the fact that the relationship between chromospheric indices and solar irradiance is really non-linear, the amplitude of solar cycle variation in the ultraviolet in the NRLSSI is weaker than in other present-day models.

Another limitation of the proxy approach is the fact that the reconstructed solar irradiance adopts the variability of the index records used in the reconstruction, along with the associated uncertainty. The uncertainty in the long-term stability of Mg II index data was argued to be the probable reason why the NRLSSI does not replicate the secular decline between the 1996 and 2008 solar cycle minima in VIRGO TSI radiometry.

The SRPM reconstruction of solar irradiance presented by \cite{fontenla11} is the only where visible flux varied in anti-phase with the solar cycle, in qualitative agreement with early SIM observations. This was achieved with modifications to the temperature stratification in the \cite{fontenla09} model atmospheres. However, the TSI from the reconstruction failed to replicate the solar cycle variation in measured TSI. The other models reviewed see visible solar irradiance varying in-phase with the solar cycle and reproduced TSI variability, including the solar cycle modulation, with reasonable success. Significantly, the OAR computation is, apart from the use of the \cite{fontenla09} model atmospheres without any modifications, largely analogous to the \cite{fontenla11} study in terms of the approach.

Considering the role of photospheric magnetism in driving variations in solar irradiance, the increase in the visible registered by SIM during its early operation, coming at a time where solar activity is declining, requires that small-scale magnetic concentrations be darker than the quiet Sun in this spectral region. However, our current understanding of the radiant properties of these solar surface features point to the converse.

Apart from the NRLSSI, the semi-empirical model SATIRE-S, recently updated by \cite{yeo14b}, gives the only other daily reconstruction of the full solar spectrum from present-day models to cover multiple solar cycles (1974 to 2013). Of the three competing and conflicting TSI composites, the model found the greatest success in replicating the solar cycle variation in the PMOD composite. The TSI reconstruction is also a good match to present-day measurements, reproducing about $95\%$ of the variability in the PMO6V record and the secular decline between the 1996 and 2008 solar cycle minima in VIRGO radiometry. The SSI reconstruction reproduced the rotational variability in UARS and SORCE spectrometry, and the cyclical variability in certain records below 240 nm. Above 240 nm, as the amplitude of solar cycle variation is poorly constrained in current measurements, SATIRE-S, as with all other models, cannot exactly replicate SORCE solar cycle variation. The amplitude of solar cycle variation between 240 and 400 nm in the reconstruction is however, matches closely to the empirical model of \cite{morrill11}, which represents an approximation of SUSIM-like SSI with the long-term stability corrected to that of the Mg II index. The model also replicates the solar cycle variation in the LASP Lyman-$\alpha$ composite, and the secular decline between the 1996 and 2008 solar cycle minima in the IUP Mg II index composite.

The intensity spectra of solar surface features employed in present-day semi-empirical models are derived from one-dimensional model atmospheres, which do not capture all the complexities of the radiant behaviour of the solar surface and atmosphere. Three-dimensional model atmospheres, though increasingly realistic, still cannot reproduce observations at all heights. Their development is impeded by the limited availability of high spatial resolution observations, and the challenge in understanding instrumental influences on apparent radiance. Current semi-empirical models also include non-LTE effects by various approximations. Constrained by observations, the intensity spectra of solar surface features generated from one-dimensional model atmospheres and present non-LTE schemes, while not exact, are sufficiently reliable for the intended purpose. This is demonstrated by the broad consistency between the various semi-empirical models, and their success in replicating measurements.

The direct observation of solar irradiance is a challenging endeavour. While the body of spaceborne measurements is still afflicted by uncertainties in the absolute radiometry, secular variation and spectral dependence of the cyclical variability, one cannot discount the considerable progress made over the past four decades with the collective effort of the community. Models of solar irradiance based on solar surface magnetism have proved to be an able complement, augmenting our understanding of the observations and the physical processes underlying solar cycle variation in solar irradiance. While open questions remain, continual observational and modeling efforts will undoubtedly see the emergence of a more cohesive picture of solar cycle variation in solar irradiance.

\chapter[Summary and outlook]{Summary and outlook}
\label{summaryoutlook}

This thesis is the compilation of four publications (each forming a chapter), detailing investigations into the nature of variations in solar irradiance over the 11-year activity cycle of the Sun. They are summarized below in the order presented in the thesis.

\paragraph{Chapter \ref{paper1}: Yeo, K. L., Solanki, S. K., Krivova, N. A., 2013, Intensity contrast of network and faculae, Astron. Astrophys., 550, A95}\mbox{}\\

In this work, we examined the relationship between the intensity contrast of network and faculae, in the continuum and core of the Fe I 6173 \AA{} line in full-disc SDO/HMI observations, with disc position and magnetogram signal. Never before has any other solar telescope been able to return simultaneous, atmospheric seeing free, full-disc measurements of intensity and magnetic flux density at similar spatial resolutions or noise level (Chap. \ref{introductionsdohmi}). So, while similar studies have been reported in the literature, the image quality of HMI data gave us the chance to look at the disc position and magnetogram signal dependence of the intensity contrast of network and faculae at unprecedented accuracy and detail, particularly in the line core (which was, up to this study, largely unstudied).

The results from this study and preceding investigations into the continuum intensity contrast of network and faculae exhibit significant discrepancies. In the case of the analogous study by \cite{ortiz02} with SoHO/MDI (the predecessor instrument to the HMI) data, this is due to the fact that we had taken care to exclude image pixels around sunspots and pores, where the apparent magnetogram signal is affected by stray light from and the magnetic canopy of these features, from consideration. Comparing the results obtained around disc centre from the various studies, we argued that the spread in apparent magnetogram signal dependence is largely due to differences in spatial resolution.

Not surprisingly, the radiant behaviour of network and faculae in the continuum and in the line core, formed at different heights, is very different. The most notable divergence is the converse centre-to-limb variation. While the intensity contrast in the continuum increases away from disc centre before decreasing again near the limb, in the line core, the intensity contrast is strongest at disc centre and declines monotonically towards the limb. This is due to the steeper temperature gradient and Zeeman splitting within magnetic concentrations, and the different mechanisms by which apparent contrast is modulated by viewing geometry at the continuum and spectral line formation heights.

Earlier works had shown, but only indirectly, that variation in solar irradiance is the sum manifestation of changes in the continuum and within spectral lines. From the data set, we derived empirical relationships relating the intensity contrast in the continuum and line core to the heliocentric angle and magnetogram signal. Using a simple model based on these relationships, we demonstrated that the variation in total solar irradiance during a three-week period in 1996, where only a single active region is visible on the solar disc, can only be explained if we consider facular contrast in both the continuum and the line core.

The study also provided further evidence that magnetic elements in quiet Sun network have a higher heating efficiency than the facular counterpart, as implied by the higher intensity contrast per unit magnetogram signal there.

Since looking at the variation in intensity contrast with magnetic field strength obviously requires co-temporal and co-spatial measurements of intensity and magnetic field, studies have largely relied on such observations from magnetographs and so mostly confined to the visible part of the spectrum. (With few exceptions, magnetographs are designed to observe the photosphere and therefore operate in the visible.) The intensity contrast of network and faculae (and therefore their contribution to variation in solar irradiance) is highest in the ultraviolet from how the Planck function responds to temperature. Yet, the relationship between the intensity contrast in the ultraviolet with magnetic field strength is not nearly as well studied as in the visible \citep[][]{berger07,viticchie10}.

In order to plug this gap in our knowledge, it will be of interest to extend this study, examining
\begin{itemize}
	\item HMI magnetograms together with the concurrent full-disc ultraviolet filtergrams from AIA \citep[Atmospheric Imaging Assembly, also onboard SDO,][]{lemen12}, and the
	\item co-temporal magnetic field and ultraviolet intensity observations from the IMaX magnetograph \citep[Imaging Magnetograph eXperiment,][]{martinezpillet11} and the SuFI filtergram instrument \citep[SUNRISE Filter Imager,][]{gandorfer11} onboard the SUNRISE balloon-borne observatory.
\end{itemize}
The AIA surveys three narrow passbands between 304 and 1700 \AA{}, and SuFI, five between 2140 and 3968 \AA{}, formed at different heights in the solar atmosphere between the photosphere and the transition region\footnote{The AIA has a total of nine passbands, the other six, located between 94 and 335 \AA{}, are formed in the corona.}. The observations from SUNRISE are not only free from atmospheric seeing but are also at the highest spatial resolutions achieved to date. They reveal the fine structure of the small-scale magnetic concentrations that make up network and faculae, largely unresolved in preceding observations, at an unsurpassed level of detail \citep{lagg10,riethmuller13}. The spatial resolution of AIA is also the highest of full-disc solar telescopes operating in a similar wavelength range. The co-temporal magnetic field and ultraviolet intensity measurements from SDO and from SUNRISE are well-suited for the intended purpose, covering a wide range of heights in the solar atmosphere at unprecedented spatial resolutions and atmospheric seeing free quality.

The complex radiant behaviour of network and faculae uncovered in this study is not completely represented in present-day semi-empirical models of solar irradiance, primarily due to the use of one-dimensional model atmospheres. Three-dimensional model atmospheres have not reached the maturity required for use in such models. As discussed in Chap. \ref{discussionsemiempirical}, while they are increasingly realistic, current three-dimensional model atmospheres cannot replicate observations at all heights. This renders the extension of this study to AIA and SUNRISE data all the more pertinent. The results of this investigation form a stringent constraint on three-dimensional model atmospheres at the HMI wavelength, 6173 \AA{}. Such an endeavour will however, also require an understanding of the effects of stray light on the apparent intensity contrast in HMI observations, investigated in the following publication.

\paragraph{Chapter \ref{paper2}: Yeo, K. L., Feller, A., Solanki, S. K., Couvidat, S., Danilovic, S., Krivova, N. A., 2014, Point spread function of SDO/HMI and the effects of stray light correction on the apparent properties of solar surface phenomena, Astron. Astrophys., 561, A22}\mbox{}\\

Here, we reported an estimate of the PSF of the HMI instrument and examined the effects of restoring HMI data for the influence of straylight by the deconvolution of this PSF. We modelled the PSF as a bivariate function of radial distance and azimuth, recovering the anisotropy of the underlying stray light behaviour. This is a departure from preceding efforts with other spaceborne solar telescopes, which had for simplicity assumed isotropy (i.e., modelling the PSF as a function of radial distance alone). The anisotropic PSF was demonstrated to perform better than an isotropic solution for stray light removal. It was also verified by comparing the image contrast in restored HMI data and in artificial solar images generated from a MHD simulation.

Restoring HMI observations for stray light with the retrieved PSF had a pronounced effect on the apparent intensity, magnetic field strength and line-of-sight velocity. Of particular relevance to solar irradiance studies is the effect on small-scale magnetic concentrations.

Image restoration decreased the surface area and increased the magnetic field strength of network and faculae. It also recovered otherwise undetectable magnetic features smeared below the noise floor by stray light. Overall, the apparent amount of magnetic flux, at least in the HMI data examined, increased by about $20\%$, most of it arising in the quiet Sun. The greater influence on quiet Sun network simply reflects the fact that the magnetic features there, generally smaller than those in active region faculae, are more affected by stray light. The observation that network elements have a higher heating efficiency than faculae and are also more obscured by stray light is not accounted in present-day models of solar irradiance that rely on full-disc imagery. The contribution by the network to variation in solar irradiance might therefore be underestimated in these models.

The PSF was recovered from observations of Venus in transit from one of the two CCDs in the instrument. It therefore represents the stray light property of this particular CCD at the time of the transit and the position of Venus in the field of view. Another important result of this study is the demonstration that the PSF reported can be applied to the entire field of view, to HMI data from either CCD and from the beginning of regular operation to the time of the study (April 2010 to June 2013) without introducing significant error. In other words, the stray light property of the instrument is sufficiently uniform across the field of view, similar between the two CCDs, and stable in time for this to be the case. Current models of solar irradiance assume that the intrinsic intensity contrast of photospheric magnetic features is invariable over time. Studies investigating if sunspot contrast varies over the solar cycle have returned conflicting results \citep[see for example,][]{mathew07,rezaei12,norton13,detoma13}. The possible time variation of network and faculae contrast is much less studied \citep{walton03,ortiz06,ermolli07}. An obstacle had been the lack of stable, extended time series of observations. The finding here that HMI data are, in terms of stray light, relatively stable over the period examined (which encompassed much of the ascending phase of solar cycle 24) demonstrates its suitability for such investigations.

HMI data products (described in Chap. \ref{introductionsdohmi}) are archived and distributed through several repositories such as JSOC and GDC-SDO\footnote{In full, the Joint Science Operations Center (http://jsoc.stanford.edu/) and the German Data Center for SDO (http://www2.mps.mpg.de/projects/seismo/GDC-SDO/).}. We plan to incorporate straylight correction by the deconvolution with the PSF reported here into the HMI data processing pipeline, with the aim to make a straylight-corrected version of the extended, near continuous time series of observations from the instrument available to the scientific community through one or more of these repositories. This will be of utility not only to investigations into the possible time variation of the intrinsic intensity contrast of solar surface features and semi-empirical models of solar irradiance such as the SATIRE-S, but any study that would benefit from accurate full-disc measurements of magnetic flux density, Doppler shift and intensity.

As stated in Chap. \ref{introductionthesisoutline}, an overall objective with the two studies discussed so far is to derive the information necessary to make quantitative comparisons of the radiant properties of small-scale magnetic concentrations in HMI data, and in artificial solar observations (of intensity and magnetic field\footnote{By generating artificial intensity images at the various HMI bandpass and polarizations (as described in Chap. \ref{gc}), and combining them together as done for HMI filtergrams to yield artificial HMI-like magnetograms.}) synthesized from three-dimensional model atmospheres based on MURaM MHD simulations (Chap. \ref{muram}). By blurring such artificial observations with the PSF of HMI, we can compare the relationship between intensity contrast, and disc position and magnetogram signal in the result with that in HMI data (obtained in the first study) directly.

\paragraph{Chapter \ref{paper3}: Yeo, K. L., Krivova, N. A., Solanki, S. K., Glassmeier, K. H., 2014, Reconstruction of total and spectral solar irradiance from 1974 to 2013 based on KPVT, SoHO/MDI and SDO/HMI observations, Astron. Astrophys., 570, A85}\mbox{}\\

We presented a daily reconstruction of total and spectral solar irradiance, extending from 1974 to 2013, based on the SATIRE-S model.

SATIRE-S had previously been applied to full-disc continuum intensity images and longitudinal magnetograms from the KPVT and SoHO/MDI to reconstruct TSI and SSI between 1974 and 2009. The parameters of the model were adjusted to the data from the two spectromagnetographs operated at the KPVT, and from MDI to combine the model output from the various data sets into a single TSI/SSI time series. This, however, still left some discrepancy in reconstructed spectra based on data from different instruments, which had to be accounted for empirically by regression. The decommissioning of MDI in 2011 also curtailed the possibility of extending these modelling efforts to the present and beyond.

In this study, we made significant improvements to the reconstruction method and extended the model to the present with similar observations from HMI. Of the improvements made, the most important is the more sophisticated cross-calibration of the model input from the various instruments. The magnetogram signal and faculae filling factor in the various data sets were brought into agreement in the periods of overlap between the different instruments (by taking into account non-linearities and variations between solar disc centre and limb) such that they yielded a consistent TSI/SSI time series as the output without the need for any additional adjustment of the reconstructed spectra.

The TSI from the reconstruction exhibited excellent agreement with the observations from the instruments in current operation, in particular, from the PMO6V radiometer on the SoHO/VIRGO experiment (which covers the period from 1996 to the present). The reconstruction reproduced most of the variability ($R^2=0.96$) including the secular decline between the 1996 and 2008 solar cycle minima. This result indicates that at least $96\%$ of the variability in solar irradiance over this period, including the secular variation, can be accounted for by photospheric magnetism alone. The reported alignment is all the more significant considering the fact that VIRGO is, of the TSI monitors sent into orbit since 1978, the only to return measurements covering an entire solar cycle minimum-to-minimum; extending all of solar cycle 23 and beyond to the current maximum.

Examining the latitudinal distribution of faculae brightening and sunspot darkening in the model at the last three solar cycle minima (1986, 1996 and 2008), we were able to illustrate, for the first time, the contribution by prevailing magnetism in the low and middle latitudes from active regions, and polar flux at high latitudes, to the minimum-to-minimum trend in TSI. The reconstruction also reproduced the solar cycle variation in the PMOD TSI composite (between 1978 and 2013) and the LASP Lyman-$\alpha$ composite (between 1974 and 2013), confirming the long-term stability of the reconstruction.

Like TSI, ultraviolet solar irradiance has been measured from space regularly since 1978. The measurements from the succession of missions, while broadly consistent at rotational timescales, exhibit discrepant solar cycle variation, especially above 240 nm. The reconstruction replicates the rotational variability in the ultraviolet solar irradiance measurements from the UARS and SORCE missions rather well but had mixed results in terms of the amplitude of solar cycle variation (which is to be expected given the gross scatter in observations). The amplitude of solar cycle variation in the reconstruction is, however, a very close match to that in the \cite{morrill11} model, even above 240 nm. This purely data-based model, extending 150 to 400 nm, is an estimation of SUSIM-equivalent SSI, with the long-term stability corrected with the Mg II index as reference. The agreement renders support to the cyclic variability in the SATIRE-S reconstruction.

The observations from SORCE/SIM represent the only extended (2003 to 2011) and continuous, radiometrically calibrated record of SSI covering the ultraviolet to the infrared (240 to 2416 nm) available. As with all other present-day models, the SATIRE-S reconstruction failed to replicate the overall trend in SIM SSI. The spectral range surveyed by SIM is accountable for more than $97\%$ of the power in solar radiation. Therefore, the integrated solar irradiance over this range should reproduce most of the variability in TSI. The total flux recorded by the instrument exhibits little resemblance to concurrent TSI measurements while the integrated solar irradiance over the spectral range of the instrument in the reconstruction replicated most of the variability in ACRIM3, VIRGO and TIM TSI radiometry ($R^2>0.88$). This suggests that the discrepancy in the overall trend between the reconstruction and SIM SSI is likely due to unresolved instrumental effects in the latter, an assertion strengthened in the review discussed next.

The reconstruction is, within uncertainty, an excellent match to observations, providing further evidence that variations in solar irradiance at timescales greater than a day is, if not solely at least predominantly driven by photospheric magnetism. The incorporation of the results from the study on the intensity contrast of network and faculae, and on the PSF of HMI, discussed earlier, into future efforts with SATIRE-S will certainly lead to even more sophisticated, reliable reconstructions of solar irradiance. (For example, the finding that network magnetic features are hotter/brighter than facular magnetic features, and the observations on the influence of sunspots/pores on the apparent magnetogram signal near these features and the effects of stray light on the manifest surface area and magnetic field strength of bright magnetic features.)

Currently, the variation in the intensity of bright magnetic features with magnetic field strength is included in SATIRE-S by the empirical relationship between the faculae filling factor and magnetogram signal described in Chap. \ref{introductionsatires}, which introduced the sole free parameter in the model, $\bsat$. This is another motivation for the studies detailed in Chaps. \ref{paper1} and \ref{paper2}, and our aim to reconcile the intensity contrast of small-scale magnetic concentrations and its magnetogram signal dependence in HMI data and in artificial observations synthesized from three-dimensional model atmospheres, discussed earlier in this chapter. As suggested by \cite{unruh09}, this would allow us to relate the intensity spectra synthesized from said three-dimensional model atmospheres to the magnetogram signal in HMI observations directly, obviating the empirical filling factor model and the associated free parameter.

Apart from the use of plane-parallel model atmospheres, another significant source of uncertainty in SATIRE-S is the assumption of local thermodynamic equilibrium, LTE (see Chap. \ref{locallte}) and the use of opacity distribution functions, ODFs (Chap. \ref{opacitysampling}) to represent spectral lines in the ATLAS9 spectral synthesis code. The detrimental effects of these simplifications are largely confined to wavelengths below 300 nm. Incidentally, this is where the variation in solar irradiance with solar activity is strongest (Fig. \ref{barchart}) and it is also a critical spectral region for climate models. So, while the empirical correction introduced in this study produced satisfactory results, going forward it is needful to tackle the LTE and ODF simplifications directly.

The approach taken to generate the intensity spectra of solar surface components in SATIRE-S has not been updated since the work of \cite{unruh99}. Since then, the application of various non-LTE spectral synthesis codes to semi-empirical modelling of solar irradiance has been demonstrated by several groups, reviewed in Chap. \ref{paper4}. These and other similar codes published in the literature differ by the approximations taken to include non-LTE effects and line opacity. A favoured alternative to the use of ODFs is the opacity sampling (OS) approach (described in Chap. \ref{opacitysampling}), which allows a more accurate account of line opacity, at the expense of computing time. It is our intention to examine the non-LTE and OS schemes reported in the literature, with the eventual aim to employ them in the generation of intensity spectra of solar surface components for solar irradiance modelling. 

In order to reconstruct solar irradiance over as long a period as possible, we employed full-disc observations from multiple instruments. Of the instruments considered, MDI has the coarsest spatial resolution and also the highest magnetogram noise level. To compose the model output based on data from the various instruments into a single, consistent time series, we had to disregard bright magnetic features in the KPVT and HMI data sets that are small and/or weak (in terms of the magnetogram signal) such that they would be undetectable at MDI's spatial resolution and magnetogram noise. Compared to other full-disc spectromagnetographs, the unsurpassed image quality of HMI observations allows prevailing photospheric magnetism to be resolved and quantified as accurately as currently achievable. On top of the proposed enhancements to the SATIRE-S model discussed, a reconstruction based on HMI data alone (optimized to the limits of the HMI instrument rather than that of MDI), though limited in timespan, will be informative of the limit to which variations in solar irradiance can be explained by photospheric magnetism.

\paragraph{Chapter \ref{paper4}: Yeo, K. L., Krivova, N. A., Solanki, S. K., 2014, Solar cycle variation in solar irradiance, Space Sci. Rev., online}\mbox{}\\

In this paper, we reviewed the current state of the measurement and modelling of solar irradiance, and the key challenges in refining present-day models.

While tremendous progress have been made over the past four decades since satellite measurements were first available, the direct observation of solar irradiance is still plagued by considerable uncertainty. Specifically, in terms of the absolute radiometry, the secular variation and the spectral dependence of the variation over the solar cycle. This is due mainly due to the immense challenge in accounting for aging and exposure degradation. For this reason, models of solar irradiance have emerged to become an important complement to measurements, augmenting our understanding not just of the apparent variability but the underlying physical processes.

We reviewed the present-day models that are able to recreate the full solar spectrum, placing the SATIRE-S reconstruction discussed above in the context of contemporary models (namely NRLSSI, SRPM, OAR and COSI).

There are two main approaches to the modelling of solar irradiance, termed proxy and semi-empirical. The models discussed, apart from NRLSSI, take the semi-empirical approach. The SATIRE-S, NRLSSI, OAR and COSI models are broadly consistent with one another, though not without disagreement. Specifically, the NRLSSI and COSI models exhibit discrepant solar cycle variation at certain spectral ranges. We showed that in the case of NRLSSI, these differences are due to shortcomings in the proxy approach, and for COSI certain limitations in the current implementation of the model.

Present-day models, with the exception of SRPM, are successful in reproducing TSI variability but not the overall trend in SIM SSI, in particular, the anti-phase (with the solar cycle) variation in the visible registered by the instrument between 2003 and 2006. The SRPM is the only model where visible solar irradiance varies in anti-phase with the solar cycle. However, the adjustments made to the model to achieve this result almost completely suppressed the solar cycle modulation in the TSI from the reconstruction. From a review of the literature and an examination of the SIM record, we summarized the evidence that the overall trend in SIM SSI is likely dominated by instrumental effects. We also argued that an inverse variation with the solar cycle in the visible is not consistent with our current knowledge of the intensity contrast of network and faculae. All these serve to refute claims that the failure of present-day models, including SATIRE-S, to replicate SIM SSI might be due to a change in the physics of the Sun or gross insufficiencies in current models of solar irradiance.

The daily reconstruction of TSI and SSI detailed in this thesis, based on the semi-empirical SATIRE-S model, was shown to be more reliable than the proxy reconstruction provided by the NRLSSI and highly consistent with measurements from multiple instruments. It also extends multiple decades, much longer than the daily reconstructions from contemporary semi-empirical models. This reconstruction is of utility not only to solar irradiance studies but also to climate models (which require solar irradiance inputs). The application of the SATIRE-S reconstruction to climate models, and a comparison of the results with that obtained with other solar irradiance models and measurements, is planned.

\bibliographystyle{thesis}
\bibliography{references}

\chapter*{Publications\markboth{Publications}{Publications}}
\addcontentsline{toc}{chapter}{Publications}

\begin{itemize}
	\item Yeo, K. L., Krivova, N. A., Solanki, S. K., Glassmeier, K. H., 2014, Reconstruction of total and spectral solar irradiance from 1974 to 2013 based on KPVT, SoHO/MDI and SDO/HMI observations, Astron. Astrophys., 570, A85
	\item Thuillier, G., Schmidtke, G., Erhardt, C., Nikutowski, B., Shapiro, A. I., Bolduc, C., Lean, J., Krivova, N. A., Charbonneau, P., Cessateur, G., Haberreiter, M., Melo, S., Delouille, V., Mampaey, B., Yeo, K. L., Schmutz, W., 2014, Solar spectral irradiance variability in November/December 2012: comparison of observations by instruments on the International Space Station and models, Sol. Phys., online
	\item Yeo, K. L., Krivova, N. A., Solanki, S. K., 2014, Solar cycle variation in solar irradiance, \ssr, online
	\item Yeo, K. L., Feller, A., Solanki, S. K., Couvidat, S., Danilovic, S., Krivova, N. A., 2014, Point spread function of SDO/HMI and the effects of stray light correction on the apparent properties of solar surface phenomena, Astron. Astrophys., 561, A22
	\item Yeo, K. L., Solanki, S. K., Krivova, N. A., 2013, Intensity contrast of network and faculae, Astron. Astrophys., 550, A95
\end{itemize}

\chapter*{Acknowledgements\markboth{Acknowledgements}{Acknowledgements}}
\addcontentsline{toc}{chapter}{Acknowledgements}

First off, I would like to thank my supervisors, Sami Solanki, Natalie Krivova and Karl-Heinz Gla\ss{}meier. Coming back to science after a six-year hiatus in industry is a challenge. Thank you, for taking a gamble on me, I hope it paid off.

Sami, after every meeting with you I am left floored by the depth and width of your knowledge and understanding, and even more by how incredibly incisive you are. What really draws my admiration though, is how no matter how busy you are, you were never sloppy in your interactions with me (or as far as I can see, with anyone else), no matter how trivial the matter. You are a role model and I consider it my privilege and a great pleasure to have come under your tutelage.

Natasha, right from the beginning, you were unwavering in your belief in my capabilities. You seem more confident in me than I myself. Thank you, for taking interest not just in my work but also my personal well-being, and for going out of your way time and again to support and set up opportunities for me.

Mention goes out to my collaborators, both here at MPS and without, who had contributed their time, effort and know how to this work; William Ball, Raymond Burston, Sebastien Couvidat, Sanja Danilovic, Ilaria Ermolli, Alex Feller, Jeff Morrill, Philip Scherrer and Jesper Schou.

Appreciation goes to the co-ordinator of the International Max Planck Research School for Solar System Science, Sonja Schuh. Your door is always open, and you take such care with all our concerns and issues, well beyond the call of duty. In less than a year, you have already made an incredible impact on the school, on each and every student.

I would have never made my way back into science without the intervention of Yvonne Unruh, my tutor back in Imperial College, and Ng Teck Khim, my former lab head at DSO. To both of you, I owe a debt of gratitude.

My life here in Germany would not have been complete without all the wonderful friends I have made. What would I do without you guys? It would take another dissertation to detail our adventures and exploits together. There is, however, one name I cannot omit here, Maria Andriopoulou. Thank you, for being there, for sharing my every joy and sorrow, quite literally from day one.

Last, but certainly not the least, I would like to thank my family, who have never wavered in their support or ever once question why I would leave home and career to pursue my interests.

This thesis is dedicated, in loving memory, to the late Dieter Schmitt.

\chapter*{Curriculum Vitae\markboth{Curriculum Vitae}{Curriculum Vitae}}
\addcontentsline{toc}{chapter}{Curriculum Vitae}

\begin{table*}[h]
\begin{tabular}{lll}
Name & \multicolumn{2}{l}{Yeo} \\
Vorname & \multicolumn{2}{l}{Kok Leng} \\
Geburtsdatum & \multicolumn{2}{l}{02.08.1979} \\
Geburtsort & \multicolumn{2}{l}{Singapur} \\
Staatsangeh\"origkeit & \multicolumn{2}{l}{singapurisch} \\
& & \\
Ausbildung & 1996-1997 & Anglo-Chinese Junior College, Singapur \\
& 2000-2004 & Imperial College London, Vereinigtes K{\"o}nigreich, \\
& & MSci Physics \\
& 2011-2014 & Doktorand am Max-Planck-Institut f{\"u}r \\
& & Sonnensystemforschung, G{\"o}ttingen und an der \\
& & Technischen Universit{\"a}t Braunschweig \\
& & \\
Wehrdienst & 1998-2000 & Singapore Armed Forces, 3rd Division Support \\
& & Command, Fernmelder \\
& & \\
Besch\"aftigungen & 2004-2010 & DSO National Laboratories, Singapur, Member \\
& & of Technical Staff \\
\end{tabular}
\end{table*}

\chapter*{Erkl\"arung zum eigenen Beitrag\markboth{Erkl\"arung zum eigenen Beitrag}{Erkl\"arung zum eigenen Beitrag}}
\addcontentsline{toc}{chapter}{Erkl\"arung zum eigenen Beitrag}

An dieser Stelle wird erkl\"art, welche Beitr\"age, der in dieser Arbeit enthaltenen wissenschaftlichen Ver\"offentlichungen, selbstst\"andig verfasst wurden. Im Folgenden werden die wissenschaftlichen Arbeiten in chronologischer Reihenfolge diskutiert.

\paragraph{Yeo, K. L., Solanki, S. K., Krivova, N. A., 2013, Intensity contrast of network and faculae, Astron. Astrophys., 550, A95}\mbox{}\\

\noindent
Diese Arbeit wurde komplett selbstst\"andig verfasst und mit Hilfe der beiden Koautoren Sami K. Solanki und Natalie A. Krivova sprachlich und inhaltlich \"uberarbeitet.

\paragraph{Yeo, K. L., Feller, A., Solanki, S. K., Couvidat, S., Danilovic, S., Krivova, N. A., 2014, Point spread function of SDO/HMI and the effects of stray light correction on the apparent properties of solar surface phenomena, Astron. Astrophys., 561, A22}\mbox{}\\

\noindent
Diese Arbeit wurde komplett selbstst\"andig verfasst und mit Hilfe der Koautoren Alex Feller, Sami K. Solanki, Sebastien Couvidat, Sanja Danilovic und Natalie A. Krivova sprachlich und inhaltlich \"uberarbeitet.

\paragraph{Yeo, K. L., Krivova, N. A., Solanki, S. K., Glassmeier, K. H., 2014, Reconstruction of total and spectral solar irradiance from 1974 to 2013 based on KPVT, SoHO/MDI and SDO/HMI observations, Astron. Astrophys., 570, A85}\mbox{}\\

\noindent
Diese Arbeit wurde komplett selbstst\"andig verfasst und mit Hilfe der Koautoren Natalie A. Krivova, Sami K. Solanki und Karl-Heinz Gla\ss{}meier sprachlich und inhaltlich \"uberarbeitet.

\paragraph{Yeo, K. L., Krivova, N. A., Solanki, S. K., 2014, Solar cycle variation in solar irradiance, Space Sci. Rev., online}\mbox{}\\

\noindent
Diese Arbeit wurde komplett selbstst\"andig verfasst und mit Hilfe der beiden Koautoren Natalie A. Krivova und Sami K. Solanki sprachlich und inhaltlich \"uberarbeitet.

\end{document}